\documentclass[aps,rmp,reprint]{revtex4-1}

\usepackage{graphicx,color}
\usepackage{amsfonts}
\usepackage[figuresright]{rotating}
\usepackage{amssymb}
\usepackage{amsmath}
\usepackage{psfrag}
\usepackage{subfigure}
\usepackage{multirow}
\usepackage{tabularx}
\usepackage{textcomp}
\usepackage{units}
\usepackage{hyperref}
\hypersetup{
 pdfnewwindow=true, colorlinks=true,
 linkcolor=blue, anchorcolor=blue,
 citecolor=blue, filecolor=blue,
 menucolor=blue, urlcolor=blue}

\usepackage{dcolumn}
\usepackage{bm}
\renewcommand{\v}[1]{\ensuremath{\mathbf{#1}}} 

\begin{document}
\title{Colloquium: Physical properties of group-IV monochalcogenide monolayers}

\author{Salvador\ \surname{Barraza-Lopez}}
\email{sbarraza@uark.edu}
\affiliation{Department of Physics, University of Arkansas, Fayetteville, AR 72701, USA}

\author{Benjamin\ \surname{M. Fregoso}}
\affiliation{Department of Physics, Kent State University, Kent, OH 44242, USA}

\author{John W.\ \surname{Villanova}}
\affiliation{Department of Physics, University of Arkansas, Fayetteville, AR 72701,{} USA}

\author{Stuart S. P.\ \surname{Parkin}}
\affiliation{Max Planck Institute of Microstructure Physics, Weinberg 2, Halle 06120, Germany}

\author{Kai\ \surname{Chang}}
\email{changkai@baqis.ac.cn}
\affiliation{Beijing Academy of Quantum Information Sciences, Beijing 100193, China}

\date{\today}

\begin{abstract}
We survey the state-of-the-art knowledge of ferroelectric and ferroelastic group-IV monochalcogenide monolayers. These semiconductors feature remarkable structural and mechanical properties, such as a {switchable} in-plane spontaneous polarization, soft elastic constants, structural degeneracies, and thermally-driven two-dimensional structural transformations. Additionally, these 2D materials also display selective valley excitations, valley Hall effects, and persistent spin helix behavior. After a description of their Raman spectra, a discussion of optical properties arising from their lack of centrosymmetry---such as an unusually strong second-harmonic intensity, large bulk photovoltaic effects, photostriction, and tunable exciton binding energies---is provided as well. The physical properties observed in these materials originate from (correlate with) their intrinsic and {switchable} electric polarization, and the physical behavior hereby reviewed could be of use in non-volatile memory, valleytronic, spintronic, and optoelectronic devices: these 2D multiferroics enrich and diversify the 2D materials toolbox.
\end{abstract}

\maketitle

\tableofcontents{}

\section{Introduction: The diversity of ultrathin ferroelectrics}

\begin{table*}
\caption{\label{tab:2dfe} Experimentally reported layered ferroelectrics, including the space group of the ferroelectric phase, its {intrinsic and switchable} polarization [in-plane (IP) and out-of-plane (OOP)], preparation methods employed, the critical temperature above which a paraelectric phase ensues, the coercive field $E_c$ (thicknesses for which $T_c$ and $E_c$ were determined are added in parenthesis), and other related properties. If the space group was not specified in the source literature, a prefix ``d-'' (for {\em distorted}) was added in front of the space group of the corresponding undistorted high-symmetry structure. ML stands for monolayer. $P_0$ stands for spontaneous polarization.}
\begin{ruledtabular}
\begin{tabular}{ccccccc}
Material & Space group & $\mathbf{P}$ direction & Preparation\footnote{MBE, molecular beam epitaxy; ME, mechanical exfoliation; PVD, physical vapor deposition}
& $T_c$ (K) & $E_c$ (kV/cm) & Other properties \\
\hline
$\alpha$-In$_2$Se$_3$ & $R3m$ & IP + OOP & ME       & 700 (4 ML) & 200 (5 nm) & $d_{33}=0.34$ pm/V\footnote{Piezoelectric coefficient} (1 ML)\\
$\beta'$-In$_2$Se$_3$ & d-$R\overline{3}m$& IP & ME & $> 473$ (100 nm) & &\\
CuInP$_2$S$_6$        & $Cc$ & IP + OOP\footnote{Vanishing in-plane polarization below a critical thickness of 90$-$100 nm.} & ME & $> 320$ (4 nm) & &\\
BA$_2$PbCl$_4$        & $Cmc2_1$ & IP & ME                   & $> 300$ (2 ML) & 13 (bulk) & $P_0\sim 13$ $\mu$C/cm$^2$\\
d1T-MoTe$_2$          & d-$P\overline{3}m$ & OOP & ME, MBE    & 330 (1 ML) & &\\
1T$'$-WTe$_2$         & $Pnm2_1$ & OOP & ME                    & 350 (2$-$3 ML) & & $P_0\neq0$ only when $\ge$ 2 ML\\
SnS odd-ML               & $P2_1mn$ & IP & MBE, PVD     & $> 300$ (1$-$15 ML) & 10.7/25 (1/9 ML) & \\
SnSe ML              & $P2_1mn$ & IP & MBE                   & 380$\sim$400 (1 ML) & 140 (1 ML) & \\
SnTe ML              & $P2_1mn$ & IP & MBE                   & 270 (1 ML) & & \\
\end{tabular}
\end{ruledtabular}
\end{table*}

Ferroelectric materials have a spontaneous, intrinsic polarization  $\mathbf{P}$ that can be switched by external electric fields. The first ferroelectric material---Rochelle salt---was discovered about a century ago \cite{rochelle_salt}. Despite a long history of applications of ferroelectrics in electric and electronic devices, the modern theory of ferroelectricity based on Berry phase---which made accurate comparisons between theory and experimental measurements possible---was not established until the 1990s \cite{King-Smith1993,Resta1994}; see Ref.~\cite{Rabe2007} for more details. From that point on, deep connections of this field with the geometry and topology of quantum mechanical wave functions have been pointed out \cite{Bernevig2013,vanderbilt2018berry}. Ferroelectric behavior is relevant from both a fundamental physical perspective as well as applications, and this Colloquium was written to highlight the physical properties of 2D ferroelectric {and ferroelastic materials} within the group-IV monochalcogenide family \cite{littlewood_jpc_1980_iv-vi_bulk}.

Researchers have always wondered whether there is a critical thickness for ferroelectric behavior below which {polarization switching becomes} suppressed \cite{Rabe2007}. Considering non-layered ferroelectric films with an out-of-plane intrinsic polarization $\mathbf{P}$, it was initially thought that the depolarization field arising from an incomplete cancellation of the space charge and an out-of-plane polarization charge at an electrode-ferroelectric interface [see, {\em e.g.,} Refs.~\cite{merz,janovec,mehta,triebwasser,black}] would raise the total energy of the system and eventually suppress the polarized state.
{Nevertheless, and as growth techniques for thin films  developed, the experimentally extracted critical thickness of ferroelectric thin films decreased from over 100 nm \cite{Feuersanger_1964} to tens of nanometers \cite{Slack_1971,YYToma_1974a,YYToma_1974b}, and eventually to only a few unit cells (u.c.s) \cite{NatureFE,Tybell_1999}. The  behavior of ultrathin ferroelectric films has been predicted to high precision by first principles calculations, suggesting critical thicknesses of several u.c.s for certain materials \cite{SBL1,Meyer_2001,Zembilgotov_2002,Wu_2004,Sai_2005,Gerra_2006}, or single-u.c.~thickness for others \cite{Sai09_PRL,ZhangYJ14_PRB,Almahmoud_2004,Almahmoud_2010}. Concurrently, sophisticated experiments on select compounds [PbTiO$_3$ (3 u.c.s) \cite{fong_pto_2004,fong_pto_2006}, BaTiO$_3$ (4 u.c.s) \cite{tenne_bto_2006,tenne_bto_2009}, PbZr$_{0.2}$Ti$_{0.8}$O$_3$ (1.5 u.c.s) \cite{gao_pzt_2017}, YMnO$_3$ (2 u.c.s) \cite{Nordlander_2019}, and BiFeO$_3$ (1 u.c.) \cite{wang_bfo_2018}] continue to push the critical thickness toward the single u.c.~limit.}

Meanwhile, a series of ultra-thin {\em layered} ferroelectric materials---especially attractive for the design and fabrication of functional (van der Waals) heterostructures---have been experimentally discovered, including In$_2$Se$_3$ \cite{ding_nc_2017_in2se3,poh_nl_2018_in2se3,in2se3_1,in2se3_3,xue_in2se3_2018_2,xue_in2se3_2018_1,in2se3_2,wan_in2se3_2018,in2se3_4}, CuInP$_2$S$_6$ \cite{cips,cips_ip_polarization_2019}, BA$_2$PbCl$_4$ \cite{ba2pbcl4_bulk,ba2pbcl4}, d1T-MoTe$_2$ \cite{yuan_nc_2019}, 1T$’$-WTe$_2$ \cite{wte2} [which is also a quantum spin Hall material \cite{song_nc_2018,asaba_sp_2018,qian_science_2014,fei_np_2017,tang_np_2017,wu_science_2018}], and of course, monolayers (MLs) of group-IV monochalcogenides like SnS, SnSe and SnTe. A brief and experimentally-driven summary of layered ferroelectrics is provided in Table \ref{tab:2dfe}.

The discovery of 2D and layered ferroelectrics facilitates the design of future non-volatile devices that are fully made of 2D material heterostructures. The experimentally verified 2D ferroelectric materials exhibit both out-of-plane and in-plane {switchable} spontaneous polarizations in few-layer films and at room temperature. Some prototype devices have also been demonstrated. For example, a ferroelectric diode in a graphene/$\alpha$-In$_2$Se$_3$ heterostructure has a relatively low coercive field of 200 kV/cm, and an electric current on/off ratio of $\sim10^5$ \cite{wan_in2se3_2018}. A d1T-MoTe$_2$ ferroelectric tunneling junction yielded an electric current on/off ratio of 1,000 \cite{yuan_nc_2019}.

Among all ultrathin ferroelectrics, a family of ferroelectric semiconductors with moderate band gaps known as group-IV monochalcogenide MLs---and referred to as $MX$ MLs henceforth---exhibit outstanding properties that are promising for many applications. By far, they are the only family of 2D ferroelectrics experimentally shown to display a robust {and switchable} in-plane spontaneous polarization at the limit of a single van der Waals ML at room temperature. Furthermore, many intriguing physical behaviors in $MX$ have been theoretically predicted in $MX$ MLs such as selective valley excitations, valley Hall effects, persistent spin helix behavior, \textit{etc.}

Nevertheless, and despite of these attractive theoretical predictions, the experimental growth and characterization remain difficult, partly because of reduced sample dimensions. Therefore, reviewing the current achievements and spurring a broader interest in this field provided the motivation to write this Colloquium. Despite the existence of several reviews focusing on the computational \cite{reviewpuru,rev}, experimental/computational \cite{AEM}, and experimental/theoretical \cite{Titova2020} aspects of 2D ferroelectrics, an all-encompassing review dedicated to the physical behavior of $MX$ MLs is still missing.

The structure of this Colloquium is as follows. The atomistic structure of O$-MX$s in the bulk and MLs is discussed in Sec.~\ref{sec:secII}. Atomistic coordination, the nature of their chemical bond, {group symmetries}, order parameters, as well as unexpected atomistic configurations experimentally obtained are covered in this Section. Sec.~\ref{sec:secIII} introduces the three members of this family (SnS, SnSe, and SnTe) that have been grown at the ML limit. Experimental characterization, including the verification of {polarization at exposed edges in ML nanoplates}, can be found there as well. The experimental ferroelectric switching of SnS, SnSe, and SnTe MLs is discussed in Sec.~\ref{sec:secIV}; novel memory concepts based on an in-plane ferroelectric switching are also introduced in that Section.

Linear elastic properties, structural degeneracies, and finite temperature thermal behavior (including phase transitions) are covered in Secs.~\ref{sec:secV}, \ref{sec:secVI}, and \ref{sec:secVII}, respectively. In a nutshell, O$-MX$ MLs are much softer than graphene, hexagonal boron nitride MLs, and transition metal dichalcogenide (TMDC) MLs. Their linear elastic properties, unusually large piezoelectric coefficients, and auxetic behavior are described in Sec.~\ref{sec:secV}. The elastic energy landscape is introduced in Sec.~\ref{sec:secVI}, which permits understanding the structural degeneracies of these 2D ferroelectrics, and the structural phase transitions that are discussed in Sec.~\ref{sec:secVII}.

Electronic and optical properties of O$-MX$ MLs are the subjects of Secs.~\ref{sec:secVIII} and \ref{sec:secIX}. The electronic properties are discussed in a gradual manner that includes band structures and valley properties without spin-orbit coupling, and a subsequent exposition of (spin-enabled) persistent spin helix behavior. Optical properties include the anisotropic absorption spectra, Raman spectra, SHG, injection and shift currents, photostriction, and excitonic effects. A summary and outlook is presented in Sec.~\ref{sec:secX}.

A unified and consistent notation has been deployed to streamline the discussion. In particular, the choice of crystallographic axes is such that orthogonal lattice vectors $\mathbf{a}_{1,0}$, $\mathbf{a}_{2,0}$, and $\mathbf{a}_{3,0}$ correspond to crystallographic vectors $\mathbf{a}$, $\mathbf{b}$, and $\mathbf{c}$ and point along the $x-$, $y-$, and $z-$direction, respectively [$\mathbf{a}_1$ ($\mathbf{a}_2$) is the so-called armchair (zigzag) direction]. These choices will lead to a modification of space group labeling, a redefinition of high-symmetry points in the electronic band structure, and to the relabeling of tensors from some of the source literature. The benefit from this effort is a self-contained discussion that is not interrupted from a lack of a standard notation. In addition, given that the structure of these materials evolves as a function of mechanical strain, temperature, electric field, and optical illumination, structural variables with a zero subindex represent their value on a ground state configuration at zero temperature and without external perturbations.

\section{Atomistic structure and chemical bonding of O$-MX$s from the bulk to MLs}\label{sec:secII}

Group-IV monochalcogenides are binary compounds with a chemical formula $MX$, where $M$ is a group IVA element and $X$ belongs to group VIA in the Periodic Table. Even though carbon, silicon, lead, oxygen, and even polonium belong to these groups, $MX$ compounds containing these elements will not be discussed here for the following reasons: SiS MLs possess a ground state structure with $Pma2$ symmetry \cite{yang_nanolett_2015_sis} which lacks a net {intrinsic electric polarization}; see Ref.~\cite{kamal_prb_2016_iv_vi_monolayers} concerning a lower-energy structure for 2D SiSe, too. Pb$X$ compounds lack a net {$\mathbf{P}$} regardless of the number of layers (more on this later). Similarly, materials such as 2D SiO, GeO, and SnO display a non-ferroelectric litharge structure \cite{lefebvre_prb_1998_sno,kamal_prb_2016_iv_vi_monolayers}. This way, $M$ will either be germanium (Ge) or tin (Sn) while $X$ represents sulphur (S), selenium (Se), or tellurium (Te) in what follows.

\begin{figure}
\includegraphics[width=\linewidth]{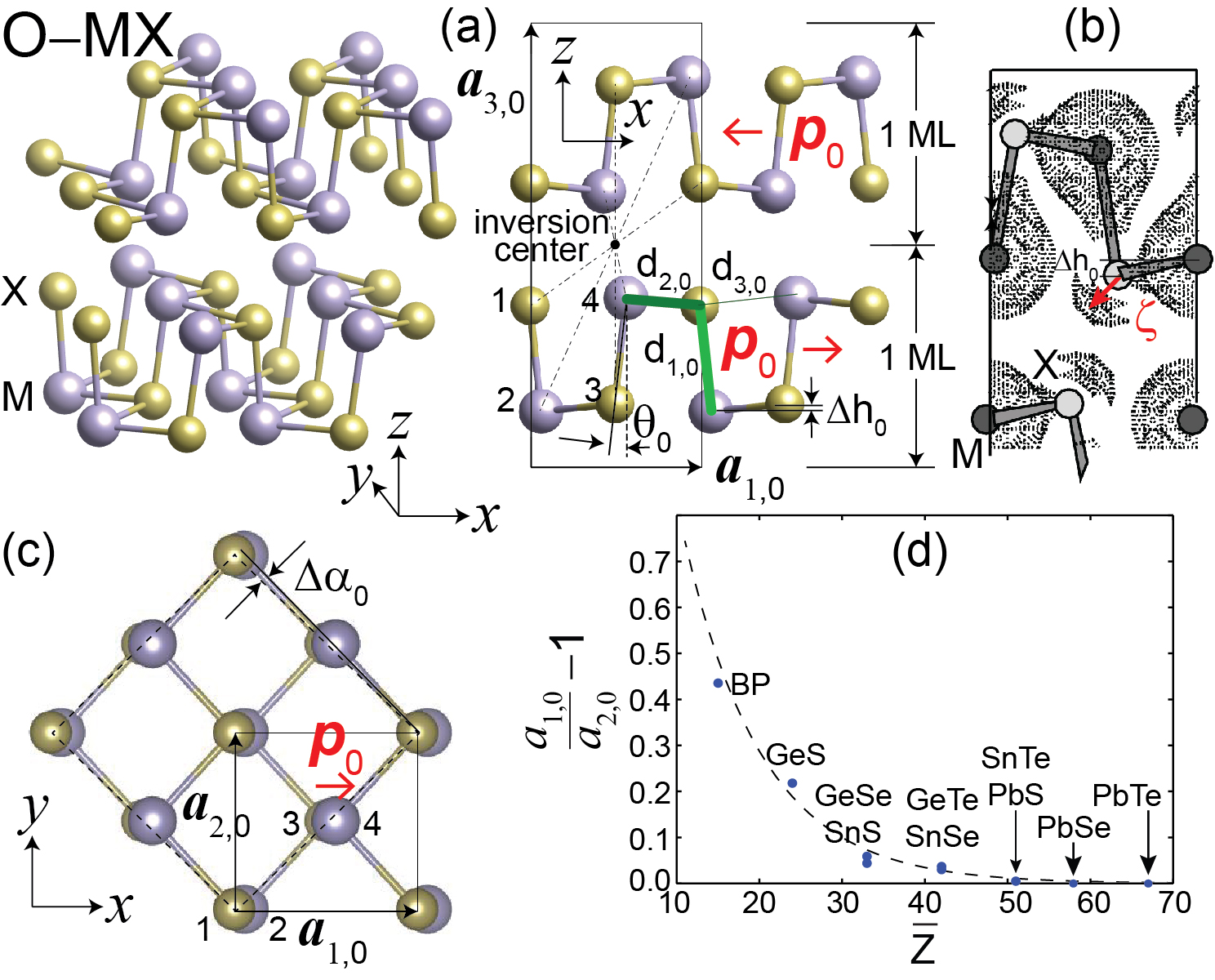}
\caption{\label{fig:figure1} (a) Structure of bulk O$-MX$s, with the $M$ atom shown in gray (big) and the $X$ atom in yellow (small) circles. Left: 3D view of two MLs with antiparallel polarization. Right: side view along the $x-z$ plane showing $\mathbf{a}_{1,0}$ and $\mathbf{a}_{3,0}$ lattice vectors ($\mathbf{a}_{2,0}$ points into the page). The inversion center swaps the direction of $\mathbf{p}_0$ at consecutive MLs (red arrows), and $p_0=|\mathbf{p}_0|\propto \theta_0$. Interatomic distances $d_{1,0}$, $d_{2,0}$, and $d_{3,0}$ as well as $\Delta h_0$ are shown, too. (b) Electronic density showing a lone pair $\zeta$, and $\Delta h_0$. Bold (shaded) bold circles stand for tin (sulfur or selenium). Adapted from Ref.~\cite{lefebvre_prb_1998_sno} with permission. Copyright, 1998, American Physical Society. (c) Top view of a O$-MX$ ML. $\Delta \alpha$ measures the deviation from 90$^{\circ}$ of the rhombus highlighted by dashed lines; $\Delta\alpha_0=0$ when $a_{1,0}=a_{2,0}$. (d) The $a_{1,0}/a_{2,0}$ ratio (proportional to $\Delta\alpha_0$) is tunable by the compound's average atomic number $\bar{Z}$. Adapted from Ref.~\cite{Mehboudi2016} with permission. Copyright, 2016, American Chemical Society.}
\end{figure}

GeTe and SnTe are rhombohedral (R$-$phase) and GeS, GeSe, SnS, and SnSe turn orthorhombic (O$-$phase) in the bulk \cite{littlewood_jpc_1980_iv-vi_bulk}. As illustrated in Fig.~\ref{fig:figure1}(a), O$-MX$ compounds have a layered structure. One ML refers to a van der Waals layer [or two atomic layers (2 ALs)] or half of an O$-MX$ u.~c.

{Strictly speaking, the intrinsic {switchable} polarization $\mathbf{P}$ should not be showcased as a vector on periodic structures. Therefore, in certain theoretical discussions, we will utilize an order parameter $\mathbf{p}_0$ (parallel to $\mathbf{P}$) that remains well-defined on periodic structures. The letter $p$ stands for {\em projection}, and this order parameter is defined in the next paragraph.

Consider the vector $\mathbf{r}_{XM}$ starting at $X$ atom 1 and ending at the nearest $M$ atom (atom 2) in the lower ML seen in Fig.~\ref{fig:figure1}(a). The positions of the remaining two atoms within the lower ML (3 and 4) are obtained by a screw operation about the $x-$axis, or by a diagonal ($n$) glide operation about the $z-$axis applied to $\mathbf{r}_{XM}$. Calling $\mathbf{r}_{XM}'$ the vector starting at ($X$) atom 3 and ending at ($M$) atom 4, we define $\mathbf{p}_0=\mathbf{r}_{XM}+\mathbf{r}_{XM}'$. The mirror symmetry along the $x-z$ plane makes $\mathbf{p}_0\cdot \hat{y}=0$, while the screw operation renders $\mathbf{p}_0\cdot \hat{z}=0$, so that $\mathbf{p}_0$ is parallel to the longer lattice vector $\mathbf{a}_{1,0}$.} These symmetries also render a zero intrinsic polarization along the $y-$ and $z-$directions. Ferroelectric O$-MX$ MLs belong to space group 31 \cite{kamal_prb_2016_iv_vi_monolayers,rodin_prb_2016_sns} [usually written as $Pnm2_1$ but labeled $P2_1mn$ for lattice vectors as drawn in Fig.~\ref{fig:figure1}(c)], and a top view of their anisotropic u.~c.~is provided in Fig.~\ref{fig:figure1}(c) within solid lines. Phonon dispersion calculations demonstrate the structural stability of these MLs \cite{singh_apl_2014_ges_gese_sns_snse}. A zero value of the {\em order parameter} $\theta$ in a ML with dissimilar lattice constants ($a_{1,0}\ne a_{2,0}$) leads onto a paraelectric structure with $\mathbf{p}=\mathbf{0}$ belonging to symmetry group 59 (
$Pmmn$).

The alternating {direction of $\mathbf{p}_0$} (or antipolar coupling) within each ML arises from the inversion center shown in the side view along the $x-z$ plane in Fig.~\ref{fig:figure1}(a) (the atoms related by inversion are joined by dash-dot lines). Bulk O$-MX$s belong to space group 62 [$Pnma$ \cite{gomes}, or $Pcmn$ with the lattice vectors employed here]. The side view of the $x-z$ plane in Fig.~\ref{fig:figure1}(a) also contains the u.~c.~boundaries in solid line, interatomic distances $d_{1,0}<d_{2,0}<d_{3,0}$, a tilt angle $\theta_0$, and the height $\Delta h_0$ of an $X$ atom relative to its nearest $M$ atom  \cite{kamal_prb_2016_iv_vi_monolayers}. A net {switchable} $\mathbf{P}_0$ ensues in binary compounds lacking inversion symmetry, which is the case for individual MLs of O$-MX$s \cite{tritsaris_jap_2013_sns,singh_apl_2014_ges_gese_sns_snse,tomanek_acsnano_2015_sis,gomes,gomes_prb_2015a_ges_gese_sns_snse,fei_apl_2015_ges_gese_sns_snse}. The atomistic structure and the in-plane $\mathbf{p}_0$ of O$-MX$ MLs, can be understood on the basis of the chemistry of black phosphorus (BP) MLs as follows.

Carbon belongs to group IVA and graphite has four valence electrons and an $sp^2$ hybridization. Phosphorus (P) belongs to group VA, and black phosphorus has five valence electrons and displays an $sp^3$ hybridization \cite{kamal_prb_2016_iv_vi_monolayers}. In (three-fold coordinated) graphite, three atoms form strong ($\sigma$) in-plane bonds and the fourth ($\pi$) electron protrudes out of plane. BP is three-fold coordinated as well, having its closest neighboring atom at a distance $d_1$ and two additional atoms located at a slightly larger distance $d_2$. Given that a phosphorus atom contains five valence electrons, such three-fold coordination requires the existence of two additional non-bonded electrons [known as a {\em lone pair} ($\zeta$)] per atom. Unlike graphene, which maintains a planar configuration with two atoms in its u.~c., lone pairs confer BP MLs with a puckered structure and a rectangular u.~c.~containing four atoms.

Similar to hexagonal boron nitride---which is made out of a group IIIA element (B) and a group VA element (N) and is isostructural to graphite---Fig.~\ref{fig:figure1}(a) indicates that O$-MX$s are isostructural to BP. Table \ref{ta:ta2} shows that bulk SnS has similar distances $d_1$ and $d_2$ for a three-fold atomistic coordination, and the same can be said of interatomic distances in bulk SnSe, also listed in the Table. The equivalent to a lone pair $\zeta$ is assigned to the more negatively charged $X$ atom in Fig.~\ref{fig:figure1}(b) \cite{lefebvre_prb_1998_sno}. The reader may notice that $\Delta h_0$ is positive in Fig.~\ref{fig:figure1}(a) and negative in Fig.~\ref{fig:figure1}(b): its sign determines certain elastic properties that will be discussed in Sec.~\ref{sec:secV}.

\begin{table}[tb]
\caption{Interatomic  distances in bulk SnS and SnSe. $N$ is  the number  of  neighbors at any given distance. Taken from Ref.~\cite{lefebvre_prb_1998_sno}.}\label{ta:ta2}
\begin{tabular}{cccc|cccc}
\hline
\hline
Material & atoms & $d$ (\AA) & $N$ & Material & atoms & $d$ (\AA) & $N$ \\
\hline
SnS      & Sn-S        & 2.63  ($d_{1,0}$)    & 1 & SnSe     & Sn-Se       & 2.74 ($d_{1,0}$)     & 1 \\
         &             & 2.66  ($d_{2,0}$)    & 2 &          &             & 2.79 ($d_{2,0}$)     & 2 \\
         &             & 3.29  ($d_{3,0}$)    & 2 &          &             & 3.34 ($d_{3,0}$)     & 2\\
         &             & 3.39      & 1 &          &             & 3.47      & 1\\
         & Sn-Sn       & 3.49      & 2 &          & Sn-Sn       & 3.55      & 2\\
         & S-S         & 3.71      & 4 &          & Se-Se       & 3.89      & 4\\
         &             & 3.90      & 2 &          &             & 3.94      & 2\\
\hline
\hline
\end{tabular}
\end{table}

The rhombic distortion angle $\Delta \alpha_0$ \cite{Kai} shown in Fig.~\ref{fig:figure1}(c) indicates the anisotropy of the u.c.~and is related to the ratio of lattice constants $a_{1,0}/a_{2,0}$ as follows \cite{other4}:
\begin{equation}\label{eq:eq1}
\frac{a_{1,0}}{a_{2,0}}=\frac{1+\sin\Delta\alpha_0}{\cos\Delta\alpha_0},
\end{equation}
or $\Delta\alpha_0\simeq \frac{a_{1,0}}{a_{2,0}}-1$ for small angles when $\Delta\alpha_0$ is expressed in radians.

Letting $Z_M$ ($Z_X$) be the atomic number of atom $M$ ($X$) and defining the average atomic number $\bar{Z}=(Z_M+Z_X)/2$, Fig.~\ref{fig:figure1}(d) illustrates a decaying exponential dependence of $\frac{a_{1,0}}{a_{2,0}}-1$ on $\bar{Z}$ \cite{Mehboudi2016}. $\frac{a_{1,0}}{a_{2,0}}-1$ has been called the {\em reversible strain} \cite{ccBP} or {\em tetragonality ratio}, and it correlates with $P_0$ in bulk ferroelectrics \cite{ref31}. Fig.~\ref{fig:figure1}(d) indicates that---at zero temperature---lattice vectors turn more equal (unequal) on heavier (lighter) $MX$ MLs. Having equal lattice vectors, Fig.~\ref{fig:figure1}(d) shows that Pb-based $MX$ MLs are paraelectric [$P_0=0$; a behavior experimentally confirmed on PbTe MLs; see Supporting Information in Ref.~\cite{Kai}] and are not discussed here for that reason. Ferroelectric O$-MX$ MLs with $a_{1,0}\ne a_{2,0}$ have similar structures and hence display similar physical behavior; this observation will permit drawing meaningful comparisons between different experimental and theoretically studied compounds within this material family.

Continuing the discussion of chemistry, one observes in Table \ref{ta:Transfer} a correlation between the charge transfer $\Delta Q$ [or {\em ionicity } \cite{littlewood_jpc_1980_iv-vi_bulk}] from the group IVA element onto the one belonging to group VIA, and Pauling's difference in electronegativity $\Delta \xi$. Although an interplay among covalent, ionic, and resonant bonding has been argued to describe $MX$s, a new type of bonding (called {\em metavalent}, and thought of as a combination of `metallic' and `covalent') has been proposed to classify these materials \cite{advmat2018,revchalcogens,advmatZanolli}. Variables employed to identify the appropriate type of bonding include the coordination number, the electronic conductivity, the dielectric constant $\epsilon_{\infty}$, the  bond polarizability, and the lattice anharmonicity. Setting up a two-dimensional map where the horizontal axis is the charge transfer $\Delta Q$  and the vertical axis (named {\em electron sharing}) is a measure of electronic exchange and correlation \cite{advmat2018}, metavalent compounds sit in between covalently-bonded and metallic materials. In the bulk, materials such as GeS, GeSe, SnS, and SnSe are assigned a covalent bonding, while GeTe can display either covalent or metavalent bonding depending on its phase [R and cubic (C) phases being metavalent and the O phase being covalent]; bulk SnTe, PbS, PbSe, and PbTe are assigned a metavalent character \cite{advmat2018,revchalcogens}.

\begin{table}[tb]
\caption{Net charge transfer $\Delta Q$ (in $e$) from atom $M$ to atom $X$ and change in electronegativity $\Delta\xi=\xi_X-\xi_M$ (in eV) for O$-MX$ MLs. Taken from Ref.~\cite{kamal_prb_2016_iv_vi_monolayers}.}\label{ta:Transfer}
\begin{tabular}{cc|cc||cc|cc}
\hline
\hline
Material & \scalebox{.85}{$\bar{Z}$} & $\Delta Q$ & $\Delta\xi$ & Material & \scalebox{.85}{$\bar{Z}$} & $\Delta Q$ & $\Delta\xi$\\
\hline
GeS ML & 24 & 0.815 & 0.57 & SnS ML  & 33 & 0.980 & 0.62\\
GeSe ML & 33 & 0.649 & 0.54 & SnSe ML & 42 & 0.855 & 0.59\\
GeTe ML & 42 & 0.372 & 0.09 & SnTe ML & 51 & 0.596 & 0.14\\
\hline
\hline
\end{tabular}
\end{table}

The in-plane u.~c.~area ({\em i.e.}, $|\mathbf{a}_{1,0}\times \mathbf{a}_{2,0}|$) of O$-MX$s is a function of the number of MLs \cite{hu_apl_2015_gese,snteapl,ourarxiv,italian,advmatZanolli}, a feature observed in BP as well \cite{Shulenburger} that is related to the thickness-dependent  spatial distribution of lone pairs. Such a dependence of $|\mathbf{a}_{1,0}\times \mathbf{a}_{2,0}|$ on thickness is not observed in more traditional 2D materials such as graphene and TMDCs.

\begin{figure}
\includegraphics[width=\linewidth]{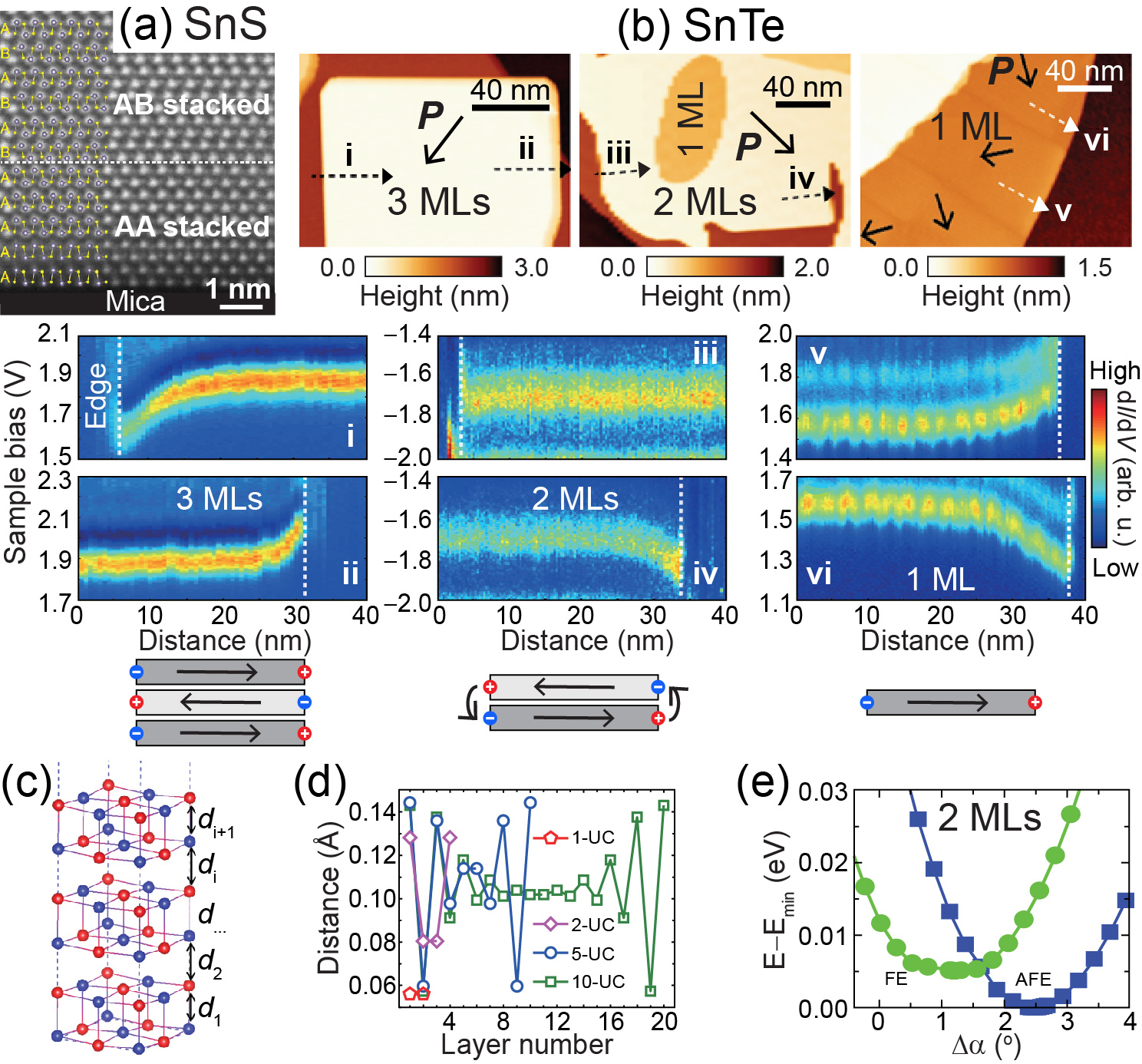}
\caption{\label{fig:figure2} (a) Cross-sectional STEM images of few-ML SnS grown on mica exhibit ferroelectric coupling for up to 15 MLs. Reproduced from Ref.~\cite{Higashitarumizu20_NC_SnS} with permission. Copyright, 2020, Springer Nature. (b) Ultrathin SnTe grown on epitaxial graphene develops an antiferroelectric coupling, as demonstrated by band bending at the exposed edges of few-layer nanoplates (subplots i through vi). Lowermost diagrams indicate electrostatic interactions at exposed edges upon antiferroelectric coupling. Reproduced from \cite{sntebl} with permission. Copyright, 2019, John Wiley and Sons. (c and d) Demonstration of layering in ferroelectrically-coupled ultrathin SnTe by the uneven distance $d_i$ among layers. Reproduced from \cite{snteapl} and \cite{liu_prl_2018_snte} with permission. Copyright, 2017, American Institute of Physics and 2018, American Physical Society. (e) A SnTe bilayer with antiferroelectric (AFE) coupling has a lower total energy when compared to a ferroelectric (FE) coupled one. Reproduced from \cite{kaloni2019} with permission. Copyright, 2019, American Physical Society.}
\end{figure}

Leaving a detailed discussion of ultrathin film creation and characterization to Sec.~\ref{sec:secIII}, Figs.~\ref{fig:figure2}(a) and \ref{fig:figure2}(b) display few-ML SnS and SnTe films and provide striking examples of unexpected structure: indeed, while bulk SnS displays the $Pcmn$ group symmetry, SnS grown on mica can take on a {\em ferroelectrically-coupled} (sometimes labeled AA) stacking sequence for up to fifteen MLs \cite{Higashitarumizu20_NC_SnS} [Fig.~\ref{fig:figure2}(a)], with a $Pcmn$ group symmetry acquired by subsequent MLs on thicker films. Non-polar, thick SnS can be switched into a ferroelectric phase by an external electric field \cite{Bao19_NL_SnS}. Additional experimental $MX$ morphologies include GeS nanowires created along an axial screw dislocation \cite{sutterNature2019} and the antiferroelectrically-coupled ultrathin SnTe grown on (metallic) epitaxial graphene \cite{sntebl} that is discussed next.

Bulk SnTe displays a metavalent, R$-$phase in the bulk. Grown on a metallic substrate, ultrathin SnTe flakes with a 3-ML, bilayer, or ML thicknesses were characterized with a scanning tunneling microscope (STM), which permits elucidating their in-plane polarization $\mathbf{P}$ {switching} from the band bending of the conduction band edge observed in Fig.~\ref{fig:figure2}(b). These STM spectra were captured along the dashed straight lines at subplots (i) through (vi) in Fig.~\ref{fig:figure2}(b) cutting through the nanoplates' edges \cite{Kai,sntebl,Chang20_arxiv_SnSe}. (Additional details on the determination of $\mathbf{P}$ will be provided in Sec.~\ref{sec:secIII}.) Band bending is almost non-existent in SnTe bilayers, which implies an antipolar coupling among MLs, and shows that the bonding of SnTe transitions from metavalent in the bulk to covalent in ultrathin films \cite{advmatZanolli}.

Three theoretical works \cite{liu_prl_2018_snte,snteapl,advmatZanolli} explain the layered nature of ultrathin SnTe. They were performed using either the local density approximation [LDA \cite{LDA2}] or the generalized gradient approximation as implemented by Perdew, Burke, and Ernzerhof [PBE \cite{PBE}] for exchange-correlation (XC) within density-functional theory \cite{martin} and assume a bulk-like ({\em i.e.,  ferroelectric}) stacking of successive MLs in freestanding SnTe configurations as the one depicted in Fig.~\ref{fig:figure2}(c).

Bulk SnTe features a Peirels distortion---a result of the competition among electron delocalization and localization \cite{advmatZanolli}---that creates a net polarization along its diagonal and distorts a cubic lattice into a rhombohedral one. As a result: (i) a bulk u.~c.~has both in-plane and an out of plane polarization and, considering two atomic layers as a ML, (ii) consecutive MLs are coupled ferroelectrically. This is different to O$-MX$s, compounds with no net out-of-plane polarization and an antipolar coupling among successive MLs; see Fig.~\ref{fig:figure1}(a). Nevertheless, the depolarization field quenches the out-of-plane polarization of SnTe films \cite{liu_prl_2018_snte}, creating an in-plane lattice expansion \cite{snteapl} and a separation between MLs resulting in the layered structure seen in Fig.~\ref{fig:figure2}(d). Freestanding SnTe films with ferroelectric coupling have an intrinsically higher $T_c$ than their bulk counterpart due to an interplay among hybridization interactions and Pauli repulsion. Additionally, electron sharing \cite{advmat2018} increases with decreasing thickness, imparting chemical bonds with a more covalent character \cite{advmatZanolli}.

Most computational works on $MX$s that employ density-functional theory make use of the PBE approximation \cite{PBE} to XC. Yet, and as seen in Fig.~\ref{fig:figure2}(e), the experimentally observed antipolar coupling and the magnitude of $\Delta \alpha$ on bilayer SnTe films is recovered when using self-consistent van der Waals [vdW-DF-cx \cite{cx}] interactions \cite{kaloni2019}. In any case, Figs.~\ref{fig:figure2}(a) and \ref{fig:figure2}(b) indicate that the details of the initial surface are crucial for the type of atomistic structure formed by ultrathin $MX$ films \cite{revchalcogens}.

\section{Experimentally available O$-MX$ MLs}\label{sec:secIII}

\subsection{SnS MLs}\label{SnS}

\begin{figure}
\includegraphics[width=\linewidth]{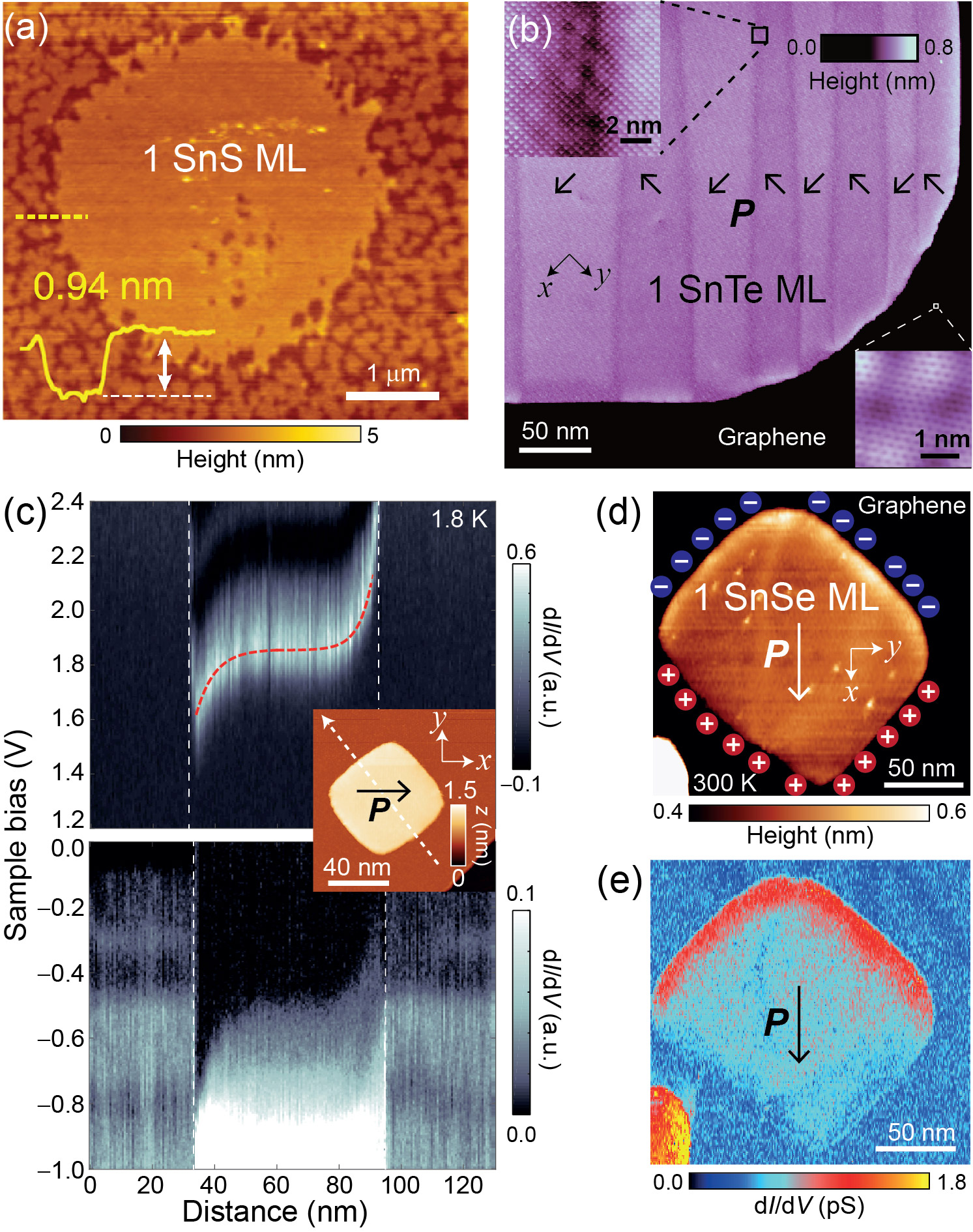}
\caption{\label{fig:figure3} (a) Atomic force microscopy topographic image of a SnS ML on mica. Reproduced from Ref.~\cite{Higashitarumizu20_NC_SnS} with permission. Copyright, 2020, Springer Nature. (b) STM topography of a SnTe ML nanoplate on epitaxial graphene, with head-to-tail 90$^\circ$ domains having an intrinsic polarization $\mathbf{P}$ indicated by arrows. Adapted from Ref.~\cite{Kai}. Copyright, 2016, American Association for the Advancement of Science. (c) Low temperature $dI/dV$ spectra across the dashed arrow in the inset for a SnSe ML nanoplate on graphene. (c and d) Room temperature STM topography and $dI/dV$: higher $dI/dV$ implies a larger electronic charge. Taken from Ref.~\cite{Chang20_arxiv_SnSe} with permission. Copyright, 2020, American Chemical Society.}
\end{figure}

Room-temperature in-plane ferroelectricity was demonstrated in few-ML SnS by a combination of piezo-force microscopy (PFM), second harmonic generation (SHG), and electric transport experiments \cite{Bao19_NL_SnS}. In order to overcome PFM's weakness in detecting the in-plane polarization of O$-MX$s, SnS films were grown by molecular beam epitaxy (MBE) on corrugated graphite substrates so that $\mathbf{P}$ was not perfectly perpendicular to the PFM tip and a finite polarization signal could be picked up. Ferroelectric domains and PFM hysteresis loops were resolved on 10 nm thick SnS films grown on mica, and a SHG signal was collected, too. The modification of film morphology illustrates the difficulties of traditional techniques such as PFM to characterize ultrathin ferroelectric films with an in-plane $\mathbf{P}$ {switching}, and the need to develop new techniques to characterize these ferroelectrics without changing morphology. Two-terminal devices were fabricated on a 15 nm thick SnS film grown on mica which was subsequently transferred onto a doped Si substrate covered by 300 nm thick SiO$_2$. Hysteresis was found in the $I-V$ curves with a coercive field of 10.7 kV/cm and a maximum $I_{on}/I_{off}$ ratio of $\sim$100. Furthermore, the remnant polarization increased when a negative gate voltage was applied \cite{Bao19_NL_SnS}.

The creation and characterization of SnS MLs has been reported subsequently \cite{Higashitarumizu20_NC_SnS}. SHG signals---a signature of lack of inversion symmetry and ferroelectricity that will be discussed from a combined theory/experiment perspective in Sec.~\ref{sec:shg}---were detected in SnS ML flakes grown {\em via} physical vapor deposition (PVD) on (insulating) mica. Two-terminal devices patterned onto these as grown flakes display hysteresis in $I-V$ loops---yet another signature of ferroelectricity that will be discussed in Sec.~\ref{sec:secIV}. Consistent with the ML arrangement depicted in Fig.~\ref{fig:figure2}(a), ferroelectricity is detected in SnS films composed of up to fifteen MLs, {\em including those composed of an even number of MLs}. Ferroelectricity is unexpected in even-ML O$-MX$ films because they are assumed to be centrosymmetric, lacking a net polarization according to the $Pcmn$ group symmetry. A coercive field of 25 kV/cm was found for 9-ML thick SnS by electric transport measurements. An apparent remnant polarization as large as $P_r\sim3$~$\mu$C/m was experimentally determined, much larger than the theoretical value of 240$-$265 pC/m listed in Table \ref{ta:ta3}, and probably an artifact due to the relatively high conductance of SnS.

\subsection{SnSe and SnTe MLs}\label{SnSeandSnTe}

The first experimentally discovered 2D ferroelectric in the O-$MX$ family is the SnTe ML grown by MBE on (metallic) graphene \cite{Kai,chang_aplm_2019} and  characterized by STM \cite{Kai,sntebl,KaiPRL} in Fig.~\ref{fig:figure3}(b). As seen in Figs.~\ref{fig:figure3}(c) and \ref{fig:figure3}(d), SnSe MLs have been grown by MBE on graphene, too \cite{Chang20_arxiv_SnSe}. Due to the metallic substrate in which these are grown, the techniques employed to characterize ultrathin SnSe and SnTe films are different and complementary to those employed for SnS. Although STM is an unconventional tool to study ferroelectrics, its extreme surface sensitivity and access to the materials' local electronic structure are advantageous for studying ultrathin ferroelectric flakes with an in-plane intrinsic polarization, where PFM lacks sensitivity and may even damage ultrathin samples. STM measurements are helped by the fact that these ultrathin films are not insulators but semiconductors, such that a tunneling current can be established into the metallic substrate~\cite{kairev}.

\begin{table}[tb]
\caption{Spontaneous in-plane polarization $P_0$ (in pC/m) as determined by DFT with the PBE XC functional.}\label{ta:ta3}
\begin{tabular}{cc|ccc||cc|ccc}
\hline
\hline
Material & \scalebox{.85}{$\bar{Z}$} & $P_0^{a}$  & $P_0^{b}$ & $P_0^{c}$ &
Material & \scalebox{.85}{$\bar{Z}$} & $P_0^{a}$  & $P_0^{b}$ & $P_0^{c}$\\
\hline
GeS ML      & 24 & 484 & 480 & 486 & SnS ML      & 33 & 260 & 240 & 265\\
GeSe  ML    & 33 & 357 & 340 & 353 & SnSe ML     & 42 & 181 & 170 & 190\\
GeTe ML     & 42 & $-$ & $-$ & 308 & SnTe ML     & 51 & $-$ & $-$ & 50\\
\hline
\hline
\end{tabular}\\
$^a$ Ref.~\cite{other3}. ${}^{b}$ Ref.~\cite{Rangel2017}.\\${}^{c}$ Our calculations.
\end{table}

Being a vector, $\mathbf{P}$ has a magnitude $P$, an orientation, and sense of direction. SHG can only tell orientation, while two-terminal electric measurements and STM [see Figs.~\ref{fig:figure2}(b-e)] can determine orientation and sense of direction. These three techniques require additional calibration to uncover the magnitude ($P$), whose calculated values are listed in Table \ref{ta:ta3}.

Similar to subplots (v) and (vi) in Fig.~\ref{fig:figure2}(b) showing band bending of SnTe MLs on STM topography images, SnSe ML nanoplates display the band bending seen in Fig.~\ref{fig:figure3}(c) as a result of bound charges accumulated at the nanoplates' edges, reflecting the in-plane polarization $\mathbf{P}$ of these 2D ferroelectrics. The direction of $\mathbf{P}$ for SnSe and SnTe MLs---shown by arrows with a $\mathbf{P}$ label in Figs.~\ref{fig:figure2}(b) and \ref{fig:figure3}(b-e)---is identified by band bending at nanoplate edges, and by the difference of lattice parameters $a_1$ and $a_2$ as extracted from atomically resolved STM images. Stripe-shaped $\sim 90^\circ$ ``head-to-tail'' domains are observed in SnTe monolayer plates in Fig.~\ref{fig:figure3}(b) \cite{Kai}, while 180$^{\circ}$ domains are formed in SnSe MLs \cite{Chang20_arxiv_SnSe}. The different type of domains formed in SnSe and SnTe MLs has to do with a lattice commensuration of SnSe MLs on graphene~\cite{Chang20_arxiv_SnSe}. A 
decrease of Sn vacancy concentration by 2$\sim$3 orders of magnitude was found in SnTe MLs with respect to bulk values \cite{Kai}. Electronic band gaps of SnSe and SnTe MLs (obtained by the determination of the valence and conduction band edges {\em via} $dI/dV$ measurements) are listed in Table \ref{ta:taGaps}.

Band bending disappears in SnTe MLs at 270 K \cite{Kai}, but it can still be observed at 300 K in SnSe MLs, implying a robust in-plane ferroelectricity at room temperature in 2D SnSe. According to variable temperature $dI/dV$ mapping experiments, $T_c$ reaches 380$-$400 K for SnSe MLs, a promising magnitude for room-temperature applications. A theoretical description of thermally-driven structural transformations can be found in Section \ref{sec:secVII}.

\section{Switching the direction of $\mathbf{P}$ on O$-MX$ MLs: demonstrating ferroelectric behavior}\label{sec:secIV}

\begin{figure}
\includegraphics[width=\linewidth]{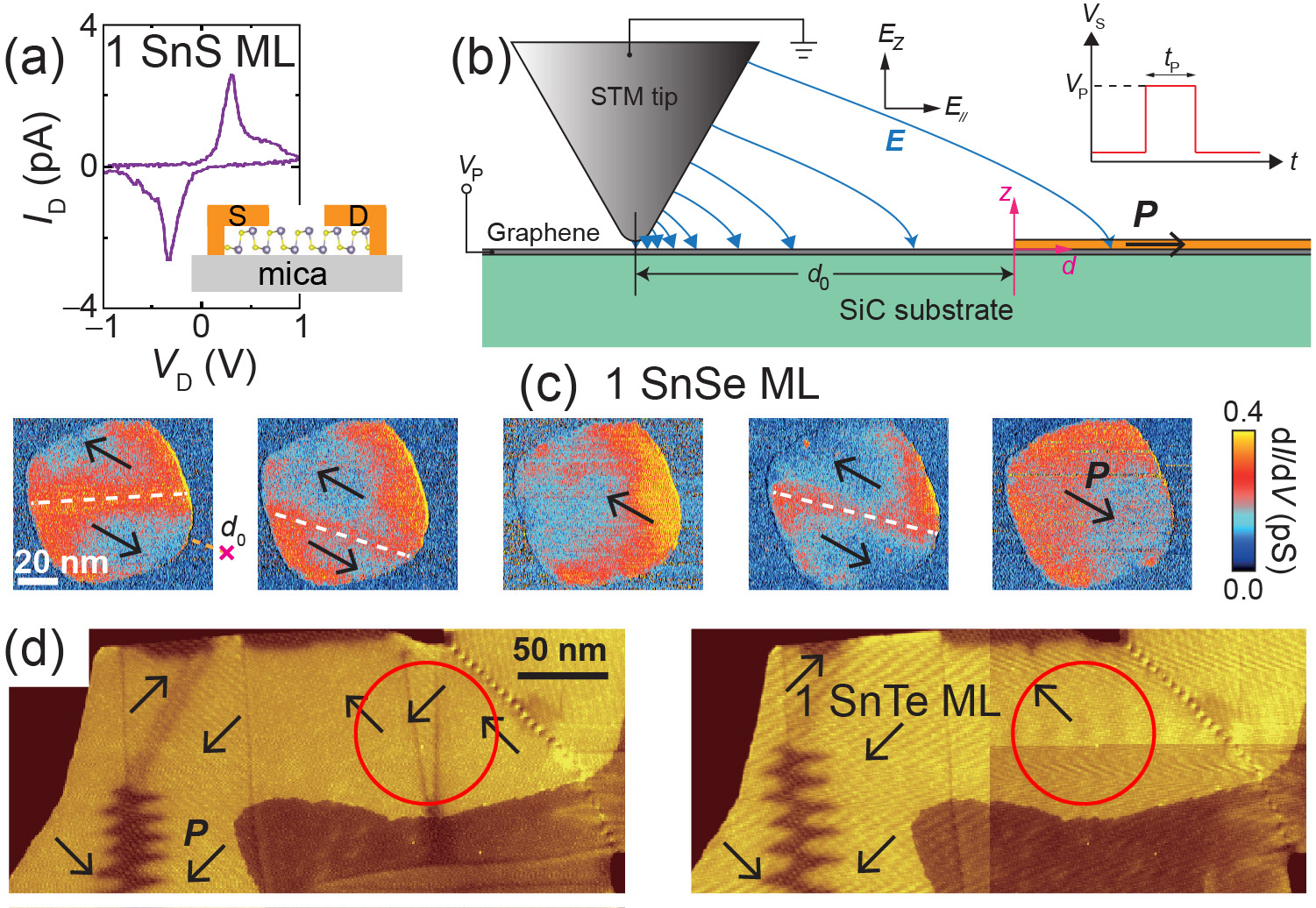}
\caption{\label{fig:figure4} (a) Ferroelectric resistive switching of a SnS ML on mica {\em via} source/drain bias. Taken from Ref.~\cite{Higashitarumizu20_NC_SnS} with permission. Copyright, 2020, Springer Nature. (b) Ferroelectric switching on a SnSe ML is achieved by bias voltage pulses $V_P$ applied on the STM tip at a point on the graphene substrate close the SnSe ML plate. (c) Consecutive $dI/dV$ images along a ferroelectric switching sequence of a SnSe ML at room temperature.  The pulses were applied at the point indicated by the cross in the first panel, and white dashed lines indicate a 180$^\circ$ domain wall. Adapted from Ref.~\cite{Chang20_arxiv_SnSe} with permission. Copyright, 2020, American Chemical Society.
(d) Ferroelectric switching of a SnTe ML by an STM tip. Taken from Ref.~\cite{Kai} with permission. Copyright, 2016, American Association for the Advancement of Science.}
\end{figure}

O$-MX$ MLs can only be considered ferroelectrics if they can be controllably switched by an external electric field. For this purpose, a two-terminal device was built by adding silver contacts to the SnS ML grown on mica, and the current $I_D$ was measured as the drain bias $V_D$ was swept from $-1$ V to $1$ V, and back to $-1$ V. The result, shown in Fig.~\ref{fig:figure4}(a), demonstrates the ferroelectric resistive switching of SnS MLs.

In addition, the domains of SnSe MLs can be switched and domain walls moved by applying bias voltage pulses onto the graphene substrate away from a SnSe nanoplate as schematically laid out in Fig.~\ref{fig:figure4}(b). Fig.~\ref{fig:figure4}(c) demonstrates the consecutive manipulation of 180$^{\circ}$ domains in a SnSe ML nanoplate. Demonstrating ferroelectric control, the polarization of the whole plate can be reversed by this approach. Statistical studies suggest a critical in-plane electric field of domain wall movement of $E_{\parallel,c}=1.4\times10^5$~V/cm. Polarization switching was also demonstrated by applying bias voltage pulses to the STM tip at the surface of SnTe MLs, which locally switches ferroelectric domains by the domain wall motion highlighted within red circles in Fig.~\ref{fig:figure4}(d).

\subsection{Polarization switching and ultrathin memories based on in-plane ferroelectric tunnel junctions}

Ferroelectrics find applications in nonvolatile memories due to their {switchable} bistable ground states \cite{Scott1400}. First-generation ferroelectric memories use the surface charge in a ferroelectric capacitor to represent data \cite{5940}. As a result, discharging the capacitor to measure the charge destroys the stored data, so the capacitor must be recharged after reading. A second generation of these memories probes the ferroelectric polarization using a tunneling-electroresistance effect \cite{Tsymbal181} within a metal-ferroelectric-metal junction in which an {\em out-of-plane} $\mathbf{P}$ exists within the ferroelectric thin film. The tunneling potential barrier is determined by the out-of-plane polarization in the ferroelectric layer.

\begin{figure}
\includegraphics[width=\linewidth]{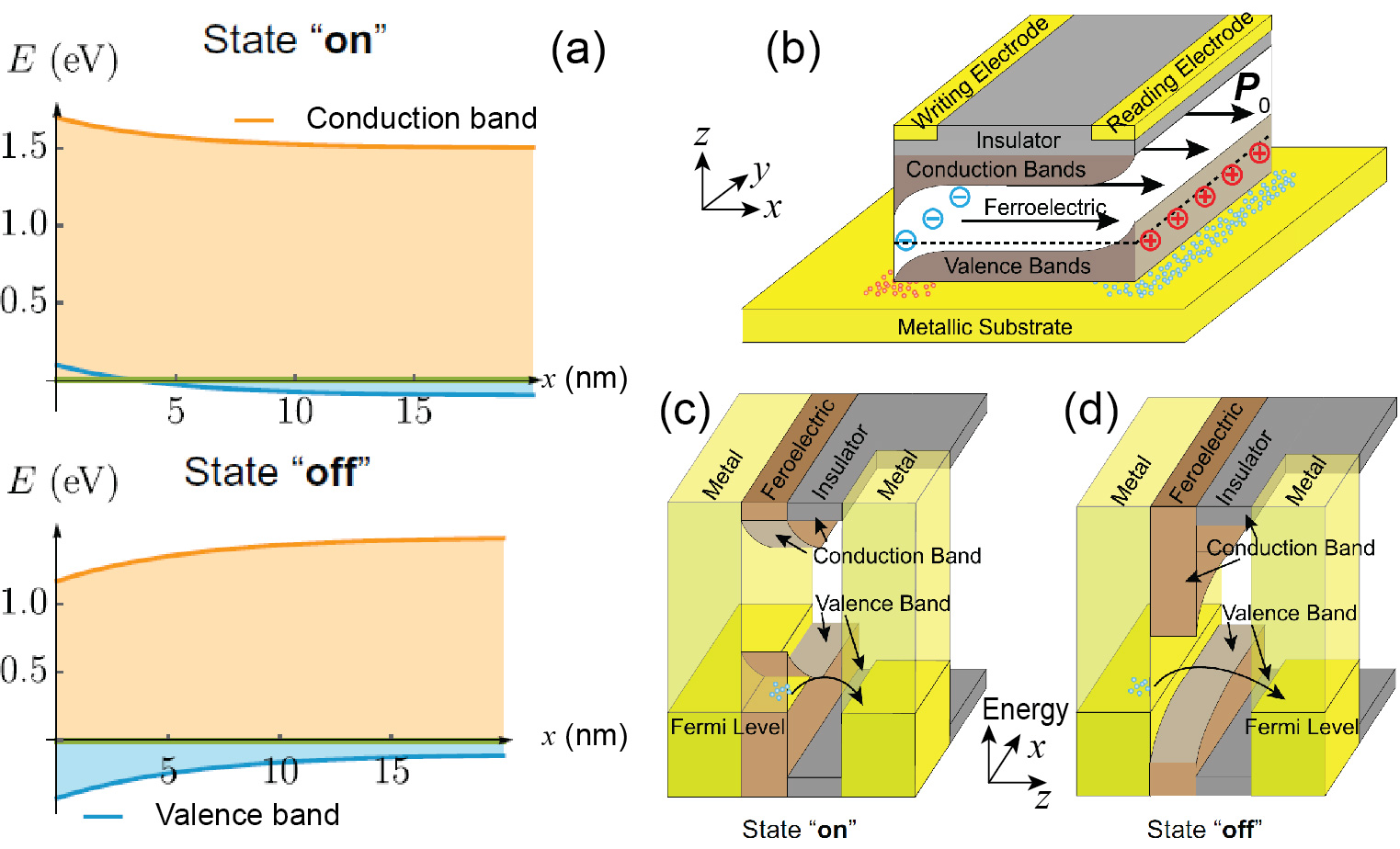}
\caption{\label{fig:figure5} (a) Schematics of upward (top) and downward (bottom) band bending near a ferroelectric's edge, for a chemical potential near the valence band edge. The Fermi energy of the electrode was set to be $E = 0$. (b)-(d) Schematic of the device (b) and band diagrams for the ``on'' (c) and ``off'' (d) sates. Adapted from Ref.~\cite{kai3}. Copyright, 2019, American Physical Society.}
\end{figure}

It may be possible to create {\em in-plane ferroelectric memories} by adding an insulator and a top gate to the two-terminal device shown in Fig.~\ref{fig:figure4}(a). Indeed, if $\mathbf{P}$ points in-plane, such as in $MX$ MLs, the band bending at the ferroelectric materials' edge can be read out with metal contacts \cite{kai3}. As depicted by the dependency of $I_D$ on $V_D$ in Fig.~\ref{fig:figure4}(a), the upward or downward band bending drawn in Fig.~\ref{fig:figure5}(a) could represent the ``on'' or ``off'' state respectively, so that the information stored is read nondestructively.

Fig.~\ref{fig:figure5}(b) shows a ferroelectric thin film sandwiched between a metallic substrate and a wide-band-gap insulator. The writing and reading electrodes are deposited at opposite edges of the top insulator. If $\mathbf{P}_0$ lies in-plane along the $+x$ direction, it will induce opposite net charges at the ferroelectric boundaries along that direction. Depending on the polarization direction (either $+x$ or $-x$), the band bending near the reading electrode could be upward or downward, leading to ``on'' and ``off'' states [Figs.~\ref{fig:figure5}(c) and Figs.~\ref{fig:figure5}(d), respectively]. Reading is nondestructive because the electric field generated by the reading voltage is perpendicular to the ferroelectric's polarization. Using Landauer's conductance formalism and suitably chosen parameters, currents of the order of $\mu A$ and $I_{on}/I_{off}$ ratios of the order of $10^{4}$ have been predicted \cite{kai3}. See Refs.~\cite{Shen19_ACSAEM} and \cite{newmemory} for additional memory devices based on O-$MX$s.

\section{Linear elastic properties, auxetic behavior, and piezoelectricity of O$-MX$ MLs}\label{sec:secV}

The physical properties of 2D materials can be tuned by strain \cite{maria,Review}. In linear elasticity theory, the strain tensor is defined as $\epsilon_{ij}\simeq \frac{1}{2}\left(\frac{\partial u_i}{\partial x_j}+\frac{\partial u_j}{\partial x_i}\right)$, where $\mathbf{u}=(u_x,u_y,u_z)$ is the displacement field  $\mathbf{u}=\mathbf{r}-\mathbf{r}_0$ away from a structural configuration that minimizes the structural energy.

The constitutive relation establishes a linear dependence among the stress tensor $\sigma_{ij}$ and $\epsilon_{ij}$: $\sigma_{ij}=C_{ijkl}\epsilon_{kl}$, where $C_{ijkl}$ is the elasticity tensor. Symmetry restrictions on O$-MX$ MLs imply that only $C_{xxxx}$, $C_{xxyy}$, $C_{yyyy}$, and $C_{xyxy}$  are non-zero \cite{fei_apl_2015_ges_gese_sns_snse,gomes_prb_2015a_ges_gese_sns_snse}; subindices $xx$ and $yy$ label compressive/tensile (normal) strain, and $xy$ is a shear strain. Using Voigt notation ($xx\to 1$, $yy \to 2$, and $xy\to 6$), these entries of the elasticity tensor are commonly written as $C_{11}$, $C_{12}$, $C_{22}$, and $C_{66}$ and their magnitudes are listed in Table \ref{ta:taElastic}.

\begin{figure}
\includegraphics[width=\linewidth]{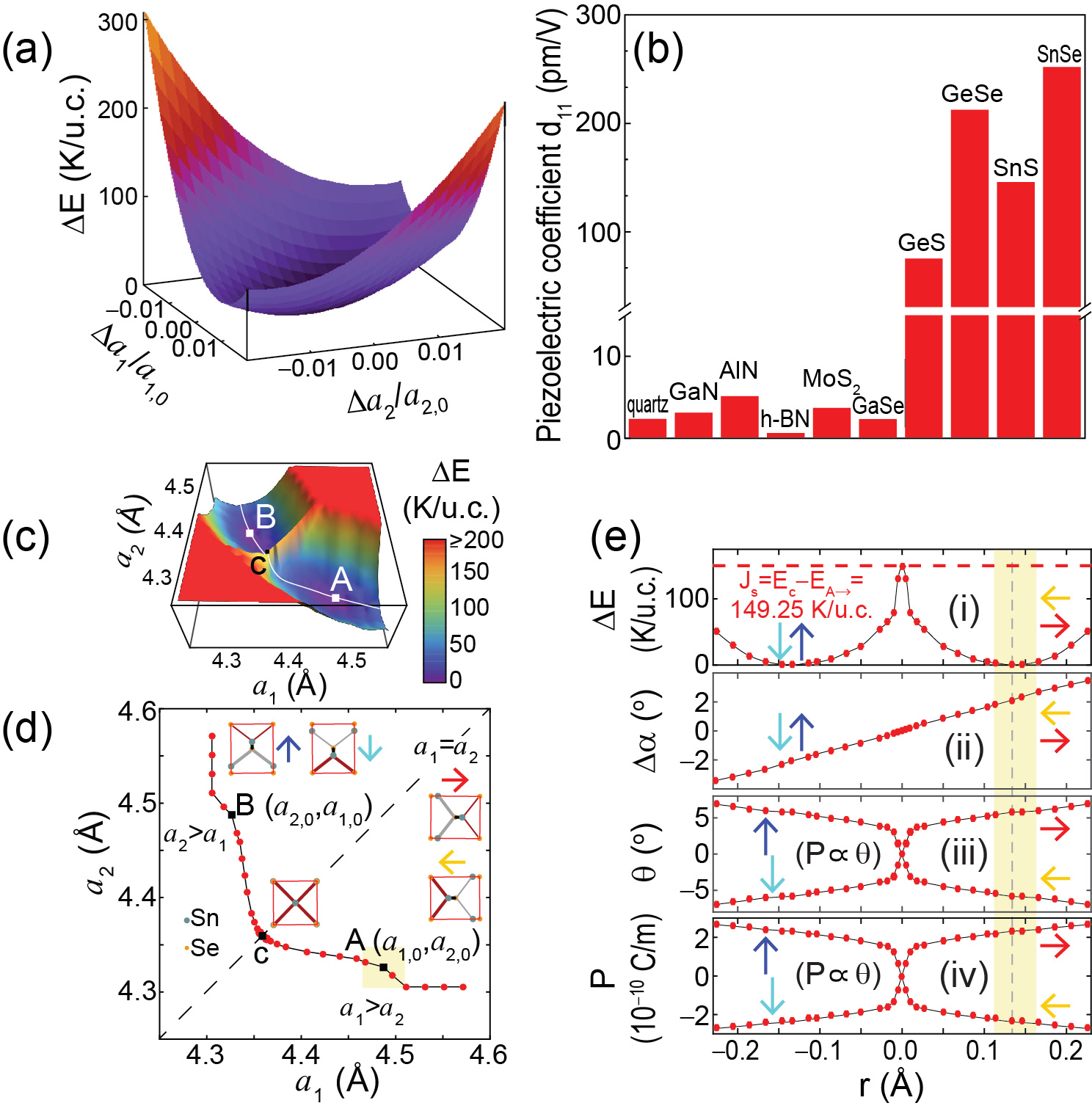}
\caption{\label{fig:figure6} (a) Elastic energy landscape of a SnSe ML (in units of K/u.c.) as a function of uniaxial strain along the $a_1$ and $a_2$ directions. Adapted from Ref.~\cite{gomes_prb_2015a_ges_gese_sns_snse} with permission. Copyright, 2015, American Physical Society. (b) Piezoelectric coefficient $d_{11}$ of GeS, GeSe, SnS and SnSe MLs and other known piezoelectric materials. Adapted from Ref.~\cite{fei_apl_2015_ges_gese_sns_snse} with permission. Copyright, 2015, American Institute of Physics. (c) Elastic energy landscape over a larger range of values for $a_1$ and $a_2$; point $A$ corresponds to the ground state unit cell shown in Fig.~\ref{fig:figure1}(c). (d) Two-dimensional lowest-energy path joining degenerate ground states $A$ and $B$; the shaded rectangle corresponds to $\pm$ 0.5\% strain. (e) Change in energy $\Delta E$, $\Delta \alpha$, $\theta$, and $P$ in going through the path shown in (c); the dashed vertical line cuts through $\Delta E=0$ and thus shows $\Delta\alpha_0$, $\theta_{0}$ and $P_0$. Subplots (c-e) are adapted from Ref.~\cite{other4} with permission. Copyright, 2018, American Physical Society.}
\end{figure}

$C_{11}$, $C_{22}$, and $C_{12}$ are obtained by fitting against the elastic energy landscape shown in Fig.~\ref{fig:figure6}(a), in which $\epsilon_1=\frac{\Delta a_1}{a_{1,0}}=\frac{a_1-a_{1,0}}{a_{1,0}}$ and $\epsilon_2=\frac{\Delta a_2}{a_{2,0}}$. Consistent with the change in area in going from to bulk to a ML, elastic constants tend to be slightly softer in MLs than in the bulk. Ref.~\cite{gomes_prb_2015a_ges_gese_sns_snse} provides the Young's modulus for GeS, GeSe, SnS and SnSe as well, which is an order of magnitude smaller that its magnitude of 340 N/m for graphene \cite{Hone}. Additionally, $C_{11}$ and $C_{22}$ are smaller when compared to their values for MoS$_2$ and GaSe MLs, listed in Table \ref{ta:taElastic} as well. The shear elastic coefficient $C_{66}$ in Table \ref{ta:taElastic} is as small as $C_{11}$: shear strain changes the magnitude of $\Delta\alpha$ in Fig.~\ref{fig:figure1}(c), implying that distortions by such an angle are as soft as a compression or elongation along the $\mathbf{a}_1$ direction. In light of Table \ref{ta:taElastic}, $MX$ MLs are {\em soft} 2D materials with anisotropic elastic properties.

The Poisson's ratio $\nu$ determines the rate of contraction in transverse directions under longitudinal uniaxial load.
Most materials have a {\em positive} Poisson's ratio but, as discussed in Refs.~\cite{gomes_prb_2015a_ges_gese_sns_snse,poisson1,poisson2,poisson3} and summarized in Table \ref{ta:ta4}, the buckled structure of O$-MX$ MLs depicted in Fig.~\ref{fig:figure1}(c) confers them with negative ratios when the out-of-plane ($z-$direction) is considered. The subindices of $\nu_{ij}$ in Table \ref{ta:ta4} indicate the (linear) Poisson's ratio along the $i$ direction due to a load along the $j$ direction as defined in Fig.~\ref{fig:figure1}. Negative values of $\nu_{ij}$ are indicative of {\em auxetic} behavior, {\em i.e.,} an elongation (compression) occurs along the $i$ direction when these 2D materials are elongated (compressed) along the $j$ direction. According to Table \ref{ta:ta4}, there is a direct correlation between a positive $\Delta h_0$ in Fig.~\ref{fig:figure1}(a) and a negative $\nu_{zx}$.

The third-order piezoelectric tensor $d_{ijk}$ links $\Delta \mathbf{P}=\mathbf{P}-\mathbf{P}_0$ with the applied strain $\epsilon_{jk}$. Using Voigt notation for the last two entries of the piezoelectric tensor and for the applied strain, Fig.~\ref{fig:figure6}(b) displays a ten times larger  magnitude of $d_{11}$ for GeS, GeSe, SnS, and SnSe when contrasted with the piezoelectric coefficients of quartz and other polar materials \cite{fei_apl_2015_ges_gese_sns_snse}.

\begin{table}[tb]
\caption{Relaxed-ion components of the elastic tensor $C_{ij}$ for $MX$ MLs in N/m. Adapted from Refs.~\cite{gomes_prb_2015a_ges_gese_sns_snse} and \cite{fei_apl_2015_ges_gese_sns_snse}.}\label{ta:taElastic}
\scalebox{.93}{
\begin{tabular}{cc|c cc cc cc}
\hline
\hline
Material & \scalebox{.85}{$\bar{Z}$}& $C_{11}$  & & $C_{22}$   & & $C_{12}$  &   & $C_{66}$ \\
\hline
GeS ML      & 24 & 15.24$-$20.87 & & 45.83$-$53.40 & & 21.62$-$22.22 & & 18.59     \\
GeSe ML     & 33 & 13.81$-$20.30 & & 46.62$-$50.16 & & 17.49$-$19.45 & & 23.19     \\
SnS ML      & 33 & 14.91$-$20.86 & & 35.97$-$43.15 & & 15.22$-$18.14 & & 19.56     \\
SnSe ML     & 42 & 19.61$-$19.88 & & 40.86$-$44.49 & & 16.36$-$18.57 & & 13.70     \\
\hline
MoS$_2$  ML $^{\text{a}}$  &   & 130        & & 130         & & 32    & & -- \\
GaSe  ML $^{\text{b}}$     &   & 83         & & 83          & & 18    & & -- \\
\hline
\hline
\end{tabular}
}\\
$^{\text{a}}$ Ref.~\cite{duerloo}. $^{\text{b}}$ Ref.~\cite{li2015}.
\end{table}

\section{Structural degeneracies and anharmonic elastic energy of O$-MX$ MLs}\label{sec:secVI}

As it turns out, the elastic energy landscape from which elastic properties were discussed in Sec.~\ref{sec:secV} is non-linear. Its non-linearity underpins the highly anharmonic vibrational properties and a propensity of O$-MX$ MLs for sudden changes in ferroelectric, structural, electronic, spin, and optical properties with temperature.

\begin{table}[tb]
\caption{Sign of $\Delta h_0$ and linear Poisson's ratio $\nu_{ij}$ of a BP ML and of group-IV monochalcogenide MLs. Negative values of $\nu_{ij}$ indicate auxetic behavior. Adapted from Refs.~\cite{poisson1,poisson2,poisson3}.}\label{ta:ta4}
\begin{tabular}{cc|c|ccccc}
\hline
\hline
Material & \scalebox{.85}{$\bar{Z}$}& $\Delta h_0$ & $\nu_{yx}$ & $\nu_{xy}$ & $\nu_{zx}$ & $\nu_{zy}$\\
\hline
BP          & 15 &   $0$        & 0.400      & 0.930      & 0.046   & $-$0.027     \\
\hline
GeS ML         & 24 & $+$       & 0.420      & 1.401      & $-0.208$& 0.411        \\
GeSe ML        & 33 & $-$       & 0.391      & 1.039      & 0.583   & $-$0.433     \\
SnS ML         & 33 & $+$       & 0.422      & 0.961      & $-$0.004& 0.404        \\
SnSe ML        & 42 & $+$       & 0.423      & 0.851      & $-$0.210& 0.352        \\
SnTe ML        & 51 & $-$       & 0.423      & 0.480      & 0.242   & 0.109        \\
\hline
\hline
\end{tabular}
\end{table}

Turning the $\Delta a_1/a_{1,0}=(a_1-a_{1,0})/a_{1,0}$ axis in Fig.~\ref{fig:figure6}(a) into $a_1$ and $\Delta a_2/a_{2,0}$ into $a_2$, and increasing the range for both $a_1$ and $a_2$ from which the structural energy $E(a_1,a_2)$ is computed, the elastic energy landscape $\Delta E(\epsilon_1,\epsilon_2)=\Delta E(a_1,a_2)=E(a_1,a_2)-E(a_{1,0},a_{2,0})$ shown in Fig.~\ref{fig:figure6}(c) ensues. Given that $E(a_{1,0},a_{2,0})=E(a_{2,0},a_{1,0})$ on 2D materials with a rectangular u.c.~($a_{1,0}>a_{2,0}$), the elastic energy landscape has two degenerate structures labeled $A$ and $B$ in Fig.~\ref{fig:figure6}(c) \cite{Mehboudi2016,other3}. O$-MX$s have eight degenerate u.c.s, occurring upon a mirror reflection with respect to the $x-z$ or $x-y$ planes, or by an exchange of $x-$ and $y-$coordinates \cite{Mehboudi2016}. Nevertheless, an inversion with respect to the $x-y$ plane does not change the orientation nor the sense of direction of $\mathbf{P}_0$ and is usually disregarded when describing degeneracies for that reason; the four remaining degenerate ground state u.c.s are displayed as an inset in Fig.~\ref{fig:figure6}(d). They have {projections $\mathbf{p}_0$} $\rightarrow$, $\uparrow$, $\leftarrow$, and $\downarrow$, that are reminiscent of discrete clock models---well known tools to discuss order-by-disorder 2D transformations in Statistical Mechanics \cite{potts} that provide important insight into the finite temperature behavior of O$-MX$ MLs \cite{Mehboudi2016}.

The saddle point $c$ in Fig.~\ref{fig:figure6} indicates the minimum elastic energy necessary to switch in between ferroelectric states $A$ and $B$. It is situated at $(a_c,a_c)$, with $a_c=\left(1-\frac{1}{\sqrt{2}}\right)a_{1,0}+\frac{a_{2,0}}{\sqrt{2}}$ \cite{ourarxiv}. The five-fold coordinated u.c.~at point $c$ is paraelectric \cite{Mehboudi2016} and it belongs to symmetry group 129 ($P4/nmm$, or $Pmm4/n$ with our choice of axes) \cite{Villanova2020PRB}. $J_s$ is the energy difference in between the five-fold coordinated paraelectric u.c. at point $c$ and any of the degenerate ferroelectric ground states ({\em i.e., } the one at point $A$ with polarization along the positive $x-$direction): $J_s=E_c-E_{A,\rightarrow}=E(a_c,a_c)-E(a_{1,0},a_{2,0})$. $J_s$ indicates the ease of a ferroelastic transformation among a pair of degenerate structures shown in Fig.~\ref{fig:figure6}(d). As it will be discussed in Sec.~\ref{sec:secVII}, it is a qualitative estimator of the critical temperature $T_c$ at which a ferroelectric to paraelectric transition takes place in these 2D materials. [$a_{1,0}=a_{2,0}=a_c$ for Pb$X$ MLs in Fig.~\ref{fig:figure1}(d), which hence have a single non-degenerate structural ground state and $J_s=0$.]

The white path $r(a_1,a_2)$ in Fig.~\ref{fig:figure6}(c) provides the lowest-energy distortion that is necessary to turn degenerate structure $A$ into $B$ elastically, and it is projected into a 2D plot in Fig.~\ref{fig:figure6}(d). For a SnSe ML, points $A$ and $B$ are located at distances $r=0.134$ \AA{} and $r=-0.134$ \AA{} along this path; point $c$ is located at $r=0$ \AA. The range of values utilized to extract linear elastic properties in Sec.~\ref{sec:secV} can be seen as a yellow rectangle in Fig.~\ref{fig:figure6}(d).

The anharmonicity of the elastic energy landscape is established by the double-well potential $\Delta E(r)$ seen as subplot (i) in Fig.~\ref{fig:figure6}(e). The magnitude of $J_s$ for a SnSe ML [as computed with the vdW-DF-CX \cite{cx} XC functional] can also be seen in that plot. The dependency of $\Delta E$ on the path coordinate $r$ is {\em bistable} ({\em i.e.,} fundamentally non-harmonic). The evolution of $\Delta \alpha$, $\theta$, and the polarization $P$ along $r$---including the four possible orientations of $\mathbf{p}$ ($\mathbf{P}$)---is displayed as subplots (ii), (iii), and (iv) in Fig.~\ref{fig:figure6}(e). The area in light yellow in Fig.~\ref{fig:figure6}(e) corresponds to the $\pm0.5\%$ strain within which $\Delta E$ can be fitted to a parabola, and where $\Delta\alpha$, $\theta$, and $P$ are linear on $r$~\cite{fei_apl_2015_ges_gese_sns_snse}.

The vertical dashed line, crossing through $\Delta E=0$ shows $\Delta\alpha_0$, $\theta_{0}$, and $P_0$; {\em i.e.}, the magnitudes of these variables in a u.~c.~like the one seen in Fig.~\ref{fig:figure1}(c). The angle $\Delta\alpha$ is positive for $r>0$ ($a_1>a_2$), zero at $r=0$ ($a_1=a_2$), and negative for $r<0$ ($a_1<a_2$). The angle $\theta$, in turn, points along the (positive or negative) $x-$direction when $r>0$, it is zero at $r=0$, and it points along $\pm y$ when $r<0$. Importantly, $\theta$ and $P$ are linearly proportional ($P\propto \theta$) and $P=0$ when $\theta=0$ and $\Delta\alpha=0$ in an elastic transformation in which lattice parameters can vary. The possibility of switching {\color{blue}$\mathbf{P}$} gives rise to a combined ferroelectricity and ferroelasticity; {\em i.e.}, to multiferroic behavior in O$-MX$ MLs \cite{ccBP,other3}.

\section{Structural phase transition and pyroelectric behavior of O$-MX$ MLs}\label{sec:secVII}

\begin{figure}
\includegraphics[width=\linewidth]{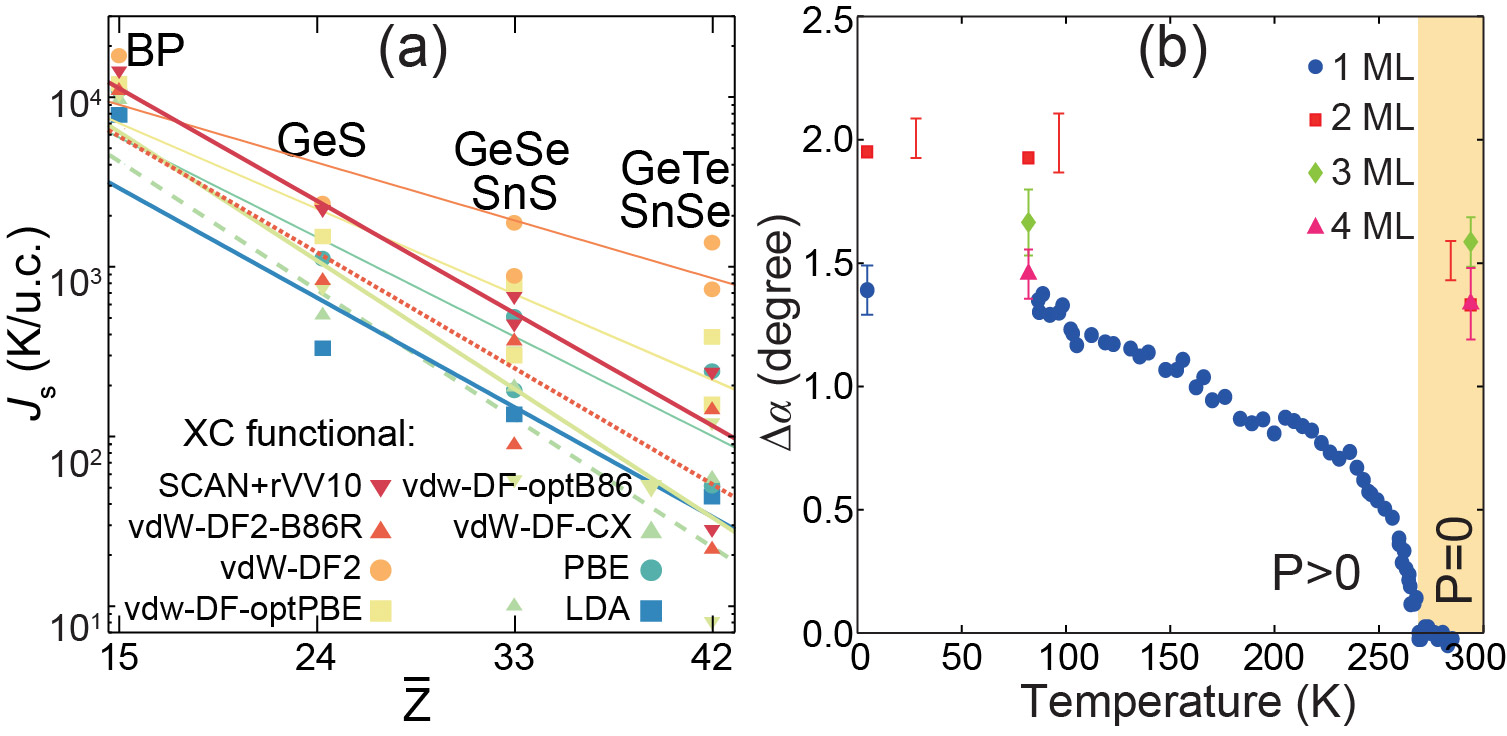}
\caption{\label{fig:figure7} (a) Exponential dependency of $J_s$ on $\bar{Z}$. This plot also shows the dependence of $J_s$ (and therefore of $a_{1,0}$, $a_{2,0}$, $\Delta \alpha_0$, $P_0$, and $\theta_0$) on the XC functional employed. Taken from Ref.~\cite{ourarxiv} with permission. Copyright, 2019, American Physical Society. (b) Experimental thermal dependency of $\Delta \alpha$ for few-layer SnTe on epitaxial graphene. A paraelectric phase (P=0) ensues at temperatures above $T_c=270$ K in SnTe MLs. Adapted from Ref.~\cite{Kai} with permission. Copyright, 2016, American Association for the Advancement of Science.}
\end{figure}

\begin{figure*}
\includegraphics[width=\linewidth]{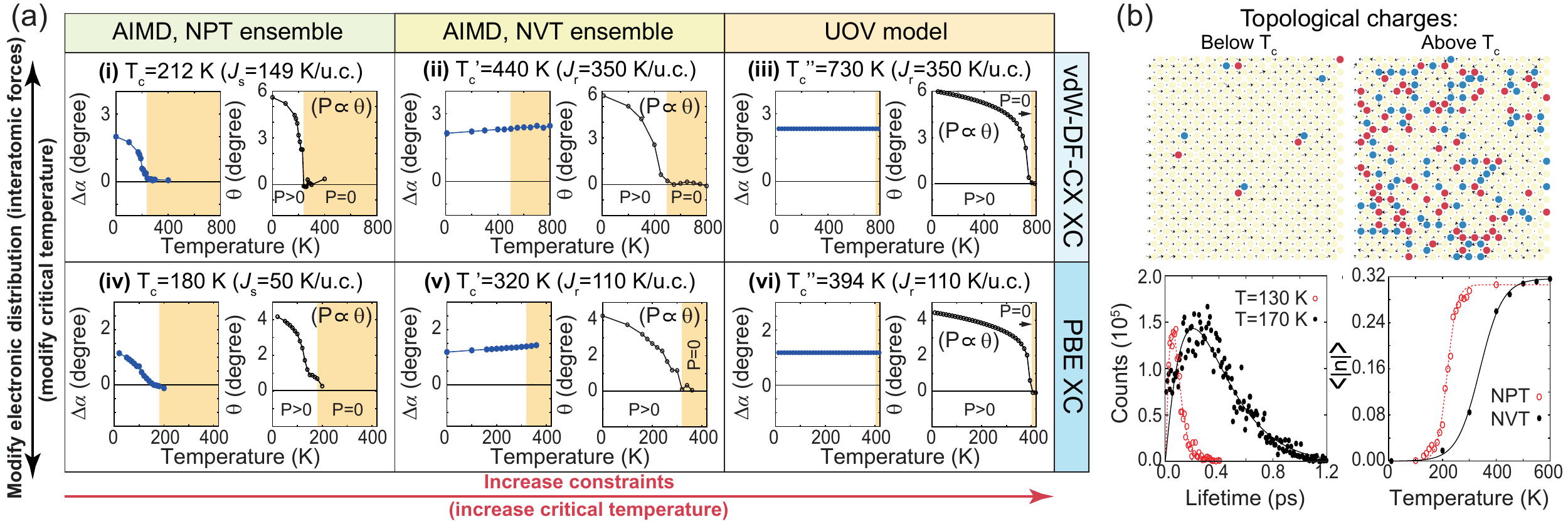}
\caption{\label{fig:figure8} (a) Temperature dependence of $\Delta \alpha$ and $\theta$ for a SnSe ML in AIMD calculations employing two XC functionals, and obtained within the NPT [subplots(i) and (iv)] and NVT ensembles [subplots(ii) and (v)]. Results from a more constrained unidirectional optical vibration (UOV) model are presented in subplots (iii) and (vi). The ferroelectric phase occurs when $\theta>0$. Note that $\Delta\alpha=0$ implies $\theta=0$ but that $\theta=0$ does not necessarily imply $\Delta\alpha=0$. (b) The structural transformations described by AIMD can be understood in terms of a fluctuating connectivity ({\em i.e.}, topology) of these 2D ferroelectrics. Pairs of topological charges have a temperature-dependent lifetime and their number saturates at temperatures within $T_c$. Adapted from Ref.~\cite{Villanova2020PRB} with permission. Copyright, 2020, American Physical Society.}
\end{figure*}

Structural degeneracies underpin strong anharmonic elastic properties, soft phonon modes, and structural phase transitions. Taking $J_s$---the relevant energy scale in the system---as an {\em ad-hoc} exchange parameter, a clock model with $r=4$ degenerate states yields the following relation among $T_c$ and $J_s$: $T_c=1.136 J_s$ \cite{potts}. The Potts model also has a prescription in case that only a subset of two degenerate states is available ({\em e.g.,} $\rightarrow$ and $\leftarrow$), which could occur in a constrained scenario in which $a_{1}$ and $a_2$ keep zero-temperature magnitudes \cite{fei_prl_2016}: calling $J_r$ the energy barrier under such constrained configuration, Potts dictates that $T_c'=2.272 J_r$ \cite{potts}. Numerical calculations indicate that $J_r \ge 1.4 J_s$, so that $T_c' \ge 2.8 T_c$. The message is that structural constraints lead to an increased $T_c$.

The ferroelectric-to-paraelectric transition temperature $T_c$ of O$-MX$ MLs calculated at the DFT level \cite{martin} has a strong dependency on the choice of XC functional, and it is unclear that the PBE XC functional ought to provide the most accurate description of the thermal behavior of O$-MX$s. The strong dependency of $T_c$ on XC functional can be already foreseen in the magnitude of $J_s$ displayed in Fig.~\ref{fig:figure7}(a), which contains predictions with LDA, PBE, multiple non-empirical van der Waals implementations \cite{reviewvdw}, and even the recent SCAN+rVV10 \cite{scan} XC functional \cite{ourarxiv}.

{\em Ab initio} molecular dynamics (AIMD) calculations performed on freestanding SnSe MLs using the NPT ensemble (in which containing walls are allowed to move to accommodate for thermal expansion) provide the following information: (i) $T_c$ is larger in GeSe MLs and bilayers than it is in SnSe MLs and bilayers, owing to the smaller $\bar{Z}$ and hence larger barrier $J_s$ [Fig.~\ref{fig:figure7}(a)] for GeSe; (ii) for a given O$-MX$, $T_c$ increases with increasing number of MLs \cite{other2,Kai}. A slight dependency of $T_c$ in the size of the simulation supercell has been documented, too \cite{other2,other4}. In agreement with $J_s$'s inverse dependency on $\bar{Z}$, experiments indicate a $T_c$  larger than 400 K for SnSe ($\bar{Z}=42$) on graphene \cite{Chang20_arxiv_SnSe}, and $T_c=$270 K for SnTe MLs ($\bar{Z}=51$) on the same substrate [Fig.~\ref{fig:figure7}(b) \cite{Kai}].

The upper row in Fig.~\ref{fig:figure8}(a) shows a progression of $T_c$ estimates for a freestanding SnSe ML that were obtained using the vdW-DF-CX XC functional \cite{cx}. From left to right, the figure displays the thermal behavior of $\Delta\alpha$ and $\theta$ when (i) using the NPT ensemble (in which containing walls move so that the material remains at atmospheric pressure), (ii) the NVT ensemble (in which the supercell volume $V$ is fixed and containing walls do not move), and (iii) a unidirectional optical vibration (UOV) model in which only one unidirectional vibrational mode---out of twelve---is employed and the containing walls do not move either \cite{fei_prl_2016}. The point is that (as already foreseen by the Potts model a few lines above) $T_c$ increases with added constraints. Energy barriers $J_s$ and $J_r$ are listed in that Figure, too.

Briefly said, AIMD calculations carried out with the NPT ensemble yield the smallest magnitude of $T_c$ (212 K). Although the compounds are not the same (a freestanding SnSe ML in calculations and a SnTe ML on graphene in experiment), the decay of $\Delta\alpha$ in the calculations seen in subplot (i) of Fig.~\ref{fig:figure8}(a) indicates a phenomenology  consistent with experiment in Fig.~\ref{fig:figure7}(b) \cite{other4,Kai}. When constraining the SnSe ML by not permitting its area to increase at finite temperature, the structural transition within the NVT ensemble necessarily requires additional (thermal) energy to take place, raising $T_c'$ up to $440$ K and displaying $\Delta\alpha>0$ at $T_c$. Despite the existence of nearly degenerate vibrational modes oscillating along both $x-$ and $y-$directions, the highly constrained UOV model only permits an optical vibration along the $x-$direction (an oscillatory mode valid only at the $\Gamma-$point) and thus yields the largest $T_c''=730$ K, still showing $\Delta\alpha>0$ at $T_c$ as $a_1$ and $a_2$ are kept fixed [Eqn.~\eqref{eq:eq1}]. $\Delta\alpha$ [Fig.~\ref{fig:figure7}(b)] is not a relevant order parameter for subplots (ii) and (iii) in Fig.~\ref{fig:figure8}(a) because $a_1$ and $a_2$ retain their zero-temperature values in these models.

The increasing sequence of critical temperatures observed with increasing mechanical constraints is independent of the XC approximation, as the lower row in Fig.~\ref{fig:figure8}(a) shows a similar phenomenology when the PBE XC functional is employed. $T_c$, $T_c'$, and $T_c''$ have smaller values than those obtained with the vdW-DF-CX XC functional \cite{Villanova2020PRB}. As illustrated in Fig.~\ref{fig:figure8}(b), the structural transition is underpinned by changes in the connectivity of the 2D lattice as the two atoms defining the angle $\theta$ in Fig.~\ref{fig:figure1}(a) rotate about the out-of-plane $z-$axis; this change in connectivity confers a topological character to the structural transformation \cite{KostRev2016,Villanova2020PRB,othertopo}.

Pyroelectricity is the creation of electricity by a temperature gradient. Given the direct proportionality between $\theta$ and $P$, the temperature derivative of $\theta$ in Fig.~\ref{fig:figure8}(a) gives a direct insight into the pyroelectric properties of O$-MX$s \cite{other2}.

The effects of substrates such as Ni, Pd, Pt, Si, Ge, CaO, and MgO on the morphology and properties of SnTe MLs (including charge transfer and atomistic distortions) have been studied \cite{substrate}. Since O$-MX$ MLs are presently grown on substrates, the effect of the substrate-$MX$ interaction on the transition temperature is an important avenue for further theory. Along these lines, the elastic energy barriers $J_s$ of GeSe, GeTe, SnS, SnSe, and SnTe have been shown to vanish under a modest hole doping of 0.2 $|e|$/u.c., where $e$ is the electron's charge \cite{doping1,doping2}.

We indicated in Sec.~\ref{sec:secII} that O$-MX$s are isostructural to BP. This makes BP MLs doubly-degenerate upon exchange of $x-$ and $y-$ coordinates, and suggests that BP MLs may also undergo a phase transition at finite temperature. Nevertheless, considering $J_s$ as an approximate measure of $T_c$, one observes $J_s>1,000$ K/u.c. for BP in Fig.~\ref{fig:figure7}(a) regardless of XC functional. Such magnitude is so large that a BP ML melts rather than undergoing a ferroelastic to paraelastic transition \cite{Mehboudi2016}, thus explaining the lack of {\em thermally-driven} 2D phase transitions in BP MLs. {[BP MLs have been shown to undergo temperature-independent 2D phase transitions by mechanical strain \cite{PhysRevLett.112.176801}.]} The propensity to undergo thermally-driven 2D transitions is a crucial aspect that sets 2D ferroelectrics apart from other 2D materials such as graphene, TMDCs with a 2H symmetry, and BP MLs.

\section{Electronic, valley and spin properties of O$-MX$ MLs}\label{sec:secVIII}
\subsection{Electronic band structure}\label{sec:secVIII-A}

\begin{figure}
\includegraphics[width=\linewidth]{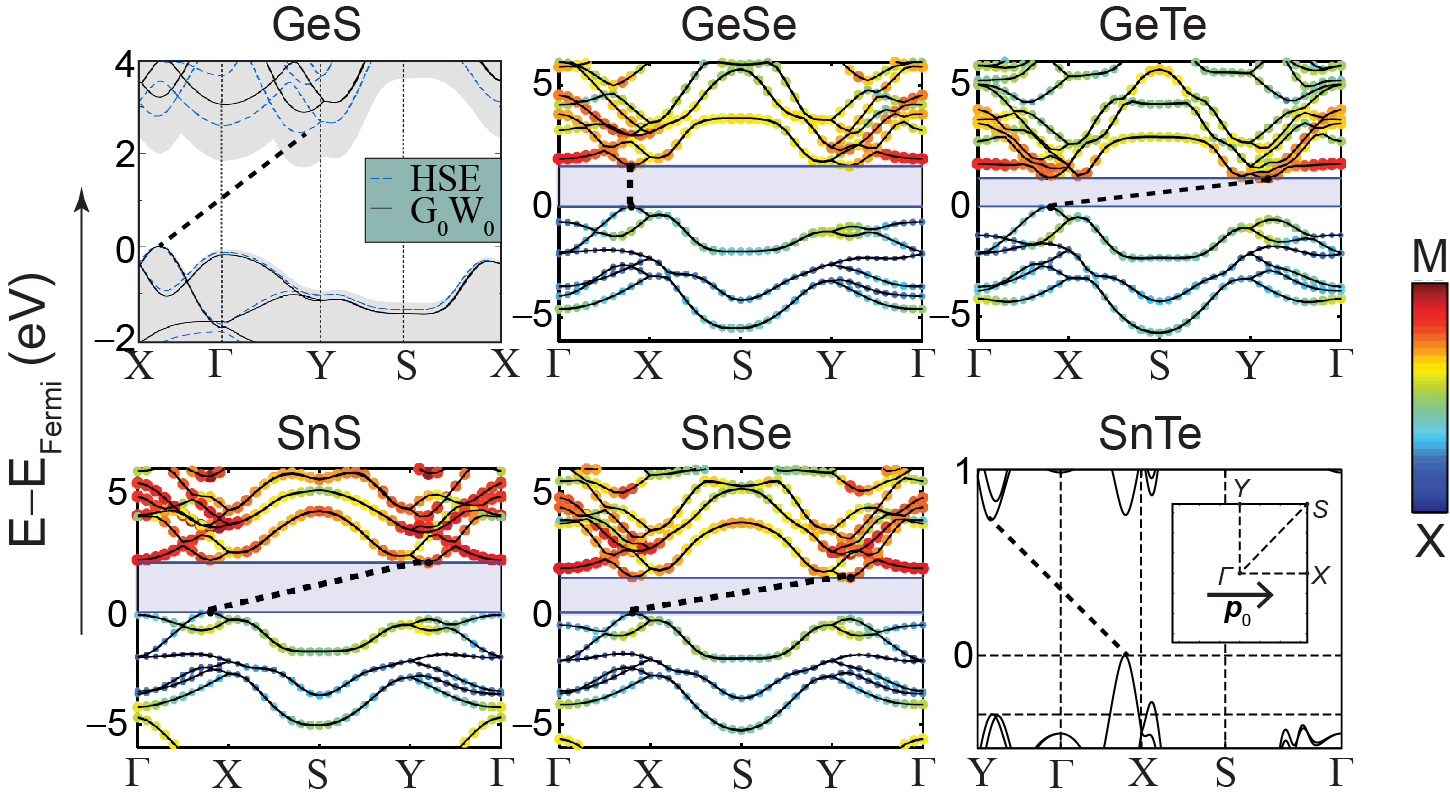}
\caption{\label{fig:figure9} Electronic band structure of O$-MX$ MLs. With the exception of the SnTe ML, these calculations were carried out with the HSE06 XC functional. GeSe MLs have direct band gaps, and the GeTe ML subplot was modified to account for an indirect band gap smaller by just 0.01 eV than the direct band gap. The inset in the SnTe subplot displays the high-symmetry points and the direction of $\mathbf{p}_0$. Adapted from Ref.~\cite{singh_apl_2014_ges_gese_sns_snse} with permission. Copyright, 2014, American Institute of Physics. The GeS subplot was taken from Ref.~\cite{gomes_prb_2016_excitons} with permission. Copyright, 2016, American Physical Society. The SnTe subplot was adapted from Ref.~\cite{KaiPRL} with permission. Copyright, 2019, American Physical Society.}
\end{figure}

TMDC MLs such as 2H-MoS$_2$ display an indirect-to-direct band gap crossover at the ML limit. As indicated in Table \ref{ta:taGaps}, the experimental band gap of MoS$_2$ increases from 1.29 eV in the bulk up to 1.90 eV in a ML due to quantum confinement \cite{mak1}, and the valence band maxima (VBM) and conduction band minima (CBM) are both located at the high-symmetry $\pm \mathbf{K}-$points in these MLs. Similarly, quantum confinement leads to an increase of the electronic band gap of BP (Table \ref{ta:taGaps}), and its electronic bands are highly anisotropic.  The VBM and CBM are both located at the $\Gamma-$point in BP MLs \cite{Tran}.

Tritsaris, Malone, and Kaxiras \cite{tritsaris_jap_2013_sns} studied the electronic properties of SnS down to the ML limit. As indicated in Table \ref{ta:taGaps}, the electronic band gap increases as these materials are thinned down, too. Nevertheless (and unlike the case for 2H-TMDs and BP MLs), the VBM and CBM are not located at high-symmetry points in the first Brillouin zone (BZ).

With reciprocal lattice vectors $\mathbf{b}_1=\frac{2\pi}{a_{1,0}}(1,0,0)$ and  $\mathbf{b}_2=\frac{2\pi}{a_{2,0}}(0,1,0)$, the high-symmetry points depicted as an inset within the SnTe subplot in Fig.~\ref{fig:figure9} are $\Gamma$, $X$ (located at $\mathbf{b}_1/2$), $Y$ (at $\mathbf{b}_2/2$), and $S$ (at $\mathbf{b}_1/2+\mathbf{b}_2/2$). O$-MX$ MLs have their VBM away from high-symmetry points, at about $\pm (0.74-0.84)X$ and their CBM at about $\pm (0.74-0.84)Y$ for an indirect band gap. The exception is GeSe, having a CBM at $\pm 0.80X$ and an direct band gap \cite{gomes,singh_apl_2014_ges_gese_sns_snse,gomes_prb_2016_excitons,shi_nl_2015_snse_gese}. Electronic band structure calculations within the GW approximation \cite{berkeleyGW} were carried out in Refs.~\cite{shi_nl_2015_snse_gese,tuttle_prb_2015_sis} and \cite{gomes_prb_2016_excitons}; their band gaps are listed in Table \ref{ta:taGaps}. The electronic band structure of O$-MX$s turns more (less) anisotropic for lighter (heavier) compounds, for which $a_{1,0}/a_{2,0}$ in Fig.~\ref{fig:figure1}(d) takes on larger (smaller) values. Going across chemical elements, the band gap for O$-MX$ MLs in Table \ref{ta:taGaps} is tunable with $\bar{Z}$: it takes its largest magnitude for lighter compounds (GeS, $\bar{Z}=24$) and it is smaller for the heaviest compound (SnTe, $\bar{Z}=51$).

\begin{table}[tb]
\caption{Electronic band gap (in eV) for BP, O$-MX$s and MoS$_2$ (ML and bulk) with PBE and HSE XC functionals \cite{scuseria}, from $GW$ calculations \cite{berkeleyGW}, or experiment. PBE values for GeTe and SnTe MLs, and for bulk SnTe were computed by us.\label{ta:taGaps}}
\scalebox{.9}{
\begin{tabular}{cc|cccc|ccc}
\hline
\hline
Material &\scalebox{.85}{$\bar{Z}$} & PBE           & HSE         & $GW$              & Exp.   &PBE           & HSE         &Exp.    \\
         &                          & (ML)          & (ML)        & (ML)              & (ML)   &(bulk)        & (bulk)      &(bulk)   \\
\hline
BP       & 15                       & 0.90$^a$      & 1.66$^a$    &   2.2$^b$         &        &0.07$^a$       & 0.39$^a$   & 0.33$^c$ \\
\hline
GeS      & 24                       & 1.65$^a$      & 2.32$^a$    &   2.85$^d$        &        &1.22$^a$       & 1.81$^a$   &  1.70-1.96$^e$\\
GeSe     & 33                       & 1.18$^a$      & 1.54$^a$    &  1.70-1.87$^{d,f}$&        &0.57$^a$       & 1.07$^a$   &  1.14$^{g}$\\
GeTe     & 42                       & 0.87          &             &                   &        &0.33$^h$       & 0.65$^h$   &  0.61$^i$\\
SnS      & 33                       & 1.38$^a$      & 1.96$^a$    &                   &        &0.82$^a$       & 1.24$^a$   &  1.20-1.37$^e$\\
SnSe     & 42                       & 0.96$^a$      & 1.44$^a$    &   1.63$^f$        & 2.1$^j$&0.54$^a$       & 1.00$^a$   &  0.90$^k$\\
SnTe     & 51                       & 0.68          &             &                   & 1.6$^l$&0.13           &            &  0.30$^m$\\
\hline
MoS$_2$  &$-$                       &  1.63$^n$     &2.11$^n$     &                   & 1.90$^o$& 0.98$^n$     &  1.46$^n$  & 1.29$^o$ \\
\hline
\hline
\end{tabular}
}\\
$^{a}$ Ref.~\cite{gomes}. $^{b}$ Ref.~\cite{BPexcitons}. $^c$ Ref.~\cite{keyes}. $^d$ Ref.~\cite{gomes_prb_2016_excitons}.\\
$^e$ Ref.~\cite{malone}.
$^f$ Ref.~\cite{shi_nl_2015_snse_gese}. $^g$ Ref.~\cite{geseexpt}.\\
$^h$ Ref.~\cite{picozzi}. $^i$ Ref.~\cite{geteexpt}.\\
$^j$ Ref.~\cite{Chang20_arxiv_SnSe}. $^k$ Ref.~\cite{exptsnse}. $^l$ Ref.~\cite{Kai}. $^m$ Ref.~\cite{bulk_SnTe_3}. $^n$ Ref.~\cite{MoS2_THEORY}. $^o$ Ref.~\cite{mak1}.

\end{table}

\begin{figure}
\includegraphics[width=\linewidth]{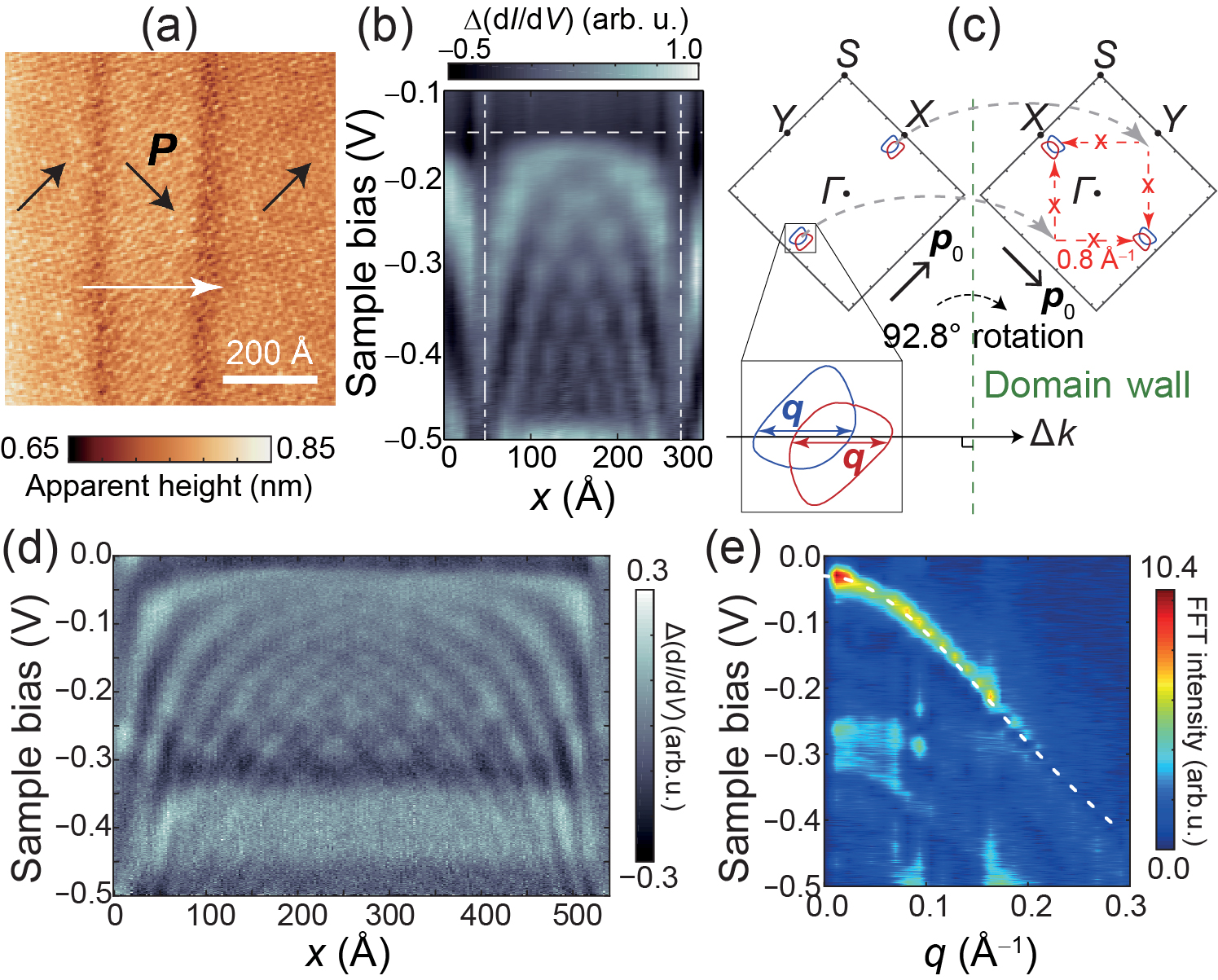}
\caption{\label{fig:figure10} (a) STM topographic image of $\sim 90^\circ$ domains on a SnTe ML. (b) $dI/dV$ spectra acquired along the white arrow in (a) at energies around the VBM. (c) Mismatched hole bands at opposite sides of a 90$^\circ$ domain wall. Constant energy contours with opposite spin components are colored in blue or red, respectively. (d) Electronic standing wave pattern across a 540~\AA{} wide domain. (e) The Fourier transform of (d) reveals the energy dispersion of scattering vectors. The dashed white curve is the $E(q)$ dispersion as obtained from the electronic band structure. Adapted from Ref.~\cite{KaiPRL} with permission. Copyright, 2019, American Physical Society.}
\end{figure}

From an experimental perspective, hole-doped SnTe MLs acquire a domain structure observed as dark vertical lines in Figs.~\ref{fig:figure3}(b) and ~\ref{fig:figure10}(a). As seen in Figs.~\ref{fig:figure10}(b) and \ref{fig:figure10}(d), the spatially resolved $dI/dV$ spectra [proportional to the sample's local density of states (LDOS)] features electronic standing wave patterns across domains for energies below the VBM that provide indirect information into these materials' electronic properties.

The  standing wave patterns observed at 4 K are induced by the electronic band mismatch at the two sides of a 90$^\circ$ domain walls [see domains with $\mathbf{P}$ forming $\sim 90^\circ$ angles in Figs.~\ref{fig:figure2}(b) and \ref{fig:figure3}(b)]. As Fig.~\ref{fig:figure10}(c) shows, the band apexes along the $\Gamma-Y$ direction are 0.3 eV below those seen along the $\Gamma-X$ direction. Such mismatch of hole momentum prevents a direct (elastic, unscattered) transmission of holes through domain walls, giving rise to a peculiar reflection resulting in standing waves \cite{KaiPRL}.

In fact, and as depicted in Fig.~\ref{fig:figure10}(c), the reflection off a domain wall occurs \textit{via} a momentum transfer $\mathbf{q}$ occurring within each hole band. This observation implies that the standing wave pattern is an indirect measure of the electronic band structure around the VBM. From the Fourier transform of the standing wave pattern in Fig.~\ref{fig:figure10}(d), a single branch of the energy dispersion with scattering vector $q$ [inset in Fig.~\ref{fig:figure10}(c)] is experimentally resolved in Fig.~\ref{fig:figure10}(e) \cite{KaiPRL}; note that $P(T=4 K)\simeq P_0$. Although spin-orbit interaction induces band splitting at the VBM, the contribution from the two spin components to the standing wave pattern are exactly the same because of time reversal symmetry.

\subsection{Valleytronics}\label{sec:secVIII-B}

\begin{figure}
\includegraphics[width=0.75\linewidth]{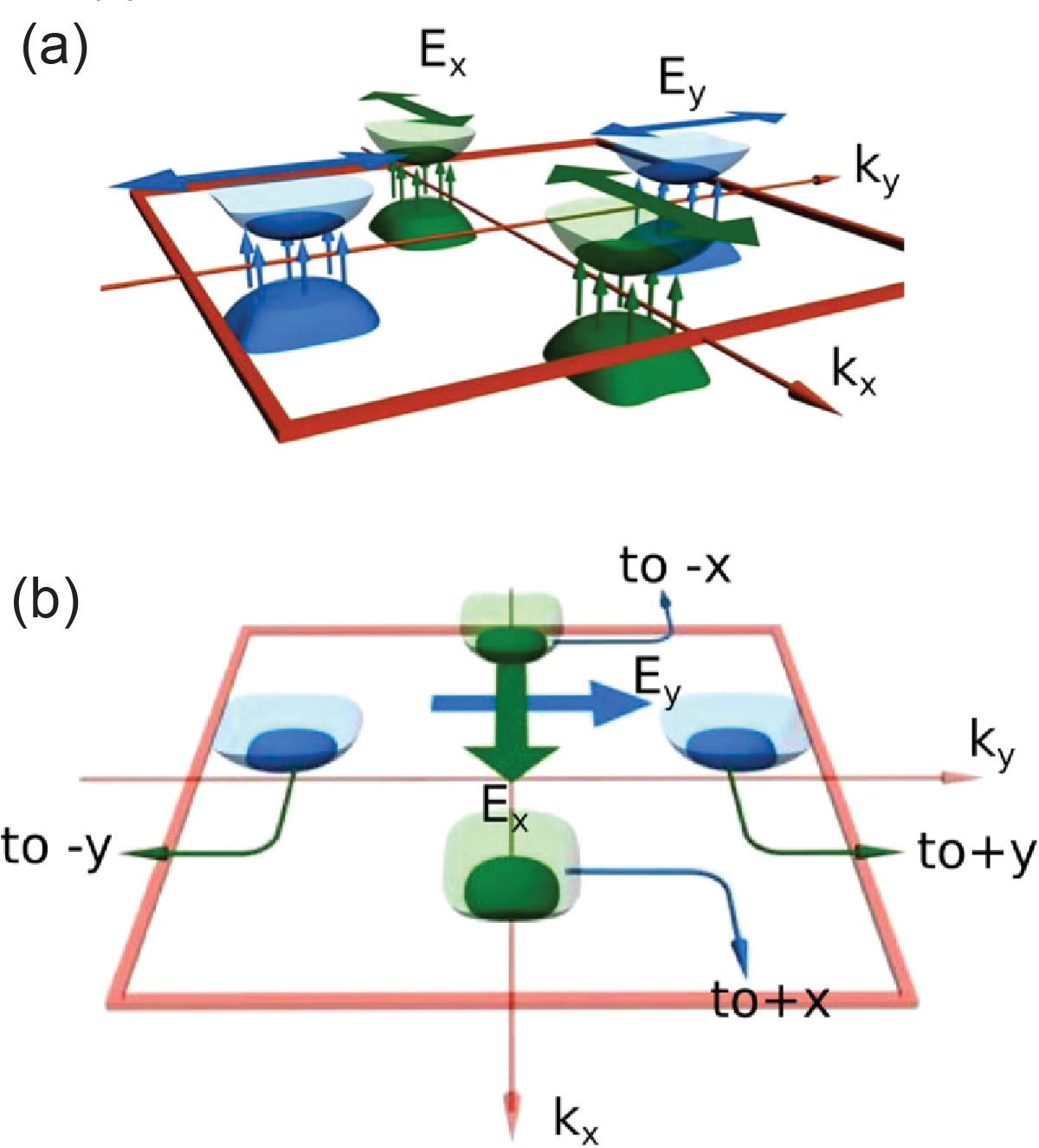}
\caption{\label{fig:figure11} (a) Valley selection under an external oscillating electric field. Different valleys are excited depending on the field's linear polarization. (b) Valley separation under a static electric field. Depending on the polarization of the field, different valleys flow along perpendicular directions. Taken from Ref.~\cite{rodin_prb_2016_sns} with permission. Copyright, 2016, American Physical Society.}
\end{figure}

The band curvature $\hbar^2/m^*$ at the VBM and CBM---where $m^*$ is the effective mass---is used to estimate hole and electron conductivities of semiconductors. When determined along orthogonal ($x-$ and $y-$) directions, it provides information about the anisotropy of the charge carriers' conductivity. The effective masses at the VBM ($m^*=m_h$) and CBM ($m^*=m_e$) for multiple O$-MX$ MLs (expressed in terms of the electron's mass $m_0$) are listed in Table \ref{ta:taMass} \cite{gomes_prb_2016_excitons}. With the exception of GeS MLs, these effective masses are smaller to those of MoS$_2$ [which range within $-$(0.44-0.48)$m_0$ for holes and 0.34-0.38$m_0$ for electrons \cite{massmos21,massmos22}], implying sharper hole/electron pockets at the VBM/CBM on O$-MX$ MLs than those existing in more traditional materials for valleytronic applications  \cite{valleytronics_review_2016}. Such sharpness of the valence (conduction) band curvature permits stating that holes (electrons) belong to a given {\em valley}.

Valleytronics refers to the use of the electron/hole pockets at the CBM or VBM  as information carriers \cite{valleytronics_review_2016}, which requires creating valley-specific gradients of charge carriers; {\em i.e.}, a {\em valley polarization}. Achieving valley polarization requires lifting the valley degeneracy; something that has been demonstrated in TMDC MLs \cite{valley_valve_grapehe_2007,TMDC_valleytronics_2012}. In these 2D materials, valleys at time-reversed states $+\v{K}$ and $-\v{K}$ in the BZ couple to the circular polarization of light so that a pseudospin (``up'' or ``down'') quantum number can be associated with each valley. Valley polarization occurs because right-hand polarized photons only excite the carriers in $+\v{K}$ valley, and left-hand polarized photons only excite those in $-\v{K}$ valley \cite{valley_polarization_2012_1,valley_polarization_2012_2,valley_polarization_2012_3,valley_polarization_2012_4}. When an in-plane electric field is applied across graphene bilayers or TMDC MLs, carriers with ``up'' and ``down'' pseudospins acquire a transverse velocity in opposite directions because of the opposite Berry curvature in $+\v{K}$ and $-\v{K}$ valleys, giving rise to a valley Hall effect \cite{valley_hall_mos2_2014,valley_hall_grapehe_2015_1,valley_hall_grapehe_2015_2}.

\begin{table}[tb]
\caption{Anisotropic effective masses of holes ($m_h/m_0$) and electrons ($m_e/m_0$) at VBM and CBM along the $x$ and $y-$directions as shown in Fig.~\ref{fig:figure1}(c) for O$-MX$ MLs.}\label{ta:taMass}
\begin{tabular}{cc|cc|cc}
\hline
\hline
Material &\scalebox{.85}{$\bar{Z}$} & $(m_h/m_0)_x$ & $(m_h/m_0)_y$ & $(m_e/m_0)_x$ & $(m_e/m_0)_y$\\
\hline
GeS ML   &24   &   $-0.26$    &  $-0.94$      &  0.24      &  0.57    \\
GeSe ML  &33   &   $-0.17$    &  $-0.32$      &  0.17      &  0.34    \\
GeTe ML  &42   &   $-0.15$    &  $-0.16$      &  0.08      &  0.32    \\
SnS ML   &33   &   $-0.24$    &  $-0.27$      &  0.20      &  0.22    \\
SnSe ML  &42   &   $-0.14$    &  $-0.14$      &  0.14      &  0.14    \\
SnTe ML  &51   &   $-0.10$    &  $-0.05$      &  0.13      &  0.14    \\
\hline
\hline
\end{tabular}
\end{table}

Unlike a BP ML, which has a single valley centered around the $\Gamma-$point, the sharp band curvature of the VBM and CBM of O$-MX$ MLs listed in Table \ref{ta:taMass} permits considering them two-valley materials, too. They feature a valley along the $\Gamma-X$ direction (the $V_x$ valley), and another valley along the $\Gamma-Y$ direction ($V_y$ valley) \cite{rodin_prb_2016_sns}, and valley-selective optical excitation can be realized in these 2D materials using linearly polarized light \cite{rodin_prb_2016_sns,hanakata_prb_2016_sns_gese,xu_prb_2017,shen_2dmater_2018_gese}.

Figure~\ref{fig:figure11}(a) shows valleys located along the $\Gamma-X$ and $\Gamma-Y$ lines in the BZ. First principles calculations and symmetry analysis show that $x$-polarized photons have a much higher probability to excite carriers in the $V_x$ valley. Similarly, carriers in the $V_y$ valley can be readily excited by $y$-polarized light almost exclusively. In other words, a specific valley can be selectively excited by controlling the polarization of the incident light [this mechanism does not distinguish sign: for instance, both $-(\Gamma-X)$ and $+(\Gamma-X)$ are both ``the $V_x$ valley'']. As an alternative mechanism to produce valley polarization by means of time-reversal symmetry, an in-plane static electric makes carriers excited from the $V_x$ valley [either located along the $-(\Gamma-X)$ or $+(\Gamma-X)$ line] bend in opposite directions, generating a valley Hall effect illustrated in Fig.~\ref{fig:figure11}(b). A similar effect occurs when exciting the $V_y$ valley. Additional transport effects arising from non-linear electric fields will be discussed in Sec.~\ref{sec:bulkphotovoltaic}.

\subsection{Persistent spin helix behavior}\label{sec:secVIII-C}

\begin{figure}
\includegraphics[width=\linewidth]{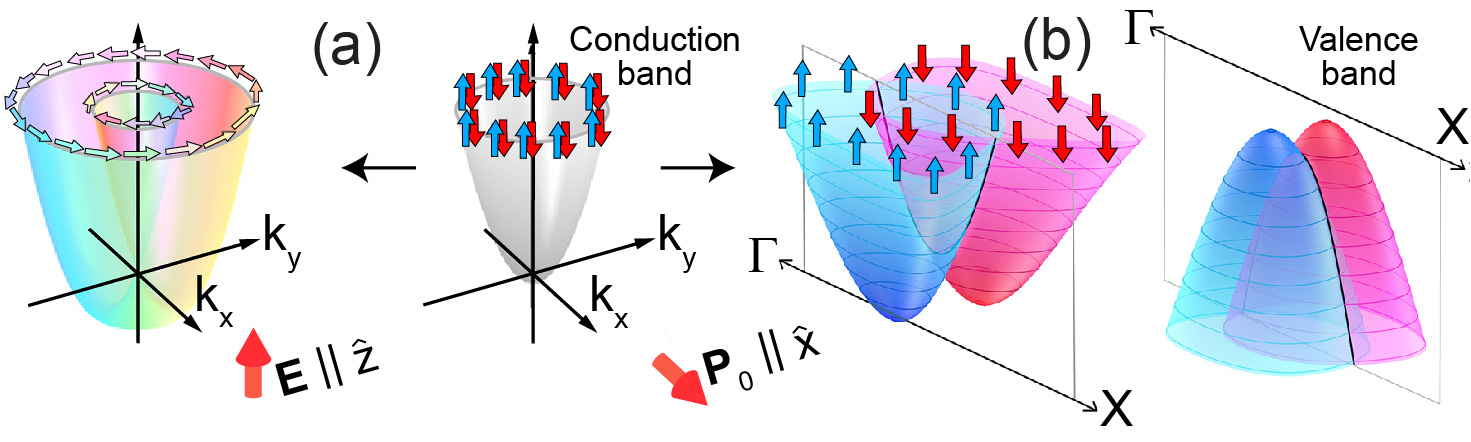}
\caption{\label{fig:figure12} (a) Conventional and (b) out-of-plane spin Rashba effects at the bottom of the conduction band. In subplot (b), the out-of-plane spin Rashba effect is realized in a SnTe ML, and both the valence and conduction bands are shown. Adapted from Ref.~\cite{HosikLee} with permission. Copyright, 2020. American Institute of Physics.}
\end{figure}

So far, we have considered the electronic properties of O$-MX$ MLs without concern for spin polarization. Spin-orbit coupling can create various types of spin splitting near the band edges, as well as spin Hall effects in O$-MX$ MLs \cite{GSHE}. {\em Zeeman-like} spin splitting is the prominent mechanism in TMDCs \cite{TMDC_valleytronics_2012}. On the other hand, a {\em Rashba-like} spin orbit coupling occurs due to the spin-orbit field $\overrightarrow{\Sigma}_{SOF}(\mathbf{k})=\alpha(\hat{\mathbf{P}}_0\times \mathbf{k})$, where $\alpha$ is the spin-orbit coupling strength, $\hat{\mathbf{P}}_0=\mathbf{P}_0/P_0$ is the direction of the intrinsic electric polarization in ferroelectrics, and $\mathbf{k}$ is the quasiparticle (electron or hole) crystal momentum [Fig.~\ref{fig:figure12}(a)].

As indicated in Sec.~\ref{sec:secII}, the out-of-plane component of $\mathbf{P}_0$ is quenched in ultrathin SnTe, making $\mathbf{P}_0=P_0\hat{x}$. Since 2D materials lack crystal momentum along the $z-$direction, $\overrightarrow{\Sigma}_{SOF}(\mathbf{k})\propto(\hat{x}\times \mathbf{k})$, with $\mathbf{k}=(k_x,k_y,0)$. Note that spin becomes degenerate along the $\Gamma-X$ ($k_y=0$) high symmetry line, that it points along the $z-$direction for $k_y\ne 0$, and it reverts direction when either $k_y$ or $\mathbf{P}_0$ change sign \cite{HosikLee}. Rotating $\mathbf{P}_0$ into the $y-$direction [see Fig.~\ref{fig:figure6}(d)] changes the orientation of the spin-split bands. The strength of the spin-orbit coupling increases with atomic number $Z$, as broadly reported for O$-MX$ MLs \cite{shi_nl_2015_snse_gese,rodin_prb_2016_sns,KaiPRL,band_splitting_2019} and other 2D ferroelectrics \cite{honeycomb_band_splitting_2015,bi_band_splitting_2018,te_band_splitting_2018,woc_band_splitting_2019}. Recalling that Pb-based $MX$ MLs lack an intrinsic polarization [$a_{1,0}=a_{2,0}$ in Fig.~\ref{fig:figure1}(d) and hence $\mathbf{P}_0=0$], the best immediate candidates for 2D ferroelectric Rashba semiconductors within $MX$ MLs are tellurides GeTe and SnTe.

Space group P$2_1$mn has the following symmetries: (i) the identity $E$; (ii) $\bar{C}_{2x}$: a two-fold rotation around the $x-$axis ($C_{2x}$), followed by a translation $\mathbf{\tau}=(\mathbf{a}_1+\mathbf{a}_2)/2$; (iii) a glide-reflection plane $\bar{M}_{xy}$: a reflection by the $xy$ plane followed by $\mathbf{\tau}$; and (iv) a reflection about the $xz$ plane ($M_{xz}$) \cite{rodin_prb_2016_sns}. Adding time-reversal symmetry as customarily defined $\hat{T}=i\sigma_yK$ (where $K$ represents complex conjugation) the following effective spin Hamiltonian applies at the top of the valence band and at the bottom of the conduction band \cite{HosikLee,absor}:
\begin{equation}\label{eq:hspin}
\hat{H}=\frac{\hbar^2}{2m^*}\left(k_x^2+k_y^2\right)+\left(\alpha k_y    +\alpha'k_x^2k_y+\alpha''k_y^3\right)\sigma_z,
\end{equation}
with $m^*$, $\alpha$, $\alpha'$, and $\alpha''$ to be fitted from band structure calculations [Fig.~\ref{fig:figure12}(b)]. This Hamiltonian does not have contributions from in-plane spin components up to third order in momentum, leading to a persistent spin Helix effect with a tunable {\em out-of-plane spin}. Eqn.~\eqref{eq:hspin} is similar to a Dresselhaus model for a bulk zinc blende crystal oriented along the [110] direction \cite{dress1}. Estimates of the spin-orbit coupling $\alpha$ in $MX$s are two to three orders of magnitude larger than those in III-V semiconductor quantum well structures \cite{HosikLee}, and their wavelength of the spin polarization $\lambda$ is smaller than that obtained for other Rashba semiconductors \cite{absor} which permits smaller lateral device dimensions. O$-MX$ MLs are a 2D platform for persistent spin helix dynamics \cite{Bernevig}.

\begin{table}[tb]
\caption{\label{ta:taSOCoperations} Transformation rules for $k_x$, $k_y$, $\sigma_x$, $\sigma_y$, and $\sigma_z$ under the symmetry operations of group P$2_1$mn. The point-group operations are defined as $\hat{C}_{2x}=i\sigma_x$, $\hat{M}_{xz}=i\sigma_y$, and $\hat{M}_{xy}=i\sigma_z$. Adapted from Ref.~\cite{absor} to reflect the choice of lattice vectors used in this work.}
\begin{tabular}{c|cc}
\hline
\hline
Symmetry operation                & $(k_x,k_y)$   & $(\sigma_x,\sigma_y,\sigma_z)$\\
\hline
$\hat{E}$                         & $(k_x,k_y)$   & $(\sigma_x,\sigma_y,\sigma_z)$\\
$\hat{C}_{2x}=i\sigma_x$          & $(k_x,-k_y)$  & $(\sigma_x,-\sigma_y,-\sigma_z)$\\
$\hat{M}_{xz}=i\sigma_y$          & $(k_x,-k_y)$  & $(-\sigma_x,\sigma_y,-\sigma_z)$\\
$\hat{M}_{xy}=i\sigma_z$          & $(k_x,k_y)$   & $(-\sigma_x,-\sigma_y,\sigma_z)$\\
\hline
$\hat{T}=i\sigma_yK$              & $(-k_x,-k_y)$ & $(-\sigma_x,-\sigma_y,-\sigma_z)$\\
\hline
\hline
\end{tabular}
\end{table}

SnTe ML spin transistors may be designed to have a channel length of $\lambda$/4 to be electrically switched in the ferroelectric channel or magnetically switched in the ferromagnetic drain \cite{HosikLee}. Another proposal is an all-in-one spin transistor based on the spin Hall effect, where the inverse spin Hall effect charge current is detuned by an out-of-plane electric field which [according to Fig.~\ref{fig:figure12}(a)] breaks the persistent spin helix state down and induces spin decoherence \cite{slawinska_2d_mat_2018_mmls}.

\section{Optical properties of O$-MX$ MLs}\label{sec:secIX}

\subsection{Optical absorption}
Optical absorption reflects the anisotropy of the electronic band structure: linearly polarized light with polarization parallel to the $x-$direction leads to a smaller absorption energy gap when contrasted with light whose polarization is parallel to the $y-$axis \cite{gomes} [this effect can be observed in Fig.~\ref{fig:figure16}, where GW corrections and Bethe-Salpeter electron-hole interactions have been added]. The symmetry imposed by the $P4/nmm$ structural transformation at $T\ge T_c$ should be reflected on a symmetric optical absorbance \cite{other2}. According to Shi and Kioupakis, the absorbance of O$-MX$ MLs is unusually strong in the visible range \cite{shi_nl_2015_snse_gese}.

\subsection{Raman spectra}
\begin{figure}
\includegraphics[width=\linewidth]{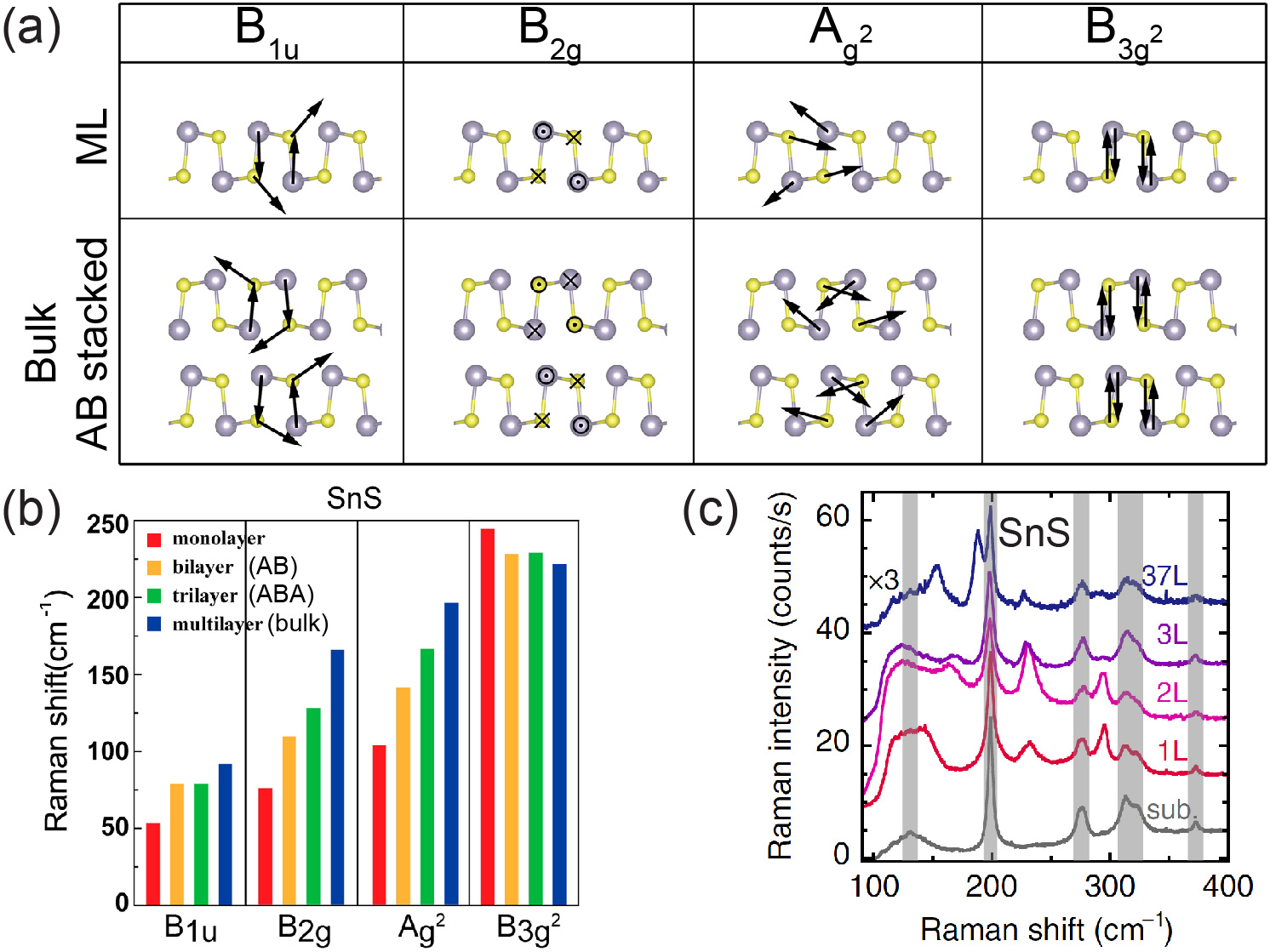}
\caption{\label{fig:figure13} (a) Relevant phonon modes for Raman spectra of ML and bulk SnS. (b) Raman shift of SnS as a function of MLs. Subplots (a) and (b) are adapted from Ref.~\cite{RamanSnS} with permission.  Copyright, 2020, Springer Nature. (c) Experimentally observed thickness dependence of Raman spectrum at 3 K. The peaks in the hatch are due to the substrate. From Ref.~\cite{Higashitarumizu20_NC_SnS} with permission. Copyright, 2020, Springer Nature.}
\end{figure}

Raman spectroscopy is employed to determine the thickness of layered materials \cite{advfunctmat,ph3}. As indicated in Sec.~\ref{sec:secII}, the atomic bonds evolve with the number of layers in O$-MX$ MLs \cite{ourarxiv,advmatZanolli}, which should leave signatures in the Raman spectra. Indeed, Raman modes $B_{1u}$, $B_{2g}$, $A_{g^2}$ and $B_{3g^2}$ are shown for monolayer and bulk SnS in Fig.~\ref{fig:figure13}(a), and shift as a function of the number of MLs is seen in Fig.~\ref{fig:figure13}(b) \cite{RamanSnS}. Experimentally-determined Raman signatures for ultrathin SnS are displayed in Fig.~\ref{fig:figure13}(c) for comparison \cite{Higashitarumizu20_NC_SnS}.

\subsection{Second harmonic generation}\label{sec:shg}

Within a semiclassical picture, the SHG originates from the non-sinusoidal motion of carriers inside crystals lacking inversion symmetry, leading to a quadratic effect in the electric field  in O-MX MLs that is forbidden in the bulk. SHG is widely utilized in applications ranging from table-top frequency multipliers, surface symmetry probes, and photon entanglement in quantum computing protocols, among others~\cite{Boyd2008}.

If the incoming electric field is homogeneous and monochromatic $E^a = E^{a}_{\omega}e^{-i\omega t}$ + c.c., the second-order polarization of the crystal oscillates at twice the driving frequency:
\begin{align}
 P^{a}_2= \sum_{bc}\chi^{abc}_2 (-2\omega;\omega,\omega) E^{b}_{\omega} E^{c}_{\omega} e^{-i 2\omega t} + \textrm{c.c.},
\label{eqn:chi_2}
\end{align}
where $\chi^{abc}_{2}(-2\omega;\omega,\omega)$ is the SHG response tensor, $a$ is the cartesian direction of the created electric field, and $b$ and $c$ are the cartesian directions of the incident electric fields. Far away from the source, the irradiated field is given by $\v{E} \sim \frac{d^2\v{P}_2}{dt^2}$ \cite{Jackson1998}.

SHG has been reported for ultrathin noncentrosymmetric samples of MoS$_2$ and h-BN with odd layer thicknesses \cite{Li2013}. The angular dependence of SHG also reveals the rotational symmetry of the crystal lattice, and can therefore be used to determine the orientation of crystallographic axes \cite{Kumar2013,Li2013,Kim2013,Janisch2014,Malard2013,Zhou2015,Attaccalite2015}. This effect has been theoretically \cite{b4,wang_nanolett_2017_gese} and experimentally \cite{Higashitarumizu20_NC_SnS} studied in O$-MX$ MLs, too.

\begin{figure}
\includegraphics[width=\linewidth]{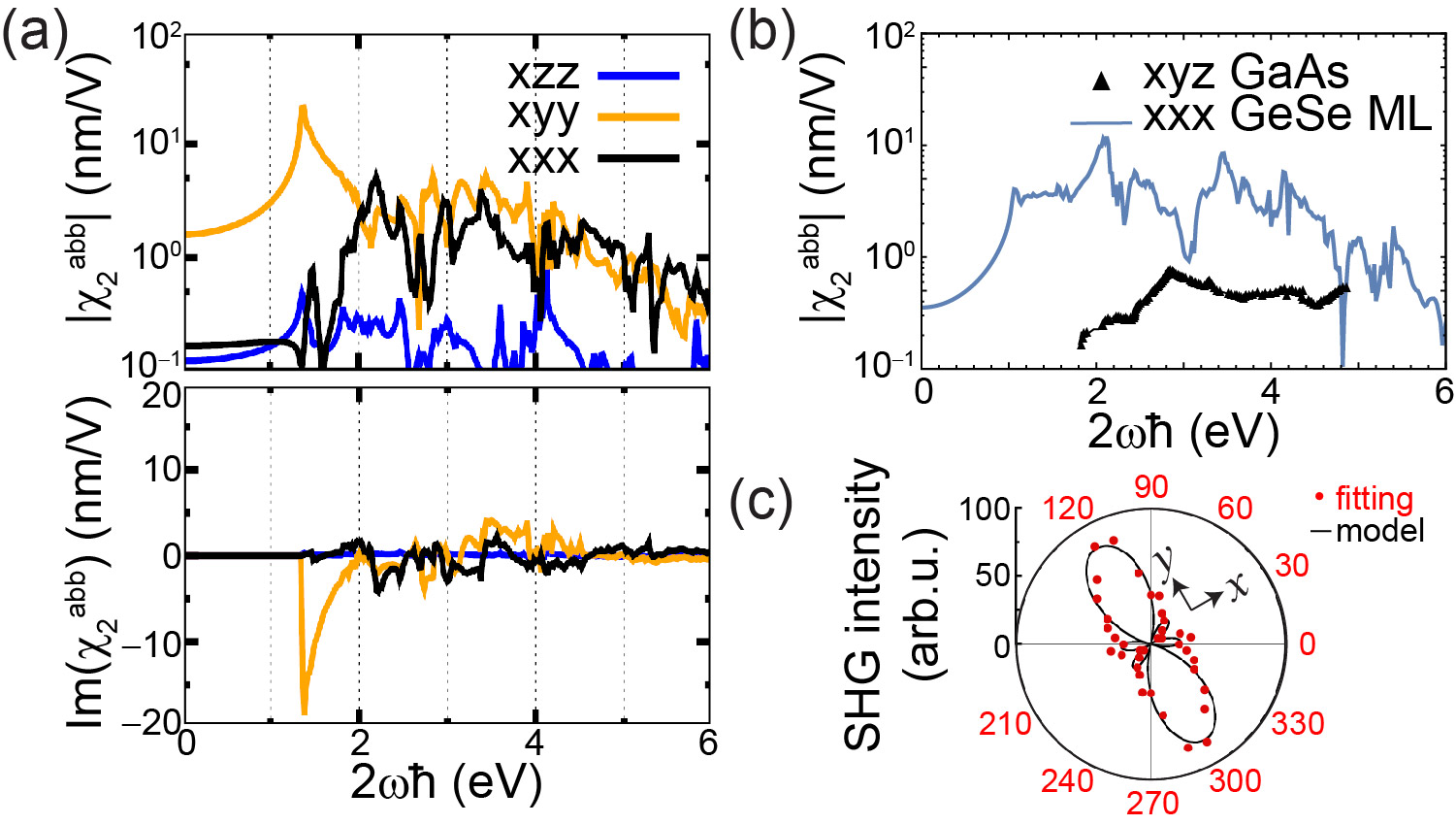}
\caption{\label{fig:figure14} (a) Absolute and imaginary SHG tensor $\chi_2^{abb}(-2\omega,\omega,\omega)$ for a SnS ML as a function of the outgoing phonon frequency $2\omega$. (b) Comparison between the experimental SHG tensor for GaAs(001) from \cite{Bergfeld2003} and the computed one for a GeSe ML, which is an order of magnitude larger. Subplots (a) and (b) are adapted from Ref.~\cite{b4} with permission. Copyright, 2017, Institute of Physics. (c) Experimental polarization dependence of SHG at 425 nm and room temperature for a SnS ML under perpendicular polarization. The inset axes show the $x-$ and $y-$directions corresponding to these defined in Fig.~\ref{fig:figure1}(c). Reproduced from Ref.~\cite{Higashitarumizu20_NC_SnS} with permission. Copyright, 2020, Springer Nature.}
\end{figure}

Following the choice of axes in Fig.~\ref{fig:figure1}(c), the O$-MX$ ML defines the $xy$-plane and the polar axis (the direction of $\mathbf{P}_0$) lies along the positive $x$-direction. Its point group only allows for non-zero $xzz$, $xyy$, $xxx$, $yyx$, $zxz$ components of $\chi^{abc}_2$ (plus the components obtained by exchanging of the last two indices, $\chi^{abc}_2=\chi^{acb}_2$). As exemplified for a SnS ML in Fig.~\ref{fig:figure14}(a), the SHG spectrum displays peak values within the visible spectrum that  can be an order of magnitude larger than those reported in GaAs ~\cite{Bergfeld2003} [Fig.~\ref{fig:figure14}(b)] or a MoS$_2$ ML \cite{Malard2013,wang_nanolett_2017_gese}. The SHG spectrum is anisotropic, and $|\chi^{xyy}|> |\chi^{xxx}| > |\chi^{xzz}|$ holds approximately true for all frequencies ~\cite{b4}. This is a counterintuitive result, as the maximum response along the polar axis occurs for incident optical fields that are perpendicularly polarized to the polar axis.

The role of the spontaneous polarization $\mathbf{P}_0$ in the large SHG response tensor and in other nonlinear responses is an active area of investigation. The large magnitude of the SHG in O$-MX$ MLs seems to be a combination of many factors, including their reduced dimensionality and in-plane polarization. Indirect evidence suggests that the in-plane $\mathbf{P}_0$ enhances the SHG by establishing mirror symmetries that strongly constrain contributions from certain regions within the BZ \cite{Panday2019}.

O$-MX$ MLs grown on insulating substrates permit performing optical experiments, and Fig.~\ref{fig:figure14}(c) is an experimental demonstration of the anisotropic behavior of the SHG of a SnS ML \cite{wang_nanolett_2017_gese} at room temperature, using an 850-nm laser as the excitation source \cite{Higashitarumizu20_NC_SnS}. The largest SHG occurs along the $y-$axis. As discussed in Sec.~\ref{sec:secIV}, the sense of direction of $\mathbf{P}$ can be set by a combination of SHG and transport measurements.

\subsection{Bulk photovoltaic effects: injection and shift currents}\label{sec:bulkphotovoltaic}
The bulk photovoltaic effect (BPVE) is the generation of a $dc$ current upon illumination in materials that lack inversion symmetry. It has been extensively studied in bulk ferroelectrics \cite{Sturman1992,ivchenko}, topological insulators \cite{Hosur2011}, 2D ferroelectrics \cite{Rangel2017,Panday2019,Kushnir2017,Kushnir2019}, Weyl semimetals \cite{Chan2017,Juan2017,Rees,Shvetsov2019}, BN nanotubes \cite{Kral2000}, among other materials. Many seemingly unrelated BPVEs have been shown to have a common origin \cite{Sipe2000,Fregoso2019}. The BPVE is much larger in 2D ferroelectrics than  in bulk ferroelectrics, potentially overcoming the low solar energy efficiency conversion found in the latter \cite{Rappe7191,rappe2,rappe3}.

\begin{figure}
\includegraphics[width=\linewidth]{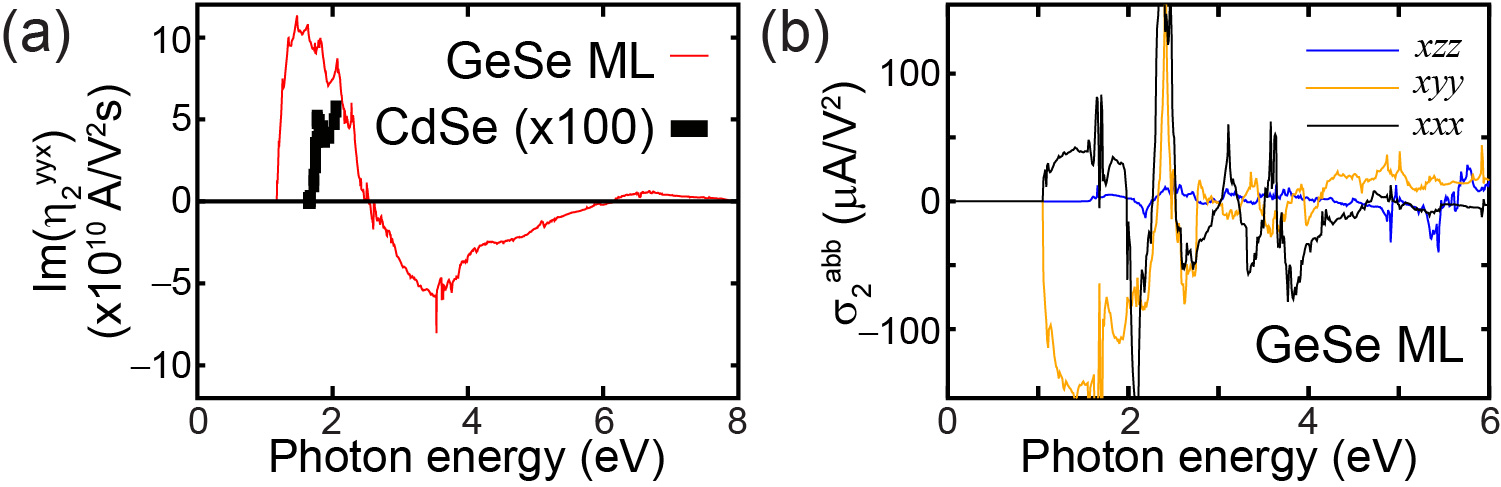}
\caption{\label{fig:figure15} (a) Injection current and (b) shift current for a GeSe ML [with axes defined as in Fig.~\ref{fig:figure1}(c)]. The injection current of CdSe is also plotted for comparison. Subplot (a) is adapted from Ref.~\cite{Panday2019}. Copyright, 2019, American Physical Society. Subplot (b) is taken from Ref.~\cite{Rangel2017}. Copyright, 2017, American Physical Society.}
\end{figure}

The BPVE differs from other photovoltaic effects in three important ways: (i) it is proportional to the intensity of the optical field; (ii) it produces large open-circuit photovoltages, {\em i.e.}, larger than the energy band gap; and (iii) it depends on the polarization state of light. These characteristics imply, respectively, (i) that the BPVE is a second order effect in the optical field, (ii) that it is an ultrafast phenomena occurring before thermalization takes place at the CBM (VBM), and (iii) that the BPVE response tensor has a real and an imaginary component. The real component ($\sigma_2$) determines the response to linearly polarized light; the imaginary component ($\eta_2$) is the response to circularly polarized light.  Denoting  an homogeneous optical field by $\v{E}$, the BPVE can then be schematically written  as \cite{Sturman1992}:
\begin{align}
\v{J}_{bpve} = \eta_2 \v{E}\times \v{E}^{*} + \sigma_2 \v{E}^2.
\label{eq:Jbpve}
\end{align}
The first term is the so-called \textit{ballistic} current, \textit{injection} current, or \textit{circular photogalvanic effect}, and it vanishes for linear polarization. The injection current is created by an unequal momentum relaxation into time reversal states \cite{Sturman1992,ivchenko} or by unequal carrier pumping rates into time-reversed states $\pm \v{k}$ \cite{Sipe2000,Fregoso2019}. In addition, and related to spin effects discussed in Sec.~\ref{sec:secVIII-C}, the chirality of circularly-polarized light couples to the spin of charge carriers to generate a spin current in spin-orbit coupled systems \cite{Sturman1992,Hosur2011,Chan2017}. Fig.~\ref{fig:figure15}(a) shows the spectrum of the injection current tensor for a GeSe ML. The only non-zero component is $\eta_2^{yyx}$ and, as a consequence, injection current can only flow perpendicularly to the polar ($x$) axis. The injection current tensor $\eta_2$ reaches peak values of $10^{11}$ A/V$^{2}$s in the visible spectrum ($1.5-3$ eV) \cite{Panday2019}, which is many orders of magnitude larger than its peak value in MoS$_2$ MLs \cite{Arzate2016}.

The second term in Eqn.~\eqref{eq:Jbpve} is the \textit{shift} current, also known as the \textit{linear photogalvanic effect}, and its microscopic interpretation is still under debate. A popular interpretation is that the shift current arises from a shift of the electron in real-space when it absorbs a photon \cite{Baltz1981}. This is reasonable since the Wannier centers of charge  are spatially separated in materials that break inversion symmetry. In a second interpretation, the quantum coherent motion of a pair of dipoles moving in $k-$space originates shift currents~\cite{Fregoso2019}, which vanish for incident circularly polarized light. Fig.~\ref{fig:figure15}(b) shows the shift-current spectra for a GeSe ML. There is a broad maximum of the order of 150 $\mu$A/V$^2$ \cite{Rangel2017} in the visible range ($1 - 3$ eV) that is larger than its magnitude in prototypical materials [{\em e.g.}, $\sigma_2\sim 0.1$~$\mu$A/V$^2$ in \textrm{BiFeO}$_3$~\cite{Young2012}]. These results demonstrate the unique potential of O$-MX$ MLs for optoelectronic applications.

\begin{figure}
\includegraphics[width=\linewidth]{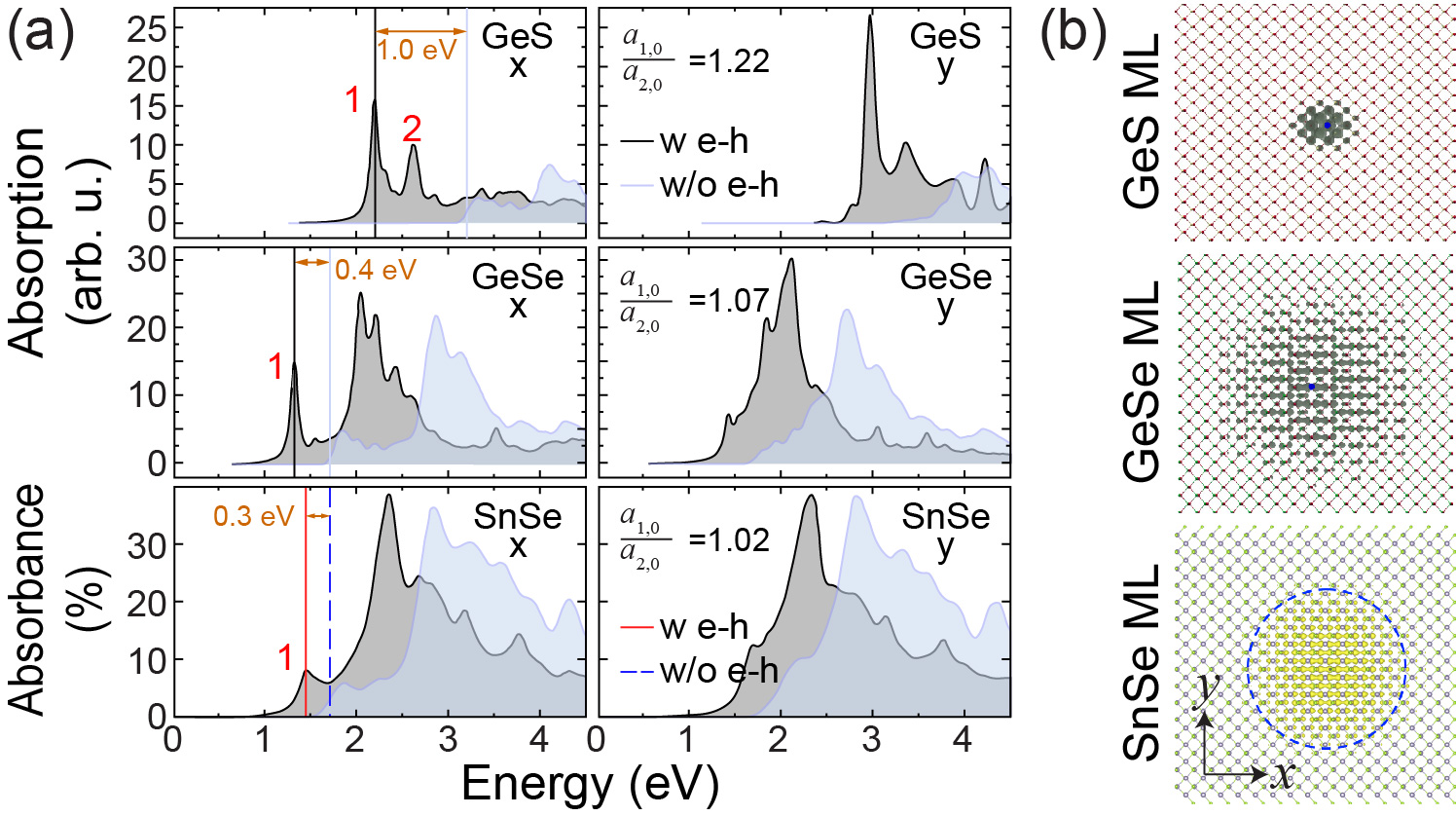}
\caption{\label{fig:figure16} (a) Absorption spectra of GeS, GeSe, and SnSe MLs with (w $e-h$) and without (w/o $e-h$) electron-hole interactions for light polarized along the $x-$direction ({\em i.e.}, parallel to $\mathbf{P}_0$ or along the $y-$direction (perpendicular to $\mathbf{P}_0$. Two peaks are identified for GeS MLs: peak 1 arises from a direct transition at the $\Gamma-$point, and peak 2 from a direct transition at the $V_x$ valley (see Figure \ref{fig:figure9}). For both GeSe and SnSe MLs, the only excitation within the band gap corresponds to a direct transition at the $V_x$ valley. (b)
Subplots for GeS and GeSe MLs were adapted from Ref.~\cite{gomes_prb_2016_excitons}. Copyright, 2016, American Physical Society. The subplots for the SnSe ML were adapted from Ref.~\cite{shi_nl_2015_snse_gese}. Copyright, 2015, American Chemical Society.}
\end{figure}

\subsection{Photostriction}

Photostriction is the structural change induced by a screened electric polarization $\mathbf{P}$ resulting from photoexcited electronic states: optical excitations will lead to a concurrent compression of lattice vector $\mathbf{a}_1$ and a comparatively smaller increase of $\mathbf{a}_2$ for an overall reduction in the unit cell area. The structural change documented for SnS and SnSe MLs is 10 times larger than that observed in bulk ferroelectric BiFeO$_3$, making O$-MX$ MLs an ultimate platform for this effect \cite{haleoot_prl_2017_sns_snse}.

\subsection{Excitons}

An exciton is an electron-hole pair hosted within a material, whose description therefore goes beyond the single-particle picture employed thus far. Excitons display a strong dependency on dimensionality, being more strongly bound in low-dimensional systems due to a reduced Coulomb screening, and hence relevant at room temperature \cite{gomes_prb_2016_excitons}.  Upon laser irradiation, a MoS$_2$ ML displays isotropic excitons with a binding energy of 0.55 eV on a graphitic substrate \cite{splendiani,Ugeda}. In turn, anisotropic excitons in BP MLs have been shown to have a larger binding energy of 0.8-0.9 eV \cite{BPexcitons,Tran,rod}. Possible optoelectronic applications of O$-MX$ MLs also call for a deep understanding of excitons \cite{gomes_prb_2016_excitons}. Given that they contribute to the dielectric environment, substrates in which 2D materials are placed may need to be accounted for when comparing experiment and theory.

The exciton binding energy has been calculated for freestanding GeS, GeSe, and SnSe MLs. It increases from less than 0.01 eV in the bulk to 1.00 eV on a GeS ML, 0.32-0.40 eV on a GeSe ML, and 0.27-0.30 eV in a SnSe ML. As seen in Fig.~\ref{fig:figure16}(a), the binding energy is larger in materials with small $\bar{Z}$, for which the absorbance is more anisotropic when shining light with polarization along the $x-$ or $y-$ directions \cite{shi_nl_2015_snse_gese,gomes_prb_2016_excitons}. Concomitant with a stronger binding and anisotropic absorbance, lighter O$-MX$ MLs host more localized and anisotropic excitons [see Fig.~\ref{fig:figure16}(b)]. An analytical, Mott-Wannier model has been employed to account for the effect of the supporting substrate---which in general lowers the exciton binding energy---in Ref.~\cite{gomes_prb_2016_excitons}.

\section{Summary and outlook}\label{sec:secX}

Two-dimensional and ultrathin ferroelectrics are gaining increased attention. They complement two-dimensional semimetal graphene, insulator hexagonal boron nitride, a large number of 2D semiconductors, and two-dimensional magnets. This Colloquium describes the structural, mechanical, electronic, and optical properties of group-IV monochalcogenide MLs in a comprehensive manner, including recent developments such as the experimental realization of SnS and SnSe MLs, and novel theoretical results such as spin helix behavior, theoretical Raman spectra, and bulk photovoltaic effects, to become the most up-to-date reference on these materials. While there are a number of challenges still to be resolved concerning chemical stability, exfoliation or growth, and their stacking into functional layered materials, these ultrathin ferroelectric and ferroelastic materials have already diversified and enriched the library of layered and two-dimensional functional materials. Their prospective use in memory, valley, and optoelectronic applications can provide the motivation and justification to drive further progress in this area.

\acknowledgments

We thank P. Kumar, L. Bellaiche, J.~E. Moore, L. V. Titova, T. Rangel and L. Fu. S.B.-L. and J.W.V. acknowledge funding from the US Department of Energy, Office of Basic Energy Sciences (Early Career Award DE-SC0016139) and DOE-NERSC contract No. DE-AC02-05CH11231. S.S.P.P. and K.C. were supported by Deutsche Forschungsgemeinschaft (DFG, German Research Foundation), Project number PA 1812/2-1.


\begin{thebibliography}{205}%
\makeatletter
\providecommand \@ifxundefined [1]{%
 \@ifx{#1\undefined}
}%
\providecommand \@ifnum [1]{%
 \ifnum #1\expandafter \@firstoftwo
 \else \expandafter \@secondoftwo
 \fi
}%
\providecommand \@ifx [1]{%
 \ifx #1\expandafter \@firstoftwo
 \else \expandafter \@secondoftwo
 \fi
}%
\providecommand \natexlab [1]{#1}%
\providecommand \enquote  [1]{``#1''}%
\providecommand \bibnamefont  [1]{#1}%
\providecommand \bibfnamefont [1]{#1}%
\providecommand \citenamefont [1]{#1}%
\providecommand \href@noop [0]{\@secondoftwo}%
\providecommand \href [0]{\begingroup \@sanitize@url \@href}%
\providecommand \@href[1]{\@@startlink{#1}\@@href}%
\providecommand \@@href[1]{\endgroup#1\@@endlink}%
\providecommand \@sanitize@url [0]{\catcode `\\12\catcode `\$12\catcode
  `\&12\catcode `\#12\catcode `\^12\catcode `\_12\catcode `\%12\relax}%
\providecommand \@@startlink[1]{}%
\providecommand \@@endlink[0]{}%
\providecommand \url  [0]{\begingroup\@sanitize@url \@url }%
\providecommand \@url [1]{\endgroup\@href {#1}{\urlprefix }}%
\providecommand \urlprefix  [0]{URL }%
\providecommand \Eprint [0]{\href }%
\providecommand \doibase [0]{http://dx.doi.org/}%
\providecommand \selectlanguage [0]{\@gobble}%
\providecommand \bibinfo  [0]{\@secondoftwo}%
\providecommand \bibfield  [0]{\@secondoftwo}%
\providecommand \translation [1]{[#1]}%
\providecommand \BibitemOpen [0]{}%
\providecommand \bibitemStop [0]{}%
\providecommand \bibitemNoStop [0]{.\EOS\space}%
\providecommand \EOS [0]{\spacefactor3000\relax}%
\providecommand \BibitemShut  [1]{\csname bibitem#1\endcsname}%
\let\auto@bib@innerbib\@empty
\bibitem [{\citenamefont {Absor}\ and\ \citenamefont {Ishii}(2019)}]{absor}%
  \BibitemOpen
  \bibfield  {author} {\bibinfo {author} {\bibnamefont {Absor}, \bibfnamefont
  {M.~A.~U.}}, \ and\ \bibinfo {author} {\bibfnamefont {F.}~\bibnamefont
  {Ishii}}} (\bibinfo {year} {2019}),\ \href {\doibase
  10.1103/PhysRevB.100.115104} {\bibfield  {journal} {\bibinfo  {journal}
  {Phys. Rev. B}\ }\textbf {\bibinfo {volume} {100}},\ \bibinfo {pages}
  {115104}}\BibitemShut {NoStop}%
\bibitem [{\citenamefont {Ai}\ \emph {et~al.}(2019)\citenamefont {Ai},
  \citenamefont {Ma}, \citenamefont {Shao}, \citenamefont {Li},\ and\
  \citenamefont {Zhao}}]{woc_band_splitting_2019}%
  \BibitemOpen
  \bibfield  {author} {\bibinfo {author} {\bibnamefont {Ai}, \bibfnamefont
  {H.}}, \bibinfo {author} {\bibfnamefont {X.}~\bibnamefont {Ma}}, \bibinfo
  {author} {\bibfnamefont {X.}~\bibnamefont {Shao}}, \bibinfo {author}
  {\bibfnamefont {W.}~\bibnamefont {Li}}, \ and\ \bibinfo {author}
  {\bibfnamefont {M.}~\bibnamefont {Zhao}}} (\bibinfo {year} {2019}),\ \href
  {\doibase 10.1103/PhysRevMaterials.3.054407} {\bibfield  {journal} {\bibinfo
  {journal} {Phys. Rev. Mater.}\ }\textbf {\bibinfo {volume} {3}},\ \bibinfo
  {pages} {054407}}\BibitemShut {NoStop}%
\bibitem [{\citenamefont {Almahmoud}\ \emph {et~al.}(2010)\citenamefont
  {Almahmoud}, \citenamefont {Kornev},\ and\ \citenamefont
  {Bellaiche}}]{Almahmoud_2010}%
  \BibitemOpen
  \bibfield  {author} {\bibinfo {author} {\bibnamefont {Almahmoud},
  \bibfnamefont {E.}}, \bibinfo {author} {\bibfnamefont {I.}~\bibnamefont
  {Kornev}}, \ and\ \bibinfo {author} {\bibfnamefont {L.}~\bibnamefont
  {Bellaiche}}} (\bibinfo {year} {2010}),\ \href {\doibase
  10.1103/PhysRevB.81.064105} {\bibfield  {journal} {\bibinfo  {journal} {Phys.
  Rev. B}\ }\textbf {\bibinfo {volume} {81}},\ \bibinfo {pages}
  {064105}}\BibitemShut {NoStop}%
\bibitem [{\citenamefont {Almahmoud}\ \emph {et~al.}(2004)\citenamefont
  {Almahmoud}, \citenamefont {Navtsenya}, \citenamefont {Kornev}, \citenamefont
  {Fu},\ and\ \citenamefont {Bellaiche}}]{Almahmoud_2004}%
  \BibitemOpen
  \bibfield  {author} {\bibinfo {author} {\bibnamefont {Almahmoud},
  \bibfnamefont {E.}}, \bibinfo {author} {\bibfnamefont {Y.}~\bibnamefont
  {Navtsenya}}, \bibinfo {author} {\bibfnamefont {I.}~\bibnamefont {Kornev}},
  \bibinfo {author} {\bibfnamefont {H.}~\bibnamefont {Fu}}, \ and\ \bibinfo
  {author} {\bibfnamefont {L.}~\bibnamefont {Bellaiche}}} (\bibinfo {year}
  {2004}),\ \href {\doibase 10.1103/PhysRevB.70.220102} {\bibfield  {journal}
  {\bibinfo  {journal} {Phys. Rev. B}\ }\textbf {\bibinfo {volume} {70}},\
  \bibinfo {pages} {220102}}\BibitemShut {NoStop}%
\bibitem [{\citenamefont {Amorim}\ \emph {et~al.}(2016)\citenamefont {Amorim},
  \citenamefont {Cortijo}, \citenamefont {{de Juan}}, \citenamefont {Grushin},
  \citenamefont {Guinea}, \citenamefont {Gutiérrez-Rubio}, \citenamefont
  {Ochoa}, \citenamefont {Parente}, \citenamefont {Roldán}, \citenamefont
  {San-Jose}, \citenamefont {Schiefele}, \citenamefont {Sturla},\ and\
  \citenamefont {Vozmediano}}]{maria}%
  \BibitemOpen
  \bibfield  {author} {\bibinfo {author} {\bibnamefont {Amorim}, \bibfnamefont
  {B.}}, \bibinfo {author} {\bibfnamefont {A.}~\bibnamefont {Cortijo}},
  \bibinfo {author} {\bibfnamefont {F.}~\bibnamefont {{de Juan}}}, \bibinfo
  {author} {\bibfnamefont {A.}~\bibnamefont {Grushin}}, \bibinfo {author}
  {\bibfnamefont {F.}~\bibnamefont {Guinea}}, \bibinfo {author} {\bibfnamefont
  {A.}~\bibnamefont {Gutiérrez-Rubio}}, \bibinfo {author} {\bibfnamefont
  {H.}~\bibnamefont {Ochoa}}, \bibinfo {author} {\bibfnamefont
  {V.}~\bibnamefont {Parente}}, \bibinfo {author} {\bibfnamefont
  {R.}~\bibnamefont {Roldán}}, \bibinfo {author} {\bibfnamefont
  {P.}~\bibnamefont {San-Jose}}, \bibinfo {author} {\bibfnamefont
  {J.}~\bibnamefont {Schiefele}}, \bibinfo {author} {\bibfnamefont
  {M.}~\bibnamefont {Sturla}}, \ and\ \bibinfo {author} {\bibfnamefont
  {M.}~\bibnamefont {Vozmediano}}} (\bibinfo {year} {2016}),\ \href {\doibase
  10.1016/j.physrep.2015.12.006} {\bibfield  {journal} {\bibinfo  {journal}
  {Phys. Rep.}\ }\textbf {\bibinfo {volume} {617}},\ \bibinfo {pages}
  {1}}\BibitemShut {NoStop}%
\bibitem [{\citenamefont {Arzate}\ \emph {et~al.}(2016)\citenamefont {Arzate},
  \citenamefont {Mendoza}, \citenamefont {V\'azquez-Nava}, \citenamefont
  {Ibarra-Borja},\ and\ \citenamefont {\'Alvarez-N\'u\~nez}}]{Arzate2016}%
  \BibitemOpen
  \bibfield  {author} {\bibinfo {author} {\bibnamefont {Arzate}, \bibfnamefont
  {N.}}, \bibinfo {author} {\bibfnamefont {B.~S.}\ \bibnamefont {Mendoza}},
  \bibinfo {author} {\bibfnamefont {R.~A.}\ \bibnamefont {V\'azquez-Nava}},
  \bibinfo {author} {\bibfnamefont {Z.}~\bibnamefont {Ibarra-Borja}}, \ and\
  \bibinfo {author} {\bibfnamefont {M.~I.}\ \bibnamefont
  {\'Alvarez-N\'u\~nez}}} (\bibinfo {year} {2016}),\ \href {\doibase
  10.1103/PhysRevB.93.115433} {\bibfield  {journal} {\bibinfo  {journal} {Phys.
  Rev. B}\ }\textbf {\bibinfo {volume} {93}},\ \bibinfo {pages}
  {115433}}\BibitemShut {NoStop}%
\bibitem [{\citenamefont {Asaba}\ \emph {et~al.}(2018)\citenamefont {Asaba},
  \citenamefont {Wang}, \citenamefont {Li}, \citenamefont {Xiang},
  \citenamefont {Tinsman}, \citenamefont {Chen}, \citenamefont {Zhou},
  \citenamefont {Zhao}, \citenamefont {Laleyan}, \citenamefont {Li},
  \citenamefont {Mi},\ and\ \citenamefont {Li}}]{asaba_sp_2018}%
  \BibitemOpen
  \bibfield  {author} {\bibinfo {author} {\bibnamefont {Asaba}, \bibfnamefont
  {T.}}, \bibinfo {author} {\bibfnamefont {Y.}~\bibnamefont {Wang}}, \bibinfo
  {author} {\bibfnamefont {G.}~\bibnamefont {Li}}, \bibinfo {author}
  {\bibfnamefont {Z.}~\bibnamefont {Xiang}}, \bibinfo {author} {\bibfnamefont
  {C.}~\bibnamefont {Tinsman}}, \bibinfo {author} {\bibfnamefont
  {L.}~\bibnamefont {Chen}}, \bibinfo {author} {\bibfnamefont {S.}~\bibnamefont
  {Zhou}}, \bibinfo {author} {\bibfnamefont {S.}~\bibnamefont {Zhao}}, \bibinfo
  {author} {\bibfnamefont {D.}~\bibnamefont {Laleyan}}, \bibinfo {author}
  {\bibfnamefont {Y.}~\bibnamefont {Li}}, \bibinfo {author} {\bibfnamefont
  {Z.}~\bibnamefont {Mi}}, \ and\ \bibinfo {author} {\bibfnamefont
  {L.}~\bibnamefont {Li}}} (\bibinfo {year} {2018}),\ \href {\doibase
  10.1038/s41598-018-24736-x} {\bibfield  {journal} {\bibinfo  {journal} {Sci.
  Rep.}\ }\textbf {\bibinfo {volume} {8}},\ \bibinfo {pages}
  {6520}}\BibitemShut {NoStop}%
\bibitem [{\citenamefont {Attaccalite}\ \emph {et~al.}(2015)\citenamefont
  {Attaccalite}, \citenamefont {Nguer}, \citenamefont {Cannuccia},\ and\
  \citenamefont {Gr{\"u}ning}}]{Attaccalite2015}%
  \BibitemOpen
  \bibfield  {author} {\bibinfo {author} {\bibnamefont {Attaccalite},
  \bibfnamefont {C.}}, \bibinfo {author} {\bibfnamefont {A.}~\bibnamefont
  {Nguer}}, \bibinfo {author} {\bibfnamefont {E.}~\bibnamefont {Cannuccia}}, \
  and\ \bibinfo {author} {\bibfnamefont {M.}~\bibnamefont {Gr{\"u}ning}}}
  (\bibinfo {year} {2015}),\ \href {\doibase 10.1039/C5CP00601E} {\bibfield
  {journal} {\bibinfo  {journal} {Phys. Chem. Chem. Phys.}\ }\textbf {\bibinfo
  {volume} {17}},\ \bibinfo {pages} {9533}}\BibitemShut {NoStop}%
\bibitem [{\citenamefont {von Baltz}\ and\ \citenamefont
  {Kraut}(1981)}]{Baltz1981}%
  \BibitemOpen
  \bibfield  {author} {\bibinfo {author} {\bibnamefont {von Baltz},
  \bibfnamefont {R.}}, \ and\ \bibinfo {author} {\bibfnamefont
  {W.}~\bibnamefont {Kraut}}} (\bibinfo {year} {1981}),\ \href {\doibase
  10.1103/PhysRevB.23.5590} {\bibfield  {journal} {\bibinfo  {journal} {Phys.
  Rev. B}\ }\textbf {\bibinfo {volume} {23}},\ \bibinfo {pages}
  {5590}}\BibitemShut {NoStop}%
\bibitem [{\citenamefont {Bao}\ \emph {et~al.}(2019)\citenamefont {Bao},
  \citenamefont {Song}, \citenamefont {Liu}, \citenamefont {Chen},
  \citenamefont {Zhu}, \citenamefont {Abdelwahab}, \citenamefont {Su},
  \citenamefont {Fu}, \citenamefont {Chi}, \citenamefont {Yu}, \citenamefont
  {Liu}, \citenamefont {Zhao}, \citenamefont {Xu}, \citenamefont {Yang},\ and\
  \citenamefont {Loh}}]{Bao19_NL_SnS}%
  \BibitemOpen
  \bibfield  {author} {\bibinfo {author} {\bibnamefont {Bao}, \bibfnamefont
  {Y.}}, \bibinfo {author} {\bibfnamefont {P.}~\bibnamefont {Song}}, \bibinfo
  {author} {\bibfnamefont {Y.}~\bibnamefont {Liu}}, \bibinfo {author}
  {\bibfnamefont {Z.}~\bibnamefont {Chen}}, \bibinfo {author} {\bibfnamefont
  {M.}~\bibnamefont {Zhu}}, \bibinfo {author} {\bibfnamefont {I.}~\bibnamefont
  {Abdelwahab}}, \bibinfo {author} {\bibfnamefont {J.}~\bibnamefont {Su}},
  \bibinfo {author} {\bibfnamefont {W.}~\bibnamefont {Fu}}, \bibinfo {author}
  {\bibfnamefont {X.}~\bibnamefont {Chi}}, \bibinfo {author} {\bibfnamefont
  {W.}~\bibnamefont {Yu}}, \bibinfo {author} {\bibfnamefont {W.}~\bibnamefont
  {Liu}}, \bibinfo {author} {\bibfnamefont {X.}~\bibnamefont {Zhao}}, \bibinfo
  {author} {\bibfnamefont {Q.-H.}\ \bibnamefont {Xu}}, \bibinfo {author}
  {\bibfnamefont {M.}~\bibnamefont {Yang}}, \ and\ \bibinfo {author}
  {\bibfnamefont {K.~P.}\ \bibnamefont {Loh}}} (\bibinfo {year} {2019}),\ \href
  {\doibase 10.1021/acs.nanolett.9b01419} {\bibfield  {journal} {\bibinfo
  {journal} {Nano Lett.}\ }\textbf {\bibinfo {volume} {19}},\ \bibinfo {pages}
  {5109}}\BibitemShut {NoStop}%
\bibitem [{\citenamefont {Barraza-Lopez}\ \emph {et~al.}(2018)\citenamefont
  {Barraza-Lopez}, \citenamefont {Kaloni}, \citenamefont {Poudel},\ and\
  \citenamefont {Kumar}}]{other4}%
  \BibitemOpen
  \bibfield  {author} {\bibinfo {author} {\bibnamefont {Barraza-Lopez},
  \bibfnamefont {S.}}, \bibinfo {author} {\bibfnamefont {T.~P.}\ \bibnamefont
  {Kaloni}}, \bibinfo {author} {\bibfnamefont {S.~P.}\ \bibnamefont {Poudel}},
  \ and\ \bibinfo {author} {\bibfnamefont {P.}~\bibnamefont {Kumar}}} (\bibinfo
  {year} {2018}),\ \href {\doibase 10.1103/PhysRevB.97.024110} {\bibfield
  {journal} {\bibinfo  {journal} {Phys. Rev. B}\ }\textbf {\bibinfo {volume}
  {97}},\ \bibinfo {pages} {024110}}\BibitemShut {NoStop}%
\bibitem [{\citenamefont {Bergfeld}\ and\ \citenamefont
  {Daum}(2003)}]{Bergfeld2003}%
  \BibitemOpen
  \bibfield  {author} {\bibinfo {author} {\bibnamefont {Bergfeld},
  \bibfnamefont {S.}}, \ and\ \bibinfo {author} {\bibfnamefont
  {W.}~\bibnamefont {Daum}}} (\bibinfo {year} {2003}),\ \href {\doibase
  10.1103/PhysRevLett.90.036801} {\bibfield  {journal} {\bibinfo  {journal}
  {Phys. Rev. Lett.}\ }\textbf {\bibinfo {volume} {90}},\ \bibinfo {pages}
  {036801}}\BibitemShut {NoStop}%
\bibitem [{\citenamefont {Berland}\ \emph {et~al.}(2015)\citenamefont
  {Berland}, \citenamefont {Cooper}, \citenamefont {Lee}, \citenamefont
  {Schr{\"o}der}, \citenamefont {Thonhauser}, \citenamefont {Hyldgaard},\ and\
  \citenamefont {Lundqvist}}]{reviewvdw}%
  \BibitemOpen
  \bibfield  {author} {\bibinfo {author} {\bibnamefont {Berland}, \bibfnamefont
  {K.}}, \bibinfo {author} {\bibfnamefont {V.~R.}\ \bibnamefont {Cooper}},
  \bibinfo {author} {\bibfnamefont {K.}~\bibnamefont {Lee}}, \bibinfo {author}
  {\bibfnamefont {E.}~\bibnamefont {Schr{\"o}der}}, \bibinfo {author}
  {\bibfnamefont {T.}~\bibnamefont {Thonhauser}}, \bibinfo {author}
  {\bibfnamefont {P.}~\bibnamefont {Hyldgaard}}, \ and\ \bibinfo {author}
  {\bibfnamefont {B.~I.}\ \bibnamefont {Lundqvist}}} (\bibinfo {year} {2015}),\
  \href {\doibase 10.1088/0034-4885/78/6/066501} {\bibfield  {journal}
  {\bibinfo  {journal} {Rep. Prog. Phys.}\ }\textbf {\bibinfo {volume} {78}},\
  \bibinfo {pages} {066501}}\BibitemShut {NoStop}%
\bibitem [{\citenamefont {Berland}\ and\ \citenamefont {Hyldgaard}(2014)}]{cx}%
  \BibitemOpen
  \bibfield  {author} {\bibinfo {author} {\bibnamefont {Berland}, \bibfnamefont
  {K.}}, \ and\ \bibinfo {author} {\bibfnamefont {P.}~\bibnamefont
  {Hyldgaard}}} (\bibinfo {year} {2014}),\ \href {\doibase
  10.1103/PhysRevB.89.035412} {\bibfield  {journal} {\bibinfo  {journal} {Phys.
  Rev. B}\ }\textbf {\bibinfo {volume} {89}},\ \bibinfo {pages}
  {035412}}\BibitemShut {NoStop}%
\bibitem [{\citenamefont {Bernevig}\ and\ \citenamefont
  {Hughes}(2013)}]{Bernevig2013}%
  \BibitemOpen
  \bibfield  {author} {\bibinfo {author} {\bibnamefont {Bernevig},
  \bibfnamefont {A.~B.}}, \ and\ \bibinfo {author} {\bibfnamefont {T.~L.}\
  \bibnamefont {Hughes}}} (\bibinfo {year} {2013}),\ \href@noop {} {\emph
  {\bibinfo {title} {Topological Insulators and Topological
  Superconductors}}},\ \bibinfo {edition} {1st}\ ed.\ (\bibinfo  {publisher}
  {Princeton U. Press},\ \bibinfo {address} {Princeton, N.J.})\BibitemShut
  {NoStop}%
\bibitem [{\citenamefont {Bernevig}\ \emph {et~al.}(2006)\citenamefont
  {Bernevig}, \citenamefont {Orenstein},\ and\ \citenamefont
  {Zhang}}]{Bernevig}%
  \BibitemOpen
  \bibfield  {author} {\bibinfo {author} {\bibnamefont {Bernevig},
  \bibfnamefont {B.~A.}}, \bibinfo {author} {\bibfnamefont {J.}~\bibnamefont
  {Orenstein}}, \ and\ \bibinfo {author} {\bibfnamefont {S.-C.}\ \bibnamefont
  {Zhang}}} (\bibinfo {year} {2006}),\ \href {\doibase
  10.1103/PhysRevLett.97.236601} {\bibfield  {journal} {\bibinfo  {journal}
  {Phys. Rev. Lett.}\ }\textbf {\bibinfo {volume} {97}},\ \bibinfo {pages}
  {236601}}\BibitemShut {NoStop}%
\bibitem [{\citenamefont {Black}\ \emph {et~al.}(1997)\citenamefont {Black},
  \citenamefont {Farrell},\ and\ \citenamefont {Licata}}]{black}%
  \BibitemOpen
  \bibfield  {author} {\bibinfo {author} {\bibnamefont {Black}, \bibfnamefont
  {C.~T.}}, \bibinfo {author} {\bibfnamefont {C.}~\bibnamefont {Farrell}}, \
  and\ \bibinfo {author} {\bibfnamefont {T.~J.}\ \bibnamefont {Licata}}}
  (\bibinfo {year} {1997}),\ \href {\doibase 10.1063/1.119781} {\bibfield
  {journal} {\bibinfo  {journal} {Appl. Phys. Lett.}\ }\textbf {\bibinfo
  {volume} {71}},\ \bibinfo {pages} {2041}}\BibitemShut {NoStop}%
\bibitem [{\citenamefont {Boyd}(2020)}]{Boyd2008}%
  \BibitemOpen
  \bibfield  {author} {\bibinfo {author} {\bibnamefont {Boyd}, \bibfnamefont
  {R.~W.}}} (\bibinfo {year} {2020}),\ \href@noop {} {\emph {\bibinfo {title}
  {Nonlinear Optics}}},\ \bibinfo {edition} {4th}\ ed.\ (\bibinfo  {publisher}
  {Academic Press},\ \bibinfo {address} {London, U.K.})\BibitemShut {NoStop}%
\bibitem [{\citenamefont {Bune}\ \emph {et~al.}(1998)\citenamefont {Bune},
  \citenamefont {Fridkin}, \citenamefont {Ducharme}, \citenamefont {Blinov},
  \citenamefont {Palto}, \citenamefont {Sorokin}, \citenamefont {Yudin},\ and\
  \citenamefont {Zlatkin}}]{NatureFE}%
  \BibitemOpen
  \bibfield  {author} {\bibinfo {author} {\bibnamefont {Bune}, \bibfnamefont
  {A.~V.}}, \bibinfo {author} {\bibfnamefont {V.~M.}\ \bibnamefont {Fridkin}},
  \bibinfo {author} {\bibfnamefont {S.}~\bibnamefont {Ducharme}}, \bibinfo
  {author} {\bibfnamefont {L.~M.}\ \bibnamefont {Blinov}}, \bibinfo {author}
  {\bibfnamefont {S.~P.}\ \bibnamefont {Palto}}, \bibinfo {author}
  {\bibfnamefont {A.~V.}\ \bibnamefont {Sorokin}}, \bibinfo {author}
  {\bibfnamefont {S.~G.}\ \bibnamefont {Yudin}}, \ and\ \bibinfo {author}
  {\bibfnamefont {A.}~\bibnamefont {Zlatkin}}} (\bibinfo {year} {1998}),\ \href
  {\doibase 10.1038/36069} {\bibfield  {journal} {\bibinfo  {journal} {Nature}\
  }\textbf {\bibinfo {volume} {391}},\ \bibinfo {pages} {874}}\BibitemShut
  {NoStop}%
\bibitem [{\citenamefont {Cao}\ \emph {et~al.}(2012)\citenamefont {Cao},
  \citenamefont {Wang}, \citenamefont {Han}, \citenamefont {Ye}, \citenamefont
  {Zhu}, \citenamefont {Shi}, \citenamefont {Niu}, \citenamefont {Tan},
  \citenamefont {Wang}, \citenamefont {Liu},\ and\ \citenamefont
  {Feng}}]{valley_polarization_2012_3}%
  \BibitemOpen
  \bibfield  {author} {\bibinfo {author} {\bibnamefont {Cao}, \bibfnamefont
  {T.}}, \bibinfo {author} {\bibfnamefont {G.}~\bibnamefont {Wang}}, \bibinfo
  {author} {\bibfnamefont {W.}~\bibnamefont {Han}}, \bibinfo {author}
  {\bibfnamefont {H.}~\bibnamefont {Ye}}, \bibinfo {author} {\bibfnamefont
  {C.}~\bibnamefont {Zhu}}, \bibinfo {author} {\bibfnamefont {J.}~\bibnamefont
  {Shi}}, \bibinfo {author} {\bibfnamefont {Q.}~\bibnamefont {Niu}}, \bibinfo
  {author} {\bibfnamefont {P.}~\bibnamefont {Tan}}, \bibinfo {author}
  {\bibfnamefont {E.}~\bibnamefont {Wang}}, \bibinfo {author} {\bibfnamefont
  {B.}~\bibnamefont {Liu}}, \ and\ \bibinfo {author} {\bibfnamefont
  {J.}~\bibnamefont {Feng}}} (\bibinfo {year} {2012}),\ \href {\doibase
  10.1038/ncomms1882} {\bibfield  {journal} {\bibinfo  {journal} {Nat.
  Commun.}\ }\textbf {\bibinfo {volume} {3}},\ \bibinfo {pages}
  {887}}\BibitemShut {NoStop}%
\bibitem [{\citenamefont {Castellanos-Gomez}\ \emph {et~al.}(2014)\citenamefont
  {Castellanos-Gomez}, \citenamefont {Vicarelli}, \citenamefont {Prada},
  \citenamefont {Island}, \citenamefont {Narasimha-Acharya}, \citenamefont
  {Blanter}, \citenamefont {Groenendijk}, \citenamefont {Buscema},
  \citenamefont {Steele}, \citenamefont {Alvarez}, \citenamefont {Zandbergen},
  \citenamefont {Palacios},\ and\ \citenamefont {van~der Zant}}]{ph3}%
  \BibitemOpen
  \bibfield  {author} {\bibinfo {author} {\bibnamefont {Castellanos-Gomez},
  \bibfnamefont {A.}}, \bibinfo {author} {\bibfnamefont {L.}~\bibnamefont
  {Vicarelli}}, \bibinfo {author} {\bibfnamefont {E.}~\bibnamefont {Prada}},
  \bibinfo {author} {\bibfnamefont {J.~O.}\ \bibnamefont {Island}}, \bibinfo
  {author} {\bibfnamefont {K.~L.}\ \bibnamefont {Narasimha-Acharya}}, \bibinfo
  {author} {\bibfnamefont {S.~I.}\ \bibnamefont {Blanter}}, \bibinfo {author}
  {\bibfnamefont {D.~J.}\ \bibnamefont {Groenendijk}}, \bibinfo {author}
  {\bibfnamefont {M.}~\bibnamefont {Buscema}}, \bibinfo {author} {\bibfnamefont
  {G.~A.}\ \bibnamefont {Steele}}, \bibinfo {author} {\bibfnamefont {J.~V.}\
  \bibnamefont {Alvarez}}, \bibinfo {author} {\bibfnamefont {H.~W.}\
  \bibnamefont {Zandbergen}}, \bibinfo {author} {\bibfnamefont {J.~J.}\
  \bibnamefont {Palacios}}, \ and\ \bibinfo {author} {\bibfnamefont {H.~S.~J.}\
  \bibnamefont {van~der Zant}}} (\bibinfo {year} {2014}),\ \href {\doibase
  10.1088/2053-1583/1/2/025001} {\bibfield  {journal} {\bibinfo  {journal} {2D
  Mater.}\ }\textbf {\bibinfo {volume} {1}},\ \bibinfo {pages}
  {025001}}\BibitemShut {NoStop}%
\bibitem [{\citenamefont {Chan}\ \emph {et~al.}(2017)\citenamefont {Chan},
  \citenamefont {Lindner}, \citenamefont {Refael},\ and\ \citenamefont
  {Lee}}]{Chan2017}%
  \BibitemOpen
  \bibfield  {author} {\bibinfo {author} {\bibnamefont {Chan}, \bibfnamefont
  {C.-K.}}, \bibinfo {author} {\bibfnamefont {N.~H.}\ \bibnamefont {Lindner}},
  \bibinfo {author} {\bibfnamefont {G.}~\bibnamefont {Refael}}, \ and\ \bibinfo
  {author} {\bibfnamefont {P.~A.}\ \bibnamefont {Lee}}} (\bibinfo {year}
  {2017}),\ \href {\doibase 10.1103/PhysRevB.95.041104} {\bibfield  {journal}
  {\bibinfo  {journal} {Phys. Rev. B}\ }\textbf {\bibinfo {volume} {95}},\
  \bibinfo {pages} {041104}}\BibitemShut {NoStop}%
\bibitem [{\citenamefont {Chang}\ \emph
  {et~al.}(2019{\natexlab{a}})\citenamefont {Chang}, \citenamefont {Kaloni},
  \citenamefont {Lin}, \citenamefont {Bedoya-Pinto}, \citenamefont {Pandeya},
  \citenamefont {Kostanovskiy}, \citenamefont {Zhao}, \citenamefont {Zhong},
  \citenamefont {Hu}, \citenamefont {Xue}, \citenamefont {Chen}, \citenamefont
  {Ji}, \citenamefont {Barraza-Lopez},\ and\ \citenamefont {Parkin}}]{sntebl}%
  \BibitemOpen
  \bibfield  {author} {\bibinfo {author} {\bibnamefont {Chang}, \bibfnamefont
  {K.}}, \bibinfo {author} {\bibfnamefont {T.~P.}\ \bibnamefont {Kaloni}},
  \bibinfo {author} {\bibfnamefont {H.}~\bibnamefont {Lin}}, \bibinfo {author}
  {\bibfnamefont {A.}~\bibnamefont {Bedoya-Pinto}}, \bibinfo {author}
  {\bibfnamefont {A.~K.}\ \bibnamefont {Pandeya}}, \bibinfo {author}
  {\bibfnamefont {I.}~\bibnamefont {Kostanovskiy}}, \bibinfo {author}
  {\bibfnamefont {K.}~\bibnamefont {Zhao}}, \bibinfo {author} {\bibfnamefont
  {Y.}~\bibnamefont {Zhong}}, \bibinfo {author} {\bibfnamefont
  {X.}~\bibnamefont {Hu}}, \bibinfo {author} {\bibfnamefont {Q.-K.}\
  \bibnamefont {Xue}}, \bibinfo {author} {\bibfnamefont {X.}~\bibnamefont
  {Chen}}, \bibinfo {author} {\bibfnamefont {S.-H.}\ \bibnamefont {Ji}},
  \bibinfo {author} {\bibfnamefont {S.}~\bibnamefont {Barraza-Lopez}}, \ and\
  \bibinfo {author} {\bibfnamefont {S.~S.~P.}\ \bibnamefont {Parkin}}}
  (\bibinfo {year} {2019}{\natexlab{a}}),\ \href {\doibase
  10.1002/adma.201804428} {\bibfield  {journal} {\bibinfo  {journal} {Adv.
  Mater.}\ }\textbf {\bibinfo {volume} {31}},\ \bibinfo {pages}
  {1804428}}\BibitemShut {NoStop}%
\bibitem [{\citenamefont {Chang}\ \emph {et~al.}(2020)\citenamefont {Chang},
  \citenamefont {K{\"u}ster}, \citenamefont {Miller}, \citenamefont {Ji},
  \citenamefont {Zhang}, \citenamefont {Sessi}, \citenamefont {Barraza-Lopez},\
  and\ \citenamefont {Parkin}}]{Chang20_arxiv_SnSe}%
  \BibitemOpen
  \bibfield  {author} {\bibinfo {author} {\bibnamefont {Chang}, \bibfnamefont
  {K.}}, \bibinfo {author} {\bibfnamefont {F.}~\bibnamefont {K{\"u}ster}},
  \bibinfo {author} {\bibfnamefont {B.~J.}\ \bibnamefont {Miller}}, \bibinfo
  {author} {\bibfnamefont {J.-R.}\ \bibnamefont {Ji}}, \bibinfo {author}
  {\bibfnamefont {J.-L.}\ \bibnamefont {Zhang}}, \bibinfo {author}
  {\bibfnamefont {P.}~\bibnamefont {Sessi}}, \bibinfo {author} {\bibfnamefont
  {S.}~\bibnamefont {Barraza-Lopez}}, \ and\ \bibinfo {author} {\bibfnamefont
  {S.~S.~P.}\ \bibnamefont {Parkin}}} (\bibinfo {year} {2020}),\ \href
  {http://arxiv.org/abs/2004.03884} {\bibinfo  {journal} {Nano Lett.
  (accepted); arXiv:2004.03884}\ }\BibitemShut {NoStop}%
\bibitem [{\citenamefont {Chang}\ \emph {et~al.}(2016)\citenamefont {Chang},
  \citenamefont {Liu}, \citenamefont {Lin}, \citenamefont {Wang}, \citenamefont
  {Zhao}, \citenamefont {Zhang}, \citenamefont {Jin}, \citenamefont {Zhong},
  \citenamefont {Hu}, \citenamefont {Duan}, \citenamefont {Zhang},
  \citenamefont {Fu}, \citenamefont {Xue}, \citenamefont {Chen},\ and\
  \citenamefont {Ji}}]{Kai}%
  \BibitemOpen
\bibfield  {journal} {  }\bibfield  {author} {\bibinfo {author} {\bibnamefont
  {Chang}, \bibfnamefont {K.}}, \bibinfo {author} {\bibfnamefont
  {J.}~\bibnamefont {Liu}}, \bibinfo {author} {\bibfnamefont {H.}~\bibnamefont
  {Lin}}, \bibinfo {author} {\bibfnamefont {N.}~\bibnamefont {Wang}}, \bibinfo
  {author} {\bibfnamefont {K.}~\bibnamefont {Zhao}}, \bibinfo {author}
  {\bibfnamefont {A.}~\bibnamefont {Zhang}}, \bibinfo {author} {\bibfnamefont
  {F.}~\bibnamefont {Jin}}, \bibinfo {author} {\bibfnamefont {Y.}~\bibnamefont
  {Zhong}}, \bibinfo {author} {\bibfnamefont {X.}~\bibnamefont {Hu}}, \bibinfo
  {author} {\bibfnamefont {W.}~\bibnamefont {Duan}}, \bibinfo {author}
  {\bibfnamefont {Q.}~\bibnamefont {Zhang}}, \bibinfo {author} {\bibfnamefont
  {L.}~\bibnamefont {Fu}}, \bibinfo {author} {\bibfnamefont {Q.-K.}\
  \bibnamefont {Xue}}, \bibinfo {author} {\bibfnamefont {X.}~\bibnamefont
  {Chen}}, \ and\ \bibinfo {author} {\bibfnamefont {S.-H.}\ \bibnamefont {Ji}}}
  (\bibinfo {year} {2016}),\ \href {\doibase 10.1126/science.aad8609}
  {\bibfield  {journal} {\bibinfo  {journal} {Science}\ }\textbf {\bibinfo
  {volume} {353}},\ \bibinfo {pages} {274}}\BibitemShut {NoStop}%
\bibitem [{\citenamefont {Chang}\ \emph
  {et~al.}(2019{\natexlab{b}})\citenamefont {Chang}, \citenamefont {Miller},
  \citenamefont {Yang}, \citenamefont {Lin}, \citenamefont {Parkin},
  \citenamefont {Barraza-Lopez}, \citenamefont {Xue}, \citenamefont {Chen},\
  and\ \citenamefont {Ji}}]{KaiPRL}%
  \BibitemOpen
  \bibfield  {author} {\bibinfo {author} {\bibnamefont {Chang}, \bibfnamefont
  {K.}}, \bibinfo {author} {\bibfnamefont {B.~J.}\ \bibnamefont {Miller}},
  \bibinfo {author} {\bibfnamefont {H.}~\bibnamefont {Yang}}, \bibinfo {author}
  {\bibfnamefont {H.}~\bibnamefont {Lin}}, \bibinfo {author} {\bibfnamefont
  {S.~S.~P.}\ \bibnamefont {Parkin}}, \bibinfo {author} {\bibfnamefont
  {S.}~\bibnamefont {Barraza-Lopez}}, \bibinfo {author} {\bibfnamefont {Q.-K.}\
  \bibnamefont {Xue}}, \bibinfo {author} {\bibfnamefont {X.}~\bibnamefont
  {Chen}}, \ and\ \bibinfo {author} {\bibfnamefont {S.-H.}\ \bibnamefont {Ji}}}
  (\bibinfo {year} {2019}{\natexlab{b}}),\ \href {\doibase
  10.1103/PhysRevLett.122.206402} {\bibfield  {journal} {\bibinfo  {journal}
  {Phys. Rev. Lett.}\ }\textbf {\bibinfo {volume} {122}},\ \bibinfo {pages}
  {206402}}\BibitemShut {NoStop}%
\bibitem [{\citenamefont {Chang}\ and\ \citenamefont
  {Parkin}(2019)}]{chang_aplm_2019}%
  \BibitemOpen
  \bibfield  {author} {\bibinfo {author} {\bibnamefont {Chang}, \bibfnamefont
  {K.}}, \ and\ \bibinfo {author} {\bibfnamefont {S.~S.~P.}\ \bibnamefont
  {Parkin}}} (\bibinfo {year} {2019}),\ \href {\doibase 10.1063/1.5091546}
  {\bibfield  {journal} {\bibinfo  {journal} {APL Mater.}\ }\textbf {\bibinfo
  {volume} {7}},\ \bibinfo {pages} {041102}}\BibitemShut {NoStop}%
\bibitem [{\citenamefont {Chang}\ and\ \citenamefont {Parkin}(2020)}]{kairev}%
  \BibitemOpen
  \bibfield  {author} {\bibinfo {author} {\bibnamefont {Chang}, \bibfnamefont
  {K.}}, \ and\ \bibinfo {author} {\bibfnamefont {S.~S.~P.}\ \bibnamefont
  {Parkin}}} (\bibinfo {year} {2020}),\ \href {\doibase 10.1063/5.0012300}
  {\bibfield  {journal} {\bibinfo  {journal} {J. Appl. Phys.}\ }\textbf
  {\bibinfo {volume} {127}},\ \bibinfo {pages} {220902}}\BibitemShut {NoStop}%
\bibitem [{\citenamefont {Cheiwchanchamnangij}\ and\ \citenamefont
  {Lambrecht}(2012)}]{massmos22}%
  \BibitemOpen
  \bibfield  {author} {\bibinfo {author} {\bibnamefont {Cheiwchanchamnangij},
  \bibfnamefont {T.}}, \ and\ \bibinfo {author} {\bibfnamefont {W.~R.~L.}\
  \bibnamefont {Lambrecht}}} (\bibinfo {year} {2012}),\ \href {\doibase
  10.1103/PhysRevB.85.205302} {\bibfield  {journal} {\bibinfo  {journal} {Phys.
  Rev. B}\ }\textbf {\bibinfo {volume} {85}},\ \bibinfo {pages}
  {205302}}\BibitemShut {NoStop}%
\bibitem [{\citenamefont {Cui}\ \emph {et~al.}(2018{\natexlab{a}})\citenamefont
  {Cui}, \citenamefont {Hu}, \citenamefont {Yan}, \citenamefont {Addiego},
  \citenamefont {Gao}, \citenamefont {Wang}, \citenamefont {Wang},
  \citenamefont {Li}, \citenamefont {Cheng}, \citenamefont {Li}, \citenamefont
  {Zhang}, \citenamefont {Alshareef}, \citenamefont {Wu}, \citenamefont {Zhu},
  \citenamefont {Pan},\ and\ \citenamefont {Li}}]{in2se3_3}%
  \BibitemOpen
  \bibfield  {author} {\bibinfo {author} {\bibnamefont {Cui}, \bibfnamefont
  {C.}}, \bibinfo {author} {\bibfnamefont {W.-J.}\ \bibnamefont {Hu}}, \bibinfo
  {author} {\bibfnamefont {X.}~\bibnamefont {Yan}}, \bibinfo {author}
  {\bibfnamefont {C.}~\bibnamefont {Addiego}}, \bibinfo {author} {\bibfnamefont
  {W.}~\bibnamefont {Gao}}, \bibinfo {author} {\bibfnamefont {Y.}~\bibnamefont
  {Wang}}, \bibinfo {author} {\bibfnamefont {Z.}~\bibnamefont {Wang}}, \bibinfo
  {author} {\bibfnamefont {L.}~\bibnamefont {Li}}, \bibinfo {author}
  {\bibfnamefont {Y.}~\bibnamefont {Cheng}}, \bibinfo {author} {\bibfnamefont
  {P.}~\bibnamefont {Li}}, \bibinfo {author} {\bibfnamefont {X.}~\bibnamefont
  {Zhang}}, \bibinfo {author} {\bibfnamefont {H.~N.}\ \bibnamefont
  {Alshareef}}, \bibinfo {author} {\bibfnamefont {T.}~\bibnamefont {Wu}},
  \bibinfo {author} {\bibfnamefont {W.}~\bibnamefont {Zhu}}, \bibinfo {author}
  {\bibfnamefont {X.}~\bibnamefont {Pan}}, \ and\ \bibinfo {author}
  {\bibfnamefont {L.-J.}\ \bibnamefont {Li}}} (\bibinfo {year}
  {2018}{\natexlab{a}}),\ \href {\doibase 10.1021/acs.nanolett.7b04852}
  {\bibfield  {journal} {\bibinfo  {journal} {Nano Lett.}\ }\textbf {\bibinfo
  {volume} {18}},\ \bibinfo {pages} {1253}}\BibitemShut {NoStop}%
\bibitem [{\citenamefont {Cui}\ \emph {et~al.}(2018{\natexlab{b}})\citenamefont
  {Cui}, \citenamefont {Xue}, \citenamefont {Hu},\ and\ \citenamefont
  {Li}}]{rev}%
  \BibitemOpen
  \bibfield  {author} {\bibinfo {author} {\bibnamefont {Cui}, \bibfnamefont
  {C.}}, \bibinfo {author} {\bibfnamefont {F.}~\bibnamefont {Xue}}, \bibinfo
  {author} {\bibfnamefont {W.-J.}\ \bibnamefont {Hu}}, \ and\ \bibinfo {author}
  {\bibfnamefont {L.-J.}\ \bibnamefont {Li}}} (\bibinfo {year}
  {2018}{\natexlab{b}}),\ \href {\doibase 10.1038/s41699-018-0063-5} {\bibfield
   {journal} {\bibinfo  {journal} {npj 2D Mat. Appl.}\ }\textbf {\bibinfo
  {volume} {2}},\ \bibinfo {pages} {18}}\BibitemShut {NoStop}%
\bibitem [{\citenamefont {Deng}\ \emph {et~al.}(2019)\citenamefont {Deng},
  \citenamefont {Liu}, \citenamefont {Li}, \citenamefont {Xu}, \citenamefont
  {Lun}, \citenamefont {Lv}, \citenamefont {Xia}, \citenamefont {Gao},
  \citenamefont {Wang},\ and\ \citenamefont
  {Hong}}]{cips_ip_polarization_2019}%
  \BibitemOpen
  \bibfield  {author} {\bibinfo {author} {\bibnamefont {Deng}, \bibfnamefont
  {J.}}, \bibinfo {author} {\bibfnamefont {Y.}~\bibnamefont {Liu}}, \bibinfo
  {author} {\bibfnamefont {M.}~\bibnamefont {Li}}, \bibinfo {author}
  {\bibfnamefont {S.}~\bibnamefont {Xu}}, \bibinfo {author} {\bibfnamefont
  {Y.}~\bibnamefont {Lun}}, \bibinfo {author} {\bibfnamefont {P.}~\bibnamefont
  {Lv}}, \bibinfo {author} {\bibfnamefont {T.}~\bibnamefont {Xia}}, \bibinfo
  {author} {\bibfnamefont {P.}~\bibnamefont {Gao}}, \bibinfo {author}
  {\bibfnamefont {X.}~\bibnamefont {Wang}}, \ and\ \bibinfo {author}
  {\bibfnamefont {J.}~\bibnamefont {Hong}}} (\bibinfo {year} {2019}),\ \href
  {\doibase 10.1002/smll.201904529} {\bibfield  {journal} {\bibinfo  {journal}
  {Small}\ }\textbf {\bibinfo {volume} {16}},\ \bibinfo {pages}
  {1904529}}\BibitemShut {NoStop}%
\bibitem [{\citenamefont {Deslippe}\ \emph {et~al.}(2012)\citenamefont
  {Deslippe}, \citenamefont {Samsonidze}, \citenamefont {Strubbe},
  \citenamefont {Jain}, \citenamefont {Cohen},\ and\ \citenamefont
  {Louie}}]{berkeleyGW}%
  \BibitemOpen
  \bibfield  {author} {\bibinfo {author} {\bibnamefont {Deslippe},
  \bibfnamefont {J.}}, \bibinfo {author} {\bibfnamefont {G.}~\bibnamefont
  {Samsonidze}}, \bibinfo {author} {\bibfnamefont {D.~A.}\ \bibnamefont
  {Strubbe}}, \bibinfo {author} {\bibfnamefont {M.}~\bibnamefont {Jain}},
  \bibinfo {author} {\bibfnamefont {M.~L.}\ \bibnamefont {Cohen}}, \ and\
  \bibinfo {author} {\bibfnamefont {S.~G.}\ \bibnamefont {Louie}}} (\bibinfo
  {year} {2012}),\ \href {\doibase 10.1016/j.cpc.2011.12.006} {\bibfield
  {journal} {\bibinfo  {journal} {Comp. Phys. Comm.}\ }\textbf {\bibinfo
  {volume} {183}},\ \bibinfo {pages} {1269}}\BibitemShut {NoStop}%
\bibitem [{\citenamefont {Dewandre}\ \emph {et~al.}(2019)\citenamefont
  {Dewandre}, \citenamefont {Verstraete}, \citenamefont {Grobert},\ and\
  \citenamefont {Zanolli}}]{italian}%
  \BibitemOpen
  \bibfield  {author} {\bibinfo {author} {\bibnamefont {Dewandre},
  \bibfnamefont {A.}}, \bibinfo {author} {\bibfnamefont {M.~J.}\ \bibnamefont
  {Verstraete}}, \bibinfo {author} {\bibfnamefont {N.}~\bibnamefont {Grobert}},
  \ and\ \bibinfo {author} {\bibfnamefont {Z.}~\bibnamefont {Zanolli}}}
  (\bibinfo {year} {2019}),\ \href {\doibase 10.1088/2515-7639/ab3513}
  {\bibfield  {journal} {\bibinfo  {journal} {J. Phys.: Mater.}\ }\textbf
  {\bibinfo {volume} {2}},\ \bibinfo {pages} {044005}}\BibitemShut {NoStop}%
\bibitem [{\citenamefont {Di~Sante}\ \emph {et~al.}(2013)\citenamefont
  {Di~Sante}, \citenamefont {Barone}, \citenamefont {Bertacco},\ and\
  \citenamefont {Picozzi}}]{picozzi}%
  \BibitemOpen
  \bibfield  {author} {\bibinfo {author} {\bibnamefont {Di~Sante},
  \bibfnamefont {D.}}, \bibinfo {author} {\bibfnamefont {P.}~\bibnamefont
  {Barone}}, \bibinfo {author} {\bibfnamefont {R.}~\bibnamefont {Bertacco}}, \
  and\ \bibinfo {author} {\bibfnamefont {S.}~\bibnamefont {Picozzi}}} (\bibinfo
  {year} {2013}),\ \href {\doibase 10.1002/adma.201203199} {\bibfield
  {journal} {\bibinfo  {journal} {Adv. Mater.}\ }\textbf {\bibinfo {volume}
  {25}},\ \bibinfo {pages} {509}}\BibitemShut {NoStop}%
\bibitem [{\citenamefont {Di~Sante}\ \emph {et~al.}(2015)\citenamefont
  {Di~Sante}, \citenamefont {Stroppa}, \citenamefont {Barone}, \citenamefont
  {Whangbo},\ and\ \citenamefont {Picozzi}}]{honeycomb_band_splitting_2015}%
  \BibitemOpen
  \bibfield  {author} {\bibinfo {author} {\bibnamefont {Di~Sante},
  \bibfnamefont {D.}}, \bibinfo {author} {\bibfnamefont {A.}~\bibnamefont
  {Stroppa}}, \bibinfo {author} {\bibfnamefont {P.}~\bibnamefont {Barone}},
  \bibinfo {author} {\bibfnamefont {M.-H.}\ \bibnamefont {Whangbo}}, \ and\
  \bibinfo {author} {\bibfnamefont {S.}~\bibnamefont {Picozzi}}} (\bibinfo
  {year} {2015}),\ \href {\doibase 10.1103/PhysRevB.91.161401} {\bibfield
  {journal} {\bibinfo  {journal} {Phys. Rev. B}\ }\textbf {\bibinfo {volume}
  {91}},\ \bibinfo {pages} {161401}}\BibitemShut {NoStop}%
\bibitem [{\citenamefont {Dimmock}\ \emph {et~al.}(1966)\citenamefont
  {Dimmock}, \citenamefont {Melngailis},\ and\ \citenamefont
  {Strauss}}]{bulk_SnTe_3}%
  \BibitemOpen
  \bibfield  {author} {\bibinfo {author} {\bibnamefont {Dimmock}, \bibfnamefont
  {J.~O.}}, \bibinfo {author} {\bibfnamefont {I.}~\bibnamefont {Melngailis}}, \
  and\ \bibinfo {author} {\bibfnamefont {A.~J.}\ \bibnamefont {Strauss}}}
  (\bibinfo {year} {1966}),\ \href {\doibase 10.1103/PhysRevLett.16.1193}
  {\bibfield  {journal} {\bibinfo  {journal} {Phys. Rev. Lett.}\ }\textbf
  {\bibinfo {volume} {16}},\ \bibinfo {pages} {1193}}\BibitemShut {NoStop}%
\bibitem [{\citenamefont {Ding}\ \emph {et~al.}(2017)\citenamefont {Ding},
  \citenamefont {Zhu}, \citenamefont {Wang}, \citenamefont {Gao}, \citenamefont
  {Xiao}, \citenamefont {Gu}, \citenamefont {Zhang},\ and\ \citenamefont
  {Zhu}}]{ding_nc_2017_in2se3}%
  \BibitemOpen
  \bibfield  {author} {\bibinfo {author} {\bibnamefont {Ding}, \bibfnamefont
  {W.}}, \bibinfo {author} {\bibfnamefont {J.}~\bibnamefont {Zhu}}, \bibinfo
  {author} {\bibfnamefont {Z.}~\bibnamefont {Wang}}, \bibinfo {author}
  {\bibfnamefont {Y.}~\bibnamefont {Gao}}, \bibinfo {author} {\bibfnamefont
  {D.}~\bibnamefont {Xiao}}, \bibinfo {author} {\bibfnamefont {Y.}~\bibnamefont
  {Gu}}, \bibinfo {author} {\bibfnamefont {Z.}~\bibnamefont {Zhang}}, \ and\
  \bibinfo {author} {\bibfnamefont {W.}~\bibnamefont {Zhu}}} (\bibinfo {year}
  {2017}),\ \href {\doibase 10.1038/ncomms14956} {\bibfield  {journal}
  {\bibinfo  {journal} {Nat. Commun.}\ }\textbf {\bibinfo {volume} {8}},\
  \bibinfo {pages} {14956}}\BibitemShut {NoStop}%
\bibitem [{\citenamefont {Dresselhaus}(1955)}]{dress1}%
  \BibitemOpen
  \bibfield  {author} {\bibinfo {author} {\bibnamefont {Dresselhaus},
  \bibfnamefont {G.}}} (\bibinfo {year} {1955}),\ \href {\doibase
  10.1103/PhysRev.100.580} {\bibfield  {journal} {\bibinfo  {journal} {Phys.
  Rev.}\ }\textbf {\bibinfo {volume} {100}},\ \bibinfo {pages}
  {580}}\BibitemShut {NoStop}%
\bibitem [{\citenamefont {Du}\ \emph {et~al.}(2020)\citenamefont {Du},
  \citenamefont {Pendergrast},\ and\ \citenamefont {Barraza-Lopez}}]{doping2}%
  \BibitemOpen
  \bibfield  {author} {\bibinfo {author} {\bibnamefont {Du}, \bibfnamefont
  {A.}}, \bibinfo {author} {\bibfnamefont {Z.}~\bibnamefont {Pendergrast}}, \
  and\ \bibinfo {author} {\bibfnamefont {S.}~\bibnamefont {Barraza-Lopez}}}
  (\bibinfo {year} {2020}),\ \href {\doibase 10.1063/5.0008502} {\bibfield
  {journal} {\bibinfo  {journal} {J. Appl. Phys.}\ }\textbf {\bibinfo {volume}
  {127}},\ \bibinfo {pages} {234103}}\BibitemShut {NoStop}%
\bibitem [{\citenamefont {Duerloo}\ \emph {et~al.}(2012)\citenamefont
  {Duerloo}, \citenamefont {Ong},\ and\ \citenamefont {Reed}}]{duerloo}%
  \BibitemOpen
  \bibfield  {author} {\bibinfo {author} {\bibnamefont {Duerloo}, \bibfnamefont
  {K.-A.~N.}}, \bibinfo {author} {\bibfnamefont {M.~T.}\ \bibnamefont {Ong}}, \
  and\ \bibinfo {author} {\bibfnamefont {E.~J.}\ \bibnamefont {Reed}}}
  (\bibinfo {year} {2012}),\ \href {\doibase 10.1021/jz3012436} {\bibfield
  {journal} {\bibinfo  {journal} {J. Phys. Chem. Lett.}\ }\textbf {\bibinfo
  {volume} {3}},\ \bibinfo {pages} {2871}}\BibitemShut {NoStop}%
\bibitem [{\citenamefont {{Evans}}\ and\ \citenamefont
  {{Womack}}(1988)}]{5940}%
  \BibitemOpen
  \bibfield  {author} {\bibinfo {author} {\bibnamefont {{Evans}}, \bibfnamefont
  {J.~T.}}, \ and\ \bibinfo {author} {\bibfnamefont {R.}~\bibnamefont
  {{Womack}}}} (\bibinfo {year} {1988}),\ \href {\doibase 10.1109/4.5940}
  {\bibfield  {journal} {\bibinfo  {journal} {IEEE J. Solid-State Circuits}\
  }\textbf {\bibinfo {volume} {23}},\ \bibinfo {pages} {1171}}\BibitemShut
  {NoStop}%
\bibitem [{\citenamefont {Fei}\ \emph {et~al.}(2016)\citenamefont {Fei},
  \citenamefont {Kang},\ and\ \citenamefont {Yang}}]{fei_prl_2016}%
  \BibitemOpen
  \bibfield  {author} {\bibinfo {author} {\bibnamefont {Fei}, \bibfnamefont
  {R.}}, \bibinfo {author} {\bibfnamefont {W.}~\bibnamefont {Kang}}, \ and\
  \bibinfo {author} {\bibfnamefont {L.}~\bibnamefont {Yang}}} (\bibinfo {year}
  {2016}),\ \href {\doibase 10.1103/PhysRevLett.117.097601} {\bibfield
  {journal} {\bibinfo  {journal} {Phys. Rev. Lett.}\ }\textbf {\bibinfo
  {volume} {117}},\ \bibinfo {pages} {097601}}\BibitemShut {NoStop}%
\bibitem [{\citenamefont {Fei}\ \emph {et~al.}(2015)\citenamefont {Fei},
  \citenamefont {Li}, \citenamefont {Li},\ and\ \citenamefont
  {Yang}}]{fei_apl_2015_ges_gese_sns_snse}%
  \BibitemOpen
  \bibfield  {author} {\bibinfo {author} {\bibnamefont {Fei}, \bibfnamefont
  {R.}}, \bibinfo {author} {\bibfnamefont {W.}~\bibnamefont {Li}}, \bibinfo
  {author} {\bibfnamefont {J.}~\bibnamefont {Li}}, \ and\ \bibinfo {author}
  {\bibfnamefont {L.}~\bibnamefont {Yang}}} (\bibinfo {year} {2015}),\ \href
  {\doibase 10.1063/1.4934750} {\bibfield  {journal} {\bibinfo  {journal}
  {Appl. Phys. Lett.}\ }\textbf {\bibinfo {volume} {107}},\ \bibinfo {pages}
  {173104}}\BibitemShut {NoStop}%
\bibitem [{\citenamefont {Fei}\ \emph {et~al.}(2017)\citenamefont {Fei},
  \citenamefont {Palomaki}, \citenamefont {Wu}, \citenamefont {Zhao},
  \citenamefont {Cai}, \citenamefont {Sun}, \citenamefont {Nguyen},
  \citenamefont {Finney}, \citenamefont {Xu},\ and\ \citenamefont
  {Cobden}}]{fei_np_2017}%
  \BibitemOpen
  \bibfield  {author} {\bibinfo {author} {\bibnamefont {Fei}, \bibfnamefont
  {Z.}}, \bibinfo {author} {\bibfnamefont {T.}~\bibnamefont {Palomaki}},
  \bibinfo {author} {\bibfnamefont {S.}~\bibnamefont {Wu}}, \bibinfo {author}
  {\bibfnamefont {W.}~\bibnamefont {Zhao}}, \bibinfo {author} {\bibfnamefont
  {X.}~\bibnamefont {Cai}}, \bibinfo {author} {\bibfnamefont {B.}~\bibnamefont
  {Sun}}, \bibinfo {author} {\bibfnamefont {P.}~\bibnamefont {Nguyen}},
  \bibinfo {author} {\bibfnamefont {J.}~\bibnamefont {Finney}}, \bibinfo
  {author} {\bibfnamefont {X.}~\bibnamefont {Xu}}, \ and\ \bibinfo {author}
  {\bibfnamefont {D.~H.}\ \bibnamefont {Cobden}}} (\bibinfo {year} {2017}),\
  \href {\doibase 10.1038/nphys4091} {\bibfield  {journal} {\bibinfo  {journal}
  {Nat. Phys.}\ }\textbf {\bibinfo {volume} {13}},\ \bibinfo {pages}
  {677}}\BibitemShut {NoStop}%
\bibitem [{\citenamefont {Fei}\ \emph {et~al.}(2018)\citenamefont {Fei},
  \citenamefont {Zhao}, \citenamefont {Palomaki}, \citenamefont {Sun},
  \citenamefont {Miller}, \citenamefont {Zhao}, \citenamefont {Yan},
  \citenamefont {Xu},\ and\ \citenamefont {Cobden}}]{wte2}%
  \BibitemOpen
  \bibfield  {author} {\bibinfo {author} {\bibnamefont {Fei}, \bibfnamefont
  {Z.}}, \bibinfo {author} {\bibfnamefont {W.}~\bibnamefont {Zhao}}, \bibinfo
  {author} {\bibfnamefont {T.~A.}\ \bibnamefont {Palomaki}}, \bibinfo {author}
  {\bibfnamefont {B.}~\bibnamefont {Sun}}, \bibinfo {author} {\bibfnamefont
  {M.~K.}\ \bibnamefont {Miller}}, \bibinfo {author} {\bibfnamefont
  {Z.}~\bibnamefont {Zhao}}, \bibinfo {author} {\bibfnamefont {J.}~\bibnamefont
  {Yan}}, \bibinfo {author} {\bibfnamefont {X.}~\bibnamefont {Xu}}, \ and\
  \bibinfo {author} {\bibfnamefont {D.~H.}\ \bibnamefont {Cobden}}} (\bibinfo
  {year} {2018}),\ \href {\doibase 10.1038/s41586-018-0336-3} {\bibfield
  {journal} {\bibinfo  {journal} {Nature}\ }\textbf {\bibinfo {volume} {560}},\
  \bibinfo {pages} {336}}\BibitemShut {NoStop}%
\bibitem [{\citenamefont {Feuersanger}\ \emph {et~al.}(1964)\citenamefont
  {Feuersanger}, \citenamefont {Hagenlocher},\ and\ \citenamefont
  {Solomon}}]{Feuersanger_1964}%
  \BibitemOpen
  \bibfield  {author} {\bibinfo {author} {\bibnamefont {Feuersanger},
  \bibfnamefont {A.~E.}}, \bibinfo {author} {\bibfnamefont {A.~K.}\
  \bibnamefont {Hagenlocher}}, \ and\ \bibinfo {author} {\bibfnamefont {A.~L.}\
  \bibnamefont {Solomon}}} (\bibinfo {year} {1964}),\ \href {\doibase
  10.1149/1.2426011} {\bibfield  {journal} {\bibinfo  {journal} {J.
  Electrochem. Soc.}\ }\textbf {\bibinfo {volume} {111}},\ \bibinfo {pages}
  {1387}}\BibitemShut {NoStop}%
\bibitem [{\citenamefont {Fong}\ \emph {et~al.}(2006)\citenamefont {Fong},
  \citenamefont {Kolpak}, \citenamefont {Eastman}, \citenamefont {Streiffer},
  \citenamefont {Fuoss}, \citenamefont {Stephenson}, \citenamefont {Thompson},
  \citenamefont {Kim}, \citenamefont {Choi}, \citenamefont {Eom}, \citenamefont
  {Grinberg},\ and\ \citenamefont {Rappe}}]{fong_pto_2006}%
  \BibitemOpen
  \bibfield  {author} {\bibinfo {author} {\bibnamefont {Fong}, \bibfnamefont
  {D.~D.}}, \bibinfo {author} {\bibfnamefont {A.~M.}\ \bibnamefont {Kolpak}},
  \bibinfo {author} {\bibfnamefont {J.~A.}\ \bibnamefont {Eastman}}, \bibinfo
  {author} {\bibfnamefont {S.~K.}\ \bibnamefont {Streiffer}}, \bibinfo {author}
  {\bibfnamefont {P.~H.}\ \bibnamefont {Fuoss}}, \bibinfo {author}
  {\bibfnamefont {G.~B.}\ \bibnamefont {Stephenson}}, \bibinfo {author}
  {\bibfnamefont {C.}~\bibnamefont {Thompson}}, \bibinfo {author}
  {\bibfnamefont {D.~M.}\ \bibnamefont {Kim}}, \bibinfo {author} {\bibfnamefont
  {K.~J.}\ \bibnamefont {Choi}}, \bibinfo {author} {\bibfnamefont {C.~B.}\
  \bibnamefont {Eom}}, \bibinfo {author} {\bibfnamefont {I.}~\bibnamefont
  {Grinberg}}, \ and\ \bibinfo {author} {\bibfnamefont {A.~M.}\ \bibnamefont
  {Rappe}}} (\bibinfo {year} {2006}),\ \href {\doibase
  10.1103/PhysRevLett.96.127601} {\bibfield  {journal} {\bibinfo  {journal}
  {Phys. Rev. Lett.}\ }\textbf {\bibinfo {volume} {96}},\ \bibinfo {pages}
  {127601}}\BibitemShut {NoStop}%
\bibitem [{\citenamefont {Fong}\ \emph {et~al.}(2004)\citenamefont {Fong},
  \citenamefont {Stephenson}, \citenamefont {Streiffer}, \citenamefont
  {Eastman}, \citenamefont {Auciello}, \citenamefont {Fuoss},\ and\
  \citenamefont {Thompson}}]{fong_pto_2004}%
  \BibitemOpen
  \bibfield  {author} {\bibinfo {author} {\bibnamefont {Fong}, \bibfnamefont
  {D.~D.}}, \bibinfo {author} {\bibfnamefont {G.~B.}\ \bibnamefont
  {Stephenson}}, \bibinfo {author} {\bibfnamefont {S.~K.}\ \bibnamefont
  {Streiffer}}, \bibinfo {author} {\bibfnamefont {J.~A.}\ \bibnamefont
  {Eastman}}, \bibinfo {author} {\bibfnamefont {O.}~\bibnamefont {Auciello}},
  \bibinfo {author} {\bibfnamefont {P.~H.}\ \bibnamefont {Fuoss}}, \ and\
  \bibinfo {author} {\bibfnamefont {C.}~\bibnamefont {Thompson}}} (\bibinfo
  {year} {2004}),\ \href {\doibase 10.1126/science.1098252} {\bibfield
  {journal} {\bibinfo  {journal} {Science}\ }\textbf {\bibinfo {volume}
  {304}},\ \bibinfo {pages} {1650}}\BibitemShut {NoStop}%
\bibitem [{\citenamefont {Fregoso}(2019)}]{Fregoso2019}%
  \BibitemOpen
  \bibfield  {author} {\bibinfo {author} {\bibnamefont {Fregoso}, \bibfnamefont
  {B.~M.}}} (\bibinfo {year} {2019}),\ \href {\doibase
  10.1103/PhysRevB.100.064301} {\bibfield  {journal} {\bibinfo  {journal}
  {Phys. Rev. B}\ }\textbf {\bibinfo {volume} {100}},\ \bibinfo {pages}
  {064301}}\BibitemShut {NoStop}%
\bibitem [{\citenamefont {Fu}\ \emph {et~al.}(2019)\citenamefont {Fu},
  \citenamefont {Liu},\ and\ \citenamefont {Yang}}]{substrate}%
  \BibitemOpen
  \bibfield  {author} {\bibinfo {author} {\bibnamefont {Fu}, \bibfnamefont
  {Z.}}, \bibinfo {author} {\bibfnamefont {M.}~\bibnamefont {Liu}}, \ and\
  \bibinfo {author} {\bibfnamefont {Z.}~\bibnamefont {Yang}}} (\bibinfo {year}
  {2019}),\ \href {\doibase 10.1103/PhysRevB.99.205425} {\bibfield  {journal}
  {\bibinfo  {journal} {Phys. Rev. B}\ }\textbf {\bibinfo {volume} {99}},\
  \bibinfo {pages} {205425}}\BibitemShut {NoStop}%
\bibitem [{\citenamefont {Gao}\ \emph {et~al.}(2017)\citenamefont {Gao},
  \citenamefont {Zhang}, \citenamefont {Li}, \citenamefont {Ishikawa},
  \citenamefont {Feng}, \citenamefont {Liu}, \citenamefont {Huang},
  \citenamefont {Shibata}, \citenamefont {Ma}, \citenamefont {Chen},
  \citenamefont {Zhang}, \citenamefont {Liu}, \citenamefont {Wang},
  \citenamefont {Yu}, \citenamefont {Liao}, \citenamefont {Chu},\ and\
  \citenamefont {Ikuhara}}]{gao_pzt_2017}%
  \BibitemOpen
  \bibfield  {author} {\bibinfo {author} {\bibnamefont {Gao}, \bibfnamefont
  {P.}}, \bibinfo {author} {\bibfnamefont {Z.}~\bibnamefont {Zhang}}, \bibinfo
  {author} {\bibfnamefont {M.}~\bibnamefont {Li}}, \bibinfo {author}
  {\bibfnamefont {R.}~\bibnamefont {Ishikawa}}, \bibinfo {author}
  {\bibfnamefont {B.}~\bibnamefont {Feng}}, \bibinfo {author} {\bibfnamefont
  {H.-J.}\ \bibnamefont {Liu}}, \bibinfo {author} {\bibfnamefont {Y.-L.}\
  \bibnamefont {Huang}}, \bibinfo {author} {\bibfnamefont {N.}~\bibnamefont
  {Shibata}}, \bibinfo {author} {\bibfnamefont {X.}~\bibnamefont {Ma}},
  \bibinfo {author} {\bibfnamefont {S.}~\bibnamefont {Chen}}, \bibinfo {author}
  {\bibfnamefont {J.}~\bibnamefont {Zhang}}, \bibinfo {author} {\bibfnamefont
  {K.}~\bibnamefont {Liu}}, \bibinfo {author} {\bibfnamefont {E.-G.}\
  \bibnamefont {Wang}}, \bibinfo {author} {\bibfnamefont {D.}~\bibnamefont
  {Yu}}, \bibinfo {author} {\bibfnamefont {L.}~\bibnamefont {Liao}}, \bibinfo
  {author} {\bibfnamefont {Y.-H.}\ \bibnamefont {Chu}}, \ and\ \bibinfo
  {author} {\bibfnamefont {Y.}~\bibnamefont {Ikuhara}}} (\bibinfo {year}
  {2017}),\ \href {\doibase 10.1038/ncomms15549} {\bibfield  {journal}
  {\bibinfo  {journal} {Nat. Commun.}\ }\textbf {\bibinfo {volume} {8}},\
  \bibinfo {pages} {15549}}\BibitemShut {NoStop}%
\bibitem [{\citenamefont {Gerra}\ \emph {et~al.}(2006)\citenamefont {Gerra},
  \citenamefont {Tagantsev}, \citenamefont {Setter},\ and\ \citenamefont
  {Parlinski}}]{Gerra_2006}%
  \BibitemOpen
  \bibfield  {author} {\bibinfo {author} {\bibnamefont {Gerra}, \bibfnamefont
  {G.}}, \bibinfo {author} {\bibfnamefont {A.~K.}\ \bibnamefont {Tagantsev}},
  \bibinfo {author} {\bibfnamefont {N.}~\bibnamefont {Setter}}, \ and\ \bibinfo
  {author} {\bibfnamefont {K.}~\bibnamefont {Parlinski}}} (\bibinfo {year}
  {2006}),\ \href {\doibase 10.1103/PhysRevLett.96.107603} {\bibfield
  {journal} {\bibinfo  {journal} {Phys. Rev. Lett.}\ }\textbf {\bibinfo
  {volume} {96}},\ \bibinfo {pages} {107603}}\BibitemShut {NoStop}%
\bibitem [{\citenamefont {Gomes}\ and\ \citenamefont {Carvalho}(2015)}]{gomes}%
  \BibitemOpen
  \bibfield  {author} {\bibinfo {author} {\bibnamefont {Gomes}, \bibfnamefont
  {L.~C.}}, \ and\ \bibinfo {author} {\bibfnamefont {A.}~\bibnamefont
  {Carvalho}}} (\bibinfo {year} {2015}),\ \href {\doibase
  10.1103/PhysRevB.92.085406} {\bibfield  {journal} {\bibinfo  {journal} {Phys.
  Rev. B}\ }\textbf {\bibinfo {volume} {92}},\ \bibinfo {pages}
  {085406}}\BibitemShut {NoStop}%
\bibitem [{\citenamefont {Gomes}\ \emph {et~al.}(2015)\citenamefont {Gomes},
  \citenamefont {Carvalho},\ and\ \citenamefont
  {Castro~Neto}}]{gomes_prb_2015a_ges_gese_sns_snse}%
  \BibitemOpen
  \bibfield  {author} {\bibinfo {author} {\bibnamefont {Gomes}, \bibfnamefont
  {L.~C.}}, \bibinfo {author} {\bibfnamefont {A.}~\bibnamefont {Carvalho}}, \
  and\ \bibinfo {author} {\bibfnamefont {A.~H.}\ \bibnamefont {Castro~Neto}}}
  (\bibinfo {year} {2015}),\ \href {\doibase 10.1103/PhysRevB.92.214103}
  {\bibfield  {journal} {\bibinfo  {journal} {Phys. Rev. B}\ }\textbf {\bibinfo
  {volume} {92}},\ \bibinfo {pages} {214103}}\BibitemShut {NoStop}%
\bibitem [{\citenamefont {Gomes}\ \emph {et~al.}(2016)\citenamefont {Gomes},
  \citenamefont {Trevisanutto}, \citenamefont {Carvalho}, \citenamefont
  {Rodin},\ and\ \citenamefont {Castro~Neto}}]{gomes_prb_2016_excitons}%
  \BibitemOpen
  \bibfield  {author} {\bibinfo {author} {\bibnamefont {Gomes}, \bibfnamefont
  {L.~C.}}, \bibinfo {author} {\bibfnamefont {P.~E.}\ \bibnamefont
  {Trevisanutto}}, \bibinfo {author} {\bibfnamefont {A.}~\bibnamefont
  {Carvalho}}, \bibinfo {author} {\bibfnamefont {A.~S.}\ \bibnamefont {Rodin}},
  \ and\ \bibinfo {author} {\bibfnamefont {A.~H.}\ \bibnamefont {Castro~Neto}}}
  (\bibinfo {year} {2016}),\ \href {\doibase 10.1103/PhysRevB.94.155428}
  {\bibfield  {journal} {\bibinfo  {journal} {Phys. Rev. B}\ }\textbf {\bibinfo
  {volume} {94}},\ \bibinfo {pages} {155428}}\BibitemShut {NoStop}%
\bibitem [{\citenamefont {Guan}\ \emph {et~al.}(2019)\citenamefont {Guan},
  \citenamefont {Hu}, \citenamefont {Shen}, \citenamefont {Xiang},
  \citenamefont {Zhong}, \citenamefont {Chu},\ and\ \citenamefont
  {Duan}}]{AEM}%
  \BibitemOpen
  \bibfield  {author} {\bibinfo {author} {\bibnamefont {Guan}, \bibfnamefont
  {Z.}}, \bibinfo {author} {\bibfnamefont {H.}~\bibnamefont {Hu}}, \bibinfo
  {author} {\bibfnamefont {X.}~\bibnamefont {Shen}}, \bibinfo {author}
  {\bibfnamefont {P.}~\bibnamefont {Xiang}}, \bibinfo {author} {\bibfnamefont
  {N.}~\bibnamefont {Zhong}}, \bibinfo {author} {\bibfnamefont
  {J.}~\bibnamefont {Chu}}, \ and\ \bibinfo {author} {\bibfnamefont
  {C.}~\bibnamefont {Duan}}} (\bibinfo {year} {2019}),\ \href {\doibase
  10.1002/aelm.201900818} {\bibfield  {journal} {\bibinfo  {journal} {Adv.
  Electron. Mater.}\ }\textbf {\bibinfo {volume} {6}},\ \bibinfo {pages}
  {1900818}}\BibitemShut {NoStop}%
\bibitem [{\citenamefont {Haleoot}\ \emph {et~al.}(2017)\citenamefont
  {Haleoot}, \citenamefont {Paillard}, \citenamefont {Kaloni}, \citenamefont
  {Mehboudi}, \citenamefont {Xu}, \citenamefont {Bellaiche},\ and\
  \citenamefont {Barraza-Lopez}}]{haleoot_prl_2017_sns_snse}%
  \BibitemOpen
  \bibfield  {author} {\bibinfo {author} {\bibnamefont {Haleoot}, \bibfnamefont
  {R.}}, \bibinfo {author} {\bibfnamefont {C.}~\bibnamefont {Paillard}},
  \bibinfo {author} {\bibfnamefont {T.~P.}\ \bibnamefont {Kaloni}}, \bibinfo
  {author} {\bibfnamefont {M.}~\bibnamefont {Mehboudi}}, \bibinfo {author}
  {\bibfnamefont {B.}~\bibnamefont {Xu}}, \bibinfo {author} {\bibfnamefont
  {L.}~\bibnamefont {Bellaiche}}, \ and\ \bibinfo {author} {\bibfnamefont
  {S.}~\bibnamefont {Barraza-Lopez}}} (\bibinfo {year} {2017}),\ \href
  {\doibase 10.1103/PhysRevLett.118.227401} {\bibfield  {journal} {\bibinfo
  {journal} {Phys. Rev. Lett.}\ }\textbf {\bibinfo {volume} {118}},\ \bibinfo
  {pages} {227401}}\BibitemShut {NoStop}%
\bibitem [{\citenamefont {Hanakata}\ \emph {et~al.}(2016)\citenamefont
  {Hanakata}, \citenamefont {Carvalho}, \citenamefont {Campbell},\ and\
  \citenamefont {Park}}]{hanakata_prb_2016_sns_gese}%
  \BibitemOpen
  \bibfield  {author} {\bibinfo {author} {\bibnamefont {Hanakata},
  \bibfnamefont {P.~Z.}}, \bibinfo {author} {\bibfnamefont {A.}~\bibnamefont
  {Carvalho}}, \bibinfo {author} {\bibfnamefont {D.~K.}\ \bibnamefont
  {Campbell}}, \ and\ \bibinfo {author} {\bibfnamefont {H.~S.}\ \bibnamefont
  {Park}}} (\bibinfo {year} {2016}),\ \href {\doibase
  10.1103/PhysRevB.94.035304} {\bibfield  {journal} {\bibinfo  {journal} {Phys.
  Rev. B}\ }\textbf {\bibinfo {volume} {94}},\ \bibinfo {pages}
  {035304}}\BibitemShut {NoStop}%
\bibitem [{\citenamefont {Heyd}\ \emph {et~al.}(2003)\citenamefont {Heyd},
  \citenamefont {Scuseria},\ and\ \citenamefont {Ernzerhof}}]{scuseria}%
  \BibitemOpen
  \bibfield  {author} {\bibinfo {author} {\bibnamefont {Heyd}, \bibfnamefont
  {J.}}, \bibinfo {author} {\bibfnamefont {G.~E.}\ \bibnamefont {Scuseria}}, \
  and\ \bibinfo {author} {\bibfnamefont {M.}~\bibnamefont {Ernzerhof}}}
  (\bibinfo {year} {2003}),\ \href {\doibase 10.1063/1.1564060} {\bibfield
  {journal} {\bibinfo  {journal} {J. Chem. Phys.}\ }\textbf {\bibinfo {volume}
  {118}},\ \bibinfo {pages} {8207}}\BibitemShut {NoStop}%
\bibitem [{\citenamefont {Higashitarumizu}\ \emph {et~al.}(2020)\citenamefont
  {Higashitarumizu}, \citenamefont {Kawamoto}, \citenamefont {Lee},
  \citenamefont {Lin}, \citenamefont {Chu}, \citenamefont {Yonemori},
  \citenamefont {Nishimura}, \citenamefont {Wakabayashi}, \citenamefont
  {Chang},\ and\ \citenamefont {Nagashio}}]{Higashitarumizu20_NC_SnS}%
  \BibitemOpen
  \bibfield  {author} {\bibinfo {author} {\bibnamefont {Higashitarumizu},
  \bibfnamefont {N.}}, \bibinfo {author} {\bibfnamefont {H.}~\bibnamefont
  {Kawamoto}}, \bibinfo {author} {\bibfnamefont {C.-J.}\ \bibnamefont {Lee}},
  \bibinfo {author} {\bibfnamefont {B.-H.}\ \bibnamefont {Lin}}, \bibinfo
  {author} {\bibfnamefont {F.-H.}\ \bibnamefont {Chu}}, \bibinfo {author}
  {\bibfnamefont {I.}~\bibnamefont {Yonemori}}, \bibinfo {author}
  {\bibfnamefont {T.}~\bibnamefont {Nishimura}}, \bibinfo {author}
  {\bibfnamefont {K.}~\bibnamefont {Wakabayashi}}, \bibinfo {author}
  {\bibfnamefont {W.-H.}\ \bibnamefont {Chang}}, \ and\ \bibinfo {author}
  {\bibfnamefont {K.}~\bibnamefont {Nagashio}}} (\bibinfo {year} {2020}),\
  \href {\doibase 10.1038/s41467-020-16291-9} {\bibfield  {journal} {\bibinfo
  {journal} {Nat. Commun.}\ }\textbf {\bibinfo {volume} {11}},\ \bibinfo
  {pages} {2428}}\BibitemShut {NoStop}%
\bibitem [{\citenamefont {Hosur}(2011)}]{Hosur2011}%
  \BibitemOpen
  \bibfield  {author} {\bibinfo {author} {\bibnamefont {Hosur}, \bibfnamefont
  {P.}}} (\bibinfo {year} {2011}),\ \href {\doibase 10.1103/PhysRevB.83.035309}
  {\bibfield  {journal} {\bibinfo  {journal} {Phys. Rev. B}\ }\textbf {\bibinfo
  {volume} {83}},\ \bibinfo {pages} {035309}}\BibitemShut {NoStop}%
\bibitem [{\citenamefont {Hu}\ \emph {et~al.}(2015)\citenamefont {Hu},
  \citenamefont {Zhang}, \citenamefont {Sun}, \citenamefont {Xie},
  \citenamefont {Cai},\ and\ \citenamefont {Zeng}}]{hu_apl_2015_gese}%
  \BibitemOpen
  \bibfield  {author} {\bibinfo {author} {\bibnamefont {Hu}, \bibfnamefont
  {Y.}}, \bibinfo {author} {\bibfnamefont {S.}~\bibnamefont {Zhang}}, \bibinfo
  {author} {\bibfnamefont {S.}~\bibnamefont {Sun}}, \bibinfo {author}
  {\bibfnamefont {M.}~\bibnamefont {Xie}}, \bibinfo {author} {\bibfnamefont
  {B.}~\bibnamefont {Cai}}, \ and\ \bibinfo {author} {\bibfnamefont
  {H.}~\bibnamefont {Zeng}}} (\bibinfo {year} {2015}),\ \href {\doibase
  10.1063/1.4931459} {\bibfield  {journal} {\bibinfo  {journal} {Appl. Phys.
  Lett.}\ }\textbf {\bibinfo {volume} {107}},\ \bibinfo {pages}
  {122107}}\BibitemShut {NoStop}%
\bibitem [{\citenamefont {Ivchenko}\ and\ \citenamefont
  {Ganichev}(2016)}]{ivchenko}%
  \BibitemOpen
  \bibfield  {author} {\bibinfo {author} {\bibnamefont {Ivchenko},
  \bibfnamefont {E.}}, \ and\ \bibinfo {author} {\bibfnamefont
  {S.}~\bibnamefont {Ganichev}}} (\bibinfo {year} {2016}),\ \enquote {\bibinfo
  {title} {Spin physics in semiconductors},}\ Chap.~\bibinfo {chapter} {9}\
  (\bibinfo  {publisher} {Springer Verlag},\ \bibinfo {address}
  {Berlin})\BibitemShut {NoStop}%
\bibitem [{\citenamefont {Jackson}(1998)}]{Jackson1998}%
  \BibitemOpen
  \bibfield  {author} {\bibinfo {author} {\bibnamefont {Jackson}, \bibfnamefont
  {J.~D.}}} (\bibinfo {year} {1998}),\ \href@noop {} {\emph {\bibinfo {title}
  {Classical Electrodynamics}}},\ \bibinfo {edition} {3rd}\ ed.\ (\bibinfo
  {publisher} {Wiley},\ \bibinfo {address} {N.Y.})\BibitemShut {NoStop}%
\bibitem [{\citenamefont {Janisch}\ \emph {et~al.}(2014)\citenamefont
  {Janisch}, \citenamefont {Wang}, \citenamefont {Ma}, \citenamefont {Mehta},
  \citenamefont {Elias}, \citenamefont {Perea-Lopez}, \citenamefont {Terrones},
  \citenamefont {Crespi},\ and\ \citenamefont {Liua}}]{Janisch2014}%
  \BibitemOpen
  \bibfield  {author} {\bibinfo {author} {\bibnamefont {Janisch}, \bibfnamefont
  {C.}}, \bibinfo {author} {\bibfnamefont {Y.}~\bibnamefont {Wang}}, \bibinfo
  {author} {\bibfnamefont {D.}~\bibnamefont {Ma}}, \bibinfo {author}
  {\bibfnamefont {N.}~\bibnamefont {Mehta}}, \bibinfo {author} {\bibfnamefont
  {A.~L.}\ \bibnamefont {Elias}}, \bibinfo {author} {\bibfnamefont
  {N.}~\bibnamefont {Perea-Lopez}}, \bibinfo {author} {\bibfnamefont
  {M.}~\bibnamefont {Terrones}}, \bibinfo {author} {\bibfnamefont
  {V.}~\bibnamefont {Crespi}}, \ and\ \bibinfo {author} {\bibfnamefont
  {Z.}~\bibnamefont {Liua}}} (\bibinfo {year} {2014}),\ \href {\doibase
  10.1038/srep05530} {\bibfield  {journal} {\bibinfo  {journal} {Sci. Rep.}\
  }\textbf {\bibinfo {volume} {4}},\ \bibinfo {pages} {5530}}\BibitemShut
  {NoStop}%
\bibitem [{\citenamefont {Janovec}(1959)}]{janovec}%
  \BibitemOpen
  \bibfield  {author} {\bibinfo {author} {\bibnamefont {Janovec}, \bibfnamefont
  {V.}}} (\bibinfo {year} {1959}),\ \href {\doibase 10.1007/BF01557932}
  {\bibfield  {journal} {\bibinfo  {journal} {Czech. J. Phys.}\ }\textbf
  {\bibinfo {volume} {9}},\ \bibinfo {pages} {468}}\BibitemShut {NoStop}%
\bibitem [{\citenamefont {Jiang}\ and\ \citenamefont {Park}(2014)}]{poisson1}%
  \BibitemOpen
  \bibfield  {author} {\bibinfo {author} {\bibnamefont {Jiang}, \bibfnamefont
  {J.}}, \ and\ \bibinfo {author} {\bibfnamefont {H.}~\bibnamefont {Park}}}
  (\bibinfo {year} {2014}),\ \href {\doibase 10.1038/ncomms5727} {\bibfield
  {journal} {\bibinfo  {journal} {Nat. Commun.}\ }\textbf {\bibinfo {volume}
  {5}},\ \bibinfo {pages} {4727}}\BibitemShut {NoStop}%
\bibitem [{\citenamefont {de~Juan}\ \emph {et~al.}(2017)\citenamefont
  {de~Juan}, \citenamefont {Grushin}, \citenamefont {Morimoto},\ and\
  \citenamefont {Moore}}]{Juan2017}%
  \BibitemOpen
  \bibfield  {author} {\bibinfo {author} {\bibnamefont {de~Juan}, \bibfnamefont
  {F.}}, \bibinfo {author} {\bibfnamefont {A.~G.}\ \bibnamefont {Grushin}},
  \bibinfo {author} {\bibfnamefont {T.}~\bibnamefont {Morimoto}}, \ and\
  \bibinfo {author} {\bibfnamefont {J.~E.}\ \bibnamefont {Moore}}} (\bibinfo
  {year} {2017}),\ \href {\doibase 10.1038/ncomms15995} {\bibfield  {journal}
  {\bibinfo  {journal} {Nat. Commun.}\ }\textbf {\bibinfo {volume} {8}},\
  \bibinfo {pages} {15995}}\BibitemShut {NoStop}%
\bibitem [{\citenamefont {Junquera}\ and\ \citenamefont {Ghosez}(2003)}]{SBL1}%
  \BibitemOpen
  \bibfield  {author} {\bibinfo {author} {\bibnamefont {Junquera},
  \bibfnamefont {J.}}, \ and\ \bibinfo {author} {\bibfnamefont
  {P.}~\bibnamefont {Ghosez}}} (\bibinfo {year} {2003}),\ \href {\doibase
  10.1038/nature01501} {\bibfield  {journal} {\bibinfo  {journal} {Nature}\
  }\textbf {\bibinfo {volume} {422}},\ \bibinfo {pages} {506}}\BibitemShut
  {NoStop}%
\bibitem [{\citenamefont {Kaloni}\ \emph {et~al.}(2019)\citenamefont {Kaloni},
  \citenamefont {Chang}, \citenamefont {Miller}, \citenamefont {Xue},
  \citenamefont {Chen}, \citenamefont {Ji}, \citenamefont {Parkin},\ and\
  \citenamefont {Barraza-Lopez}}]{kaloni2019}%
  \BibitemOpen
  \bibfield  {author} {\bibinfo {author} {\bibnamefont {Kaloni}, \bibfnamefont
  {T.~P.}}, \bibinfo {author} {\bibfnamefont {K.}~\bibnamefont {Chang}},
  \bibinfo {author} {\bibfnamefont {B.~J.}\ \bibnamefont {Miller}}, \bibinfo
  {author} {\bibfnamefont {Q.-K.}\ \bibnamefont {Xue}}, \bibinfo {author}
  {\bibfnamefont {X.}~\bibnamefont {Chen}}, \bibinfo {author} {\bibfnamefont
  {S.-H.}\ \bibnamefont {Ji}}, \bibinfo {author} {\bibfnamefont {S.~S.~P.}\
  \bibnamefont {Parkin}}, \ and\ \bibinfo {author} {\bibfnamefont
  {S.}~\bibnamefont {Barraza-Lopez}}} (\bibinfo {year} {2019}),\ \href
  {\doibase 10.1103/PhysRevB.99.134108} {\bibfield  {journal} {\bibinfo
  {journal} {Phys. Rev. B}\ }\textbf {\bibinfo {volume} {99}},\ \bibinfo
  {pages} {134108}}\BibitemShut {NoStop}%
\bibitem [{\citenamefont {Kamal}\ \emph {et~al.}(2016)\citenamefont {Kamal},
  \citenamefont {Chakrabarti},\ and\ \citenamefont
  {Ezawa}}]{kamal_prb_2016_iv_vi_monolayers}%
  \BibitemOpen
  \bibfield  {author} {\bibinfo {author} {\bibnamefont {Kamal}, \bibfnamefont
  {C.}}, \bibinfo {author} {\bibfnamefont {A.}~\bibnamefont {Chakrabarti}}, \
  and\ \bibinfo {author} {\bibfnamefont {M.}~\bibnamefont {Ezawa}}} (\bibinfo
  {year} {2016}),\ \href {\doibase 10.1103/PhysRevB.93.125428} {\bibfield
  {journal} {\bibinfo  {journal} {Phys. Rev. B}\ }\textbf {\bibinfo {volume}
  {93}},\ \bibinfo {pages} {125428}}\BibitemShut {NoStop}%
\bibitem [{\citenamefont {Keyes}(1953)}]{keyes}%
  \BibitemOpen
  \bibfield  {author} {\bibinfo {author} {\bibnamefont {Keyes}, \bibfnamefont
  {R.~W.}}} (\bibinfo {year} {1953}),\ \href {\doibase 10.1103/PhysRev.92.580}
  {\bibfield  {journal} {\bibinfo  {journal} {Phys. Rev.}\ }\textbf {\bibinfo
  {volume} {92}},\ \bibinfo {pages} {580}}\BibitemShut {NoStop}%
\bibitem [{\citenamefont {Kim}\ \emph {et~al.}(2013)\citenamefont {Kim},
  \citenamefont {Brown}, \citenamefont {Graham}, \citenamefont {Hovden},
  \citenamefont {Havener}, \citenamefont {McEuen}, \citenamefont {Muller},\
  and\ \citenamefont {Park}}]{Kim2013}%
  \BibitemOpen
  \bibfield  {author} {\bibinfo {author} {\bibnamefont {Kim}, \bibfnamefont
  {C.-J.}}, \bibinfo {author} {\bibfnamefont {L.}~\bibnamefont {Brown}},
  \bibinfo {author} {\bibfnamefont {M.~W.}\ \bibnamefont {Graham}}, \bibinfo
  {author} {\bibfnamefont {R.}~\bibnamefont {Hovden}}, \bibinfo {author}
  {\bibfnamefont {R.~W.}\ \bibnamefont {Havener}}, \bibinfo {author}
  {\bibfnamefont {P.~L.}\ \bibnamefont {McEuen}}, \bibinfo {author}
  {\bibfnamefont {D.~A.}\ \bibnamefont {Muller}}, \ and\ \bibinfo {author}
  {\bibfnamefont {J.}~\bibnamefont {Park}}} (\bibinfo {year} {2013}),\ \href
  {\doibase 10.1021/nl403328s} {\bibfield  {journal} {\bibinfo  {journal} {Nano
  Lett.}\ }\textbf {\bibinfo {volume} {13}},\ \bibinfo {pages}
  {5660}}\BibitemShut {NoStop}%
\bibitem [{\citenamefont {King-Smith}\ and\ \citenamefont
  {Vanderbilt}(1993)}]{King-Smith1993}%
  \BibitemOpen
  \bibfield  {author} {\bibinfo {author} {\bibnamefont {King-Smith},
  \bibfnamefont {R.~D.}}, \ and\ \bibinfo {author} {\bibfnamefont
  {D.}~\bibnamefont {Vanderbilt}}} (\bibinfo {year} {1993}),\ \href@noop {}
  {\bibfield  {journal} {\bibinfo  {journal} {Phys. Rev. B}\ }\textbf {\bibinfo
  {volume} {47}},\ \bibinfo {pages} {1651}}\BibitemShut {NoStop}%
\bibitem [{\citenamefont {Kong}\ \emph {et~al.}(2018)\citenamefont {Kong},
  \citenamefont {Deng}, \citenamefont {Li}, \citenamefont {Liu}, \citenamefont
  {Ding}, \citenamefont {Sun},\ and\ \citenamefont {Liu}}]{poisson2}%
  \BibitemOpen
  \bibfield  {author} {\bibinfo {author} {\bibnamefont {Kong}, \bibfnamefont
  {X.}}, \bibinfo {author} {\bibfnamefont {J.}~\bibnamefont {Deng}}, \bibinfo
  {author} {\bibfnamefont {L.}~\bibnamefont {Li}}, \bibinfo {author}
  {\bibfnamefont {Y.}~\bibnamefont {Liu}}, \bibinfo {author} {\bibfnamefont
  {X.}~\bibnamefont {Ding}}, \bibinfo {author} {\bibfnamefont {J.}~\bibnamefont
  {Sun}}, \ and\ \bibinfo {author} {\bibfnamefont {J.~Z.}\ \bibnamefont {Liu}}}
  (\bibinfo {year} {2018}),\ \href {\doibase 10.1103/PhysRevB.98.184104}
  {\bibfield  {journal} {\bibinfo  {journal} {Phys. Rev. B}\ }\textbf {\bibinfo
  {volume} {98}},\ \bibinfo {pages} {184104}}\BibitemShut {NoStop}%
\bibitem [{\citenamefont {Kooi}\ and\ \citenamefont
  {Wuttig}(2020)}]{revchalcogens}%
  \BibitemOpen
  \bibfield  {author} {\bibinfo {author} {\bibnamefont {Kooi}, \bibfnamefont
  {B.~J.}}, \ and\ \bibinfo {author} {\bibfnamefont {M.}~\bibnamefont
  {Wuttig}}} (\bibinfo {year} {2020}),\ \href {\doibase 10.1002/adma.201908302}
  {\bibfield  {journal} {\bibinfo  {journal} {Adv. Mater.}\ }\textbf {\bibinfo
  {volume} {32}},\ \bibinfo {pages} {1908302}}\BibitemShut {NoStop}%
\bibitem [{\citenamefont {Kosterlitz}(2016)}]{KostRev2016}%
  \BibitemOpen
  \bibfield  {author} {\bibinfo {author} {\bibnamefont {Kosterlitz},
  \bibfnamefont {J.~M.}}} (\bibinfo {year} {2016}),\ \href {\doibase
  10.1088/0034-4885/79/2/026001} {\bibfield  {journal} {\bibinfo  {journal}
  {Rep. Prog. Phys.}\ }\textbf {\bibinfo {volume} {79}},\ \bibinfo {pages}
  {026001}}\BibitemShut {NoStop}%
\bibitem [{\citenamefont {Kou}\ \emph {et~al.}(2018)\citenamefont {Kou},
  \citenamefont {Fu}, \citenamefont {Ma}, \citenamefont {Yan}, \citenamefont
  {Ting~Liao},\ and\ \citenamefont {Chen}}]{bi_band_splitting_2018}%
  \BibitemOpen
  \bibfield  {author} {\bibinfo {author} {\bibnamefont {Kou}, \bibfnamefont
  {L.}}, \bibinfo {author} {\bibfnamefont {H.}~\bibnamefont {Fu}}, \bibinfo
  {author} {\bibfnamefont {Y.}~\bibnamefont {Ma}}, \bibinfo {author}
  {\bibfnamefont {B.}~\bibnamefont {Yan}}, \bibinfo {author} {\bibfnamefont
  {A.~D.}\ \bibnamefont {Ting~Liao}}, \ and\ \bibinfo {author} {\bibfnamefont
  {C.}~\bibnamefont {Chen}}} (\bibinfo {year} {2018}),\ \href {\doibase
  10.1103/PhysRevB.97.075429} {\bibfield  {journal} {\bibinfo  {journal} {Phys.
  Rev. B}\ }\textbf {\bibinfo {volume} {97}},\ \bibinfo {pages}
  {075429}}\BibitemShut {NoStop}%
\bibitem [{\citenamefont {Kr\'al}\ \emph {et~al.}(2000)\citenamefont {Kr\'al},
  \citenamefont {Mele},\ and\ \citenamefont {Tom\'anek}}]{Kral2000}%
  \BibitemOpen
  \bibfield  {author} {\bibinfo {author} {\bibnamefont {Kr\'al}, \bibfnamefont
  {P.}}, \bibinfo {author} {\bibfnamefont {E.~J.}\ \bibnamefont {Mele}}, \ and\
  \bibinfo {author} {\bibfnamefont {D.}~\bibnamefont {Tom\'anek}}} (\bibinfo
  {year} {2000}),\ \href {\doibase 10.1103/PhysRevLett.85.1512} {\bibfield
  {journal} {\bibinfo  {journal} {Phys. Rev. Lett.}\ }\textbf {\bibinfo
  {volume} {85}},\ \bibinfo {pages} {1512}}\BibitemShut {NoStop}%
\bibitem [{\citenamefont {Kumar}\ \emph {et~al.}(2013)\citenamefont {Kumar},
  \citenamefont {Najmaei}, \citenamefont {Cui}, \citenamefont {Ceballos},
  \citenamefont {Ajayan}, \citenamefont {Lou},\ and\ \citenamefont
  {Zhao}}]{Kumar2013}%
  \BibitemOpen
  \bibfield  {author} {\bibinfo {author} {\bibnamefont {Kumar}, \bibfnamefont
  {N.}}, \bibinfo {author} {\bibfnamefont {S.}~\bibnamefont {Najmaei}},
  \bibinfo {author} {\bibfnamefont {Q.}~\bibnamefont {Cui}}, \bibinfo {author}
  {\bibfnamefont {F.}~\bibnamefont {Ceballos}}, \bibinfo {author}
  {\bibfnamefont {P.~M.}\ \bibnamefont {Ajayan}}, \bibinfo {author}
  {\bibfnamefont {J.}~\bibnamefont {Lou}}, \ and\ \bibinfo {author}
  {\bibfnamefont {H.}~\bibnamefont {Zhao}}} (\bibinfo {year} {2013}),\ \href
  {\doibase 10.1103/PhysRevB.87.161403} {\bibfield  {journal} {\bibinfo
  {journal} {Phys. Rev. B}\ }\textbf {\bibinfo {volume} {87}},\ \bibinfo
  {pages} {161403}}\BibitemShut {NoStop}%
\bibitem [{\citenamefont {Kushnir}\ \emph {et~al.}(2019)\citenamefont
  {Kushnir}, \citenamefont {Qin}, \citenamefont {Shen}, \citenamefont {Li},
  \citenamefont {Fregoso}, \citenamefont {Tongay},\ and\ \citenamefont
  {Titova}}]{Kushnir2019}%
  \BibitemOpen
  \bibfield  {author} {\bibinfo {author} {\bibnamefont {Kushnir}, \bibfnamefont
  {K.}}, \bibinfo {author} {\bibfnamefont {Y.}~\bibnamefont {Qin}}, \bibinfo
  {author} {\bibfnamefont {Y.}~\bibnamefont {Shen}}, \bibinfo {author}
  {\bibfnamefont {G.}~\bibnamefont {Li}}, \bibinfo {author} {\bibfnamefont
  {B.~M.}\ \bibnamefont {Fregoso}}, \bibinfo {author} {\bibfnamefont
  {S.}~\bibnamefont {Tongay}}, \ and\ \bibinfo {author} {\bibfnamefont {L.~V.}\
  \bibnamefont {Titova}}} (\bibinfo {year} {2019}),\ \href {\doibase
  10.1021/acsami.8b17225} {\bibfield  {journal} {\bibinfo  {journal} {ACS Appl.
  Mater. Interfaces}\ }\textbf {\bibinfo {volume} {11}},\ \bibinfo {pages}
  {5492}}\BibitemShut {NoStop}%
\bibitem [{\citenamefont {Kushnir}\ \emph {et~al.}(2017)\citenamefont
  {Kushnir}, \citenamefont {Wang}, \citenamefont {Fitzgerald}, \citenamefont
  {Koski},\ and\ \citenamefont {Titova}}]{Kushnir2017}%
  \BibitemOpen
  \bibfield  {author} {\bibinfo {author} {\bibnamefont {Kushnir}, \bibfnamefont
  {K.}}, \bibinfo {author} {\bibfnamefont {M.}~\bibnamefont {Wang}}, \bibinfo
  {author} {\bibfnamefont {P.~D.}\ \bibnamefont {Fitzgerald}}, \bibinfo
  {author} {\bibfnamefont {K.~J.}\ \bibnamefont {Koski}}, \ and\ \bibinfo
  {author} {\bibfnamefont {L.~V.}\ \bibnamefont {Titova}}} (\bibinfo {year}
  {2017}),\ \href {\doibase 10.1021/acsenergylett.7b00330} {\bibfield
  {journal} {\bibinfo  {journal} {ACS Energy Lett.}\ }\textbf {\bibinfo
  {volume} {2}},\ \bibinfo {pages} {1429}}\BibitemShut {NoStop}%
\bibitem [{\citenamefont {Kwon}\ \emph {et~al.}(2020)\citenamefont {Kwon},
  \citenamefont {Zhang}, \citenamefont {Wang}, \citenamefont {Yu},
  \citenamefont {Wang}, \citenamefont {Park}, \citenamefont {Choi},
  \citenamefont {Ma}, \citenamefont {Zhu}, \citenamefont {Tian}, \citenamefont
  {Su},\ and\ \citenamefont {Loh}}]{newmemory}%
  \BibitemOpen
  \bibfield  {author} {\bibinfo {author} {\bibnamefont {Kwon}, \bibfnamefont
  {K.~C.}}, \bibinfo {author} {\bibfnamefont {Y.}~\bibnamefont {Zhang}},
  \bibinfo {author} {\bibfnamefont {L.}~\bibnamefont {Wang}}, \bibinfo {author}
  {\bibfnamefont {W.}~\bibnamefont {Yu}}, \bibinfo {author} {\bibfnamefont
  {X.}~\bibnamefont {Wang}}, \bibinfo {author} {\bibfnamefont {I.-H.}\
  \bibnamefont {Park}}, \bibinfo {author} {\bibfnamefont {H.~S.}\ \bibnamefont
  {Choi}}, \bibinfo {author} {\bibfnamefont {T.}~\bibnamefont {Ma}}, \bibinfo
  {author} {\bibfnamefont {Z.}~\bibnamefont {Zhu}}, \bibinfo {author}
  {\bibfnamefont {B.}~\bibnamefont {Tian}}, \bibinfo {author} {\bibfnamefont
  {C.}~\bibnamefont {Su}}, \ and\ \bibinfo {author} {\bibfnamefont {K.~P.}\
  \bibnamefont {Loh}}} (\bibinfo {year} {2020}),\ \href {\doibase
  10.1021/acsnano.0c03869} {\bibfield  {journal} {\bibinfo  {journal} {ACS
  Nano}\ }\textbf {\bibinfo {volume} {14}},\ \bibinfo {pages}
  {7628}}\BibitemShut {NoStop}%
\bibitem [{\citenamefont {Lee}\ \emph {et~al.}(2008)\citenamefont {Lee},
  \citenamefont {Wei}, \citenamefont {Kysar},\ and\ \citenamefont
  {Hone}}]{Hone}%
  \BibitemOpen
  \bibfield  {author} {\bibinfo {author} {\bibnamefont {Lee}, \bibfnamefont
  {C.}}, \bibinfo {author} {\bibfnamefont {X.}~\bibnamefont {Wei}}, \bibinfo
  {author} {\bibfnamefont {J.~W.}\ \bibnamefont {Kysar}}, \ and\ \bibinfo
  {author} {\bibfnamefont {J.}~\bibnamefont {Hone}}} (\bibinfo {year} {2008}),\
  \href {\doibase 10.1126/science.1157996} {\bibfield  {journal} {\bibinfo
  {journal} {Science}\ }\textbf {\bibinfo {volume} {321}},\ \bibinfo {pages}
  {385}}\BibitemShut {NoStop}%
\bibitem [{\citenamefont {Lee}\ \emph {et~al.}(2020)\citenamefont {Lee},
  \citenamefont {Im},\ and\ \citenamefont {Jin}}]{HosikLee}%
  \BibitemOpen
  \bibfield  {author} {\bibinfo {author} {\bibnamefont {Lee}, \bibfnamefont
  {H.}}, \bibinfo {author} {\bibfnamefont {J.}~\bibnamefont {Im}}, \ and\
  \bibinfo {author} {\bibfnamefont {H.}~\bibnamefont {Jin}}} (\bibinfo {year}
  {2020}),\ \href {\doibase 10.1063/1.5137753} {\bibfield  {journal} {\bibinfo
  {journal} {Appl. Phys. Lett.}\ }\textbf {\bibinfo {volume} {116}},\ \bibinfo
  {pages} {022411}}\BibitemShut {NoStop}%
\bibitem [{\citenamefont {Lefebvre}\ \emph {et~al.}(1998)\citenamefont
  {Lefebvre}, \citenamefont {Szymanski}, \citenamefont {Olivier-Fourcade},\
  and\ \citenamefont {Jumas}}]{lefebvre_prb_1998_sno}%
  \BibitemOpen
  \bibfield  {author} {\bibinfo {author} {\bibnamefont {Lefebvre},
  \bibfnamefont {I.}}, \bibinfo {author} {\bibfnamefont {M.~A.}\ \bibnamefont
  {Szymanski}}, \bibinfo {author} {\bibfnamefont {J.}~\bibnamefont
  {Olivier-Fourcade}}, \ and\ \bibinfo {author} {\bibfnamefont {J.~C.}\
  \bibnamefont {Jumas}}} (\bibinfo {year} {1998}),\ \href {\doibase
  10.1103/PhysRevB.58.1896} {\bibfield  {journal} {\bibinfo  {journal} {Phys.
  Rev. B}\ }\textbf {\bibinfo {volume} {58}},\ \bibinfo {pages}
  {1896}}\BibitemShut {NoStop}%
\bibitem [{\citenamefont {Li}\ \emph {et~al.}(2012)\citenamefont {Li},
  \citenamefont {Zhang}, \citenamefont {Yap}, \citenamefont {Tay},
  \citenamefont {Edwin}, \citenamefont {Olivier},\ and\ \citenamefont
  {Baillargeat}}]{advfunctmat}%
  \BibitemOpen
  \bibfield  {author} {\bibinfo {author} {\bibnamefont {Li}, \bibfnamefont
  {H.}}, \bibinfo {author} {\bibfnamefont {Q.}~\bibnamefont {Zhang}}, \bibinfo
  {author} {\bibfnamefont {C.~C.~R.}\ \bibnamefont {Yap}}, \bibinfo {author}
  {\bibfnamefont {B.~K.}\ \bibnamefont {Tay}}, \bibinfo {author} {\bibfnamefont
  {T.~H.~T.}\ \bibnamefont {Edwin}}, \bibinfo {author} {\bibfnamefont
  {A.}~\bibnamefont {Olivier}}, \ and\ \bibinfo {author} {\bibfnamefont
  {D.}~\bibnamefont {Baillargeat}}} (\bibinfo {year} {2012}),\ \href {\doibase
  10.1002/adfm.201102111} {\bibfield  {journal} {\bibinfo  {journal} {Adv.
  Funct. Mater.}\ }\textbf {\bibinfo {volume} {22}},\ \bibinfo {pages}
  {1385}}\BibitemShut {NoStop}%
\bibitem [{\citenamefont {Li}\ and\ \citenamefont {Li}(2015)}]{li2015}%
  \BibitemOpen
  \bibfield  {author} {\bibinfo {author} {\bibnamefont {Li}, \bibfnamefont
  {W.}}, \ and\ \bibinfo {author} {\bibfnamefont {J.}~\bibnamefont {Li}}}
  (\bibinfo {year} {2015}),\ \href {\doibase 10.1007/s12274-015-0878-8}
  {\bibfield  {journal} {\bibinfo  {journal} {Nano Res.}\ }\textbf {\bibinfo
  {volume} {8}},\ \bibinfo {pages} {3796}}\BibitemShut {NoStop}%
\bibitem [{\citenamefont {Li}\ \emph {et~al.}(2013)\citenamefont {Li},
  \citenamefont {Rao}, \citenamefont {Mak}, \citenamefont {You}, \citenamefont
  {Wang}, \citenamefont {Dean},\ and\ \citenamefont {Heinz}}]{Li2013}%
  \BibitemOpen
  \bibfield  {author} {\bibinfo {author} {\bibnamefont {Li}, \bibfnamefont
  {Y.}}, \bibinfo {author} {\bibfnamefont {Y.}~\bibnamefont {Rao}}, \bibinfo
  {author} {\bibfnamefont {K.~F.}\ \bibnamefont {Mak}}, \bibinfo {author}
  {\bibfnamefont {Y.}~\bibnamefont {You}}, \bibinfo {author} {\bibfnamefont
  {S.}~\bibnamefont {Wang}}, \bibinfo {author} {\bibfnamefont {C.~R.}\
  \bibnamefont {Dean}}, \ and\ \bibinfo {author} {\bibfnamefont {T.~F.}\
  \bibnamefont {Heinz}}} (\bibinfo {year} {2013}),\ \href {\doibase
  10.1021/nl401561r} {\bibfield  {journal} {\bibinfo  {journal} {Nano Lett.}\
  }\textbf {\bibinfo {volume} {13}},\ \bibinfo {pages} {3329}}\BibitemShut
  {NoStop}%
\bibitem [{\citenamefont {Liao}\ \emph {et~al.}(2015)\citenamefont {Liao},
  \citenamefont {Zhang}, \citenamefont {Hu}, \citenamefont {Mao}, \citenamefont
  {Ye}, \citenamefont {Li}, \citenamefont {Huang},\ and\ \citenamefont
  {Xiong}}]{ba2pbcl4_bulk}%
  \BibitemOpen
  \bibfield  {author} {\bibinfo {author} {\bibnamefont {Liao}, \bibfnamefont
  {W.-Q.}}, \bibinfo {author} {\bibfnamefont {Y.}~\bibnamefont {Zhang}},
  \bibinfo {author} {\bibfnamefont {C.-L.}\ \bibnamefont {Hu}}, \bibinfo
  {author} {\bibfnamefont {J.-G.}\ \bibnamefont {Mao}}, \bibinfo {author}
  {\bibfnamefont {H.-Y.}\ \bibnamefont {Ye}}, \bibinfo {author} {\bibfnamefont
  {P.-F.}\ \bibnamefont {Li}}, \bibinfo {author} {\bibfnamefont {S.~D.}\
  \bibnamefont {Huang}}, \ and\ \bibinfo {author} {\bibfnamefont {R.-G.}\
  \bibnamefont {Xiong}}} (\bibinfo {year} {2015}),\ \href {\doibase
  10.1038/ncomms8338} {\bibfield  {journal} {\bibinfo  {journal} {Nat.
  Commun.}\ }\textbf {\bibinfo {volume} {6}},\ \bibinfo {pages}
  {7338}}\BibitemShut {NoStop}%
\bibitem [{\citenamefont {Lichtensteiger}\ \emph {et~al.}(2005)\citenamefont
  {Lichtensteiger}, \citenamefont {Triscone}, \citenamefont {Junquera},\ and\
  \citenamefont {Ghosez}}]{ref31}%
  \BibitemOpen
  \bibfield  {author} {\bibinfo {author} {\bibnamefont {Lichtensteiger},
  \bibfnamefont {C.}}, \bibinfo {author} {\bibfnamefont {J.-M.}\ \bibnamefont
  {Triscone}}, \bibinfo {author} {\bibfnamefont {J.}~\bibnamefont {Junquera}},
  \ and\ \bibinfo {author} {\bibfnamefont {P.}~\bibnamefont {Ghosez}}}
  (\bibinfo {year} {2005}),\ \href {\doibase 10.1103/PhysRevLett.94.047603}
  {\bibfield  {journal} {\bibinfo  {journal} {Phys. Rev. Lett.}\ }\textbf
  {\bibinfo {volume} {94}},\ \bibinfo {pages} {047603}}\BibitemShut {NoStop}%
\bibitem [{\citenamefont {Littlewood}(1980)}]{littlewood_jpc_1980_iv-vi_bulk}%
  \BibitemOpen
  \bibfield  {author} {\bibinfo {author} {\bibnamefont {Littlewood},
  \bibfnamefont {P.~B.}}} (\bibinfo {year} {1980}),\ \href {\doibase
  10.1088/0022-3719/13/26/009} {\bibfield  {journal} {\bibinfo  {journal} {J.
  Phys. C: Solid State Phys.}\ }\textbf {\bibinfo {volume} {13}},\ \bibinfo
  {pages} {4855}}\BibitemShut {NoStop}%
\bibitem [{\citenamefont {Liu}\ \emph {et~al.}(2019{\natexlab{a}})\citenamefont
  {Liu}, \citenamefont {Niu}, \citenamefont {Fu}, \citenamefont {Xi},
  \citenamefont {Lei},\ and\ \citenamefont {Quhe}}]{poisson3}%
  \BibitemOpen
  \bibfield  {author} {\bibinfo {author} {\bibnamefont {Liu}, \bibfnamefont
  {B.}}, \bibinfo {author} {\bibfnamefont {M.}~\bibnamefont {Niu}}, \bibinfo
  {author} {\bibfnamefont {J.}~\bibnamefont {Fu}}, \bibinfo {author}
  {\bibfnamefont {Z.}~\bibnamefont {Xi}}, \bibinfo {author} {\bibfnamefont
  {M.}~\bibnamefont {Lei}}, \ and\ \bibinfo {author} {\bibfnamefont
  {R.}~\bibnamefont {Quhe}}} (\bibinfo {year} {2019}{\natexlab{a}}),\ \href
  {\doibase 10.1103/PhysRevMaterials.3.054002} {\bibfield  {journal} {\bibinfo
  {journal} {Phys. Rev. Mater.}\ }\textbf {\bibinfo {volume} {3}},\ \bibinfo
  {pages} {054002}}\BibitemShut {NoStop}%
\bibitem [{\citenamefont {Liu}\ \emph {et~al.}(2016)\citenamefont {Liu},
  \citenamefont {You}, \citenamefont {Seyler}, \citenamefont {Li},
  \citenamefont {Yu}, \citenamefont {Lin}, \citenamefont {Wang}, \citenamefont
  {Zhou}, \citenamefont {Wang}, \citenamefont {He}, \citenamefont {Pantelides},
  \citenamefont {Zhou}, \citenamefont {Sharma}, \citenamefont {Xu},
  \citenamefont {Ajayan}, \citenamefont {Wang},\ and\ \citenamefont
  {Liu}}]{cips}%
  \BibitemOpen
  \bibfield  {author} {\bibinfo {author} {\bibnamefont {Liu}, \bibfnamefont
  {F.}}, \bibinfo {author} {\bibfnamefont {L.}~\bibnamefont {You}}, \bibinfo
  {author} {\bibfnamefont {K.~L.}\ \bibnamefont {Seyler}}, \bibinfo {author}
  {\bibfnamefont {X.}~\bibnamefont {Li}}, \bibinfo {author} {\bibfnamefont
  {P.}~\bibnamefont {Yu}}, \bibinfo {author} {\bibfnamefont {J.}~\bibnamefont
  {Lin}}, \bibinfo {author} {\bibfnamefont {X.}~\bibnamefont {Wang}}, \bibinfo
  {author} {\bibfnamefont {J.}~\bibnamefont {Zhou}}, \bibinfo {author}
  {\bibfnamefont {H.}~\bibnamefont {Wang}}, \bibinfo {author} {\bibfnamefont
  {H.}~\bibnamefont {He}}, \bibinfo {author} {\bibfnamefont {S.~T.}\
  \bibnamefont {Pantelides}}, \bibinfo {author} {\bibfnamefont
  {W.}~\bibnamefont {Zhou}}, \bibinfo {author} {\bibfnamefont {P.}~\bibnamefont
  {Sharma}}, \bibinfo {author} {\bibfnamefont {X.}~\bibnamefont {Xu}}, \bibinfo
  {author} {\bibfnamefont {P.~M.}\ \bibnamefont {Ajayan}}, \bibinfo {author}
  {\bibfnamefont {J.}~\bibnamefont {Wang}}, \ and\ \bibinfo {author}
  {\bibfnamefont {Z.}~\bibnamefont {Liu}}} (\bibinfo {year} {2016}),\ \href
  {\doibase 10.1038/ncomms12357} {\bibfield  {journal} {\bibinfo  {journal}
  {Nat. Commun.}\ }\textbf {\bibinfo {volume} {7}},\ \bibinfo {pages}
  {12357}}\BibitemShut {NoStop}%
\bibitem [{\citenamefont {Liu}\ \emph {et~al.}(2018)\citenamefont {Liu},
  \citenamefont {Lu}, \citenamefont {Picozzi}, \citenamefont {Bellaiche},\ and\
  \citenamefont {Xiang}}]{liu_prl_2018_snte}%
  \BibitemOpen
  \bibfield  {author} {\bibinfo {author} {\bibnamefont {Liu}, \bibfnamefont
  {K.}}, \bibinfo {author} {\bibfnamefont {J.}~\bibnamefont {Lu}}, \bibinfo
  {author} {\bibfnamefont {S.}~\bibnamefont {Picozzi}}, \bibinfo {author}
  {\bibfnamefont {L.}~\bibnamefont {Bellaiche}}, \ and\ \bibinfo {author}
  {\bibfnamefont {H.}~\bibnamefont {Xiang}}} (\bibinfo {year} {2018}),\ \href
  {\doibase 10.1103/PhysRevLett.121.027601} {\bibfield  {journal} {\bibinfo
  {journal} {Phys. Rev. Lett.}\ }\textbf {\bibinfo {volume} {121}},\ \bibinfo
  {pages} {027601}}\BibitemShut {NoStop}%
\bibitem [{\citenamefont {Liu}\ \emph {et~al.}(2019{\natexlab{b}})\citenamefont
  {Liu}, \citenamefont {Luo}, \citenamefont {Ji}, \citenamefont {Barone},
  \citenamefont {Picozzi},\ and\ \citenamefont {Xiang}}]{band_splitting_2019}%
  \BibitemOpen
  \bibfield  {author} {\bibinfo {author} {\bibnamefont {Liu}, \bibfnamefont
  {K.}}, \bibinfo {author} {\bibfnamefont {W.}~\bibnamefont {Luo}}, \bibinfo
  {author} {\bibfnamefont {J.}~\bibnamefont {Ji}}, \bibinfo {author}
  {\bibfnamefont {P.}~\bibnamefont {Barone}}, \bibinfo {author} {\bibfnamefont
  {S.}~\bibnamefont {Picozzi}}, \ and\ \bibinfo {author} {\bibfnamefont
  {H.}~\bibnamefont {Xiang}}} (\bibinfo {year} {2019}{\natexlab{b}}),\ \href
  {\doibase 10.1038/s41467-019-13197-z} {\bibfield  {journal} {\bibinfo
  {journal} {Nat. Commun.}\ }\textbf {\bibinfo {volume} {10}},\ \bibinfo
  {pages} {5144}}\BibitemShut {NoStop}%
\bibitem [{\citenamefont {Mak}\ \emph {et~al.}(2012)\citenamefont {Mak},
  \citenamefont {He}, \citenamefont {Shan},\ and\ \citenamefont
  {Heinz}}]{valley_polarization_2012_1}%
  \BibitemOpen
  \bibfield  {author} {\bibinfo {author} {\bibnamefont {Mak}, \bibfnamefont
  {K.~F.}}, \bibinfo {author} {\bibfnamefont {K.}~\bibnamefont {He}}, \bibinfo
  {author} {\bibfnamefont {J.}~\bibnamefont {Shan}}, \ and\ \bibinfo {author}
  {\bibfnamefont {T.~F.}\ \bibnamefont {Heinz}}} (\bibinfo {year} {2012}),\
  \href {\doibase 10.1038/nnano.2012.96} {\bibfield  {journal} {\bibinfo
  {journal} {Nat. Nanotechnol.}\ }\textbf {\bibinfo {volume} {7}},\ \bibinfo
  {pages} {494}}\BibitemShut {NoStop}%
\bibitem [{\citenamefont {Mak}\ \emph {et~al.}(2010)\citenamefont {Mak},
  \citenamefont {Lee}, \citenamefont {Hone}, \citenamefont {Shan},\ and\
  \citenamefont {Heinz}}]{mak1}%
  \BibitemOpen
  \bibfield  {author} {\bibinfo {author} {\bibnamefont {Mak}, \bibfnamefont
  {K.~F.}}, \bibinfo {author} {\bibfnamefont {C.}~\bibnamefont {Lee}}, \bibinfo
  {author} {\bibfnamefont {J.}~\bibnamefont {Hone}}, \bibinfo {author}
  {\bibfnamefont {J.}~\bibnamefont {Shan}}, \ and\ \bibinfo {author}
  {\bibfnamefont {T.~F.}\ \bibnamefont {Heinz}}} (\bibinfo {year} {2010}),\
  \href {\doibase 10.1103/PhysRevLett.105.136805} {\bibfield  {journal}
  {\bibinfo  {journal} {Phys. Rev. Lett.}\ }\textbf {\bibinfo {volume} {105}},\
  \bibinfo {pages} {136805}}\BibitemShut {NoStop}%
\bibitem [{\citenamefont {Mak}\ \emph {et~al.}(2014)\citenamefont {Mak},
  \citenamefont {McGill}, \citenamefont {Park},\ and\ \citenamefont
  {McEuen}}]{valley_hall_mos2_2014}%
  \BibitemOpen
  \bibfield  {author} {\bibinfo {author} {\bibnamefont {Mak}, \bibfnamefont
  {K.~F.}}, \bibinfo {author} {\bibfnamefont {K.~L.}\ \bibnamefont {McGill}},
  \bibinfo {author} {\bibfnamefont {J.}~\bibnamefont {Park}}, \ and\ \bibinfo
  {author} {\bibfnamefont {P.~L.}\ \bibnamefont {McEuen}}} (\bibinfo {year}
  {2014}),\ \href {\doibase 10.1126/science.1250140} {\bibfield  {journal}
  {\bibinfo  {journal} {Science}\ }\textbf {\bibinfo {volume} {344}},\ \bibinfo
  {pages} {1489}}\BibitemShut {NoStop}%
\bibitem [{\citenamefont {Malard}\ \emph {et~al.}(2013)\citenamefont {Malard},
  \citenamefont {Alencar}, \citenamefont {Barboza}, \citenamefont {Mak},\ and\
  \citenamefont {de~Paula}}]{Malard2013}%
  \BibitemOpen
  \bibfield  {author} {\bibinfo {author} {\bibnamefont {Malard}, \bibfnamefont
  {L.~M.}}, \bibinfo {author} {\bibfnamefont {T.~V.}\ \bibnamefont {Alencar}},
  \bibinfo {author} {\bibfnamefont {A.~P.~M.}\ \bibnamefont {Barboza}},
  \bibinfo {author} {\bibfnamefont {K.~F.}\ \bibnamefont {Mak}}, \ and\
  \bibinfo {author} {\bibfnamefont {A.~M.}\ \bibnamefont {de~Paula}}} (\bibinfo
  {year} {2013}),\ \href {\doibase 10.1103/PhysRevB.87.201401} {\bibfield
  {journal} {\bibinfo  {journal} {Phys. Rev. B}\ }\textbf {\bibinfo {volume}
  {87}},\ \bibinfo {pages} {201401}}\BibitemShut {NoStop}%
\bibitem [{\citenamefont {Malone}\ and\ \citenamefont
  {Kaxiras}(2013)}]{malone}%
  \BibitemOpen
  \bibfield  {author} {\bibinfo {author} {\bibnamefont {Malone}, \bibfnamefont
  {B.~D.}}, \ and\ \bibinfo {author} {\bibfnamefont {E.}~\bibnamefont
  {Kaxiras}}} (\bibinfo {year} {2013}),\ \href {\doibase
  10.1103/PhysRevB.87.245312} {\bibfield  {journal} {\bibinfo  {journal} {Phys.
  Rev. B}\ }\textbf {\bibinfo {volume} {87}},\ \bibinfo {pages}
  {245312}}\BibitemShut {NoStop}%
\bibitem [{\citenamefont {Martin}(2004)}]{martin}%
  \BibitemOpen
  \bibfield  {author} {\bibinfo {author} {\bibnamefont {Martin}, \bibfnamefont
  {R.~M.}}} (\bibinfo {year} {2004}),\ \href {\doibase
  10.1017/CBO9780511805769} {\emph {\bibinfo {title} {{Electronic Structure:
  Basic Theory and Practical Methods}}}},\ \bibinfo {edition} {1st}\ ed.\
  (\bibinfo  {publisher} {Cambridge U. Press},\ \bibinfo {address} {Cambdridge,
  UK})\BibitemShut {NoStop}%
\bibitem [{\citenamefont {Mehboudi}\ \emph
  {et~al.}(2016{\natexlab{a}})\citenamefont {Mehboudi}, \citenamefont {Dorio},
  \citenamefont {Zhu}, \citenamefont {van~der Zande}, \citenamefont
  {Churchill}, \citenamefont {Pacheco-Sanjuan}, \citenamefont {Harriss},
  \citenamefont {Kumar},\ and\ \citenamefont {Barraza-Lopez}}]{Mehboudi2016}%
  \BibitemOpen
  \bibfield  {author} {\bibinfo {author} {\bibnamefont {Mehboudi},
  \bibfnamefont {M.}}, \bibinfo {author} {\bibfnamefont {A.~M.}\ \bibnamefont
  {Dorio}}, \bibinfo {author} {\bibfnamefont {W.}~\bibnamefont {Zhu}}, \bibinfo
  {author} {\bibfnamefont {A.}~\bibnamefont {van~der Zande}}, \bibinfo {author}
  {\bibfnamefont {H.~O.~H.}\ \bibnamefont {Churchill}}, \bibinfo {author}
  {\bibfnamefont {A.~A.}\ \bibnamefont {Pacheco-Sanjuan}}, \bibinfo {author}
  {\bibfnamefont {E.~O.}\ \bibnamefont {Harriss}}, \bibinfo {author}
  {\bibfnamefont {P.}~\bibnamefont {Kumar}}, \ and\ \bibinfo {author}
  {\bibfnamefont {S.}~\bibnamefont {Barraza-Lopez}}} (\bibinfo {year}
  {2016}{\natexlab{a}}),\ \href {\doibase 10.1021/acs.nanolett.5b04613}
  {\bibfield  {journal} {\bibinfo  {journal} {Nano Lett.}\ }\textbf {\bibinfo
  {volume} {16}},\ \bibinfo {pages} {1704}}\BibitemShut {NoStop}%
\bibitem [{\citenamefont {Mehboudi}\ \emph
  {et~al.}(2016{\natexlab{b}})\citenamefont {Mehboudi}, \citenamefont
  {Fregoso}, \citenamefont {Yang}, \citenamefont {Zhu}, \citenamefont {van~der
  Zande}, \citenamefont {Ferrer}, \citenamefont {Bellaiche}, \citenamefont
  {Kumar},\ and\ \citenamefont {Barraza-Lopez}}]{other2}%
  \BibitemOpen
  \bibfield  {author} {\bibinfo {author} {\bibnamefont {Mehboudi},
  \bibfnamefont {M.}}, \bibinfo {author} {\bibfnamefont {B.~M.}\ \bibnamefont
  {Fregoso}}, \bibinfo {author} {\bibfnamefont {Y.}~\bibnamefont {Yang}},
  \bibinfo {author} {\bibfnamefont {W.}~\bibnamefont {Zhu}}, \bibinfo {author}
  {\bibfnamefont {A.}~\bibnamefont {van~der Zande}}, \bibinfo {author}
  {\bibfnamefont {J.}~\bibnamefont {Ferrer}}, \bibinfo {author} {\bibfnamefont
  {L.}~\bibnamefont {Bellaiche}}, \bibinfo {author} {\bibfnamefont
  {P.}~\bibnamefont {Kumar}}, \ and\ \bibinfo {author} {\bibfnamefont
  {S.}~\bibnamefont {Barraza-Lopez}}} (\bibinfo {year} {2016}{\natexlab{b}}),\
  \href {\doibase 10.1103/PhysRevLett.117.246802} {\bibfield  {journal}
  {\bibinfo  {journal} {Phys. Rev. Lett.}\ }\textbf {\bibinfo {volume} {117}},\
  \bibinfo {pages} {246802}}\BibitemShut {NoStop}%
\bibitem [{\citenamefont {Mehta}\ \emph {et~al.}(1973)\citenamefont {Mehta},
  \citenamefont {Silverman},\ and\ \citenamefont {Jacobs}}]{mehta}%
  \BibitemOpen
  \bibfield  {author} {\bibinfo {author} {\bibnamefont {Mehta}, \bibfnamefont
  {R.~R.}}, \bibinfo {author} {\bibfnamefont {B.~D.}\ \bibnamefont
  {Silverman}}, \ and\ \bibinfo {author} {\bibfnamefont {J.~T.}\ \bibnamefont
  {Jacobs}}} (\bibinfo {year} {1973}),\ \href {\doibase 10.1063/1.1662770}
  {\bibfield  {journal} {\bibinfo  {journal} {J. Appl. Phys.}\ }\textbf
  {\bibinfo {volume} {44}},\ \bibinfo {pages} {3379}}\BibitemShut {NoStop}%
\bibitem [{\citenamefont {Merz}(1956)}]{merz}%
  \BibitemOpen
  \bibfield  {author} {\bibinfo {author} {\bibnamefont {Merz}, \bibfnamefont
  {W.~J.}}} (\bibinfo {year} {1956}),\ \href {\doibase 10.1063/1.1722518}
  {\bibfield  {journal} {\bibinfo  {journal} {J. Appl. Phys.}\ }\textbf
  {\bibinfo {volume} {27}},\ \bibinfo {pages} {938}}\BibitemShut {NoStop}%
\bibitem [{\citenamefont {Meyer}\ and\ \citenamefont
  {Vanderbilt}(2001)}]{Meyer_2001}%
  \BibitemOpen
  \bibfield  {author} {\bibinfo {author} {\bibnamefont {Meyer}, \bibfnamefont
  {B.}}, \ and\ \bibinfo {author} {\bibfnamefont {D.}~\bibnamefont
  {Vanderbilt}}} (\bibinfo {year} {2001}),\ \href {\doibase
  10.1103/PhysRevB.63.205426} {\bibfield  {journal} {\bibinfo  {journal} {Phys.
  Rev. B}\ }\textbf {\bibinfo {volume} {63}},\ \bibinfo {pages}
  {205426}}\BibitemShut {NoStop}%
\bibitem [{\citenamefont {Naumis}\ \emph {et~al.}(2017)\citenamefont {Naumis},
  \citenamefont {Barraza-Lopez}, \citenamefont {Oliva-Leyva},\ and\
  \citenamefont {Terrones}}]{Review}%
  \BibitemOpen
  \bibfield  {author} {\bibinfo {author} {\bibnamefont {Naumis}, \bibfnamefont
  {G.~G.}}, \bibinfo {author} {\bibfnamefont {S.}~\bibnamefont
  {Barraza-Lopez}}, \bibinfo {author} {\bibfnamefont {M.}~\bibnamefont
  {Oliva-Leyva}}, \ and\ \bibinfo {author} {\bibfnamefont {H.}~\bibnamefont
  {Terrones}}} (\bibinfo {year} {2017}),\ \href {\doibase
  https://doi.org/10.1088/1361-6633/aa74ef} {\bibfield  {journal} {\bibinfo
  {journal} {Rep. Prog. Phys.}\ }\textbf {\bibinfo {volume} {80}},\ \bibinfo
  {pages} {096501}}\BibitemShut {NoStop}%
\bibitem [{\citenamefont {Nordlander}\ \emph {et~al.}(2019)\citenamefont
  {Nordlander}, \citenamefont {Campanini}, \citenamefont {Rossell},
  \citenamefont {Erni}, \citenamefont {Meier}, \citenamefont {Cano},
  \citenamefont {Spaldin}, \citenamefont {Fiebig},\ and\ \citenamefont
  {Trassin}}]{Nordlander_2019}%
  \BibitemOpen
  \bibfield  {author} {\bibinfo {author} {\bibnamefont {Nordlander},
  \bibfnamefont {J.}}, \bibinfo {author} {\bibfnamefont {M.}~\bibnamefont
  {Campanini}}, \bibinfo {author} {\bibfnamefont {M.~D.}\ \bibnamefont
  {Rossell}}, \bibinfo {author} {\bibfnamefont {R.}~\bibnamefont {Erni}},
  \bibinfo {author} {\bibfnamefont {Q.~N.}\ \bibnamefont {Meier}}, \bibinfo
  {author} {\bibfnamefont {A.}~\bibnamefont {Cano}}, \bibinfo {author}
  {\bibfnamefont {N.~A.}\ \bibnamefont {Spaldin}}, \bibinfo {author}
  {\bibfnamefont {M.}~\bibnamefont {Fiebig}}, \ and\ \bibinfo {author}
  {\bibfnamefont {M.}~\bibnamefont {Trassin}}} (\bibinfo {year} {2019}),\ \href
  {\doibase 10.1038/s41467-019-13474-x} {\bibfield  {journal} {\bibinfo
  {journal} {Nat. Commun.}\ }\textbf {\bibinfo {volume} {10}},\ \bibinfo
  {pages} {5591}}\BibitemShut {NoStop}%
\bibitem [{\citenamefont {Panday}\ \emph {et~al.}(2019)\citenamefont {Panday},
  \citenamefont {Barraza-Lopez}, \citenamefont {Rangel},\ and\ \citenamefont
  {Fregoso}}]{Panday2019}%
  \BibitemOpen
  \bibfield  {author} {\bibinfo {author} {\bibnamefont {Panday}, \bibfnamefont
  {S.~R.}}, \bibinfo {author} {\bibfnamefont {S.}~\bibnamefont
  {Barraza-Lopez}}, \bibinfo {author} {\bibfnamefont {T.}~\bibnamefont
  {Rangel}}, \ and\ \bibinfo {author} {\bibfnamefont {B.~M.}\ \bibnamefont
  {Fregoso}}} (\bibinfo {year} {2019}),\ \href {\doibase
  10.1103/PhysRevB.100.195305} {\bibfield  {journal} {\bibinfo  {journal}
  {Phys. Rev. B}\ }\textbf {\bibinfo {volume} {100}},\ \bibinfo {pages}
  {195305}}\BibitemShut {NoStop}%
\bibitem [{\citenamefont {Panday}\ and\ \citenamefont {Fregoso}(2017)}]{b4}%
  \BibitemOpen
  \bibfield  {author} {\bibinfo {author} {\bibnamefont {Panday}, \bibfnamefont
  {S.~R.}}, \ and\ \bibinfo {author} {\bibfnamefont {B.~M.}\ \bibnamefont
  {Fregoso}}} (\bibinfo {year} {2017}),\ \href
  {http://stacks.iop.org/0953-8984/29/i=43/a=43LT01} {\bibfield  {journal}
  {\bibinfo  {journal} {J. Phys.: Condens. Matter}\ }\textbf {\bibinfo {volume}
  {29}},\ \bibinfo {pages} {43LT01}}\BibitemShut {NoStop}%
\bibitem [{\citenamefont {Parenteau}\ and\ \citenamefont
  {Carlone}(1990)}]{exptsnse}%
  \BibitemOpen
  \bibfield  {author} {\bibinfo {author} {\bibnamefont {Parenteau},
  \bibfnamefont {M.}}, \ and\ \bibinfo {author} {\bibfnamefont
  {C.}~\bibnamefont {Carlone}}} (\bibinfo {year} {1990}),\ \href {\doibase
  10.1103/PhysRevB.41.5227} {\bibfield  {journal} {\bibinfo  {journal} {Phys.
  Rev. B}\ }\textbf {\bibinfo {volume} {41}},\ \bibinfo {pages}
  {5227}}\BibitemShut {NoStop}%
\bibitem [{\citenamefont {Park}\ \emph {et~al.}(2009)\citenamefont {Park},
  \citenamefont {Eom}, \citenamefont {Lee}, \citenamefont {Da~Silva},
  \citenamefont {Kang}, \citenamefont {Lee},\ and\ \citenamefont
  {Khang}}]{geteexpt}%
  \BibitemOpen
  \bibfield  {author} {\bibinfo {author} {\bibnamefont {Park}, \bibfnamefont
  {J.-W.}}, \bibinfo {author} {\bibfnamefont {S.~H.}\ \bibnamefont {Eom}},
  \bibinfo {author} {\bibfnamefont {H.}~\bibnamefont {Lee}}, \bibinfo {author}
  {\bibfnamefont {J.~L.~F.}\ \bibnamefont {Da~Silva}}, \bibinfo {author}
  {\bibfnamefont {Y.-S.}\ \bibnamefont {Kang}}, \bibinfo {author}
  {\bibfnamefont {T.-Y.}\ \bibnamefont {Lee}}, \ and\ \bibinfo {author}
  {\bibfnamefont {Y.~H.}\ \bibnamefont {Khang}}} (\bibinfo {year} {2009}),\
  \href {\doibase 10.1103/PhysRevB.80.115209} {\bibfield  {journal} {\bibinfo
  {journal} {Phys. Rev. B}\ }\textbf {\bibinfo {volume} {80}},\ \bibinfo
  {pages} {115209}}\BibitemShut {NoStop}%
\bibitem [{\citenamefont {Park}\ \emph {et~al.}(2019)\citenamefont {Park},
  \citenamefont {Choi},\ and\ \citenamefont {Lee}}]{RamanSnS}%
  \BibitemOpen
  \bibfield  {author} {\bibinfo {author} {\bibnamefont {Park}, \bibfnamefont
  {M.}}, \bibinfo {author} {\bibfnamefont {J.~S.}\ \bibnamefont {Choi}}, \ and\
  \bibinfo {author} {\bibfnamefont {H.}~\bibnamefont {Lee}}} (\bibinfo {year}
  {2019}),\ \href {\doibase 10.1038/s41598-019-55577-x} {\bibfield  {journal}
  {\bibinfo  {journal} {Sci. Rep.}\ }\textbf {\bibinfo {volume} {9}},\ \bibinfo
  {pages} {19826}}\BibitemShut {NoStop}%
\bibitem [{\citenamefont {Peelaers}\ and\ \citenamefont {Van~de
  Walle}(2012)}]{massmos21}%
  \BibitemOpen
  \bibfield  {author} {\bibinfo {author} {\bibnamefont {Peelaers},
  \bibfnamefont {H.}}, \ and\ \bibinfo {author} {\bibfnamefont {C.~G.}\
  \bibnamefont {Van~de Walle}}} (\bibinfo {year} {2012}),\ \href {\doibase
  10.1103/PhysRevB.86.241401} {\bibfield  {journal} {\bibinfo  {journal} {Phys.
  Rev. B}\ }\textbf {\bibinfo {volume} {86}},\ \bibinfo {pages}
  {241401}}\BibitemShut {NoStop}%
\bibitem [{\citenamefont {Peng}\ \emph {et~al.}(2016)\citenamefont {Peng},
  \citenamefont {Yang}, \citenamefont {Perdew},\ and\ \citenamefont
  {Sun}}]{scan}%
  \BibitemOpen
  \bibfield  {author} {\bibinfo {author} {\bibnamefont {Peng}, \bibfnamefont
  {H.}}, \bibinfo {author} {\bibfnamefont {Z.-H.}\ \bibnamefont {Yang}},
  \bibinfo {author} {\bibfnamefont {J.~P.}\ \bibnamefont {Perdew}}, \ and\
  \bibinfo {author} {\bibfnamefont {J.}~\bibnamefont {Sun}}} (\bibinfo {year}
  {2016}),\ \href {\doibase 10.1103/PhysRevX.6.041005} {\bibfield  {journal}
  {\bibinfo  {journal} {Phys. Rev. X}\ }\textbf {\bibinfo {volume} {6}},\
  \bibinfo {pages} {041005}}\BibitemShut {NoStop}%
\bibitem [{\citenamefont {Perdew}\ \emph {et~al.}(1996)\citenamefont {Perdew},
  \citenamefont {Burke},\ and\ \citenamefont {Ernzerhof}}]{PBE}%
  \BibitemOpen
  \bibfield  {author} {\bibinfo {author} {\bibnamefont {Perdew}, \bibfnamefont
  {J.~P.}}, \bibinfo {author} {\bibfnamefont {K.}~\bibnamefont {Burke}}, \ and\
  \bibinfo {author} {\bibfnamefont {M.}~\bibnamefont {Ernzerhof}}} (\bibinfo
  {year} {1996}),\ \href {\doibase 10.1103/PhysRevLett.77.3865} {\bibfield
  {journal} {\bibinfo  {journal} {Phys. Rev. Lett.}\ }\textbf {\bibinfo
  {volume} {77}},\ \bibinfo {pages} {3865}}\BibitemShut {NoStop}%
\bibitem [{\citenamefont {Perdew}\ and\ \citenamefont {Zunger}(1981)}]{LDA2}%
  \BibitemOpen
  \bibfield  {author} {\bibinfo {author} {\bibnamefont {Perdew}, \bibfnamefont
  {J.~P.}}, \ and\ \bibinfo {author} {\bibfnamefont {A.}~\bibnamefont
  {Zunger}}} (\bibinfo {year} {1981}),\ \href {\doibase
  10.1103/PhysRevB.23.5048} {\bibfield  {journal} {\bibinfo  {journal} {Phys.
  Rev. B}\ }\textbf {\bibinfo {volume} {23}},\ \bibinfo {pages}
  {5048}}\BibitemShut {NoStop}%
\bibitem [{\citenamefont {Poh}\ \emph {et~al.}(2018)\citenamefont {Poh},
  \citenamefont {Tan}, \citenamefont {Wang}, \citenamefont {Song},
  \citenamefont {Abidi}, \citenamefont {Zhao}, \citenamefont {Dan},
  \citenamefont {Chen}, \citenamefont {Luo}, \citenamefont {Pennycook},
  \citenamefont {Castro-Neto},\ and\ \citenamefont {Loh}}]{poh_nl_2018_in2se3}%
  \BibitemOpen
  \bibfield  {author} {\bibinfo {author} {\bibnamefont {Poh}, \bibfnamefont
  {S.~M.}}, \bibinfo {author} {\bibfnamefont {S.~J.~R.}\ \bibnamefont {Tan}},
  \bibinfo {author} {\bibfnamefont {H.}~\bibnamefont {Wang}}, \bibinfo {author}
  {\bibfnamefont {P.}~\bibnamefont {Song}}, \bibinfo {author} {\bibfnamefont
  {I.~H.}\ \bibnamefont {Abidi}}, \bibinfo {author} {\bibfnamefont
  {X.}~\bibnamefont {Zhao}}, \bibinfo {author} {\bibfnamefont {J.}~\bibnamefont
  {Dan}}, \bibinfo {author} {\bibfnamefont {J.}~\bibnamefont {Chen}}, \bibinfo
  {author} {\bibfnamefont {Z.}~\bibnamefont {Luo}}, \bibinfo {author}
  {\bibfnamefont {S.~J.}\ \bibnamefont {Pennycook}}, \bibinfo {author}
  {\bibfnamefont {A.~H.}\ \bibnamefont {Castro-Neto}}, \ and\ \bibinfo {author}
  {\bibfnamefont {K.~P.}\ \bibnamefont {Loh}}} (\bibinfo {year} {2018}),\ \href
  {\doibase 10.1021/acs.nanolett.8b02688} {\bibfield  {journal} {\bibinfo
  {journal} {Nano Lett.}\ }\textbf {\bibinfo {volume} {10}},\ \bibinfo {pages}
  {6340}}\BibitemShut {NoStop}%
\bibitem [{\citenamefont {Potts}(1952)}]{potts}%
  \BibitemOpen
  \bibfield  {author} {\bibinfo {author} {\bibnamefont {Potts}, \bibfnamefont
  {R.~B.}}} (\bibinfo {year} {1952}),\ \href {\doibase
  10.1017/S0305004100027419} {\bibfield  {journal} {\bibinfo  {journal} {Math.
  Proc. Cambridge Philos. Soc.}\ }\textbf {\bibinfo {volume} {48}},\ \bibinfo
  {pages} {106}}\BibitemShut {NoStop}%
\bibitem [{\citenamefont {Poudel}\ \emph {et~al.}(2019)\citenamefont {Poudel},
  \citenamefont {Villanova},\ and\ \citenamefont {Barraza-Lopez}}]{ourarxiv}%
  \BibitemOpen
  \bibfield  {author} {\bibinfo {author} {\bibnamefont {Poudel}, \bibfnamefont
  {S.~P.}}, \bibinfo {author} {\bibfnamefont {J.~W.}\ \bibnamefont
  {Villanova}}, \ and\ \bibinfo {author} {\bibfnamefont {S.}~\bibnamefont
  {Barraza-Lopez}}} (\bibinfo {year} {2019}),\ \href {\doibase
  10.1103/PhysRevMaterials.3.124004} {\bibfield  {journal} {\bibinfo  {journal}
  {Phys. Rev. Mater.}\ }\textbf {\bibinfo {volume} {3}},\ \bibinfo {pages}
  {124004}}\BibitemShut {NoStop}%
\bibitem [{\citenamefont {Qian}\ \emph {et~al.}(2014)\citenamefont {Qian},
  \citenamefont {Liu}, \citenamefont {Fu},\ and\ \citenamefont
  {Li}}]{qian_science_2014}%
  \BibitemOpen
  \bibfield  {author} {\bibinfo {author} {\bibnamefont {Qian}, \bibfnamefont
  {X.}}, \bibinfo {author} {\bibfnamefont {J.}~\bibnamefont {Liu}}, \bibinfo
  {author} {\bibfnamefont {L.}~\bibnamefont {Fu}}, \ and\ \bibinfo {author}
  {\bibfnamefont {J.}~\bibnamefont {Li}}} (\bibinfo {year} {2014}),\ \href
  {\doibase 10.1126/science.1256815} {\bibfield  {journal} {\bibinfo  {journal}
  {Science}\ }\textbf {\bibinfo {volume} {346}},\ \bibinfo {pages}
  {1344}}\BibitemShut {NoStop}%
\bibitem [{\citenamefont {Rabe}\ \emph {et~al.}(2007)\citenamefont {Rabe},
  \citenamefont {Ahn},\ and\ \citenamefont {Triscone}}]{Rabe2007}%
  \BibitemOpen
  \bibinfo {editor} {\bibnamefont {Rabe}, \bibfnamefont {K.~M.}}, \bibinfo
  {editor} {\bibfnamefont {C.~H.}\ \bibnamefont {Ahn}}, \ and\ \bibinfo
  {editor} {\bibfnamefont {J.-M.}\ \bibnamefont {Triscone}},\ Eds. (\bibinfo
  {year} {2007}),\ \href@noop {} {\emph {\bibinfo {title} {Physics of
  Ferroelectrics: A Modern Perspective}}},\ \bibinfo {edition} {1st}\ ed.\
  (\bibinfo  {publisher} {Springer-Verlag Berlin Heidelberg},\ \bibinfo
  {address} {Berlin})\BibitemShut {NoStop}%
\bibitem [{\citenamefont {Rangel}\ \emph {et~al.}(2017)\citenamefont {Rangel},
  \citenamefont {Fregoso}, \citenamefont {Mendoza}, \citenamefont {Morimoto},
  \citenamefont {Moore},\ and\ \citenamefont {Neaton}}]{Rangel2017}%
  \BibitemOpen
  \bibfield  {author} {\bibinfo {author} {\bibnamefont {Rangel}, \bibfnamefont
  {T.}}, \bibinfo {author} {\bibfnamefont {B.~M.}\ \bibnamefont {Fregoso}},
  \bibinfo {author} {\bibfnamefont {B.~S.}\ \bibnamefont {Mendoza}}, \bibinfo
  {author} {\bibfnamefont {T.}~\bibnamefont {Morimoto}}, \bibinfo {author}
  {\bibfnamefont {J.~E.}\ \bibnamefont {Moore}}, \ and\ \bibinfo {author}
  {\bibfnamefont {J.~B.}\ \bibnamefont {Neaton}}} (\bibinfo {year} {2017}),\
  \href {\doibase 10.1103/PhysRevLett.119.067402} {\bibfield  {journal}
  {\bibinfo  {journal} {Phys. Rev. Lett.}\ }\textbf {\bibinfo {volume} {119}},\
  \bibinfo {pages} {067402}}\BibitemShut {NoStop}%
\bibitem [{\citenamefont {Rappe}\ \emph {et~al.}(2017)\citenamefont {Rappe},
  \citenamefont {Grinberg},\ and\ \citenamefont {Spanier}}]{Rappe7191}%
  \BibitemOpen
  \bibfield  {author} {\bibinfo {author} {\bibnamefont {Rappe}, \bibfnamefont
  {A.~M.}}, \bibinfo {author} {\bibfnamefont {I.}~\bibnamefont {Grinberg}}, \
  and\ \bibinfo {author} {\bibfnamefont {J.~E.}\ \bibnamefont {Spanier}}}
  (\bibinfo {year} {2017}),\ \href {\doibase 10.1073/pnas.1708154114}
  {\bibfield  {journal} {\bibinfo  {journal} {Proc. Natl. Acad. Sci. (USA)}\
  }\textbf {\bibinfo {volume} {114}}~(\bibinfo {number} {28}),\ \bibinfo
  {pages} {7191}}\BibitemShut {NoStop}%
\bibitem [{\citenamefont {Raty}\ \emph {et~al.}(2019)\citenamefont {Raty},
  \citenamefont {Schumacher}, \citenamefont {Golub}, \citenamefont {Deringer},
  \citenamefont {Gatti},\ and\ \citenamefont {Wuttig}}]{advmat2018}%
  \BibitemOpen
  \bibfield  {author} {\bibinfo {author} {\bibnamefont {Raty}, \bibfnamefont
  {J.-Y.}}, \bibinfo {author} {\bibfnamefont {M.}~\bibnamefont {Schumacher}},
  \bibinfo {author} {\bibfnamefont {P.}~\bibnamefont {Golub}}, \bibinfo
  {author} {\bibfnamefont {V.~L.}\ \bibnamefont {Deringer}}, \bibinfo {author}
  {\bibfnamefont {C.}~\bibnamefont {Gatti}}, \ and\ \bibinfo {author}
  {\bibfnamefont {M.}~\bibnamefont {Wuttig}}} (\bibinfo {year} {2019}),\ \href
  {\doibase 10.1002/adma.201806280} {\bibfield  {journal} {\bibinfo  {journal}
  {Adv. Mater.}\ }\textbf {\bibinfo {volume} {31}},\ \bibinfo {pages}
  {1806280}}\BibitemShut {NoStop}%
\bibitem [{\citenamefont {Rees}\ \emph {et~al.}(2019)\citenamefont {Rees},
  \citenamefont {Manna}, \citenamefont {Lu}, \citenamefont {Morimoto},
  \citenamefont {Borrmann}, \citenamefont {Felser}, \citenamefont {Moore},
  \citenamefont {Torchinsky},\ and\ \citenamefont {Orenstein}}]{Rees}%
  \BibitemOpen
  \bibfield  {author} {\bibinfo {author} {\bibnamefont {Rees}, \bibfnamefont
  {D.}}, \bibinfo {author} {\bibfnamefont {K.}~\bibnamefont {Manna}}, \bibinfo
  {author} {\bibfnamefont {B.}~\bibnamefont {Lu}}, \bibinfo {author}
  {\bibfnamefont {T.}~\bibnamefont {Morimoto}}, \bibinfo {author}
  {\bibfnamefont {H.}~\bibnamefont {Borrmann}}, \bibinfo {author}
  {\bibfnamefont {C.}~\bibnamefont {Felser}}, \bibinfo {author} {\bibfnamefont
  {J.}~\bibnamefont {Moore}}, \bibinfo {author} {\bibfnamefont {D.~H.}\
  \bibnamefont {Torchinsky}}, \ and\ \bibinfo {author} {\bibfnamefont
  {J.}~\bibnamefont {Orenstein}}} (\bibinfo {year} {2019}),\ \href@noop {}
  {\enquote {\bibinfo {title} {Quantized photocurrents in the chiral multifold
  fermion system {RhSi}},}\ }\bibinfo {note} {ArXiv:1902.03230}\BibitemShut
  {NoStop}%
\bibitem [{\citenamefont {Resta}(1994)}]{Resta1994}%
  \BibitemOpen
  \bibfield  {author} {\bibinfo {author} {\bibnamefont {Resta}, \bibfnamefont
  {R.}}} (\bibinfo {year} {1994}),\ \href {\doibase 10.1103/RevModPhys.66.899}
  {\bibfield  {journal} {\bibinfo  {journal} {Rev. Mod. Phys.}\ }\textbf
  {\bibinfo {volume} {66}},\ \bibinfo {pages} {899}}\BibitemShut {NoStop}%
\bibitem [{\citenamefont {Rodin}\ \emph
  {et~al.}(2014{\natexlab{a}})\citenamefont {Rodin}, \citenamefont {Carvalho},\
  and\ \citenamefont {Castro~Neto}}]{rod}%
  \BibitemOpen
  \bibfield  {author} {\bibinfo {author} {\bibnamefont {Rodin}, \bibfnamefont
  {A.~S.}}, \bibinfo {author} {\bibfnamefont {A.}~\bibnamefont {Carvalho}}, \
  and\ \bibinfo {author} {\bibfnamefont {A.~H.}\ \bibnamefont {Castro~Neto}}}
  (\bibinfo {year} {2014}{\natexlab{a}}),\ \href {\doibase
  10.1103/PhysRevB.90.075429} {\bibfield  {journal} {\bibinfo  {journal} {Phys.
  Rev. B}\ }\textbf {\bibinfo {volume} {90}},\ \bibinfo {pages}
  {075429}}\BibitemShut {NoStop}%
\bibitem [{\citenamefont {Rodin}\ \emph
  {et~al.}(2014{\natexlab{b}})\citenamefont {Rodin}, \citenamefont {Carvalho},\
  and\ \citenamefont {Castro~Neto}}]{PhysRevLett.112.176801}%
  \BibitemOpen
  \bibfield  {author} {\bibinfo {author} {\bibnamefont {Rodin}, \bibfnamefont
  {A.~S.}}, \bibinfo {author} {\bibfnamefont {A.}~\bibnamefont {Carvalho}}, \
  and\ \bibinfo {author} {\bibfnamefont {A.~H.}\ \bibnamefont {Castro~Neto}}}
  (\bibinfo {year} {2014}{\natexlab{b}}),\ \href {\doibase
  10.1103/PhysRevLett.112.176801} {\bibfield  {journal} {\bibinfo  {journal}
  {Phys. Rev. Lett.}\ }\textbf {\bibinfo {volume} {112}},\ \bibinfo {pages}
  {176801}}\BibitemShut {NoStop}%
\bibitem [{\citenamefont {Rodin}\ \emph {et~al.}(2016)\citenamefont {Rodin},
  \citenamefont {Gomes}, \citenamefont {Carvalho},\ and\ \citenamefont
  {Castro~Neto}}]{rodin_prb_2016_sns}%
  \BibitemOpen
  \bibfield  {author} {\bibinfo {author} {\bibnamefont {Rodin}, \bibfnamefont
  {A.~S.}}, \bibinfo {author} {\bibfnamefont {L.~C.}\ \bibnamefont {Gomes}},
  \bibinfo {author} {\bibfnamefont {A.}~\bibnamefont {Carvalho}}, \ and\
  \bibinfo {author} {\bibfnamefont {A.~H.}\ \bibnamefont {Castro~Neto}}}
  (\bibinfo {year} {2016}),\ \href {\doibase 10.1103/PhysRevB.93.045431}
  {\bibfield  {journal} {\bibinfo  {journal} {Phys. Rev. B}\ }\textbf {\bibinfo
  {volume} {93}},\ \bibinfo {pages} {045431}}\BibitemShut {NoStop}%
\bibitem [{\citenamefont {Ronneberger}\ \emph {et~al.}(2020)\citenamefont
  {Ronneberger}, \citenamefont {Zanolli}, \citenamefont {Wuttig},\ and\
  \citenamefont {Mazzarello}}]{advmatZanolli}%
  \BibitemOpen
  \bibfield  {author} {\bibinfo {author} {\bibnamefont {Ronneberger},
  \bibfnamefont {I.}}, \bibinfo {author} {\bibfnamefont {Z.}~\bibnamefont
  {Zanolli}}, \bibinfo {author} {\bibfnamefont {M.}~\bibnamefont {Wuttig}}, \
  and\ \bibinfo {author} {\bibfnamefont {R.}~\bibnamefont {Mazzarello}}}
  (\bibinfo {year} {2020}),\ \href {\doibase 10.1002/adma.202001033} {\bibfield
   {journal} {\bibinfo  {journal} {Adv. Mater.}\ }\textbf {\bibinfo {volume}
  {32}},\ \bibinfo {pages} {2001033}}\BibitemShut {NoStop}%
\bibitem [{\citenamefont {Rycerz}\ \emph {et~al.}(2007)\citenamefont {Rycerz},
  \citenamefont {Tworzydlo},\ and\ \citenamefont
  {Beenakker}}]{valley_valve_grapehe_2007}%
  \BibitemOpen
  \bibfield  {author} {\bibinfo {author} {\bibnamefont {Rycerz}, \bibfnamefont
  {A.}}, \bibinfo {author} {\bibfnamefont {J.}~\bibnamefont {Tworzydlo}}, \
  and\ \bibinfo {author} {\bibfnamefont {C.~W.~J.}\ \bibnamefont {Beenakker}}}
  (\bibinfo {year} {2007}),\ \href {\doibase 10.1038/nphys547} {\bibfield
  {journal} {\bibinfo  {journal} {Nat. Phys.}\ }\textbf {\bibinfo {volume}
  {3}},\ \bibinfo {pages} {172}}\BibitemShut {NoStop}%
\bibitem [{\citenamefont {Sai}\ \emph {et~al.}(2009)\citenamefont {Sai},
  \citenamefont {Fennie},\ and\ \citenamefont {Demkov}}]{Sai09_PRL}%
  \BibitemOpen
  \bibfield  {author} {\bibinfo {author} {\bibnamefont {Sai}, \bibfnamefont
  {N.}}, \bibinfo {author} {\bibfnamefont {C.~J.}\ \bibnamefont {Fennie}}, \
  and\ \bibinfo {author} {\bibfnamefont {A.~A.}\ \bibnamefont {Demkov}}}
  (\bibinfo {year} {2009}),\ \href {\doibase 10.1103/PhysRevLett.102.107601}
  {\bibfield  {journal} {\bibinfo  {journal} {Phys. Rev. Lett.}\ }\textbf
  {\bibinfo {volume} {102}},\ \bibinfo {pages} {107601}}\BibitemShut {NoStop}%
\bibitem [{\citenamefont {Sai}\ \emph {et~al.}(2005)\citenamefont {Sai},
  \citenamefont {Kolpak},\ and\ \citenamefont {Rappe}}]{Sai_2005}%
  \BibitemOpen
  \bibfield  {author} {\bibinfo {author} {\bibnamefont {Sai}, \bibfnamefont
  {N.}}, \bibinfo {author} {\bibfnamefont {A.~M.}\ \bibnamefont {Kolpak}}, \
  and\ \bibinfo {author} {\bibfnamefont {A.~M.}\ \bibnamefont {Rappe}}}
  (\bibinfo {year} {2005}),\ \href {\doibase 10.1103/PhysRevB.72.020101}
  {\bibfield  {journal} {\bibinfo  {journal} {Phys. Rev. B}\ }\textbf {\bibinfo
  {volume} {72}},\ \bibinfo {pages} {020101}}\BibitemShut {NoStop}%
\bibitem [{\citenamefont {Sallen}\ \emph {et~al.}(2012)\citenamefont {Sallen},
  \citenamefont {Bouet}, \citenamefont {Marie}, \citenamefont {Wang},
  \citenamefont {Zhu}, \citenamefont {Han}, \citenamefont {Lu}, \citenamefont
  {Tan}, \citenamefont {Amand}, \citenamefont {Liu},\ and\ \citenamefont
  {Urbaszek}}]{valley_polarization_2012_4}%
  \BibitemOpen
  \bibfield  {author} {\bibinfo {author} {\bibnamefont {Sallen}, \bibfnamefont
  {G.}}, \bibinfo {author} {\bibfnamefont {L.}~\bibnamefont {Bouet}}, \bibinfo
  {author} {\bibfnamefont {X.}~\bibnamefont {Marie}}, \bibinfo {author}
  {\bibfnamefont {G.}~\bibnamefont {Wang}}, \bibinfo {author} {\bibfnamefont
  {C.~R.}\ \bibnamefont {Zhu}}, \bibinfo {author} {\bibfnamefont {W.~P.}\
  \bibnamefont {Han}}, \bibinfo {author} {\bibfnamefont {Y.}~\bibnamefont
  {Lu}}, \bibinfo {author} {\bibfnamefont {P.~H.}\ \bibnamefont {Tan}},
  \bibinfo {author} {\bibfnamefont {T.}~\bibnamefont {Amand}}, \bibinfo
  {author} {\bibfnamefont {B.~L.}\ \bibnamefont {Liu}}, \ and\ \bibinfo
  {author} {\bibfnamefont {B.}~\bibnamefont {Urbaszek}}} (\bibinfo {year}
  {2012}),\ \href {\doibase 10.1103/PhysRevB.86.081301} {\bibfield  {journal}
  {\bibinfo  {journal} {Phys. Rev. B}\ }\textbf {\bibinfo {volume} {86}},\
  \bibinfo {pages} {081301}}\BibitemShut {NoStop}%
\bibitem [{\citenamefont {Schaibley}\ \emph {et~al.}(2016)\citenamefont
  {Schaibley}, \citenamefont {Yu}, \citenamefont {Clark}, \citenamefont
  {Rivera}, \citenamefont {Ross}, \citenamefont {Seyler}, \citenamefont {Yao},\
  and\ \citenamefont {Xu}}]{valleytronics_review_2016}%
  \BibitemOpen
  \bibfield  {author} {\bibinfo {author} {\bibnamefont {Schaibley},
  \bibfnamefont {J.~R.}}, \bibinfo {author} {\bibfnamefont {H.}~\bibnamefont
  {Yu}}, \bibinfo {author} {\bibfnamefont {G.}~\bibnamefont {Clark}}, \bibinfo
  {author} {\bibfnamefont {P.}~\bibnamefont {Rivera}}, \bibinfo {author}
  {\bibfnamefont {J.~S.}\ \bibnamefont {Ross}}, \bibinfo {author}
  {\bibfnamefont {K.~L.}\ \bibnamefont {Seyler}}, \bibinfo {author}
  {\bibfnamefont {W.}~\bibnamefont {Yao}}, \ and\ \bibinfo {author}
  {\bibfnamefont {X.}~\bibnamefont {Xu}}} (\bibinfo {year} {2016}),\ \href
  {\doibase 10.1038/natrevmats.2016.55} {\bibfield  {journal} {\bibinfo
  {journal} {Nat. Rev. Mater.}\ }\textbf {\bibinfo {volume} {1}},\ \bibinfo
  {pages} {16055}}\BibitemShut {NoStop}%
\bibitem [{\citenamefont {Scott}\ and\ \citenamefont {Paz~de
  Araujo}(1989)}]{Scott1400}%
  \BibitemOpen
  \bibfield  {author} {\bibinfo {author} {\bibnamefont {Scott}, \bibfnamefont
  {J.~F.}}, \ and\ \bibinfo {author} {\bibfnamefont {C.~A.}\ \bibnamefont
  {Paz~de Araujo}}} (\bibinfo {year} {1989}),\ \href {\doibase
  10.1126/science.246.4936.1400} {\bibfield  {journal} {\bibinfo  {journal}
  {Science}\ }\textbf {\bibinfo {volume} {246}},\ \bibinfo {pages}
  {1400}}\BibitemShut {NoStop}%
\bibitem [{\citenamefont {Shen}\ \emph
  {et~al.}(2019{\natexlab{a}})\citenamefont {Shen}, \citenamefont {Liu},
  \citenamefont {Chang},\ and\ \citenamefont {Fu}}]{kai3}%
  \BibitemOpen
  \bibfield  {author} {\bibinfo {author} {\bibnamefont {Shen}, \bibfnamefont
  {H.}}, \bibinfo {author} {\bibfnamefont {J.}~\bibnamefont {Liu}}, \bibinfo
  {author} {\bibfnamefont {K.}~\bibnamefont {Chang}}, \ and\ \bibinfo {author}
  {\bibfnamefont {L.}~\bibnamefont {Fu}}} (\bibinfo {year}
  {2019}{\natexlab{a}}),\ \href {\doibase 10.1103/PhysRevApplied.11.024048}
  {\bibfield  {journal} {\bibinfo  {journal} {Phys. Rev. Appl.}\ }\textbf
  {\bibinfo {volume} {11}},\ \bibinfo {pages} {024048}}\BibitemShut {NoStop}%
\bibitem [{\citenamefont {Shen}\ \emph
  {et~al.}(2019{\natexlab{b}})\citenamefont {Shen}, \citenamefont {Fang},
  \citenamefont {Tian},\ and\ \citenamefont {Duan}}]{Shen19_ACSAEM}%
  \BibitemOpen
  \bibfield  {author} {\bibinfo {author} {\bibnamefont {Shen}, \bibfnamefont
  {X.-W.}}, \bibinfo {author} {\bibfnamefont {Y.-W.}\ \bibnamefont {Fang}},
  \bibinfo {author} {\bibfnamefont {B.-B.}\ \bibnamefont {Tian}}, \ and\
  \bibinfo {author} {\bibfnamefont {C.-G.}\ \bibnamefont {Duan}}} (\bibinfo
  {year} {2019}{\natexlab{b}}),\ \href {\doibase 10.1021/acsaelm.9b00146}
  {\bibfield  {journal} {\bibinfo  {journal} {ACS Appl. Electron. Mater.}\
  }\textbf {\bibinfo {volume} {1}},\ \bibinfo {pages} {1133}}\BibitemShut
  {NoStop}%
\bibitem [{\citenamefont {Shen}\ \emph {et~al.}(2017)\citenamefont {Shen},
  \citenamefont {Tong}, \citenamefont {Gong},\ and\ \citenamefont
  {Duan}}]{shen_2dmater_2018_gese}%
  \BibitemOpen
  \bibfield  {author} {\bibinfo {author} {\bibnamefont {Shen}, \bibfnamefont
  {X.-W.}}, \bibinfo {author} {\bibfnamefont {W.-Y.}\ \bibnamefont {Tong}},
  \bibinfo {author} {\bibfnamefont {S.-J.}\ \bibnamefont {Gong}}, \ and\
  \bibinfo {author} {\bibfnamefont {C.-G.}\ \bibnamefont {Duan}}} (\bibinfo
  {year} {2017}),\ \href {\doibase 10.1088/2053-1583/aa8d3b} {\bibfield
  {journal} {\bibinfo  {journal} {2D Mater.}\ }\textbf {\bibinfo {volume}
  {5}},\ \bibinfo {pages} {011001}}\BibitemShut {NoStop}%
\bibitem [{\citenamefont {Shi}\ and\ \citenamefont
  {Kioupakis}(2015)}]{shi_nl_2015_snse_gese}%
  \BibitemOpen
  \bibfield  {author} {\bibinfo {author} {\bibnamefont {Shi}, \bibfnamefont
  {G.}}, \ and\ \bibinfo {author} {\bibfnamefont {E.}~\bibnamefont
  {Kioupakis}}} (\bibinfo {year} {2015}),\ \href {\doibase
  10.1021/acs.nanolett.5b02861} {\bibfield  {journal} {\bibinfo  {journal}
  {Nano Lett.}\ }\textbf {\bibinfo {volume} {15}},\ \bibinfo {pages}
  {6926}}\BibitemShut {NoStop}%
\bibitem [{\citenamefont {Shi}\ \emph {et~al.}(2017)\citenamefont {Shi},
  \citenamefont {Li}, \citenamefont {Xiu}, \citenamefont {Liu}, \citenamefont
  {Zhang}, \citenamefont {Liu}, \citenamefont {Li},\ and\ \citenamefont
  {Dong}}]{MoS2_THEORY}%
  \BibitemOpen
  \bibfield  {author} {\bibinfo {author} {\bibnamefont {Shi}, \bibfnamefont
  {L.-B.}}, \bibinfo {author} {\bibfnamefont {M.-B.}\ \bibnamefont {Li}},
  \bibinfo {author} {\bibfnamefont {X.-M.}\ \bibnamefont {Xiu}}, \bibinfo
  {author} {\bibfnamefont {X.-Y.}\ \bibnamefont {Liu}}, \bibinfo {author}
  {\bibfnamefont {K.-C.}\ \bibnamefont {Zhang}}, \bibinfo {author}
  {\bibfnamefont {Y.-H.}\ \bibnamefont {Liu}}, \bibinfo {author} {\bibfnamefont
  {C.-R.}\ \bibnamefont {Li}}, \ and\ \bibinfo {author} {\bibfnamefont {H.-K.}\
  \bibnamefont {Dong}}} (\bibinfo {year} {2017}),\ \href {\doibase
  10.1063/1.4983815} {\bibfield  {journal} {\bibinfo  {journal} {J. Appl.
  Phys.}\ }\textbf {\bibinfo {volume} {121}},\ \bibinfo {pages}
  {205305}}\BibitemShut {NoStop}%
\bibitem [{\citenamefont {Shimazaki}\ \emph {et~al.}(2015)\citenamefont
  {Shimazaki}, \citenamefont {Yamamoto}, \citenamefont {Borzenets},
  \citenamefont {K.Watanabe}, \citenamefont {Taniguchi},\ and\ \citenamefont
  {Tarucha}}]{valley_hall_grapehe_2015_1}%
  \BibitemOpen
  \bibfield  {author} {\bibinfo {author} {\bibnamefont {Shimazaki},
  \bibfnamefont {Y.}}, \bibinfo {author} {\bibfnamefont {M.}~\bibnamefont
  {Yamamoto}}, \bibinfo {author} {\bibfnamefont {I.~V.}\ \bibnamefont
  {Borzenets}}, \bibinfo {author} {\bibnamefont {K.Watanabe}}, \bibinfo
  {author} {\bibfnamefont {T.}~\bibnamefont {Taniguchi}}, \ and\ \bibinfo
  {author} {\bibfnamefont {S.}~\bibnamefont {Tarucha}}} (\bibinfo {year}
  {2015}),\ \href {\doibase 10.1038/NPHYS3551} {\bibfield  {journal} {\bibinfo
  {journal} {Nat. Phys.}\ }\textbf {\bibinfo {volume} {11}},\ \bibinfo {pages}
  {1032}}\BibitemShut {NoStop}%
\bibitem [{\citenamefont {Shulenburger}\ \emph {et~al.}(2015)\citenamefont
  {Shulenburger}, \citenamefont {Baczewski}, \citenamefont {Zhu}, \citenamefont
  {Guan},\ and\ \citenamefont {Tománek}}]{Shulenburger}%
  \BibitemOpen
  \bibfield  {author} {\bibinfo {author} {\bibnamefont {Shulenburger},
  \bibfnamefont {L.}}, \bibinfo {author} {\bibfnamefont {A.}~\bibnamefont
  {Baczewski}}, \bibinfo {author} {\bibfnamefont {Z.}~\bibnamefont {Zhu}},
  \bibinfo {author} {\bibfnamefont {J.}~\bibnamefont {Guan}}, \ and\ \bibinfo
  {author} {\bibfnamefont {D.}~\bibnamefont {Tománek}}} (\bibinfo {year}
  {2015}),\ \href {\doibase 10.1021/acs.nanolett.5b03615} {\bibfield  {journal}
  {\bibinfo  {journal} {Nano Lett.}\ }\textbf {\bibinfo {volume} {15}},\
  \bibinfo {pages} {8170}}\BibitemShut {NoStop}%
\bibitem [{\citenamefont {Shvetsov}\ \emph {et~al.}(2019)\citenamefont
  {Shvetsov}, \citenamefont {Esin}, \citenamefont {Timonina}, \citenamefont
  {Kolesnikov},\ and\ \citenamefont {Deviatov}}]{Shvetsov2019}%
  \BibitemOpen
  \bibfield  {author} {\bibinfo {author} {\bibnamefont {Shvetsov},
  \bibfnamefont {O.~O.}}, \bibinfo {author} {\bibfnamefont {V.~D.}\
  \bibnamefont {Esin}}, \bibinfo {author} {\bibfnamefont {A.~V.}\ \bibnamefont
  {Timonina}}, \bibinfo {author} {\bibfnamefont {N.~N.}\ \bibnamefont
  {Kolesnikov}}, \ and\ \bibinfo {author} {\bibfnamefont {E.~V.}\ \bibnamefont
  {Deviatov}}} (\bibinfo {year} {2019}),\ \href {\doibase
  10.1134/S0021364019110018} {\bibfield  {journal} {\bibinfo  {journal} {JETP
  Lett.}\ }\textbf {\bibinfo {volume} {109}},\ \bibinfo {pages}
  {715}}\BibitemShut {NoStop}%
\bibitem [{\citenamefont {Singh}\ and\ \citenamefont
  {Hennig}(2014)}]{singh_apl_2014_ges_gese_sns_snse}%
  \BibitemOpen
  \bibfield  {author} {\bibinfo {author} {\bibnamefont {Singh}, \bibfnamefont
  {A.~K.}}, \ and\ \bibinfo {author} {\bibfnamefont {R.~G.}\ \bibnamefont
  {Hennig}}} (\bibinfo {year} {2014}),\ \href {\doibase 10.1063/1.4891230}
  {\bibfield  {journal} {\bibinfo  {journal} {Appl. Phys. Lett.}\ }\textbf
  {\bibinfo {volume} {105}},\ \bibinfo {pages} {042103}}\BibitemShut {NoStop}%
\bibitem [{\citenamefont {Sipe}\ and\ \citenamefont
  {Shkrebtii}(2000)}]{Sipe2000}%
  \BibitemOpen
  \bibfield  {author} {\bibinfo {author} {\bibnamefont {Sipe}, \bibfnamefont
  {J.~E.}}, \ and\ \bibinfo {author} {\bibfnamefont {A.~I.}\ \bibnamefont
  {Shkrebtii}}} (\bibinfo {year} {2000}),\ \href {\doibase
  10.1103/PhysRevB.61.5337} {\bibfield  {journal} {\bibinfo  {journal} {Phys.
  Rev. B}\ }\textbf {\bibinfo {volume} {61}},\ \bibinfo {pages}
  {5337}}\BibitemShut {NoStop}%
\bibitem [{\citenamefont {Slack}\ and\ \citenamefont
  {Burfoot}(1971)}]{Slack_1971}%
  \BibitemOpen
  \bibfield  {author} {\bibinfo {author} {\bibnamefont {Slack}, \bibfnamefont
  {J.~R.}}, \ and\ \bibinfo {author} {\bibfnamefont {J.~C.}\ \bibnamefont
  {Burfoot}}} (\bibinfo {year} {1971}),\ \href {\doibase
  10.1088/0022-3719/4/8/016} {\bibfield  {journal} {\bibinfo  {journal} {J.
  Phys. C: Solid State Phys.}\ }\textbf {\bibinfo {volume} {4}},\ \bibinfo
  {pages} {898}}\BibitemShut {NoStop}%
\bibitem [{\citenamefont {S{\l}awi{\'{n}}ska}\ \emph
  {et~al.}(2019)\citenamefont {S{\l}awi{\'{n}}ska}, \citenamefont {Cerasoli},
  \citenamefont {Wang}, \citenamefont {Postorino}, \citenamefont {Supka},
  \citenamefont {Curtarolo}, \citenamefont {Fornari},\ and\ \citenamefont
  {Nardelli}}]{GSHE}%
  \BibitemOpen
  \bibfield  {author} {\bibinfo {author} {\bibnamefont {S{\l}awi{\'{n}}ska},
  \bibfnamefont {J.}}, \bibinfo {author} {\bibfnamefont {F.~T.}\ \bibnamefont
  {Cerasoli}}, \bibinfo {author} {\bibfnamefont {H.}~\bibnamefont {Wang}},
  \bibinfo {author} {\bibfnamefont {S.}~\bibnamefont {Postorino}}, \bibinfo
  {author} {\bibfnamefont {A.}~\bibnamefont {Supka}}, \bibinfo {author}
  {\bibfnamefont {S.}~\bibnamefont {Curtarolo}}, \bibinfo {author}
  {\bibfnamefont {M.}~\bibnamefont {Fornari}}, \ and\ \bibinfo {author}
  {\bibfnamefont {M.~B.}\ \bibnamefont {Nardelli}}} (\bibinfo {year} {2019}),\
  \href {\doibase 10.1088/2053-1583/ab0146} {\bibfield  {journal} {\bibinfo
  {journal} {2D Mater.}\ }\textbf {\bibinfo {volume} {6}},\ \bibinfo {pages}
  {025012}}\BibitemShut {NoStop}%
\bibitem [{\citenamefont {Slawinska}\ \emph {et~al.}(2019)\citenamefont
  {Slawinska}, \citenamefont {Cerasoli}, \citenamefont {Wang}, \citenamefont
  {Postorino}, \citenamefont {Supka}, \citenamefont {Curtarolo}, \citenamefont
  {Fornari},\ and\ \citenamefont {Nardelli}}]{slawinska_2d_mat_2018_mmls}%
  \BibitemOpen
  \bibfield  {author} {\bibinfo {author} {\bibnamefont {Slawinska},
  \bibfnamefont {J.}}, \bibinfo {author} {\bibfnamefont {F.~T.}\ \bibnamefont
  {Cerasoli}}, \bibinfo {author} {\bibfnamefont {H.}~\bibnamefont {Wang}},
  \bibinfo {author} {\bibfnamefont {S.}~\bibnamefont {Postorino}}, \bibinfo
  {author} {\bibfnamefont {A.}~\bibnamefont {Supka}}, \bibinfo {author}
  {\bibfnamefont {S.}~\bibnamefont {Curtarolo}}, \bibinfo {author}
  {\bibfnamefont {M.}~\bibnamefont {Fornari}}, \ and\ \bibinfo {author}
  {\bibfnamefont {M.~B.}\ \bibnamefont {Nardelli}}} (\bibinfo {year} {2019}),\
  \href {\doibase 10.1088/2053-1583/ab0146} {\bibfield  {journal} {\bibinfo
  {journal} {2D Mater.}\ }\textbf {\bibinfo {volume} {6}},\ \bibinfo {pages}
  {025012}}\BibitemShut {NoStop}%
\bibitem [{\citenamefont {Song}\ \emph {et~al.}(2018)\citenamefont {Song},
  \citenamefont {Jia}, \citenamefont {Zhang}, \citenamefont {Zhu},
  \citenamefont {Shi}, \citenamefont {Wang}, \citenamefont {Zhu}, \citenamefont
  {Yuan}, \citenamefont {Zhang}, \citenamefont {Xing},\ and\ \citenamefont
  {Li}}]{song_nc_2018}%
  \BibitemOpen
  \bibfield  {author} {\bibinfo {author} {\bibnamefont {Song}, \bibfnamefont
  {Y.-H.}}, \bibinfo {author} {\bibfnamefont {Z.-Y.}\ \bibnamefont {Jia}},
  \bibinfo {author} {\bibfnamefont {D.}~\bibnamefont {Zhang}}, \bibinfo
  {author} {\bibfnamefont {X.-Y.}\ \bibnamefont {Zhu}}, \bibinfo {author}
  {\bibfnamefont {Z.-Q.}\ \bibnamefont {Shi}}, \bibinfo {author} {\bibfnamefont
  {H.}~\bibnamefont {Wang}}, \bibinfo {author} {\bibfnamefont {L.}~\bibnamefont
  {Zhu}}, \bibinfo {author} {\bibfnamefont {Q.-Q.}\ \bibnamefont {Yuan}},
  \bibinfo {author} {\bibfnamefont {H.}~\bibnamefont {Zhang}}, \bibinfo
  {author} {\bibfnamefont {D.-Y.}\ \bibnamefont {Xing}}, \ and\ \bibinfo
  {author} {\bibfnamefont {S.-C.}\ \bibnamefont {Li}}} (\bibinfo {year}
  {2018}),\ \href {\doibase 10.1038/s41467-018-06635-x} {\bibfield  {journal}
  {\bibinfo  {journal} {Nat. Phys.}\ }\textbf {\bibinfo {volume} {9}},\
  \bibinfo {pages} {4071}}\BibitemShut {NoStop}%
\bibitem [{\citenamefont {Spanier}\ \emph {et~al.}(2016)\citenamefont
  {Spanier}, \citenamefont {Fridkin}, \citenamefont {Rappe}, \citenamefont
  {Akbashev}, \citenamefont {Polemi}, \citenamefont {Qi}, \citenamefont {Gu},
  \citenamefont {Young}, \citenamefont {Hawley}, \citenamefont {Imbrenda},
  \citenamefont {Xiao}, \citenamefont {Bennett-Jackson},\ and\ \citenamefont
  {Johnson}}]{rappe2}%
  \BibitemOpen
  \bibfield  {author} {\bibinfo {author} {\bibnamefont {Spanier}, \bibfnamefont
  {J.~E.}}, \bibinfo {author} {\bibfnamefont {V.~M.}\ \bibnamefont {Fridkin}},
  \bibinfo {author} {\bibfnamefont {A.~M.}\ \bibnamefont {Rappe}}, \bibinfo
  {author} {\bibfnamefont {A.~R.}\ \bibnamefont {Akbashev}}, \bibinfo {author}
  {\bibfnamefont {A.}~\bibnamefont {Polemi}}, \bibinfo {author} {\bibfnamefont
  {Y.}~\bibnamefont {Qi}}, \bibinfo {author} {\bibfnamefont {Z.}~\bibnamefont
  {Gu}}, \bibinfo {author} {\bibfnamefont {S.~M.}\ \bibnamefont {Young}},
  \bibinfo {author} {\bibfnamefont {C.~J.}\ \bibnamefont {Hawley}}, \bibinfo
  {author} {\bibfnamefont {D.}~\bibnamefont {Imbrenda}}, \bibinfo {author}
  {\bibfnamefont {G.}~\bibnamefont {Xiao}}, \bibinfo {author} {\bibfnamefont
  {A.~L.}\ \bibnamefont {Bennett-Jackson}}, \ and\ \bibinfo {author}
  {\bibfnamefont {C.~L.}\ \bibnamefont {Johnson}}} (\bibinfo {year} {2016}),\
  \href {\doibase 10.1038/nphoton.2016.143} {\bibfield  {journal} {\bibinfo
  {journal} {Nat. Photon.}\ }\textbf {\bibinfo {volume} {10}},\ \bibinfo
  {pages} {611}}\BibitemShut {NoStop}%
\bibitem [{\citenamefont {Splendiani}\ \emph {et~al.}(2010)\citenamefont
  {Splendiani}, \citenamefont {Sun}, \citenamefont {Zhang}, \citenamefont {Li},
  \citenamefont {Kim}, \citenamefont {Chim}, \citenamefont {Galli},\ and\
  \citenamefont {Wang}}]{splendiani}%
  \BibitemOpen
  \bibfield  {author} {\bibinfo {author} {\bibnamefont {Splendiani},
  \bibfnamefont {A.}}, \bibinfo {author} {\bibfnamefont {L.}~\bibnamefont
  {Sun}}, \bibinfo {author} {\bibfnamefont {Y.}~\bibnamefont {Zhang}}, \bibinfo
  {author} {\bibfnamefont {T.}~\bibnamefont {Li}}, \bibinfo {author}
  {\bibfnamefont {J.}~\bibnamefont {Kim}}, \bibinfo {author} {\bibfnamefont
  {C.-Y.}\ \bibnamefont {Chim}}, \bibinfo {author} {\bibfnamefont
  {G.}~\bibnamefont {Galli}}, \ and\ \bibinfo {author} {\bibfnamefont
  {F.}~\bibnamefont {Wang}}} (\bibinfo {year} {2010}),\ \href {\doibase
  10.1021/nl903868w} {\bibfield  {journal} {\bibinfo  {journal} {Nano Lett.}\
  }\textbf {\bibinfo {volume} {10}},\ \bibinfo {pages} {1271}}\BibitemShut
  {NoStop}%
\bibitem [{\citenamefont {Sturman}\ and\ \citenamefont
  {Sturman}(1992)}]{Sturman1992}%
  \BibitemOpen
  \bibfield  {author} {\bibinfo {author} {\bibnamefont {Sturman}, \bibfnamefont
  {B.~I.}}, \ and\ \bibinfo {author} {\bibfnamefont {P.~J.}\ \bibnamefont
  {Sturman}}} (\bibinfo {year} {1992}),\ \href@noop {} {\emph {\bibinfo {title}
  {Photovoltaic and Photo-refractive Effects in Noncentrosymmetric
  Materials}}},\ \bibinfo {edition} {1st}\ ed.,\ \bibinfo {series}
  {ferroelectricity and related phenomena}, Vol.~\bibinfo {volume} {8}\
  (\bibinfo  {publisher} {Routledge})\BibitemShut {NoStop}%
\bibitem [{\citenamefont {Sui}\ \emph {et~al.}(2015)\citenamefont {Sui},
  \citenamefont {Chen}, \citenamefont {Ma}, \citenamefont {Shan}, \citenamefont
  {Tian}, \citenamefont {Watanabe}, \citenamefont {Taniguchi}, \citenamefont
  {Jin}, \citenamefont {Yao}, \citenamefont {Xiao},\ and\ \citenamefont
  {Zhang}}]{valley_hall_grapehe_2015_2}%
  \BibitemOpen
  \bibfield  {author} {\bibinfo {author} {\bibnamefont {Sui}, \bibfnamefont
  {M.}}, \bibinfo {author} {\bibfnamefont {G.}~\bibnamefont {Chen}}, \bibinfo
  {author} {\bibfnamefont {L.}~\bibnamefont {Ma}}, \bibinfo {author}
  {\bibfnamefont {W.-Y.}\ \bibnamefont {Shan}}, \bibinfo {author}
  {\bibfnamefont {D.}~\bibnamefont {Tian}}, \bibinfo {author} {\bibfnamefont
  {K.}~\bibnamefont {Watanabe}}, \bibinfo {author} {\bibfnamefont
  {T.}~\bibnamefont {Taniguchi}}, \bibinfo {author} {\bibfnamefont
  {X.}~\bibnamefont {Jin}}, \bibinfo {author} {\bibfnamefont {W.}~\bibnamefont
  {Yao}}, \bibinfo {author} {\bibfnamefont {D.}~\bibnamefont {Xiao}}, \ and\
  \bibinfo {author} {\bibfnamefont {Y.}~\bibnamefont {Zhang}}} (\bibinfo {year}
  {2015}),\ \href {\doibase 10.1038/NPHYS3485} {\bibfield  {journal} {\bibinfo
  {journal} {Nat. Phys.}\ }\textbf {\bibinfo {volume} {11}},\ \bibinfo {pages}
  {1027}}\BibitemShut {NoStop}%
\bibitem [{\citenamefont {Sutter}\ \emph {et~al.}(2019)\citenamefont {Sutter},
  \citenamefont {Wimer},\ and\ \citenamefont {Sutter}}]{sutterNature2019}%
  \BibitemOpen
  \bibfield  {author} {\bibinfo {author} {\bibnamefont {Sutter}, \bibfnamefont
  {P.}}, \bibinfo {author} {\bibfnamefont {S.}~\bibnamefont {Wimer}}, \ and\
  \bibinfo {author} {\bibfnamefont {E.}~\bibnamefont {Sutter}}} (\bibinfo
  {year} {2019}),\ \href {\doibase 10.1038/s41586-019-1147-x} {\bibfield
  {journal} {\bibinfo  {journal} {Nature}\ }\textbf {\bibinfo {volume} {570}},\
  \bibinfo {pages} {354}}\BibitemShut {NoStop}%
\bibitem [{\citenamefont {Tan}\ \emph {et~al.}(2016)\citenamefont {Tan},
  \citenamefont {Zheng}, \citenamefont {Young}, \citenamefont {Wang},
  \citenamefont {Liu},\ and\ \citenamefont {Rappe}}]{rappe3}%
  \BibitemOpen
  \bibfield  {author} {\bibinfo {author} {\bibnamefont {Tan}, \bibfnamefont
  {L.~Z.}}, \bibinfo {author} {\bibfnamefont {F.}~\bibnamefont {Zheng}},
  \bibinfo {author} {\bibfnamefont {S.~M.}\ \bibnamefont {Young}}, \bibinfo
  {author} {\bibfnamefont {F.}~\bibnamefont {Wang}}, \bibinfo {author}
  {\bibfnamefont {S.}~\bibnamefont {Liu}}, \ and\ \bibinfo {author}
  {\bibfnamefont {A.~M.}\ \bibnamefont {Rappe}}} (\bibinfo {year} {2016}),\
  \href {\doibase 10.1038/npjcompumats.2016.26} {\bibfield  {journal} {\bibinfo
   {journal} {npj Comp. Mater.}\ }\textbf {\bibinfo {volume} {2}},\ \bibinfo
  {pages} {16026}}\BibitemShut {NoStop}%
\bibitem [{\citenamefont {Tang}\ \emph {et~al.}(2017)\citenamefont {Tang},
  \citenamefont {Zhang}, \citenamefont {Wong}, \citenamefont {Pedramrazi},
  \citenamefont {Tsai}, \citenamefont {Jia}, \citenamefont {Moritz},
  \citenamefont {Claassen}, \citenamefont {Ryu}, \citenamefont {Kahn},
  \citenamefont {Jiang}, \citenamefont {Yan}, \citenamefont {Hashimoto},
  \citenamefont {Lu}, \citenamefont {Moore}, \citenamefont {Hwang},
  \citenamefont {Hwang}, \citenamefont {Hussain}, \citenamefont {Chen},
  \citenamefont {Ugeda}, \citenamefont {Liu}, \citenamefont {Xie},
  \citenamefont {Devereaux}, \citenamefont {Crommie}, \citenamefont {Mo},\ and\
  \citenamefont {Shen}}]{tang_np_2017}%
  \BibitemOpen
  \bibfield  {author} {\bibinfo {author} {\bibnamefont {Tang}, \bibfnamefont
  {S.}}, \bibinfo {author} {\bibfnamefont {C.}~\bibnamefont {Zhang}}, \bibinfo
  {author} {\bibfnamefont {D.}~\bibnamefont {Wong}}, \bibinfo {author}
  {\bibfnamefont {Z.}~\bibnamefont {Pedramrazi}}, \bibinfo {author}
  {\bibfnamefont {H.-Z.}\ \bibnamefont {Tsai}}, \bibinfo {author}
  {\bibfnamefont {C.}~\bibnamefont {Jia}}, \bibinfo {author} {\bibfnamefont
  {B.}~\bibnamefont {Moritz}}, \bibinfo {author} {\bibfnamefont
  {M.}~\bibnamefont {Claassen}}, \bibinfo {author} {\bibfnamefont
  {H.}~\bibnamefont {Ryu}}, \bibinfo {author} {\bibfnamefont {S.}~\bibnamefont
  {Kahn}}, \bibinfo {author} {\bibfnamefont {J.}~\bibnamefont {Jiang}},
  \bibinfo {author} {\bibfnamefont {H.}~\bibnamefont {Yan}}, \bibinfo {author}
  {\bibfnamefont {M.}~\bibnamefont {Hashimoto}}, \bibinfo {author}
  {\bibfnamefont {D.}~\bibnamefont {Lu}}, \bibinfo {author} {\bibfnamefont
  {R.~G.}\ \bibnamefont {Moore}}, \bibinfo {author} {\bibfnamefont {C.-C.}\
  \bibnamefont {Hwang}}, \bibinfo {author} {\bibfnamefont {C.}~\bibnamefont
  {Hwang}}, \bibinfo {author} {\bibfnamefont {Z.}~\bibnamefont {Hussain}},
  \bibinfo {author} {\bibfnamefont {Y.}~\bibnamefont {Chen}}, \bibinfo {author}
  {\bibfnamefont {M.~M.}\ \bibnamefont {Ugeda}}, \bibinfo {author}
  {\bibfnamefont {Z.}~\bibnamefont {Liu}}, \bibinfo {author} {\bibfnamefont
  {X.}~\bibnamefont {Xie}}, \bibinfo {author} {\bibfnamefont {T.~P.}\
  \bibnamefont {Devereaux}}, \bibinfo {author} {\bibfnamefont {M.~F.}\
  \bibnamefont {Crommie}}, \bibinfo {author} {\bibfnamefont {S.-K.}\
  \bibnamefont {Mo}}, \ and\ \bibinfo {author} {\bibfnamefont {Z.-X.}\
  \bibnamefont {Shen}}} (\bibinfo {year} {2017}),\ \href {\doibase
  10.1038/nphys4174} {\bibfield  {journal} {\bibinfo  {journal} {Nat. Phys.}\
  }\textbf {\bibinfo {volume} {13}},\ \bibinfo {pages} {683}}\BibitemShut
  {NoStop}%
\bibitem [{\citenamefont {Tenne}\ \emph {et~al.}(2006)\citenamefont {Tenne},
  \citenamefont {Bruchhausen}, \citenamefont {Lanzillotti-Kimura},
  \citenamefont {Fainstein}, \citenamefont {Katiyar}, \citenamefont
  {Cantarero}, \citenamefont {Soukiassian}, \citenamefont {Vaithyanathan},
  \citenamefont {Haeni}, \citenamefont {Tian}, \citenamefont {Schlom},
  \citenamefont {Choi}, \citenamefont {Kim}, \citenamefont {Eom}, \citenamefont
  {Sun}, \citenamefont {Pan}, \citenamefont {Li}, \citenamefont {Chen},
  \citenamefont {Jia}, \citenamefont {Nakhmanson}, \citenamefont {Rabe},\ and\
  \citenamefont {Xi}}]{tenne_bto_2006}%
  \BibitemOpen
  \bibfield  {author} {\bibinfo {author} {\bibnamefont {Tenne}, \bibfnamefont
  {D.~A.}}, \bibinfo {author} {\bibfnamefont {A.}~\bibnamefont {Bruchhausen}},
  \bibinfo {author} {\bibfnamefont {N.~D.}\ \bibnamefont {Lanzillotti-Kimura}},
  \bibinfo {author} {\bibfnamefont {A.}~\bibnamefont {Fainstein}}, \bibinfo
  {author} {\bibfnamefont {R.~S.}\ \bibnamefont {Katiyar}}, \bibinfo {author}
  {\bibfnamefont {A.}~\bibnamefont {Cantarero}}, \bibinfo {author}
  {\bibfnamefont {A.}~\bibnamefont {Soukiassian}}, \bibinfo {author}
  {\bibfnamefont {V.}~\bibnamefont {Vaithyanathan}}, \bibinfo {author}
  {\bibfnamefont {J.~H.}\ \bibnamefont {Haeni}}, \bibinfo {author}
  {\bibfnamefont {W.}~\bibnamefont {Tian}}, \bibinfo {author} {\bibfnamefont
  {D.~G.}\ \bibnamefont {Schlom}}, \bibinfo {author} {\bibfnamefont {K.~J.}\
  \bibnamefont {Choi}}, \bibinfo {author} {\bibfnamefont {D.~M.}\ \bibnamefont
  {Kim}}, \bibinfo {author} {\bibfnamefont {C.~B.}\ \bibnamefont {Eom}},
  \bibinfo {author} {\bibfnamefont {H.~P.}\ \bibnamefont {Sun}}, \bibinfo
  {author} {\bibfnamefont {X.~Q.}\ \bibnamefont {Pan}}, \bibinfo {author}
  {\bibfnamefont {Y.~L.}\ \bibnamefont {Li}}, \bibinfo {author} {\bibfnamefont
  {L.~Q.}\ \bibnamefont {Chen}}, \bibinfo {author} {\bibfnamefont {Q.~X.}\
  \bibnamefont {Jia}}, \bibinfo {author} {\bibfnamefont {S.~M.}\ \bibnamefont
  {Nakhmanson}}, \bibinfo {author} {\bibfnamefont {K.~M.}\ \bibnamefont
  {Rabe}}, \ and\ \bibinfo {author} {\bibfnamefont {X.~X.}\ \bibnamefont {Xi}}}
  (\bibinfo {year} {2006}),\ \href {\doibase 10.1126/science.1130306}
  {\bibfield  {journal} {\bibinfo  {journal} {Science}\ }\textbf {\bibinfo
  {volume} {313}},\ \bibinfo {pages} {1614}}\BibitemShut {NoStop}%
\bibitem [{\citenamefont {Tenne}\ \emph {et~al.}(2009)\citenamefont {Tenne},
  \citenamefont {Turner}, \citenamefont {Schmidt}, \citenamefont {Biegalski},
  \citenamefont {Li}, \citenamefont {Chen}, \citenamefont {Soukiassian},
  \citenamefont {Trolier-McKinstry}, \citenamefont {Schlom}, \citenamefont
  {Xi}, \citenamefont {Fong}, \citenamefont {Fuoss}, \citenamefont {Eastman},
  \citenamefont {Stephenson}, \citenamefont {Thompson},\ and\ \citenamefont
  {Streiffer}}]{tenne_bto_2009}%
  \BibitemOpen
  \bibfield  {author} {\bibinfo {author} {\bibnamefont {Tenne}, \bibfnamefont
  {D.~A.}}, \bibinfo {author} {\bibfnamefont {P.}~\bibnamefont {Turner}},
  \bibinfo {author} {\bibfnamefont {J.~D.}\ \bibnamefont {Schmidt}}, \bibinfo
  {author} {\bibfnamefont {M.}~\bibnamefont {Biegalski}}, \bibinfo {author}
  {\bibfnamefont {Y.~L.}\ \bibnamefont {Li}}, \bibinfo {author} {\bibfnamefont
  {L.~Q.}\ \bibnamefont {Chen}}, \bibinfo {author} {\bibfnamefont
  {A.}~\bibnamefont {Soukiassian}}, \bibinfo {author} {\bibfnamefont
  {S.}~\bibnamefont {Trolier-McKinstry}}, \bibinfo {author} {\bibfnamefont
  {D.~G.}\ \bibnamefont {Schlom}}, \bibinfo {author} {\bibfnamefont {X.~X.}\
  \bibnamefont {Xi}}, \bibinfo {author} {\bibfnamefont {D.~D.}\ \bibnamefont
  {Fong}}, \bibinfo {author} {\bibfnamefont {P.~H.}\ \bibnamefont {Fuoss}},
  \bibinfo {author} {\bibfnamefont {J.~A.}\ \bibnamefont {Eastman}}, \bibinfo
  {author} {\bibfnamefont {G.~B.}\ \bibnamefont {Stephenson}}, \bibinfo
  {author} {\bibfnamefont {C.}~\bibnamefont {Thompson}}, \ and\ \bibinfo
  {author} {\bibfnamefont {S.~K.}\ \bibnamefont {Streiffer}}} (\bibinfo {year}
  {2009}),\ \href {\doibase 10.1103/PhysRevLett.103.177601} {\bibfield
  {journal} {\bibinfo  {journal} {Phys. Rev. Lett.}\ }\textbf {\bibinfo
  {volume} {103}},\ \bibinfo {pages} {177601}}\BibitemShut {NoStop}%
\bibitem [{\citenamefont {Titova}\ \emph {et~al.}(2020)\citenamefont {Titova},
  \citenamefont {Fregoso},\ and\ \citenamefont {Grimm}}]{Titova2020}%
  \BibitemOpen
  \bibfield  {author} {\bibinfo {author} {\bibnamefont {Titova}, \bibfnamefont
  {L.~V.}}, \bibinfo {author} {\bibfnamefont {B.~M.}\ \bibnamefont {Fregoso}},
  \ and\ \bibinfo {author} {\bibfnamefont {R.~L.}\ \bibnamefont {Grimm}}}
  (\bibinfo {year} {2020}),\ in\ \href@noop {} {\emph {\bibinfo {booktitle}
  {Chalcogenide}}},\ \bibinfo {series and number} {Woodhead Publishing Series
  in Electronic and Optical Materials},\ \bibinfo {editor} {edited by\ \bibinfo
  {editor} {\bibfnamefont {X.}~\bibnamefont {Liu}}, \bibinfo {editor}
  {\bibfnamefont {S.}~\bibnamefont {Lee}}, \bibinfo {editor} {\bibfnamefont
  {J.~K.}\ \bibnamefont {Furdyna}}, \bibinfo {editor} {\bibfnamefont
  {T.}~\bibnamefont {Luo}}, \ and\ \bibinfo {editor} {\bibfnamefont {Y.-H.}\
  \bibnamefont {Zhang}}}\ (\bibinfo  {publisher} {Woodhead
  Publishing})\BibitemShut {NoStop}%
\bibitem [{\citenamefont {Tomashpolski}(1974)}]{YYToma_1974a}%
  \BibitemOpen
  \bibfield  {author} {\bibinfo {author} {\bibnamefont {Tomashpolski},
  \bibfnamefont {Y.~Y.}}} (\bibinfo {year} {1974}),\ \href {\doibase
  10.1080/00150197408238012} {\bibfield  {journal} {\bibinfo  {journal}
  {Ferroelectrics}\ }\textbf {\bibinfo {volume} {7}},\ \bibinfo {pages}
  {253}}\BibitemShut {NoStop}%
\bibitem [{\citenamefont {Tomashpolski}\ \emph {et~al.}(1974)\citenamefont
  {Tomashpolski}, \citenamefont {Sevostianov}, \citenamefont {Pentegova},
  \citenamefont {Sorokina},\ and\ \citenamefont {Venevtsev}}]{YYToma_1974b}%
  \BibitemOpen
  \bibfield  {author} {\bibinfo {author} {\bibnamefont {Tomashpolski},
  \bibfnamefont {Y.~Y.}}, \bibinfo {author} {\bibfnamefont {M.~A.}\
  \bibnamefont {Sevostianov}}, \bibinfo {author} {\bibfnamefont {M.~V.}\
  \bibnamefont {Pentegova}}, \bibinfo {author} {\bibfnamefont {L.~A.}\
  \bibnamefont {Sorokina}}, \ and\ \bibinfo {author} {\bibfnamefont {Y.~N.}\
  \bibnamefont {Venevtsev}}} (\bibinfo {year} {1974}),\ \href {\doibase
  10.1080/00150197408238013} {\bibfield  {journal} {\bibinfo  {journal}
  {Ferroelectrics}\ }\textbf {\bibinfo {volume} {7}},\ \bibinfo {pages}
  {257}}\BibitemShut {NoStop}%
\bibitem [{\citenamefont {Tran}\ \emph {et~al.}(2014)\citenamefont {Tran},
  \citenamefont {Soklaski}, \citenamefont {Liang},\ and\ \citenamefont
  {Yang}}]{Tran}%
  \BibitemOpen
  \bibfield  {author} {\bibinfo {author} {\bibnamefont {Tran}, \bibfnamefont
  {V.}}, \bibinfo {author} {\bibfnamefont {R.}~\bibnamefont {Soklaski}},
  \bibinfo {author} {\bibfnamefont {Y.}~\bibnamefont {Liang}}, \ and\ \bibinfo
  {author} {\bibfnamefont {L.}~\bibnamefont {Yang}}} (\bibinfo {year} {2014}),\
  \href {\doibase 10.1103/PhysRevB.89.235319} {\bibfield  {journal} {\bibinfo
  {journal} {Phys. Rev. B}\ }\textbf {\bibinfo {volume} {89}},\ \bibinfo
  {pages} {235319}}\BibitemShut {NoStop}%
\bibitem [{\citenamefont {Triebwasser}(1960)}]{triebwasser}%
  \BibitemOpen
  \bibfield  {author} {\bibinfo {author} {\bibnamefont {Triebwasser},
  \bibfnamefont {S.}}} (\bibinfo {year} {1960}),\ \href {\doibase
  10.1103/PhysRev.118.100} {\bibfield  {journal} {\bibinfo  {journal} {Phys.
  Rev.}\ }\textbf {\bibinfo {volume} {118}},\ \bibinfo {pages}
  {100}}\BibitemShut {NoStop}%
\bibitem [{\citenamefont {Tritsaris}\ \emph {et~al.}(2013)\citenamefont
  {Tritsaris}, \citenamefont {Malone},\ and\ \citenamefont
  {Kaxiras}}]{tritsaris_jap_2013_sns}%
  \BibitemOpen
  \bibfield  {author} {\bibinfo {author} {\bibnamefont {Tritsaris},
  \bibfnamefont {G.}}, \bibinfo {author} {\bibfnamefont {B.}~\bibnamefont
  {Malone}}, \ and\ \bibinfo {author} {\bibfnamefont {E.}~\bibnamefont
  {Kaxiras}}} (\bibinfo {year} {2013}),\ \href {\doibase 10.1063/1.4811455}
  {\bibfield  {journal} {\bibinfo  {journal} {J. Appl. Phys.}\ }\textbf
  {\bibinfo {volume} {113}},\ \bibinfo {pages} {233507}}\BibitemShut {NoStop}%
\bibitem [{\citenamefont {Tsymbal}\ and\ \citenamefont
  {Kohlstedt}(2006)}]{Tsymbal181}%
  \BibitemOpen
  \bibfield  {author} {\bibinfo {author} {\bibnamefont {Tsymbal}, \bibfnamefont
  {E.~Y.}}, \ and\ \bibinfo {author} {\bibfnamefont {H.}~\bibnamefont
  {Kohlstedt}}} (\bibinfo {year} {2006}),\ \href {\doibase
  10.1126/science.1126230} {\bibfield  {journal} {\bibinfo  {journal}
  {Science}\ }\textbf {\bibinfo {volume} {313}},\ \bibinfo {pages}
  {181}}\BibitemShut {NoStop}%
\bibitem [{\citenamefont {Tuttle}\ \emph {et~al.}(2015)\citenamefont {Tuttle},
  \citenamefont {Alhassan},\ and\ \citenamefont
  {Pantelides}}]{tuttle_prb_2015_sis}%
  \BibitemOpen
  \bibfield  {author} {\bibinfo {author} {\bibnamefont {Tuttle}, \bibfnamefont
  {B.~R.}}, \bibinfo {author} {\bibfnamefont {S.~M.}\ \bibnamefont {Alhassan}},
  \ and\ \bibinfo {author} {\bibfnamefont {S.~T.}\ \bibnamefont {Pantelides}}}
  (\bibinfo {year} {2015}),\ \href {\doibase 10.1103/PhysRevB.92.235405}
  {\bibfield  {journal} {\bibinfo  {journal} {Phys. Rev. B}\ }\textbf {\bibinfo
  {volume} {92}},\ \bibinfo {pages} {235405}}\BibitemShut {NoStop}%
\bibitem [{\citenamefont {Tybell}\ \emph {et~al.}(1999)\citenamefont {Tybell},
  \citenamefont {Ahn},\ and\ \citenamefont {Triscone}}]{Tybell_1999}%
  \BibitemOpen
  \bibfield  {author} {\bibinfo {author} {\bibnamefont {Tybell}, \bibfnamefont
  {T.}}, \bibinfo {author} {\bibfnamefont {C.~H.}\ \bibnamefont {Ahn}}, \ and\
  \bibinfo {author} {\bibfnamefont {J.-M.}\ \bibnamefont {Triscone}}} (\bibinfo
  {year} {1999}),\ \href {\doibase 10.1063/1.124536} {\bibfield  {journal}
  {\bibinfo  {journal} {Appl. Phys. Lett.}\ }\textbf {\bibinfo {volume} {75}},\
  \bibinfo {pages} {856}}\BibitemShut {NoStop}%
\bibitem [{\citenamefont {Ugeda}\ \emph {et~al.}(2014)\citenamefont {Ugeda},
  \citenamefont {Bradley}, \citenamefont {Shi}, \citenamefont {da~Jornada},
  \citenamefont {Zhang}, \citenamefont {Qiu}, \citenamefont {Ruan},
  \citenamefont {Mo}, \citenamefont {Hussain}, \citenamefont {Shen},
  \citenamefont {Wang}, \citenamefont {G.},\ and\ \citenamefont
  {Crommie}}]{Ugeda}%
  \BibitemOpen
  \bibfield  {author} {\bibinfo {author} {\bibnamefont {Ugeda}, \bibfnamefont
  {M.~M.}}, \bibinfo {author} {\bibfnamefont {A.~J.}\ \bibnamefont {Bradley}},
  \bibinfo {author} {\bibfnamefont {S.-F.}\ \bibnamefont {Shi}}, \bibinfo
  {author} {\bibfnamefont {F.~H.}\ \bibnamefont {da~Jornada}}, \bibinfo
  {author} {\bibfnamefont {Y.}~\bibnamefont {Zhang}}, \bibinfo {author}
  {\bibfnamefont {D.~Y.}\ \bibnamefont {Qiu}}, \bibinfo {author} {\bibfnamefont
  {W.}~\bibnamefont {Ruan}}, \bibinfo {author} {\bibfnamefont {S.-K.}\
  \bibnamefont {Mo}}, \bibinfo {author} {\bibfnamefont {Z.}~\bibnamefont
  {Hussain}}, \bibinfo {author} {\bibfnamefont {Z.-X.}\ \bibnamefont {Shen}},
  \bibinfo {author} {\bibfnamefont {F.}~\bibnamefont {Wang}}, \bibinfo {author}
  {\bibfnamefont {L.~S.}\ \bibnamefont {G.}}, \ and\ \bibinfo {author}
  {\bibfnamefont {M.~F.}\ \bibnamefont {Crommie}}} (\bibinfo {year} {2014}),\
  \href {\doibase 10.1038/nmat4061} {\bibfield  {journal} {\bibinfo  {journal}
  {Nat. Mater.}\ }\textbf {\bibinfo {volume} {13}},\ \bibinfo {pages}
  {1091}}\BibitemShut {NoStop}%
\bibitem [{\citenamefont {Valasek}(1921)}]{rochelle_salt}%
  \BibitemOpen
  \bibfield  {author} {\bibinfo {author} {\bibnamefont {Valasek}, \bibfnamefont
  {J.}}} (\bibinfo {year} {1921}),\ \href {\doibase 10.1103/PhysRev.17.475}
  {\bibfield  {journal} {\bibinfo  {journal} {Phys. Rev.}\ }\textbf {\bibinfo
  {volume} {17}},\ \bibinfo {pages} {475}}\BibitemShut {NoStop}%
\bibitem [{\citenamefont {Vanderbilt}(2018)}]{vanderbilt2018berry}%
  \BibitemOpen
  \bibfield  {author} {\bibinfo {author} {\bibnamefont {Vanderbilt},
  \bibfnamefont {D.}}} (\bibinfo {year} {2018}),\ \href@noop {} {\emph
  {\bibinfo {title} {Berry Phases in Electronic Structure Theory: Electric
  Polarization, Orbital Magnetization and Topological Insulators}}}\ (\bibinfo
  {publisher} {Cambridge University Press},\ \bibinfo {address} {Cambridge,
  U.K.})\BibitemShut {NoStop}%
\bibitem [{\citenamefont {Vaughn}\ \emph {et~al.}(2010)\citenamefont {Vaughn},
  \citenamefont {Patel}, \citenamefont {Hickner},\ and\ \citenamefont
  {Schaak}}]{geseexpt}%
  \BibitemOpen
  \bibfield  {author} {\bibinfo {author} {\bibnamefont {Vaughn}, \bibfnamefont
  {D.~D.}}, \bibinfo {author} {\bibfnamefont {R.~J.}\ \bibnamefont {Patel}},
  \bibinfo {author} {\bibfnamefont {M.~A.}\ \bibnamefont {Hickner}}, \ and\
  \bibinfo {author} {\bibfnamefont {R.~E.}\ \bibnamefont {Schaak}}} (\bibinfo
  {year} {2010}),\ \href {\doibase 10.1021/ja107520b} {\bibfield  {journal}
  {\bibinfo  {journal} {J. Am. Chem. Soc.}\ }\textbf {\bibinfo {volume}
  {132}},\ \bibinfo {pages} {15170}}\BibitemShut {NoStop}%
\bibitem [{\citenamefont {Villanova}\ \emph {et~al.}(2020)\citenamefont
  {Villanova}, \citenamefont {Kumar},\ and\ \citenamefont
  {Barraza-Lopez}}]{Villanova2020PRB}%
  \BibitemOpen
  \bibfield  {author} {\bibinfo {author} {\bibnamefont {Villanova},
  \bibfnamefont {J.~W.}}, \bibinfo {author} {\bibfnamefont {P.}~\bibnamefont
  {Kumar}}, \ and\ \bibinfo {author} {\bibfnamefont {S.}~\bibnamefont
  {Barraza-Lopez}}} (\bibinfo {year} {2020}),\ \href {\doibase
  10.1103/PhysRevB.101.184101} {\bibfield  {journal} {\bibinfo  {journal}
  {Phys. Rev. B}\ }\textbf {\bibinfo {volume} {101}},\ \bibinfo {pages}
  {184101}}\BibitemShut {NoStop}%
\bibitem [{\citenamefont {Wan}\ \emph {et~al.}(2018)\citenamefont {Wan},
  \citenamefont {Li}, \citenamefont {Li}, \citenamefont {Mao}, \citenamefont
  {Zhu},\ and\ \citenamefont {Zeng}}]{wan_in2se3_2018}%
  \BibitemOpen
  \bibfield  {author} {\bibinfo {author} {\bibnamefont {Wan}, \bibfnamefont
  {S.}}, \bibinfo {author} {\bibfnamefont {Y.}~\bibnamefont {Li}}, \bibinfo
  {author} {\bibfnamefont {W.}~\bibnamefont {Li}}, \bibinfo {author}
  {\bibfnamefont {X.}~\bibnamefont {Mao}}, \bibinfo {author} {\bibfnamefont
  {W.}~\bibnamefont {Zhu}}, \ and\ \bibinfo {author} {\bibfnamefont
  {H.}~\bibnamefont {Zeng}}} (\bibinfo {year} {2018}),\ \href {\doibase
  10.1039/c8nr04422h} {\bibfield  {journal} {\bibinfo  {journal} {Nanoscale}\
  }\textbf {\bibinfo {volume} {10}},\ \bibinfo {pages} {14885}}\BibitemShut
  {NoStop}%
\bibitem [{\citenamefont {Wang}\ \emph
  {et~al.}(2018{\natexlab{a}})\citenamefont {Wang}, \citenamefont {Liu},
  \citenamefont {Yoong}, \citenamefont {Paudel}, \citenamefont {Xiao},
  \citenamefont {Guo}, \citenamefont {Lin}, \citenamefont {Yang}, \citenamefont
  {Wang}, \citenamefont {Chow}, \citenamefont {Venkatesan}, \citenamefont
  {Tsymbal}, \citenamefont {Tian},\ and\ \citenamefont {Chen}}]{wang_bfo_2018}%
  \BibitemOpen
  \bibfield  {author} {\bibinfo {author} {\bibnamefont {Wang}, \bibfnamefont
  {H.}}, \bibinfo {author} {\bibfnamefont {Z.~R.}\ \bibnamefont {Liu}},
  \bibinfo {author} {\bibfnamefont {H.~Y.}\ \bibnamefont {Yoong}}, \bibinfo
  {author} {\bibfnamefont {T.~R.}\ \bibnamefont {Paudel}}, \bibinfo {author}
  {\bibfnamefont {J.~X.}\ \bibnamefont {Xiao}}, \bibinfo {author}
  {\bibfnamefont {R.}~\bibnamefont {Guo}}, \bibinfo {author} {\bibfnamefont
  {W.~N.}\ \bibnamefont {Lin}}, \bibinfo {author} {\bibfnamefont
  {P.}~\bibnamefont {Yang}}, \bibinfo {author} {\bibfnamefont {J.}~\bibnamefont
  {Wang}}, \bibinfo {author} {\bibfnamefont {G.~M.}\ \bibnamefont {Chow}},
  \bibinfo {author} {\bibfnamefont {T.}~\bibnamefont {Venkatesan}}, \bibinfo
  {author} {\bibfnamefont {E.~Y.}\ \bibnamefont {Tsymbal}}, \bibinfo {author}
  {\bibfnamefont {H.}~\bibnamefont {Tian}}, \ and\ \bibinfo {author}
  {\bibfnamefont {J.~S.}\ \bibnamefont {Chen}}} (\bibinfo {year}
  {2018}{\natexlab{a}}),\ \href {\doibase 10.1038/s41467-018-05662-y}
  {\bibfield  {journal} {\bibinfo  {journal} {Nat. Commun.}\ }\textbf {\bibinfo
  {volume} {9}},\ \bibinfo {pages} {3319}}\BibitemShut {NoStop}%
\bibitem [{\citenamefont {Wang}\ and\ \citenamefont
  {Qian}(2017{\natexlab{a}})}]{wang_nanolett_2017_gese}%
  \BibitemOpen
  \bibfield  {author} {\bibinfo {author} {\bibnamefont {Wang}, \bibfnamefont
  {H.}}, \ and\ \bibinfo {author} {\bibfnamefont {X.}~\bibnamefont {Qian}}}
  (\bibinfo {year} {2017}{\natexlab{a}}),\ \href {\doibase
  10.1021/acs.nanolett.7b02268} {\bibfield  {journal} {\bibinfo  {journal}
  {Nano Lett.}\ }\textbf {\bibinfo {volume} {17}},\ \bibinfo {pages}
  {5027}}\BibitemShut {NoStop}%
\bibitem [{\citenamefont {Wang}\ and\ \citenamefont
  {Qian}(2017{\natexlab{b}})}]{other3}%
  \BibitemOpen
  \bibfield  {author} {\bibinfo {author} {\bibnamefont {Wang}, \bibfnamefont
  {H.}}, \ and\ \bibinfo {author} {\bibfnamefont {X.}~\bibnamefont {Qian}}}
  (\bibinfo {year} {2017}{\natexlab{b}}),\ \href {\doibase
  10.1088/2053-1583/4/1/015042} {\bibfield  {journal} {\bibinfo  {journal} {2D
  Mater.}\ }\textbf {\bibinfo {volume} {4}},\ \bibinfo {pages}
  {015042}}\BibitemShut {NoStop}%
\bibitem [{\citenamefont {Wang}\ \emph {et~al.}(2015)\citenamefont {Wang},
  \citenamefont {Jones}, \citenamefont {Seyler}, \citenamefont {Tran},
  \citenamefont {Jia}, \citenamefont {Zhao}, \citenamefont {Wang},
  \citenamefont {Yang}, \citenamefont {Xu},\ and\ \citenamefont
  {Xia}}]{BPexcitons}%
  \BibitemOpen
  \bibfield  {author} {\bibinfo {author} {\bibnamefont {Wang}, \bibfnamefont
  {X.}}, \bibinfo {author} {\bibfnamefont {A.~M.}\ \bibnamefont {Jones}},
  \bibinfo {author} {\bibfnamefont {K.~L.}\ \bibnamefont {Seyler}}, \bibinfo
  {author} {\bibfnamefont {V.}~\bibnamefont {Tran}}, \bibinfo {author}
  {\bibfnamefont {Y.}~\bibnamefont {Jia}}, \bibinfo {author} {\bibfnamefont
  {H.}~\bibnamefont {Zhao}}, \bibinfo {author} {\bibfnamefont {H.}~\bibnamefont
  {Wang}}, \bibinfo {author} {\bibfnamefont {L.}~\bibnamefont {Yang}}, \bibinfo
  {author} {\bibfnamefont {X.}~\bibnamefont {Xu}}, \ and\ \bibinfo {author}
  {\bibfnamefont {F.}~\bibnamefont {Xia}}} (\bibinfo {year} {2015}),\ \href
  {\doibase 10.1038/nnano.2015.71} {\bibfield  {journal} {\bibinfo  {journal}
  {Nat. Nanotechnol.}\ }\textbf {\bibinfo {volume} {10}},\ \bibinfo {pages}
  {517}}\BibitemShut {NoStop}%
\bibitem [{\citenamefont {Wang}\ \emph
  {et~al.}(2018{\natexlab{b}})\citenamefont {Wang}, \citenamefont {Xiao},
  \citenamefont {Chen}, \citenamefont {Hua}, \citenamefont {Zou}, \citenamefont
  {Wu}, \citenamefont {Jiang}, \citenamefont {Yang}, \citenamefont {Lu},\ and\
  \citenamefont {Ji}}]{te_band_splitting_2018}%
  \BibitemOpen
  \bibfield  {author} {\bibinfo {author} {\bibnamefont {Wang}, \bibfnamefont
  {Y.}}, \bibinfo {author} {\bibfnamefont {C.}~\bibnamefont {Xiao}}, \bibinfo
  {author} {\bibfnamefont {M.}~\bibnamefont {Chen}}, \bibinfo {author}
  {\bibfnamefont {C.}~\bibnamefont {Hua}}, \bibinfo {author} {\bibfnamefont
  {J.}~\bibnamefont {Zou}}, \bibinfo {author} {\bibfnamefont {C.}~\bibnamefont
  {Wu}}, \bibinfo {author} {\bibfnamefont {J.}~\bibnamefont {Jiang}}, \bibinfo
  {author} {\bibfnamefont {S.~A.}\ \bibnamefont {Yang}}, \bibinfo {author}
  {\bibfnamefont {Y.}~\bibnamefont {Lu}}, \ and\ \bibinfo {author}
  {\bibfnamefont {W.}~\bibnamefont {Ji}}} (\bibinfo {year}
  {2018}{\natexlab{b}}),\ \href {\doibase 10.1039/c8mh00082d} {\bibfield
  {journal} {\bibinfo  {journal} {Mater. Horiz.}\ }\textbf {\bibinfo {volume}
  {5}},\ \bibinfo {pages} {521}}\BibitemShut {NoStop}%
\bibitem [{\citenamefont {Wu}\ and\ \citenamefont {Jena}(2018)}]{reviewpuru}%
  \BibitemOpen
  \bibfield  {author} {\bibinfo {author} {\bibnamefont {Wu}, \bibfnamefont
  {M.}}, \ and\ \bibinfo {author} {\bibfnamefont {P.}~\bibnamefont {Jena}}}
  (\bibinfo {year} {2018}),\ \href {\doibase 10.1002/wcms.1365} {\bibfield
  {journal} {\bibinfo  {journal} {WIREs Comput. Mol. Sci.}\ }\textbf {\bibinfo
  {volume} {8}},\ \bibinfo {pages} {e1365}}\BibitemShut {NoStop}%
\bibitem [{\citenamefont {Wu}\ and\ \citenamefont {Zeng}(2016)}]{ccBP}%
  \BibitemOpen
  \bibfield  {author} {\bibinfo {author} {\bibnamefont {Wu}, \bibfnamefont
  {M.}}, \ and\ \bibinfo {author} {\bibfnamefont {X.~C.}\ \bibnamefont {Zeng}}}
  (\bibinfo {year} {2016}),\ \href {\doibase 10.1021/acs.nanolett.6b00726}
  {\bibfield  {journal} {\bibinfo  {journal} {Nano Lett.}\ }\textbf {\bibinfo
  {volume} {16}},\ \bibinfo {pages} {3236}}\BibitemShut {NoStop}%
\bibitem [{\citenamefont {Wu}\ \emph {et~al.}(2018)\citenamefont {Wu},
  \citenamefont {Fatemi}, \citenamefont {Gibson}, \citenamefont {Watanabe},
  \citenamefont {Taniguchi}, \citenamefont {Cava},\ and\ \citenamefont
  {Jarillo-Herrero}}]{wu_science_2018}%
  \BibitemOpen
  \bibfield  {author} {\bibinfo {author} {\bibnamefont {Wu}, \bibfnamefont
  {S.}}, \bibinfo {author} {\bibfnamefont {V.}~\bibnamefont {Fatemi}}, \bibinfo
  {author} {\bibfnamefont {Q.~D.}\ \bibnamefont {Gibson}}, \bibinfo {author}
  {\bibfnamefont {K.}~\bibnamefont {Watanabe}}, \bibinfo {author}
  {\bibfnamefont {T.}~\bibnamefont {Taniguchi}}, \bibinfo {author}
  {\bibfnamefont {R.~J.}\ \bibnamefont {Cava}}, \ and\ \bibinfo {author}
  {\bibfnamefont {P.}~\bibnamefont {Jarillo-Herrero}}} (\bibinfo {year}
  {2018}),\ \href {\doibase 10.1126/science.aan6003} {\bibfield  {journal}
  {\bibinfo  {journal} {Science}\ }\textbf {\bibinfo {volume} {359}},\ \bibinfo
  {pages} {76}}\BibitemShut {NoStop}%
\bibitem [{\citenamefont {Wu}\ \emph {et~al.}(2004)\citenamefont {Wu},
  \citenamefont {Huang}, \citenamefont {Liu}, \citenamefont {Wu}, \citenamefont
  {Duan}, \citenamefont {Gu},\ and\ \citenamefont {Zhang}}]{Wu_2004}%
  \BibitemOpen
  \bibfield  {author} {\bibinfo {author} {\bibnamefont {Wu}, \bibfnamefont
  {Z.}}, \bibinfo {author} {\bibfnamefont {N.}~\bibnamefont {Huang}}, \bibinfo
  {author} {\bibfnamefont {Z.}~\bibnamefont {Liu}}, \bibinfo {author}
  {\bibfnamefont {J.}~\bibnamefont {Wu}}, \bibinfo {author} {\bibfnamefont
  {W.}~\bibnamefont {Duan}}, \bibinfo {author} {\bibfnamefont {B.-L.}\
  \bibnamefont {Gu}}, \ and\ \bibinfo {author} {\bibfnamefont {X.-W.}\
  \bibnamefont {Zhang}}} (\bibinfo {year} {2004}),\ \href {\doibase
  10.1103/PhysRevB.70.104108} {\bibfield  {journal} {\bibinfo  {journal} {Phys.
  Rev. B}\ }\textbf {\bibinfo {volume} {70}},\ \bibinfo {pages}
  {104108}}\BibitemShut {NoStop}%
\bibitem [{\citenamefont {Xiao}\ \emph {et~al.}(2012)\citenamefont {Xiao},
  \citenamefont {Liu}, \citenamefont {Feng}, \citenamefont {Xu},\ and\
  \citenamefont {Yao}}]{TMDC_valleytronics_2012}%
  \BibitemOpen
  \bibfield  {author} {\bibinfo {author} {\bibnamefont {Xiao}, \bibfnamefont
  {D.}}, \bibinfo {author} {\bibfnamefont {G.-B.}\ \bibnamefont {Liu}},
  \bibinfo {author} {\bibfnamefont {W.}~\bibnamefont {Feng}}, \bibinfo {author}
  {\bibfnamefont {X.}~\bibnamefont {Xu}}, \ and\ \bibinfo {author}
  {\bibfnamefont {W.}~\bibnamefont {Yao}}} (\bibinfo {year} {2012}),\ \href
  {\doibase 10.1103/PhysRevLett.108.196802} {\bibfield  {journal} {\bibinfo
  {journal} {Phys. Rev. Lett.}\ }\textbf {\bibinfo {volume} {108}},\ \bibinfo
  {pages} {196802}}\BibitemShut {NoStop}%
\bibitem [{\citenamefont {Xiao}\ \emph {et~al.}(2018)\citenamefont {Xiao},
  \citenamefont {Zhu}, \citenamefont {Wang}, \citenamefont {Feng},
  \citenamefont {Hu}, \citenamefont {Dasgupta}, \citenamefont {Han},
  \citenamefont {Wang}, \citenamefont {Muller}, \citenamefont {Martin},
  \citenamefont {Hu},\ and\ \citenamefont {Zhang}}]{in2se3_2}%
  \BibitemOpen
  \bibfield  {author} {\bibinfo {author} {\bibnamefont {Xiao}, \bibfnamefont
  {J.}}, \bibinfo {author} {\bibfnamefont {H.}~\bibnamefont {Zhu}}, \bibinfo
  {author} {\bibfnamefont {Y.}~\bibnamefont {Wang}}, \bibinfo {author}
  {\bibfnamefont {W.}~\bibnamefont {Feng}}, \bibinfo {author} {\bibfnamefont
  {Y.}~\bibnamefont {Hu}}, \bibinfo {author} {\bibfnamefont {A.}~\bibnamefont
  {Dasgupta}}, \bibinfo {author} {\bibfnamefont {Y.}~\bibnamefont {Han}},
  \bibinfo {author} {\bibfnamefont {Y.}~\bibnamefont {Wang}}, \bibinfo {author}
  {\bibfnamefont {D.~A.}\ \bibnamefont {Muller}}, \bibinfo {author}
  {\bibfnamefont {L.~W.}\ \bibnamefont {Martin}}, \bibinfo {author}
  {\bibfnamefont {P.}~\bibnamefont {Hu}}, \ and\ \bibinfo {author}
  {\bibfnamefont {X.}~\bibnamefont {Zhang}}} (\bibinfo {year} {2018}),\ \href
  {\doibase 10.1103/PhysRevLett.120.227601} {\bibfield  {journal} {\bibinfo
  {journal} {Phys. Rev. Lett.}\ }\textbf {\bibinfo {volume} {120}},\ \bibinfo
  {pages} {227601}}\BibitemShut {NoStop}%
\bibitem [{\citenamefont {Xu}\ \emph {et~al.}(2020)\citenamefont {Xu},
  \citenamefont {Nahas}, \citenamefont {Prokhorenko}, \citenamefont {Xiang},\
  and\ \citenamefont {Bellaiche}}]{othertopo}%
  \BibitemOpen
  \bibfield  {author} {\bibinfo {author} {\bibnamefont {Xu}, \bibfnamefont
  {C.}}, \bibinfo {author} {\bibfnamefont {Y.}~\bibnamefont {Nahas}}, \bibinfo
  {author} {\bibfnamefont {S.}~\bibnamefont {Prokhorenko}}, \bibinfo {author}
  {\bibfnamefont {H.}~\bibnamefont {Xiang}}, \ and\ \bibinfo {author}
  {\bibfnamefont {L.}~\bibnamefont {Bellaiche}}} (\bibinfo {year} {2020}),\
  \href {\doibase 10.1103/PhysRevB.101.241402} {\bibfield  {journal} {\bibinfo
  {journal} {Phys. Rev. B}\ }\textbf {\bibinfo {volume} {101}},\ \bibinfo
  {pages} {241402}}\BibitemShut {NoStop}%
\bibitem [{\citenamefont {Xu}\ \emph {et~al.}(2017)\citenamefont {Xu},
  \citenamefont {Yang}, \citenamefont {Wang},\ and\ \citenamefont
  {Feng}}]{xu_prb_2017}%
  \BibitemOpen
  \bibfield  {author} {\bibinfo {author} {\bibnamefont {Xu}, \bibfnamefont
  {L.}}, \bibinfo {author} {\bibfnamefont {M.}~\bibnamefont {Yang}}, \bibinfo
  {author} {\bibfnamefont {S.~J.}\ \bibnamefont {Wang}}, \ and\ \bibinfo
  {author} {\bibfnamefont {Y.~P.}\ \bibnamefont {Feng}}} (\bibinfo {year}
  {2017}),\ \href {\doibase 10.1103/PhysRevB.95.235434} {\bibfield  {journal}
  {\bibinfo  {journal} {Phys. Rev. B}\ }\textbf {\bibinfo {volume} {95}},\
  \bibinfo {pages} {235434}}\BibitemShut {NoStop}%
\bibitem [{\citenamefont {Xue}\ \emph {et~al.}(2018{\natexlab{a}})\citenamefont
  {Xue}, \citenamefont {Hu}, \citenamefont {Lee}, \citenamefont {Lu},
  \citenamefont {Zhang}, \citenamefont {Tang}, \citenamefont {Han},
  \citenamefont {Hsu}, \citenamefont {Tu}, \citenamefont {Chang}, \citenamefont
  {Lien}, \citenamefont {He}, \citenamefont {Zhang}, \citenamefont {Li},\ and\
  \citenamefont {Zhang}}]{xue_in2se3_2018_1}%
  \BibitemOpen
  \bibfield  {author} {\bibinfo {author} {\bibnamefont {Xue}, \bibfnamefont
  {F.}}, \bibinfo {author} {\bibfnamefont {W.}~\bibnamefont {Hu}}, \bibinfo
  {author} {\bibfnamefont {K.-C.}\ \bibnamefont {Lee}}, \bibinfo {author}
  {\bibfnamefont {L.-S.}\ \bibnamefont {Lu}}, \bibinfo {author} {\bibfnamefont
  {J.}~\bibnamefont {Zhang}}, \bibinfo {author} {\bibfnamefont {H.-L.}\
  \bibnamefont {Tang}}, \bibinfo {author} {\bibfnamefont {A.}~\bibnamefont
  {Han}}, \bibinfo {author} {\bibfnamefont {W.-T.}\ \bibnamefont {Hsu}},
  \bibinfo {author} {\bibfnamefont {S.}~\bibnamefont {Tu}}, \bibinfo {author}
  {\bibfnamefont {W.-H.}\ \bibnamefont {Chang}}, \bibinfo {author}
  {\bibfnamefont {C.-H.}\ \bibnamefont {Lien}}, \bibinfo {author}
  {\bibfnamefont {J.-H.}\ \bibnamefont {He}}, \bibinfo {author} {\bibfnamefont
  {Z.}~\bibnamefont {Zhang}}, \bibinfo {author} {\bibfnamefont {L.-J.}\
  \bibnamefont {Li}}, \ and\ \bibinfo {author} {\bibfnamefont {X.}~\bibnamefont
  {Zhang}}} (\bibinfo {year} {2018}{\natexlab{a}}),\ \href {\doibase
  0.1002/adfm.201803738} {\bibfield  {journal} {\bibinfo  {journal} {Adv.
  Funct. Mater.}\ }\textbf {\bibinfo {volume} {28}},\ \bibinfo {pages}
  {183738}}\BibitemShut {NoStop}%
\bibitem [{\citenamefont {Xue}\ \emph {et~al.}(2018{\natexlab{b}})\citenamefont
  {Xue}, \citenamefont {Zhang}, \citenamefont {Hu}, \citenamefont {Hsu},
  \citenamefont {Han}, \citenamefont {Leung}, \citenamefont {Huang},
  \citenamefont {Wan}, \citenamefont {Liu}, \citenamefont {Zhang},
  \citenamefont {He}, \citenamefont {Chang}, \citenamefont {Wang},
  \citenamefont {Zhang},\ and\ \citenamefont {Li}}]{xue_in2se3_2018_2}%
  \BibitemOpen
  \bibfield  {author} {\bibinfo {author} {\bibnamefont {Xue}, \bibfnamefont
  {F.}}, \bibinfo {author} {\bibfnamefont {J.}~\bibnamefont {Zhang}}, \bibinfo
  {author} {\bibfnamefont {W.}~\bibnamefont {Hu}}, \bibinfo {author}
  {\bibfnamefont {W.-T.}\ \bibnamefont {Hsu}}, \bibinfo {author} {\bibfnamefont
  {A.}~\bibnamefont {Han}}, \bibinfo {author} {\bibfnamefont {S.-F.}\
  \bibnamefont {Leung}}, \bibinfo {author} {\bibfnamefont {J.-K.}\ \bibnamefont
  {Huang}}, \bibinfo {author} {\bibfnamefont {Y.}~\bibnamefont {Wan}}, \bibinfo
  {author} {\bibfnamefont {S.}~\bibnamefont {Liu}}, \bibinfo {author}
  {\bibfnamefont {J.}~\bibnamefont {Zhang}}, \bibinfo {author} {\bibfnamefont
  {J.-H.}\ \bibnamefont {He}}, \bibinfo {author} {\bibfnamefont {W.-H.}\
  \bibnamefont {Chang}}, \bibinfo {author} {\bibfnamefont {Z.~L.}\ \bibnamefont
  {Wang}}, \bibinfo {author} {\bibfnamefont {X.}~\bibnamefont {Zhang}}, \ and\
  \bibinfo {author} {\bibfnamefont {L.-J.}\ \bibnamefont {Li}}} (\bibinfo
  {year} {2018}{\natexlab{b}}),\ \href {\doibase 10.1021/acsnano.8b02152}
  {\bibfield  {journal} {\bibinfo  {journal} {ACS Nano}\ }\textbf {\bibinfo
  {volume} {12}},\ \bibinfo {pages} {4976}}\BibitemShut {NoStop}%
\bibitem [{\citenamefont {Yang}\ \emph {et~al.}(2018)\citenamefont {Yang},
  \citenamefont {Liu}, \citenamefont {Tang}, \citenamefont {Wang},\ and\
  \citenamefont {Hong}}]{snteapl}%
  \BibitemOpen
  \bibfield  {author} {\bibinfo {author} {\bibnamefont {Yang}, \bibfnamefont
  {C.}}, \bibinfo {author} {\bibfnamefont {Y.}~\bibnamefont {Liu}}, \bibinfo
  {author} {\bibfnamefont {G.}~\bibnamefont {Tang}}, \bibinfo {author}
  {\bibfnamefont {X.}~\bibnamefont {Wang}}, \ and\ \bibinfo {author}
  {\bibfnamefont {J.}~\bibnamefont {Hong}}} (\bibinfo {year} {2018}),\ \href
  {\doibase 10.1063/1.5040671} {\bibfield  {journal} {\bibinfo  {journal}
  {Appl. Phys. Lett.}\ }\textbf {\bibinfo {volume} {113}},\ \bibinfo {pages}
  {082905}}\BibitemShut {NoStop}%
\bibitem [{\citenamefont {Yang}\ \emph {et~al.}(2016)\citenamefont {Yang},
  \citenamefont {Zhang}, \citenamefont {Yin}, \citenamefont {Gong},
  \citenamefont {Yakobson},\ and\ \citenamefont
  {Wei}}]{yang_nanolett_2015_sis}%
  \BibitemOpen
  \bibfield  {author} {\bibinfo {author} {\bibnamefont {Yang}, \bibfnamefont
  {J.-H.}}, \bibinfo {author} {\bibfnamefont {Y.}~\bibnamefont {Zhang}},
  \bibinfo {author} {\bibfnamefont {W.-J.}\ \bibnamefont {Yin}}, \bibinfo
  {author} {\bibfnamefont {X.~G.}\ \bibnamefont {Gong}}, \bibinfo {author}
  {\bibfnamefont {B.~I.}\ \bibnamefont {Yakobson}}, \ and\ \bibinfo {author}
  {\bibfnamefont {S.-H.}\ \bibnamefont {Wei}}} (\bibinfo {year} {2016}),\ \href
  {\doibase 10.1021/acs.nanolett.5b04341} {\bibfield  {journal} {\bibinfo
  {journal} {Nano Lett.}\ }\textbf {\bibinfo {volume} {16}},\ \bibinfo {pages}
  {1110}}\BibitemShut {NoStop}%
\bibitem [{\citenamefont {You}\ \emph {et~al.}(2018)\citenamefont {You},
  \citenamefont {Liu}, \citenamefont {Li}, \citenamefont {Hu}, \citenamefont
  {Zhou}, \citenamefont {Chang}, \citenamefont {Zhou}, \citenamefont {Fu},
  \citenamefont {Yuan}, \citenamefont {Dong}, \citenamefont {Fan},
  \citenamefont {Gruverman}, \citenamefont {Liu},\ and\ \citenamefont
  {Wang}}]{ba2pbcl4}%
  \BibitemOpen
  \bibfield  {author} {\bibinfo {author} {\bibnamefont {You}, \bibfnamefont
  {L.}}, \bibinfo {author} {\bibfnamefont {F.}~\bibnamefont {Liu}}, \bibinfo
  {author} {\bibfnamefont {H.}~\bibnamefont {Li}}, \bibinfo {author}
  {\bibfnamefont {Y.}~\bibnamefont {Hu}}, \bibinfo {author} {\bibfnamefont
  {S.}~\bibnamefont {Zhou}}, \bibinfo {author} {\bibfnamefont {L.}~\bibnamefont
  {Chang}}, \bibinfo {author} {\bibfnamefont {Y.}~\bibnamefont {Zhou}},
  \bibinfo {author} {\bibfnamefont {Q.}~\bibnamefont {Fu}}, \bibinfo {author}
  {\bibfnamefont {G.}~\bibnamefont {Yuan}}, \bibinfo {author} {\bibfnamefont
  {S.}~\bibnamefont {Dong}}, \bibinfo {author} {\bibfnamefont {H.~J.}\
  \bibnamefont {Fan}}, \bibinfo {author} {\bibfnamefont {A.}~\bibnamefont
  {Gruverman}}, \bibinfo {author} {\bibfnamefont {Z.}~\bibnamefont {Liu}}, \
  and\ \bibinfo {author} {\bibfnamefont {J.}~\bibnamefont {Wang}}} (\bibinfo
  {year} {2018}),\ \href {\doibase 10.1002/adma.201803249} {\bibfield
  {journal} {\bibinfo  {journal} {Adv. Mater.}\ }\textbf {\bibinfo {volume}
  {30}},\ \bibinfo {pages} {1803249}}\BibitemShut {NoStop}%
\bibitem [{\citenamefont {Young}\ and\ \citenamefont
  {Rappe}(2012)}]{Young2012}%
  \BibitemOpen
  \bibfield  {author} {\bibinfo {author} {\bibnamefont {Young}, \bibfnamefont
  {S.~M.}}, \ and\ \bibinfo {author} {\bibfnamefont {A.~M.}\ \bibnamefont
  {Rappe}}} (\bibinfo {year} {2012}),\ \href {\doibase
  10.1103/PhysRevLett.109.116601} {\bibfield  {journal} {\bibinfo  {journal}
  {Phys. Rev. Lett.}\ }\textbf {\bibinfo {volume} {109}},\ \bibinfo {pages}
  {116601}}\BibitemShut {NoStop}%
\bibitem [{\citenamefont {Yuan}\ \emph {et~al.}(2019)\citenamefont {Yuan},
  \citenamefont {Luo}, \citenamefont {Chan}, \citenamefont {Xiao},
  \citenamefont {Dai}, \citenamefont {Xie},\ and\ \citenamefont
  {Hao}}]{yuan_nc_2019}%
  \BibitemOpen
  \bibfield  {author} {\bibinfo {author} {\bibnamefont {Yuan}, \bibfnamefont
  {S.}}, \bibinfo {author} {\bibfnamefont {X.}~\bibnamefont {Luo}}, \bibinfo
  {author} {\bibfnamefont {H.~L.}\ \bibnamefont {Chan}}, \bibinfo {author}
  {\bibfnamefont {C.}~\bibnamefont {Xiao}}, \bibinfo {author} {\bibfnamefont
  {Y.}~\bibnamefont {Dai}}, \bibinfo {author} {\bibfnamefont {M.}~\bibnamefont
  {Xie}}, \ and\ \bibinfo {author} {\bibfnamefont {J.}~\bibnamefont {Hao}}}
  (\bibinfo {year} {2019}),\ \href {\doibase 10.1038/s41467-019-09669-x}
  {\bibfield  {journal} {\bibinfo  {journal} {Nat. Commun.}\ }\textbf {\bibinfo
  {volume} {10}},\ \bibinfo {pages} {1775}}\BibitemShut {NoStop}%
\bibitem [{\citenamefont {Zembilgotov}\ \emph {et~al.}(2002)\citenamefont
  {Zembilgotov}, \citenamefont {Pertsev}, \citenamefont {Kohlstedt},\ and\
  \citenamefont {Waser}}]{Zembilgotov_2002}%
  \BibitemOpen
  \bibfield  {author} {\bibinfo {author} {\bibnamefont {Zembilgotov},
  \bibfnamefont {A.~G.}}, \bibinfo {author} {\bibfnamefont {N.~A.}\
  \bibnamefont {Pertsev}}, \bibinfo {author} {\bibfnamefont {H.}~\bibnamefont
  {Kohlstedt}}, \ and\ \bibinfo {author} {\bibfnamefont {R.}~\bibnamefont
  {Waser}}} (\bibinfo {year} {2002}),\ \href {\doibase 10.1063/1.1427406}
  {\bibfield  {journal} {\bibinfo  {journal} {J. Appl. Phys.}\ }\textbf
  {\bibinfo {volume} {91}},\ \bibinfo {pages} {2247}}\BibitemShut {NoStop}%
\bibitem [{\citenamefont {Zeng}\ \emph {et~al.}(2012)\citenamefont {Zeng},
  \citenamefont {Dai}, \citenamefont {Yao}, \citenamefont {Xiao},\ and\
  \citenamefont {Cui}}]{valley_polarization_2012_2}%
  \BibitemOpen
  \bibfield  {author} {\bibinfo {author} {\bibnamefont {Zeng}, \bibfnamefont
  {H.}}, \bibinfo {author} {\bibfnamefont {J.}~\bibnamefont {Dai}}, \bibinfo
  {author} {\bibfnamefont {W.}~\bibnamefont {Yao}}, \bibinfo {author}
  {\bibfnamefont {D.}~\bibnamefont {Xiao}}, \ and\ \bibinfo {author}
  {\bibfnamefont {X.}~\bibnamefont {Cui}}} (\bibinfo {year} {2012}),\ \href
  {\doibase 10.1038/nnano.2012.95} {\bibfield  {journal} {\bibinfo  {journal}
  {Nat. Nanotechnol.}\ }\textbf {\bibinfo {volume} {7}},\ \bibinfo {pages}
  {490}}\BibitemShut {NoStop}%
\bibitem [{\citenamefont {Zhang}\ \emph {et~al.}(2014)\citenamefont {Zhang},
  \citenamefont {Li}, \citenamefont {Shimada}, \citenamefont {Wang},\ and\
  \citenamefont {Kitamura}}]{ZhangYJ14_PRB}%
  \BibitemOpen
  \bibfield  {author} {\bibinfo {author} {\bibnamefont {Zhang}, \bibfnamefont
  {Y.}}, \bibinfo {author} {\bibfnamefont {G.-P.}\ \bibnamefont {Li}}, \bibinfo
  {author} {\bibfnamefont {T.}~\bibnamefont {Shimada}}, \bibinfo {author}
  {\bibfnamefont {J.}~\bibnamefont {Wang}}, \ and\ \bibinfo {author}
  {\bibfnamefont {T.}~\bibnamefont {Kitamura}}} (\bibinfo {year} {2014}),\
  \href {\doibase 10.1103/PhysRevB.90.184107} {\bibfield  {journal} {\bibinfo
  {journal} {Phys. Rev. B}\ }\textbf {\bibinfo {volume} {90}},\ \bibinfo
  {pages} {184107}}\BibitemShut {NoStop}%
\bibitem [{\citenamefont {Zheng}\ \emph {et~al.}(2018)\citenamefont {Zheng},
  \citenamefont {Yu}, \citenamefont {Zhu}, \citenamefont {Collins},
  \citenamefont {Kim}, \citenamefont {Lou}, \citenamefont {Xu}, \citenamefont
  {Li}, \citenamefont {Wei}, \citenamefont {Zhang}, \citenamefont {Edmonds},
  \citenamefont {Li}, \citenamefont {Seidel}, \citenamefont {Zhu},
  \citenamefont {Liu}, \citenamefont {Tang},\ and\ \citenamefont
  {Fuhrer}}]{in2se3_4}%
  \BibitemOpen
  \bibfield  {author} {\bibinfo {author} {\bibnamefont {Zheng}, \bibfnamefont
  {C.}}, \bibinfo {author} {\bibfnamefont {L.}~\bibnamefont {Yu}}, \bibinfo
  {author} {\bibfnamefont {L.}~\bibnamefont {Zhu}}, \bibinfo {author}
  {\bibfnamefont {J.~L.}\ \bibnamefont {Collins}}, \bibinfo {author}
  {\bibfnamefont {D.}~\bibnamefont {Kim}}, \bibinfo {author} {\bibfnamefont
  {Y.}~\bibnamefont {Lou}}, \bibinfo {author} {\bibfnamefont {C.}~\bibnamefont
  {Xu}}, \bibinfo {author} {\bibfnamefont {M.}~\bibnamefont {Li}}, \bibinfo
  {author} {\bibfnamefont {Z.}~\bibnamefont {Wei}}, \bibinfo {author}
  {\bibfnamefont {Y.}~\bibnamefont {Zhang}}, \bibinfo {author} {\bibfnamefont
  {M.~T.}\ \bibnamefont {Edmonds}}, \bibinfo {author} {\bibfnamefont
  {S.}~\bibnamefont {Li}}, \bibinfo {author} {\bibfnamefont {J.}~\bibnamefont
  {Seidel}}, \bibinfo {author} {\bibfnamefont {Y.}~\bibnamefont {Zhu}},
  \bibinfo {author} {\bibfnamefont {J.~Z.}\ \bibnamefont {Liu}}, \bibinfo
  {author} {\bibfnamefont {W.-X.}\ \bibnamefont {Tang}}, \ and\ \bibinfo
  {author} {\bibfnamefont {M.~S.}\ \bibnamefont {Fuhrer}}} (\bibinfo {year}
  {2018}),\ \href {\doibase 10.1126/sciadv.aar7720} {\bibfield  {journal}
  {\bibinfo  {journal} {Sci. Adv.}\ }\textbf {\bibinfo {volume} {4}},\ \bibinfo
  {pages} {eaar7220}}\BibitemShut {NoStop}%
\bibitem [{\citenamefont {Zhou}\ \emph {et~al.}(2015)\citenamefont {Zhou},
  \citenamefont {Cheng}, \citenamefont {Zhou}, \citenamefont {Cao},
  \citenamefont {Hong}, \citenamefont {Liao}, \citenamefont {Wu}, \citenamefont
  {Peng}, \citenamefont {Liu},\ and\ \citenamefont {Yu}}]{Zhou2015}%
  \BibitemOpen
  \bibfield  {author} {\bibinfo {author} {\bibnamefont {Zhou}, \bibfnamefont
  {X.}}, \bibinfo {author} {\bibfnamefont {J.}~\bibnamefont {Cheng}}, \bibinfo
  {author} {\bibfnamefont {Y.}~\bibnamefont {Zhou}}, \bibinfo {author}
  {\bibfnamefont {T.}~\bibnamefont {Cao}}, \bibinfo {author} {\bibfnamefont
  {H.}~\bibnamefont {Hong}}, \bibinfo {author} {\bibfnamefont {Z.}~\bibnamefont
  {Liao}}, \bibinfo {author} {\bibfnamefont {S.}~\bibnamefont {Wu}}, \bibinfo
  {author} {\bibfnamefont {H.}~\bibnamefont {Peng}}, \bibinfo {author}
  {\bibfnamefont {K.}~\bibnamefont {Liu}}, \ and\ \bibinfo {author}
  {\bibfnamefont {D.}~\bibnamefont {Yu}}} (\bibinfo {year} {2015}),\ \href
  {\doibase 10.1021/jacs.5b04305} {\bibfield  {journal} {\bibinfo  {journal}
  {J. Am. Chem. Soc.}\ }\textbf {\bibinfo {volume} {137}},\ \bibinfo {pages}
  {7994}}\BibitemShut {NoStop}%
\bibitem [{\citenamefont {Zhou}\ \emph {et~al.}(2017)\citenamefont {Zhou},
  \citenamefont {Wu}, \citenamefont {Zhu}, \citenamefont {Cho}, \citenamefont
  {He}, \citenamefont {Yang}, \citenamefont {Herrera}, \citenamefont {Chu},
  \citenamefont {Han}, \citenamefont {Downer}, \citenamefont {Peng},\ and\
  \citenamefont {Lai}}]{in2se3_1}%
  \BibitemOpen
  \bibfield  {author} {\bibinfo {author} {\bibnamefont {Zhou}, \bibfnamefont
  {Y.}}, \bibinfo {author} {\bibfnamefont {D.}~\bibnamefont {Wu}}, \bibinfo
  {author} {\bibfnamefont {Y.}~\bibnamefont {Zhu}}, \bibinfo {author}
  {\bibfnamefont {Y.}~\bibnamefont {Cho}}, \bibinfo {author} {\bibfnamefont
  {Q.}~\bibnamefont {He}}, \bibinfo {author} {\bibfnamefont {X.}~\bibnamefont
  {Yang}}, \bibinfo {author} {\bibfnamefont {K.}~\bibnamefont {Herrera}},
  \bibinfo {author} {\bibfnamefont {Z.}~\bibnamefont {Chu}}, \bibinfo {author}
  {\bibfnamefont {Y.}~\bibnamefont {Han}}, \bibinfo {author} {\bibfnamefont
  {M.~C.}\ \bibnamefont {Downer}}, \bibinfo {author} {\bibfnamefont
  {H.}~\bibnamefont {Peng}}, \ and\ \bibinfo {author} {\bibfnamefont
  {K.}~\bibnamefont {Lai}}} (\bibinfo {year} {2017}),\ \href {\doibase
  10.1021/acs.nanolett.7b02198} {\bibfield  {journal} {\bibinfo  {journal}
  {Nano Lett.}\ }\textbf {\bibinfo {volume} {17}},\ \bibinfo {pages}
  {5508}}\BibitemShut {NoStop}%
\bibitem [{\citenamefont {Zhu}\ \emph {et~al.}(2020)\citenamefont {Zhu},
  \citenamefont {Lu},\ and\ \citenamefont {Wang}}]{doping1}%
  \BibitemOpen
  \bibfield  {author} {\bibinfo {author} {\bibnamefont {Zhu}, \bibfnamefont
  {L.}}, \bibinfo {author} {\bibfnamefont {Y.}~\bibnamefont {Lu}}, \ and\
  \bibinfo {author} {\bibfnamefont {L.}~\bibnamefont {Wang}}} (\bibinfo {year}
  {2020}),\ \href {\doibase 10.1063/1.5123296} {\bibfield  {journal} {\bibinfo
  {journal} {J. Appl. Phys.}\ }\textbf {\bibinfo {volume} {127}},\ \bibinfo
  {pages} {014101}}\BibitemShut {NoStop}%
\bibitem [{\citenamefont {Zhu}\ \emph {et~al.}(2015)\citenamefont {Zhu},
  \citenamefont {Guan}, \citenamefont {Liu},\ and\ \citenamefont
  {Tom{\'a}nek}}]{tomanek_acsnano_2015_sis}%
  \BibitemOpen
  \bibfield  {author} {\bibinfo {author} {\bibnamefont {Zhu}, \bibfnamefont
  {Z.}}, \bibinfo {author} {\bibfnamefont {J.}~\bibnamefont {Guan}}, \bibinfo
  {author} {\bibfnamefont {D.}~\bibnamefont {Liu}}, \ and\ \bibinfo {author}
  {\bibfnamefont {D.}~\bibnamefont {Tom{\'a}nek}}} (\bibinfo {year} {2015}),\
  \href {\doibase 10.1021/acsnano.5b02742} {\bibfield  {journal} {\bibinfo
  {journal} {ACS Nano}\ }\textbf {\bibinfo {volume} {9}},\ \bibinfo {pages}
  {8284}}\BibitemShut {NoStop}%
\end{thebibliography}

%

\end{document}